\documentclass{aa}  

\usepackage{graphicx}
\usepackage{xcolor}
\usepackage{txfonts}

\usepackage{subfig}
\usepackage{pdflscape}
\usepackage[export]{adjustbox}

\newcommand{\kms }{km s$^{-1}$}

\newcommand{ \Mstar } {$M_{\star}$}
\newcommand{ \Msun } {M$_{\odot}$}

\newcommand{ \Herschel } {\textit{Herschel}}

\newcommand{ \Ha} {H$\alpha$}
\newcommand{ \Hb} {H$\beta$}
\newcommand{ \OIII} {[O\,\textsc{iii}]}
\newcommand{ \NII} {[N\,\textsc{ii}]}
\newcommand{ \SII} {[S\,\textsc{ii}]}
\newcommand{ \uvmodelfit}{{\tt uvmodelfit}}
\newcommand{ \sersic}{S{\'e}rsic}
\newcommand{ \XN}{X\_N\_81\_44}
\newcommand{ \casa}{{\tt CASA}}
\newcommand{ \cigale}{{\tt CIGALE}}
\newcommand{ \gaia}{\textit{Gaia}}

\newcommand{ \curvefit}{{\tt scipy.optimize.curve\_fit}}
\newcommand{ \micron}{$\mu$m}

\usepackage[colorlinks=True, citecolor=blue, linkcolor=blue]{hyperref}
\begin{document}

   \title{SUPER V. ALMA continuum observations of z$\sim$2 AGN and the elusive evidence of outflows influencing star formation}
   \titlerunning{SUPER V: obscured star formation and outflows}

    \author{ \normalsize \parbox{0.9\textwidth}{
    I. Lamperti \inst{1,2,3}\thanks{E-mail: isabellalamperti@gmail.com} \and
    C. M. Harrison\inst{4}\thanks{E-mail: christopher.harrison@newcastle.ac.uk} \and
    V. Mainieri\inst{1} \and
    D. Kakkad\inst{5,6} \and
    M. Perna\inst{3, 7} \and
    C. Circosta\inst{2} \and
    J. Scholtz\inst{8} \and
    S. Carniani\inst{9} \and 
    C. Cicone\inst{10} \and 
    D. M. Alexander\inst{11} \and
    M. Bischetti\inst{12} \and
    G. Calistro Rivera\inst{1} \and
    C.-C. Chen\inst{13} \and
    G. Cresci\inst{7} \and
    C. Feruglio\inst{12} \and
    F. Fiore\inst{12} \and
    F. Mannucci\inst{7} \and
    A. Marconi\inst{14,7} \and
    L. N. Martínez-Ramírez\inst{1,15} \and
    H. Netzer\inst{16} \and
    E. Piconcelli\inst{17} \and
    A. Puglisi\inst{11} \and
    D. J. Rosario\inst{11} \and
    M. Schramm\inst{18} \and
    G. Vietri\inst{19} \and
    C. Vignali\inst{20, 21} \and
    L. Zappacosta\inst{17}
      }}
\authorrunning{I. Lamperti et al.}

\institute{
   European Southern Observatory, Karl-Schwarzschild-Str. 2, 85748 Garching bei M{\"u}nchen, Germany
    \and Department of Physics \& Astronomy, University College London, Gower Street, London WC1E 6BT, UK
    \and Centro de Astrobiología (CAB, CSIC–INTA), Departamento de Astrofísica, Ctra. de Ajalvir Km. 4, 28850 Torrejón de Ardoz, Madrid, Spain
    \and
    School of Mathematics, Statistics and Physics, Herschel Building, Newcastle University, Newcastle upon Tyne NE1 7RU, UK
    \and European Southern Observatory, Alonso de Cordova 3107, Casilla 19, Santiago 19001, Chile
    \and Department of Physics, University of Oxford, Denys Wilkinson Building, Keble Road, Oxford, OX1 3RH, UK
     \and INAF - Osservatorio Astrofisico di Arcetri, Largo E. Fermi 5, 50125 Firenze, Italy
    \and  Chalmers University of Technology, Department of Earth and Space Sciences, Onsala Space Observatory, 43992, Onsala, Sweden
    \and Scuola Normale Superiore, Piazza dei Cavalieri 7, I-56126 Pisa, Italy
    \and Institute of Theoretical Astrophysics, University of Oslo, P.O. Box 1029, Blindern, 0315 Oslo, Norway
    \and Centre for Extragalactic Astronomy, Department of Physics, Durham University, South Road, Durham DH1 3LE, UK
    \and INAF - Osservatorio Astronomico di Trieste, via G.B. Tiepolo 11, 34143 Trieste, Italy
    \and Academia Sinica Institute of Astronomy and Astrophysics (ASIAA), No. 1, Sec. 4, Roosevelt Rd., Taipei 10617, Taiwan
    \and Dipartimento di Fisica e Astronomia, Università di Firenze, Via G. Sansone 1, 50019 Sesto Fiorentino (Firenze), Italy
    \and Escuela de Física - Universidad Industrial de Santander, 680002 Bucaramanga, Colombia
    \and School of Physics and Astronomy, Tel-Aviv University, Tel-Aviv 69978, Israel
    \and INAF – Osservatorio Astronomico di Roma, via di Frascati 33, 00078 Monte Porzio Catone, Italy
    \and Graduate school of Science and Engineering, Saitama Univ. 255 Shimo-Okubo, Sakura-ku, Saitama City, Saitama 338-8570, Japan
    \and INAF IASF-Milano, Via Alfonso Corti 12, 20133 Milano, Italy
    \and Dipartimento di Fisica e Astronomia, Alma Mater Studiorum, Università degli Studi di Bologna, Via Gobetti 93/2, I-40129 Bologna, Italy
    \and INAF – Osservatorio di Astrofisica e Scienza dello Spazio di Bologna, Via Gobetti 93/3, I-40129 Bologna, Italy
    }


 
  \abstract
    {
We study the impact of active galactic nuclei (AGN) ionised outflows on star formation in high-redshift AGN host galaxies, by combining near-infrared integral field spectroscopic (IFS) observations, mapping the \Ha\ emission and \OIII $\lambda5007$ outflows, with matched-resolution observations of the rest-frame far-infrared (FIR) emission.
We present high-resolution ALMA Band 7 observations of eight X-ray selected AGN ($L_{2-10~keV}=10^{43.8}-10^{45.2}$~erg~s$^{-1}$) at $z\sim2$ from the SUPER (SINFONI Survey for Unveiling the Physics and Effect of Radiative feedback) sample, targeting the observed-frame 870~\micron\ (rest-frame $\sim 260$~\micron) continuum at $\sim2$~kpc (0.2'') spatial resolution. The targets were selected among the SUPER AGN with an \OIII\ detection in the IFS maps and with a detection in the FIR photometry. 
We detected six out of eight targets with signal-to-noise ratio S/N$\gtrsim10$ in the ALMA maps, from which we measured  continuum flux densities in the range $0.27-2.58$~mJy and FIR half-light radii ($R_e$) in the range $0.8-2.1$~kpc. The other two targets were detected with S/N of 3.6 and 5.9, which are insufficient for spatially resolved analysis. 
The FIR $R_e$ of our sample are comparable to other AGN and star-forming galaxies at a similar redshift from the literature.
 However, combining our sample with the literature samples, we find that the mean FIR size in X-ray AGN ($R_e = 1.16\pm 0.11$~kpc) is slightly smaller than in non-AGN ($R_e = 1.69\pm 0.13$~kpc).
 From spectral energy distribution (SED) fitting, we find that the main contribution to the 260~\micron\  flux density is dust heated by star formation, with $\leq 4\%$ contribution from AGN-heated dust and $\leq 1$\% from synchrotron emission.
The majority of our sample show different morphologies for the FIR (mostly due to reprocessed stellar emission) and the ionised gas emission (\Ha\ and \OIII, mostly due to AGN emission). This could be due to the different locations of dust and ionised gas, the different sources of the emission (stars and AGN), or the effect of dust obscuration. We are unable to identify any residual \Ha\ emission, above that dominated by AGN, that could be attributed to star formation.
Under the assumption that the FIR emission is a reliable tracer of obscured star formation, we find that the obscured star formation activity in these AGN host galaxies is not clearly affected by the ionised outflows. However, we cannot rule out that star formation suppression is happening on smaller spatial scales than the ones we probe with our observations ($< 2$~kpc) or on different timescales. 
}
 


   \maketitle
%

\section{Introduction}
Super-massive black holes (SMBHs) in the centre of galaxies have phases of intense accretion of material, during which they become very luminous and are identified as active galactic nuclei (AGN). 
AGN release large amounts of energy into the host galaxies \citep[$\sim$ 10\% of the rest-mass energy of the accreted material, ][]{Shapiro1983, Marconi2004} 
and therefore have the potential to influence the properties of the interstellar medium (ISM)  and ultimately its star formation rate (SFR). This phenomenon is called 
 AGN feedback \citep[e.g. ][]{Fabian2012, King2015a, Harrison2017}.
Simulations of galaxy formation require AGN feedback in massive galaxies (stellar masses $> 10^{10}$ \Msun) in order to match many observables, such as the evolution of galaxy sizes \citep[e.g,][]{Choi2018}, the galaxy colour bi-modality \citep{Trayford2016}, and the lack of very massive galaxies in the most massive galaxy haloes \citep{Somerville2008, Behroozi2013}.

A viable AGN feedback mechanism is represented by fast winds ($> 1000$~\kms) originating from the AGN accretion disk, or by collimated radio jets, both of which  can impact the surrounding gas, generating outflows which propagate to large distances into the host galaxy \citep[e.g. ][]{King2003,Mukherjee2018,Costa2020}. These outflows can potentially heat the ISM and also expel gas from the host galaxy \citep[][]{Ishibashi2016, Zubovas2018}.

AGN outflows have been detected at different scales, extending from a few parsecs to kiloparsecs \citep[e.g][]{Storchi-Bergmann2010, Veilleux2013, Cresci2015, Feruglio2015, Kakkad2016, McElroy2016, Rupke2017, Jarvis2019, Marasco2020, Kakkad2020, Vietri2018, Vietri2020}. 
In addition, outflows have been observed in  multiple gas phases: neutral and ionised as well as atomic and molecular \citep{Cicone2018}.

Understanding whether these outflows are able to have an impact on star formation is crucial for testing and refining the aforementioned simulations, which include a variety of different models of `AGN feedback'.
This impact can be particularly important for galaxies at redshift $z\sim2$, which corresponds to the peak of star formation and AGN activity in the Universe \citep[e.g. ][]{Shankar2009, Madau2014, Tacconi2020}.

To study the impact of AGN outflows on star formation, one possibility is to use spatially resolved observations to map the distribution of both the outflows and the star-forming regions and look for  spatial correlations or anti-correlations, which can indicate signatures of star formation enhancement (`positive feedback') or suppression (`negative feedback') \citep[][and references therein]{Cresci2018}.
Using integral field spectroscopy (IFS) at rest-frame optical wavelengths, several studies investigate the impact of ionised outflows, traced by \OIII$\lambda5007$ (\OIII\ hereafter), on star formation in AGN host galaxies and found evidence of both positive and negative feedback \citep[e.g. ][]{Cano-Diaz2012, Cresci2015, Carniani2016, Maiolino2017, Gallagher2019, Perna2020}. 
However, the interpretation of some of these observations may not be trivial \citep[see][]{Scholtz2020, Scholtz2021}. 
For example, assessing the impact of outflows on star formation can be complicated by the fact that common SFR tracers (e.g. the \Ha\ emission line) are affected by dust extinction \citep{Madau1996, Casey2014, Whitaker2014}. In some objects, the UV and \Ha\ emission could be completely hidden by the dust \citep[e.g. ][]{Hodge2016, Chen2017, Puglisi2017}.
 Thus, to have a complete view of the star formation in the host galaxy, it is crucial to also have information about the dust-obscured star formation traced by the far-infrared (FIR) emission \citep[e.g. ][]{Whitaker2014, Brusa2018, Scholtz2020, Bouwens2020}.

High-resolution  observations from the Atacama Large Millimeter/submillimeter Array (ALMA) have been used to measure the FIR sizes of high-redshift star-forming galaxies \citep[e.g.][]{Ikarashi2015, Simpson2015, Hodge2016, Gullberg2019, Puglisi2019, Cheng2020}. Several studies found the rest-frame FIR sizes of  star-forming galaxies at $z>1$ to be smaller than the rest-frame optical sizes  \citep[e.g. ][]{Chen2015, Barro2016, Tadaki2017a, Fujimoto2017, Fujimoto2018, Elbaz2018, Calistro2018, Lang2019}. A possible interpretation is that the central compact dusty star-forming component is related to the formation of the bulge  \citep{Fujimoto2017}. 
 Alternatively, \citet{Popping2021} suggested, based on simulations, that the larger rest-frame optical sizes could be due to higher dust-obscuration in the central region of galaxies, which artificially increases the ratio between the derived optical and FIR sizes.
It is still unclear whether the presence of an AGN can affect (or is related to) the FIR size of the host galaxy. For example, \citet{Chang2020} studied a sample of seven AGN and 20 non-AGN at $z\sim1$ and found that obscured IR-selected AGN have smaller FIR sizes (median size $R_e = 0.76\pm0.48$~kpc) than non-AGN (median size $R_e = 1.62\pm0.50$~kpc)  at the same redshift and stellar mass. On the other hand, \citet{Harrison2016b} measured the FIR sizes of a sample of five X-ray selected AGN at $z\sim1.5-4.5$ and found that their FIR sizes (FWHM size $1-3$~kpc, median 1.8~kpc) are comparable to the sizes of sub-millimetre galaxies (SMGs) at the same redshift (median FWHM size $2.4\pm0.2$~kpc).

This paper is part of a series of papers from  SUPER\footnote{http://www.super-survey.org} \citep[SINFONI Survey for Unveiling the Physics and Effects of Radiative feedback,][]{Circosta2018}, an ESO large programme which aims to investigate the outflows properties of $z\sim2$ X-ray AGN. The high resolution ($\sim 0.2-0.3''$, corresponding to $\sim 2$~kpc) of the IFS SINFONI observations is critical to resolve the morphology of the ionised outflow \citep{Kakkad2020} and connect it with the star formation properties of the host galaxy. 
In this work, we study the impact of AGN ionised outflows on both obscured and unobscured star formation in a sub-sample of eight SUPER AGN.
We combine SINFONI IFS observations, to map the \Ha\ emission and \OIII\ outflows, with matched-resolution ALMA observations of the continuum FIR emission at $\lambda_{obs} \sim 870$ \micron\ ($\lambda_{rest} \sim 260$ \micron).

The paper is organised as follows.
In Section~\ref{sec:sample}, we describe the selection criteria and the general properties of the  sample. In Section~\ref{sec:ALMA_data}, we present the ALMA data and the analysis to extract the information about the FIR sizes and flux densities. Section~\ref{sec:SINFONI} describes the SINFONI \Ha\ and \OIII\ data. In Section~\ref{sec:results} we present our results. First, we investigate the origin of the FIR emission in our targets (Sec.~\ref{sec:origin}). Then, we compare the FIR sizes with other samples from the literature (Sec.~\ref{sec:FIR_sizes_lit}). Finally,  we compare the spatial distribution of the FIR continuum with the \Ha\ and \OIII\ emission, as well as with the ionised outflows (Sec.~\ref{sec:FIR_Ha_OIII} and \ref{sec:outflows}).
In Section~\ref{sec:conclusions}, we summarise the main results and our conclusions.

Throughout this work, we assume wavelength in vacuum and a cosmological model with $\Omega_{\lambda} = 0.7$, $\Omega_{\text{M}}= 0.3$ , and $H_0 = 70$ km s$^{-1}$ Mpc$^{-1}$.

\section{Sample}
\label{sec:sample}

The sample studied in this paper consists of eight AGN with ALMA Band 7 continuum observations, which have been selected from the SUPER parent sample.
 The sample selection of SUPER is described in detail in \cite{Circosta2018}. In brief, the 39 SUPER AGN were selected in the X-rays from various surveys, by adopting as a threshold an absorption-corrected X-ray luminosity of $L_X \geq 10^{42}$ erg s$^{-1}$, to exclude sources where the X-ray emission may come from star formation \citep{Aird2017}. The SUPER sample covers the redshift range $z=2-2.5$ and spans a wide range of AGN bolometric luminosities\footnote{AGN bolometric luminosities are derived from the fit of the multi-wavelength spectral energy distributions (SEDs) \citep[for details, see][]{Circosta2018}.} ($\sim 10^{44} -10^{48}$ erg s$^{-1}$)  and X-ray luminosities ($L_{2-10keV}\sim 10^{43} -10^{46}$ erg s$^{-1}$).
To select the sub-sample to be observed with ALMA, we considered all the SUPER objects with at least one photometric detection,  based on the photometry we had in hand
\citep[i.e. a signal-to-noise S/N $>3$ in the catalogue from ][]{Circosta2018} in the FIR (i.e. observed wavelength in the range $70-870$~\micron), to be able to properly assess the emission process at 870~\micron\ (observed wavelength), and with \OIII\ detections in the completed IFS data at the time of the proposal (April 2018), to be able to investigate the impact of \OIII\ outflows on the dust-obscured star formation. The eight SUPER targets that satisfied these criteria were selected for the ALMA observations.

 Our sample includes five AGN from the COSMOS-Legacy survey \citep{Civano2016, Marchesi2016a}, two from the \textit{Chandra} Deep Field South \citep[CDF-S,][]{Luo2017}, and one from the wide-area XMM-Newton XXL survey \citep{Pierre2016}. 
A summary of the sample characteristics is provided in Table \ref{tab:sample}.

\begin{table*}

\centering
\caption{Properties of the sample. All quantities are taken from \citet{Circosta2018}, except for the SFRs and stellar masses which have been derived from the SED fitting including the Band 7 fluxes presented in this paper (see Section~\ref{sec:ALMA_SED_fit}). 
 (1) Field where the targets are located; (2) Source identification number from the catalogues corresponding to each field; 
(3) RA and (4) Dec of the optical counterpart;
(5) Spectroscopic redshift; 
(6) AGN classification as broad line (BL) or narrow line (NL) objects according to the optical spectra;
(7) Host galaxy stellar mass and 1$\sigma$ error; 
(8) SFR from the 8-1000\micron\ FIR luminosity and 1$\sigma$ error; 
(9) AGN bolometric luminosity and 1$\sigma$ error, derived from SED fitting; 
(10) Absorption-corrected X-ray luminosity in the hard band (2-10 keV) and 90\% confidence level error; 
(11) Absorbing hydrogen column density and 90\%
 confidence level error, derived from either X-ray
 spectral fitting or hardness-ratio \citep[see][]{Circosta2018}.}
\setlength{\tabcolsep}{3pt}
\begin{tabular}{llccccccccc}
\hline
Field & ID & RA  & Dec & z & AGN &  log M$_*$ &  SFR & log L$_\text{bol}$ & log L$_\text{2-10keV}$ & log N$_\text{H}$    \\ 
   &  &[deg] & [deg] & & type &[log \Msun] & [ \Msun\ yr$^{-1}$]  &  [erg s$^{-1}$] & [erg s$^{-1}$] & [cm$^{-2}$] \\
 (1) & (2)  & (3) & (4) &(5) & (6) & (7) & (8) & (9) & (10) & (11)\\ 
  \hline \hline
XMM-XXL & X\_N\_81\_44 & 34.378942 & -4.306572 & 2.311 & BL & 10.89$\pm$0.25 & 173$\pm$92 & 46.80$\pm$0.03& 44.77$^{+0.07}_{-0.09}$ &$< 21.86$ \\
CDF-S & XID36 & 52.961556 & -27.784281 & 2.259 & NL & 10.72$\pm$0.08 & 186$\pm$9 & 45.70$\pm$0.06& 43.84$^{+0.31}_{-0.63}$ &$> 24.10$ \\
CDF-S & XID419 & 53.097649 & -27.715270 & 2.145 & NL & 10.89$\pm$0.02 & 44$\pm$2 & 45.54$\pm$0.05& 43.84$^{+0.29}_{-0.44}$ &24.28$^{+0.19}_{-0.31}$ \\
COSMOS & cid\_1057 & 149.812485 & 2.111014 & 2.214 & NL & 10.99$\pm$0.08 & 64$\pm$7 & 45.91$\pm$0.06& 44.53$^{+0.26}_{-0.30}$ &23.98$^{+0.24}_{-0.28}$ \\
COSMOS & cid\_346 & 149.930878 & 2.118734 & 2.219 & BL & 10.67$\pm$0.37 & 380$\pm$29 & 46.66$\pm$0.02& 44.47$^{+0.08}_{-0.09}$ &23.05$^{+0.17}_{-0.19}$ \\
COSMOS & cid\_451 & 150.002533 & 2.258629 & 2.450 & NL & 11.21$\pm$0.05 & 48$\pm$19 & 46.44$\pm$0.07& 45.18$^{+0.23}_{-0.19}$ &23.87$^{+0.19}_{-0.15}$ \\
COSMOS & cid\_1205 & 150.010696 & 2.332968 & 2.255 &  BL$^{a}$ & 11.16$\pm$0.22 & 88$\pm$16 & 45.75$\pm$0.17& 44.25$^{+0.21}_{-0.23}$ &23.50$^{+0.27}_{-0.27}$ \\
COSMOS & cid\_1143 & 150.036819 & 2.257776 & 2.492 & NL & 10.82$\pm$0.04 & 8$\pm$2 & 44.85$\pm$0.12& 44.83$^{+0.43}_{-0.36}$ &24.01$^{+0.77}_{-0.29}$ \\

\hline
\end{tabular}
$^{(a)}$ cid\_1205 was previously reported as a NL target in \citet{Circosta2018}, but it is now classified as a BL target based on the presence of BLR emission in the \Ha\ line in the K-band SINFONI spectrum \citep{Kakkad2020}.
\label{tab:sample}
\end{table*}

\begin{figure*}
\centering

\includegraphics[width=0.45\textwidth]{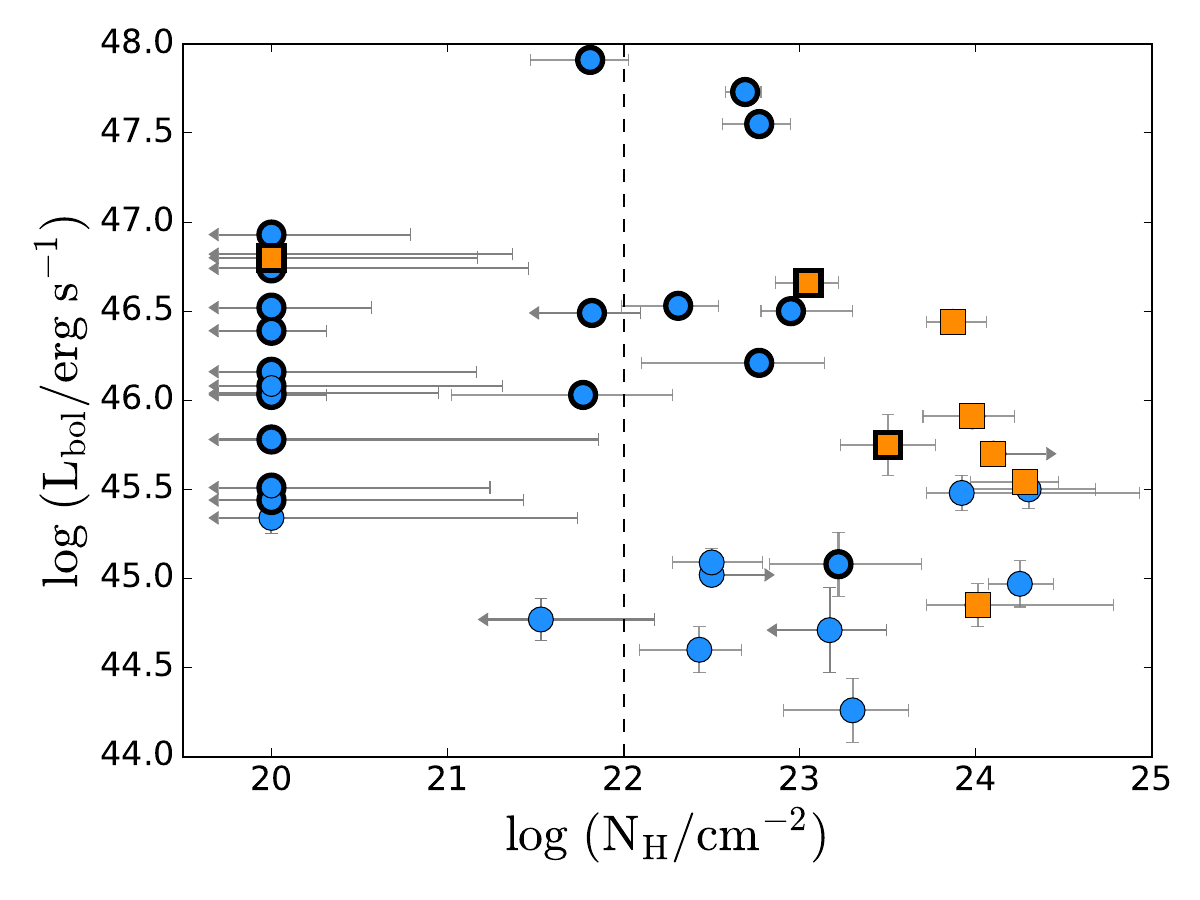}
\includegraphics[width=0.45\textwidth]{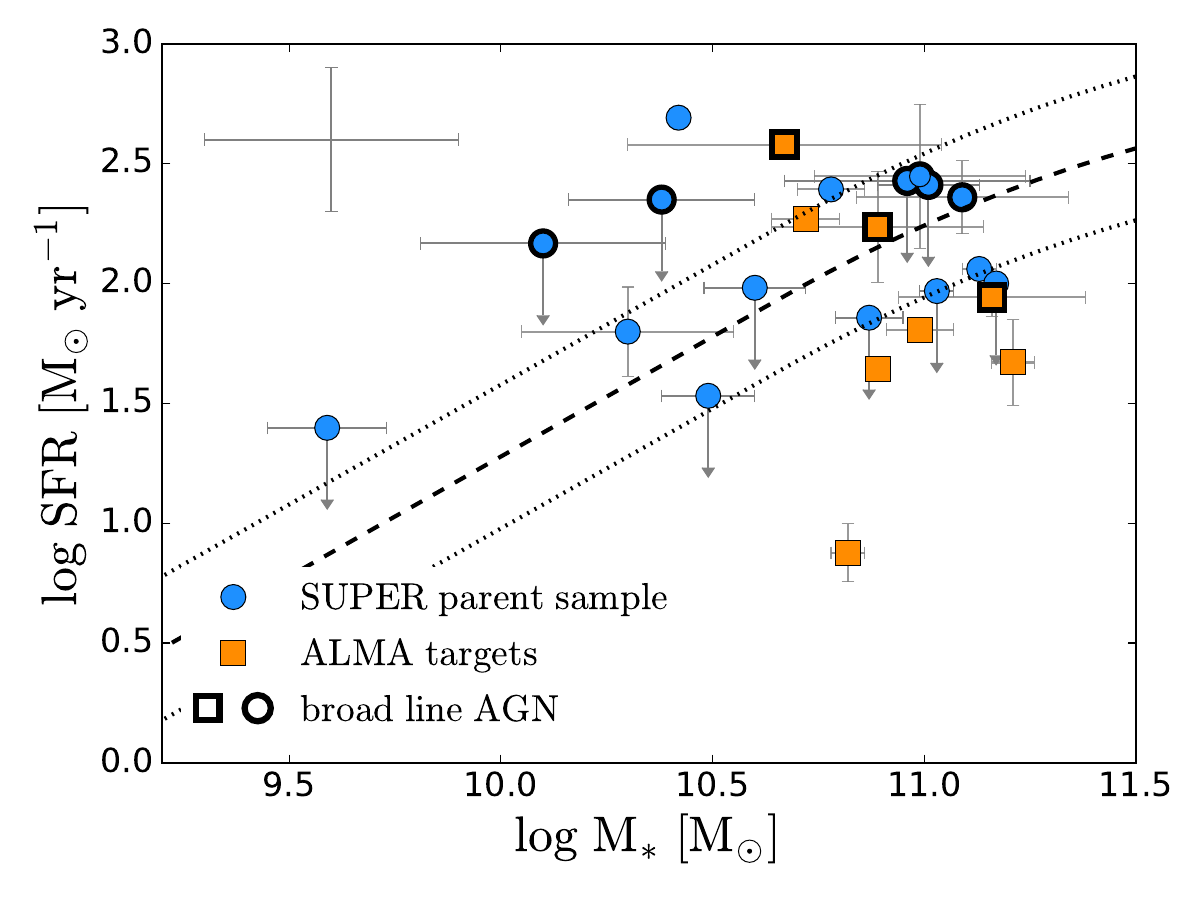}
\caption{\textit{Left:} AGN bolometric luminosity  versus hydrogen column density ($N_H$) for the SUPER sample. ALMA targets from this work are marked with orange squares. Targets classified as broad line AGN in the optical are marked with black contours.  The dashed line at $\log(N_H/$cm$^{-2})=22$ marks the  separation between X-ray unobscured and obscured AGN.
\textit{Right:} SUPER sample in the star formation rate (SFR) vs. stellar mass ($M_*$) plane. Only the 24/39 objects with SFR and stellar mass constraints are shown. The size of the systematic uncertainties is shown on the top left corner.
The black dashed curve shows the main sequence (MS) of star-forming galaxies from Schreiber et al. (2015) at the average redshift of our target sample (z$\sim$2.3), the 0.3 dex dispersion is showed by dotted lines. Most of the ALMA targets are consistent with being on the MS, with the exception of cid\_1143, which lies  below.}
\label{fig:SUPER_sample}
\end{figure*}
 
Given the selection criteria based on the FIR and \OIII\ properties described above, it is important to consider how this sub-sample is representative of the parent population of the SUPER survey.
 Figure~\ref{fig:SUPER_sample} shows the AGN bolometric luminosities ($L_{bol}$) and hydrogen column densities ($N_{H}$) for the total SUPER sample, with the ALMA targets highlighted in orange. Bolometric luminosities are derived from  SED fitting and $N_H$  from the analysis of the X-ray spectra \citep[for more details see][]{Circosta2018}.
The ALMA targets have bolometric luminosities in the range $\log (L_{bol}$/[erg s$^{-1}])= 44.9-46.8$, with  median $\log (L_{bol}$/[erg s$^{-1}])= 45.8$. They are therefore representative in terms of bolometric luminosities of the parent SUPER sample (median $\log L_{bol}$/(erg s$^{-1})= 46.0$ and interquartile range  $\log (L_{bol}$/[erg s$^{-1}])=45.4-46.5$).

Most of the ALMA targets have relatively high hydrogen column densities  ($\log( N_{H}$/cm$^{-2}) = 23.1-24.3$), with the exception of \XN, which has an upper limit ($\log (N_{H}$/cm$^{-2}) < 21.9$).
The three ALMA targets with the lowest $N_{H}$ are classified as optical broad line (Type 1) AGN.
 Our ALMA sample includes 7/8 (88\%) obscured sources \citep[based on $\log( N_{H}/\text{cm}^{-2})>22$;][]{Mainieri2002, Szokoly2004}, compared to 54\% in the parent sample, and 5/8 (63\%) optical Type 2 (narrow line) AGN, compared to 41\% in the parent sample. 
However, if we consider only SUPER targets with photometric coverage in the FIR, 18/26 (70\%) have $\log N_{H}$/cm$^{-2}>22$ and 15/26 (58\%) are classified as optical Type 2 AGN. 
 The targets with FIR observations are biased against unobscured (Type 1) AGN. This is due to the fact that Type 1 targets in SUPER are mostly located in the XMM-XXL, WISSH and Stripe82X fields, which do not have a good FIR coverage:  WISSH and Stripe82X do not have FIR photometric observations available and XMM-XXL has FIR coverage for only some of the targets.  
 On the contrary, the CDF-S and COSMOS fields have good FIR coverage, and the SUPER targets in those fields are biased towards Type 2 objects (73\% of the objects, compared to 41\% in the total SUPER sample) which is expected given that these two fields have the deepest X-ray observations \citep[e.g. ][]{Mainieri2002}.
We conclude that the high $N_{H}$ values of our sample are likely due to the fact that obscured targets are preferentially located in the two fields (COSMOS and CDF-S) with good FIR photometry.

In summary, although the ALMA sample presented in this paper has bolometric luminosities which are representative of the parent sample, we are biased towards heavily obscured (high $N_{H}$) and Type 2 targets.
We discuss how representative our ALMA sample is in terms of SFRs and stellar masses in section~\ref{sec:ALMA_SED_fit}.

\section{ALMA observations and analyses}
\label{sec:ALMA_data}

We use FIR continuum observations from ALMA with the goal of tracing the dust obscured star formation at the same resolution ($\sim$2~kpc) as the \OIII\ and \Ha\ emission line maps extracted from our SINFONI IFS data.
 The ALMA Band 7 continuum observations used in this work belong to the Cycle 6 observing programme ID 2018.1.00992.S (P.I: C. M. Harrison). 
 
 Observations were taken between 18 October and 5 November 2018.
The on-source integration time for each target is between 14 and 70 minutes. The observations were performed with $45-49$ antennas, with baselines in the range $15.1-2500$~m. 
The Band 7 870$\mu$m continuum observations correspond to rest-frame 250-277~\micron\ for our sources. 
The requested angular resolution was 0.15-0.30'', to match the resolution of the SINFONI observations. 
The maximum recoverable scales are in the range $1.9-2.8$''. The field of view is $\sim17$''.

\subsection{Data reduction and imaging}

We reduced the data using the Common Astronomy Software Application (\casa) version 5.4.0. The calibrated measurement sets were created using the standard ALMA pipeline provided with the raw data in the ALMA archive.
We used the \casa\ task {\tt fixvis} to  phase-centre the calibrated measurement set on the central position of the source, determined by fitting a 2D Gaussian profile on the reduced image provided by the ALMA data-reduction pipeline.

We produced an image of the calibrated measurement set using the  \casa\ task {\tt tclean}.
We measured the RMS (root mean square) noise from the dirty image (obtained without applying the primary-beam correction), excluding the emission from the sources.
 We set the cleaning threshold of {\tt tclean} to 2$\times$RMS and  the pixel size to be one fifth of the beam size. 
In three cases (cid\_451, cid\_1205 and cid\_1143), there is a secondary source in the field of view. 
The secondary sources are at a distance of $\sim2.5$", $8$" and $9$" from the primary target for cid\_1205, cid\_451 and cid\_1143, respectively.
We modelled the secondary source with a 2D Gaussian and then we removed it from the $uv$-data using the \casa\ task {\tt uvsub}. We note that in all cases the presence of the secondary source does not affect significantly our size  measurements. We also tested that fitting the two sources simultaneously in the $uv$-plane gives consistent results.

One of our aims is to compare location, sizes and morphologies between the FIR emission, measured from the ALMA maps, and the ionised gas emission, measured with SINFONI. Thus, we want to create ALMA maps with a similar resolution to the SINFONI IFS maps (FWHM PSF$\sim$ 0.3").
 We generated the maps using the Briggs weighting scheme with robust parameter $=2$ (corresponding to natural weighting). The beam sizes obtained using natural weighting are in the range 0.17-0.25'' and closely match the resolution of the SINFONI maps. 
We also created maps using the Briggs weighting scheme with robust $=0.5$ (higher resolution, beam size 0.16-0.20'') and we found no appreciable difference in the morphology and size of the FIR emission that would affect our conclusions. Therefore we decided to use  the maps obtained with natural weighting  for the rest of the analysis. 
We show these final ALMA maps in Figure~\ref{fig:ALMA_maps}. 

We re-measured the RMS from the `clean' maps produced with natural weighting. The RMS  are in the range 15-31 $\mu$Jy beam$^{-1}$.
We measured the peak signal-to-noise ratio (S/N) of the sources by dividing the peak flux density by the RMS. The S/N values vary between 3.6 and 29.
 Two of the targets (cid\_1057 and cid\_451) have low S/N ($<<10$), 
 and therefore we cannot derive reliable size measurements for these objects \citep{Simpson2015}. We highlight these two targets with low S/N in all the relevant figures. The beam size, RMS, and peak signal-to-noise of the ALMA images  are summarised in Table \ref{tab:ALMA_obs}.

\begin{figure*}
\centering

\includegraphics[width=0.3\textwidth]{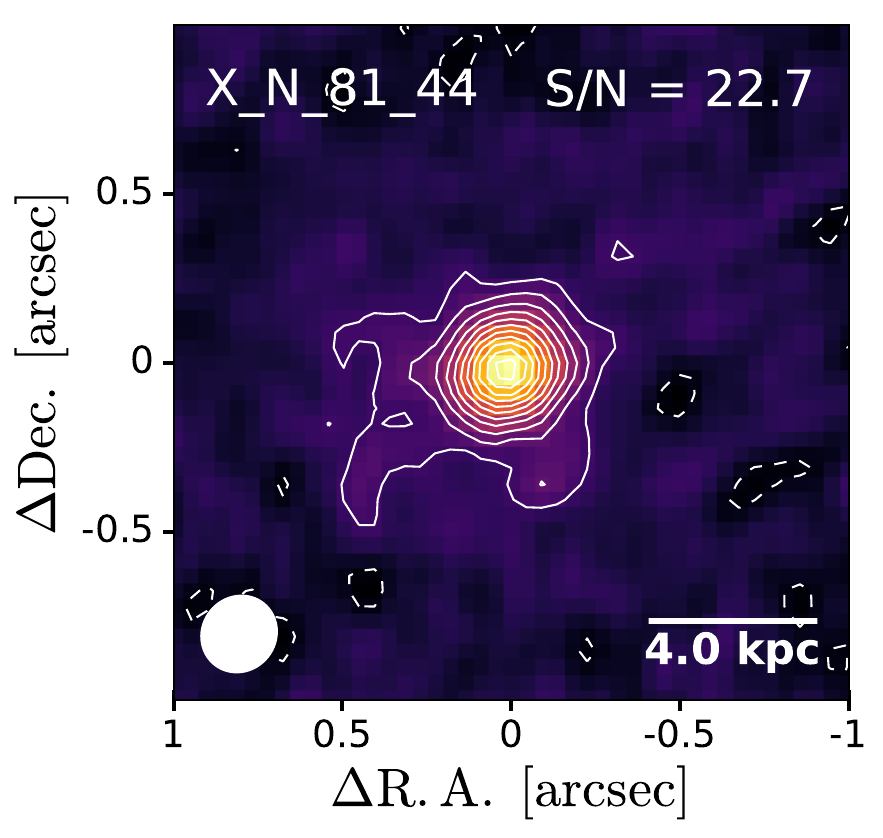}
\includegraphics[width=0.38\textwidth]{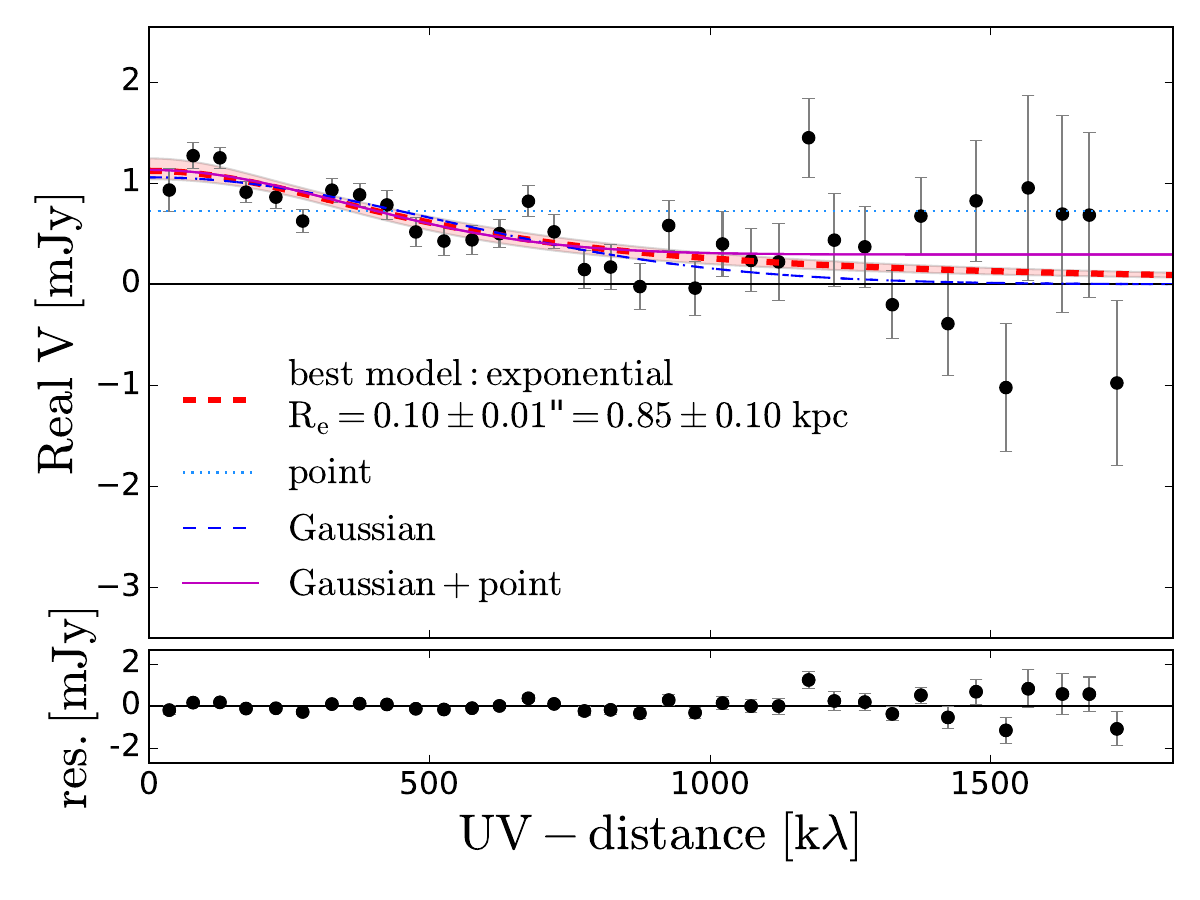}

\includegraphics[width=0.3\textwidth]{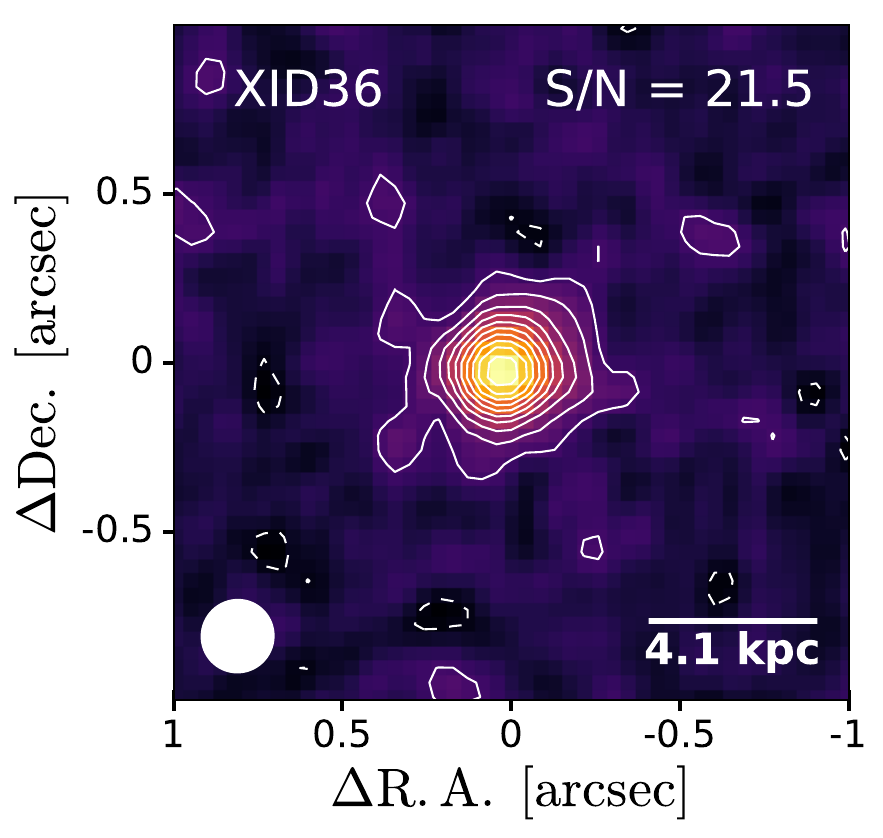}
\includegraphics[width=0.38\textwidth]{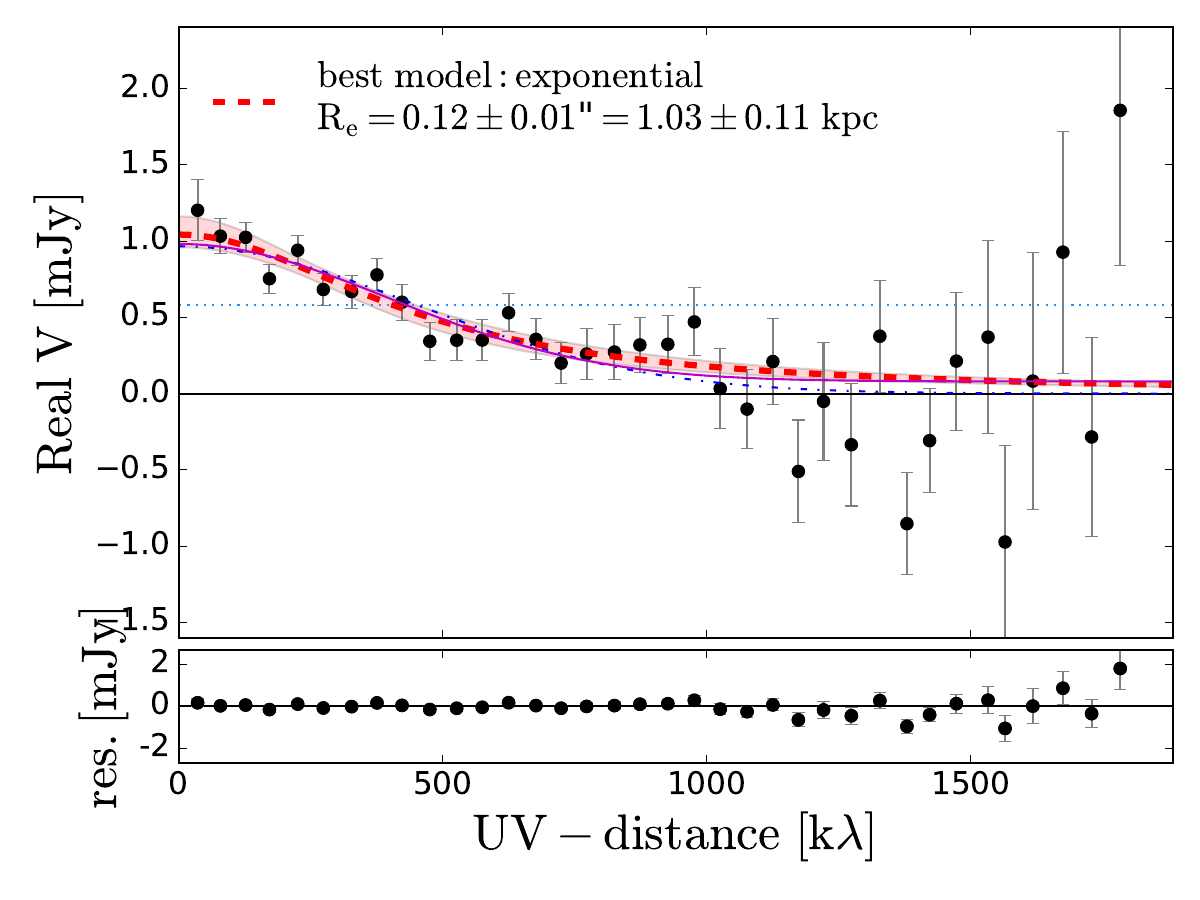}

\includegraphics[width=0.3\textwidth]{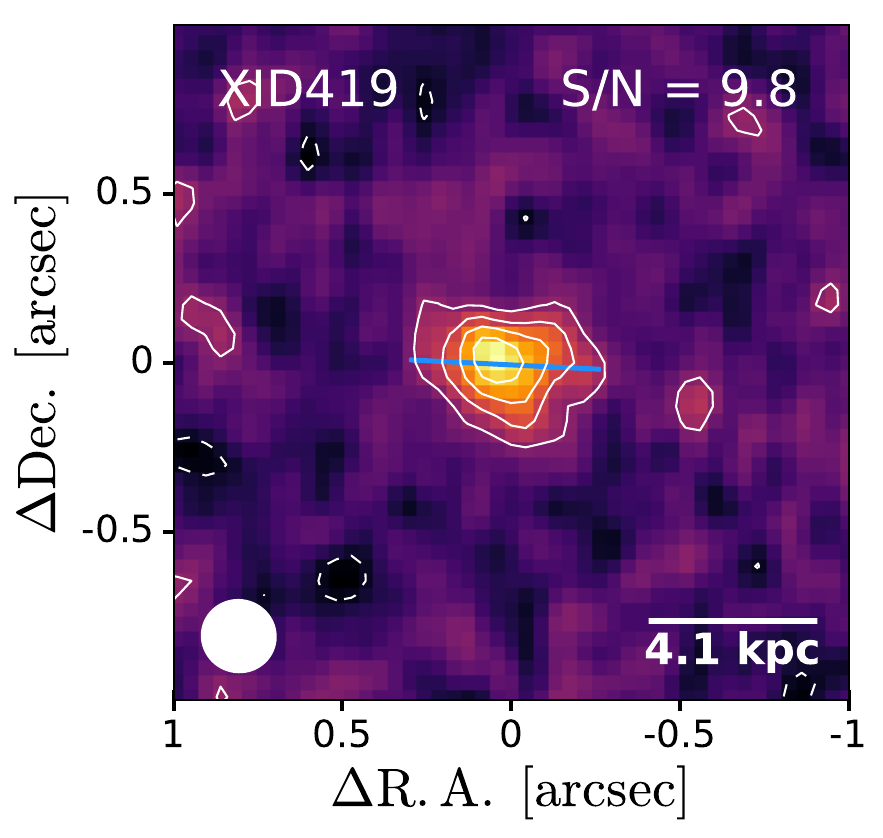}
\includegraphics[width=0.38\textwidth]{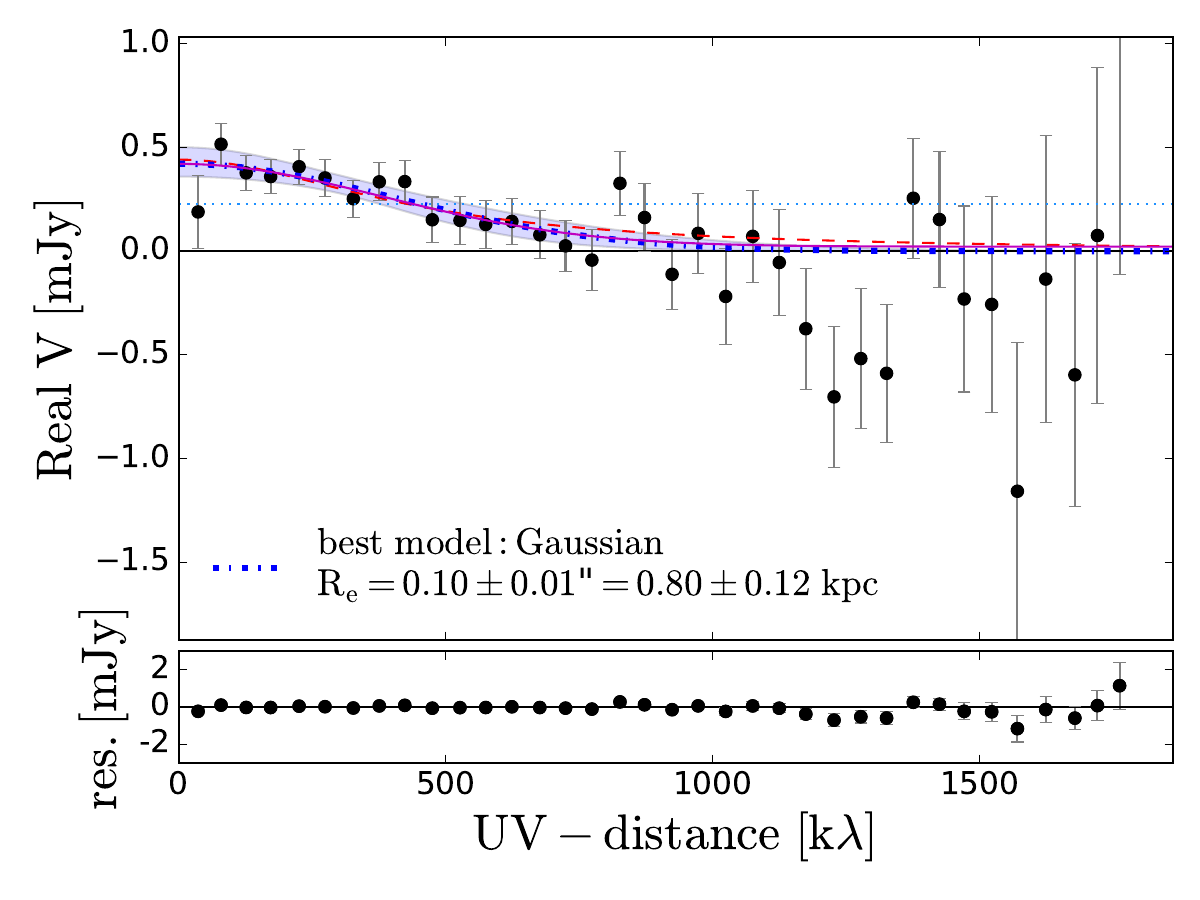}

\includegraphics[width=0.3\textwidth]{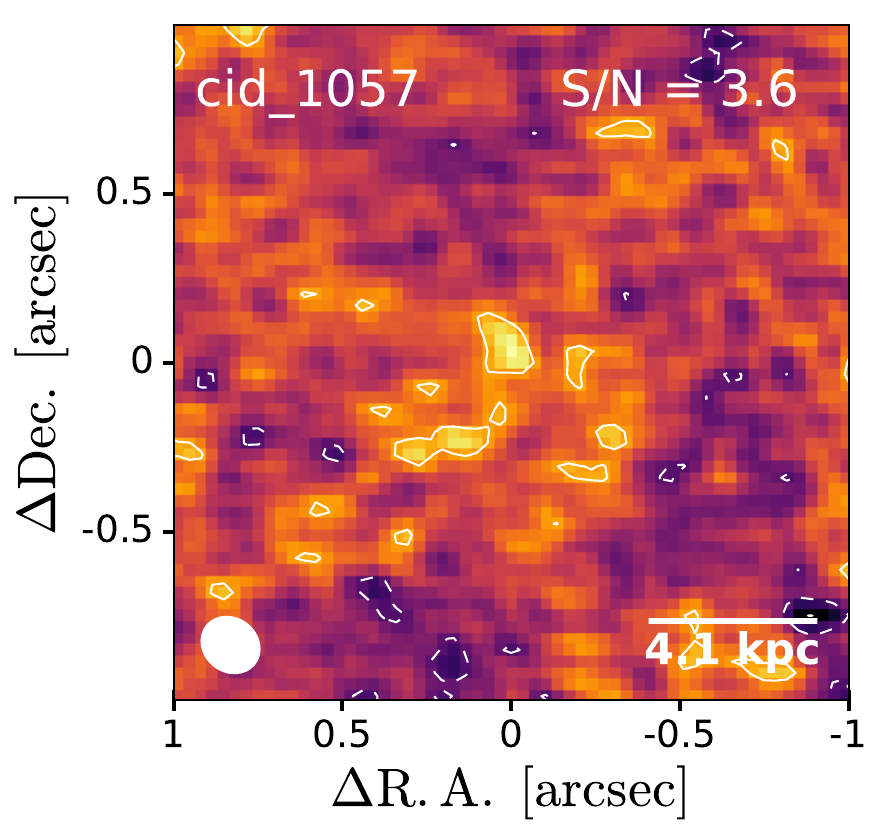}
\includegraphics[width=0.38\textwidth]{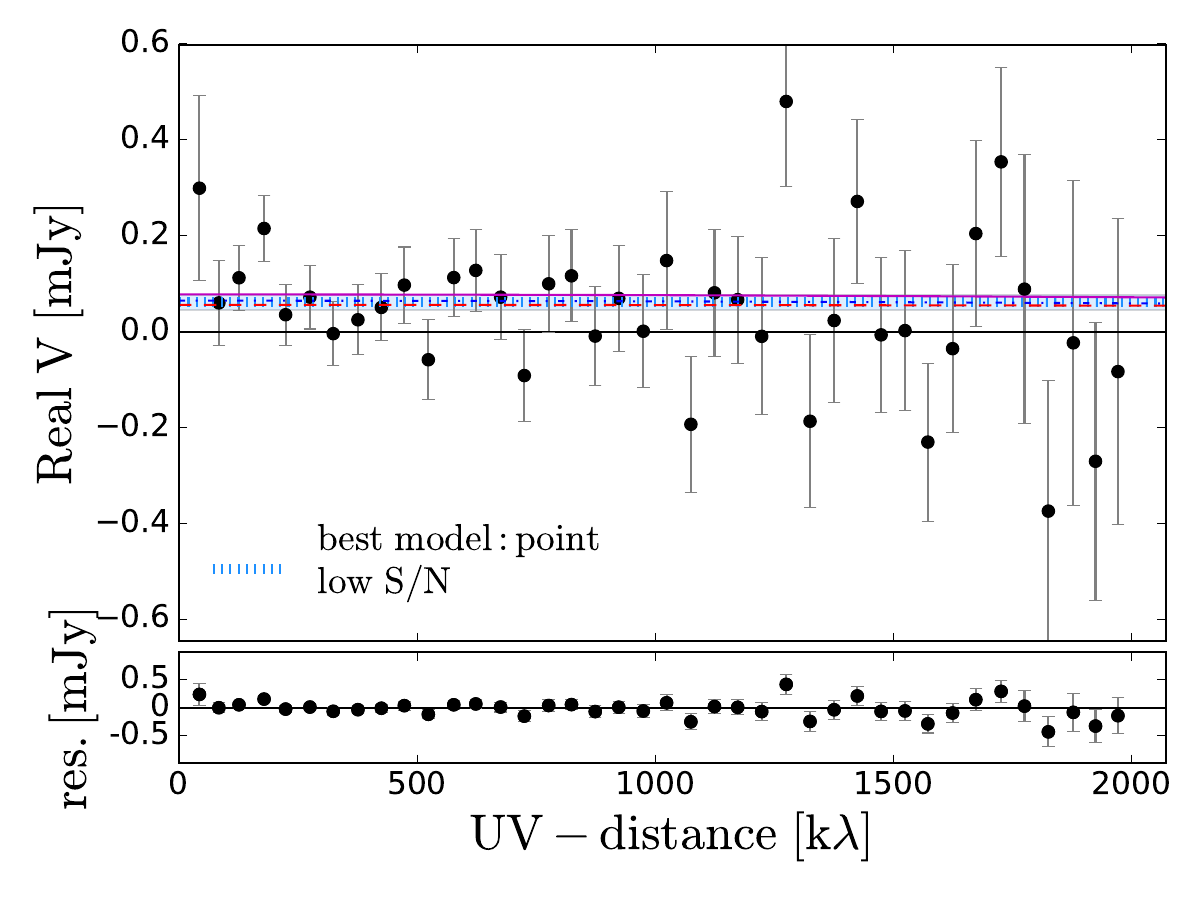}

\caption{\small \textit{Left column:} Maps of the ALMA continuum. 
The white ellipse represents the ALMA beam. The peak S/N is also shown. Contours start at 2$\sigma$ and increase at intervals of 2$\sigma$. Dashed contours are negative contours at -2$\sigma$. The lightblue line indicates the position angle along the major axis of the FIR emission, when it can be reliably determined (see Section~\ref{sec:FIR_size_morpho}).
\textit{Right column:} Real part of the visibilities versus $uv$-distance. The overlaid straight and curved lines show the fit using different models: point source (dotted lightblue line), exponential (dashed red curve), Gaussian (dashed-dotted blue curve), and Gaussian profile plus point source (solid magenta curve). The best model according the Bayesian Information Criterion (BIC) is labelled in each case. The shaded area shows the $1\sigma$ uncertainties on the best fit model. The lower panel shows the residuals with respect to the best-fit model. 
} 
\label{fig:ALMA_maps}
\end{figure*}

\begin{figure*}\ContinuedFloat
\centering

\includegraphics[width=0.3\textwidth]{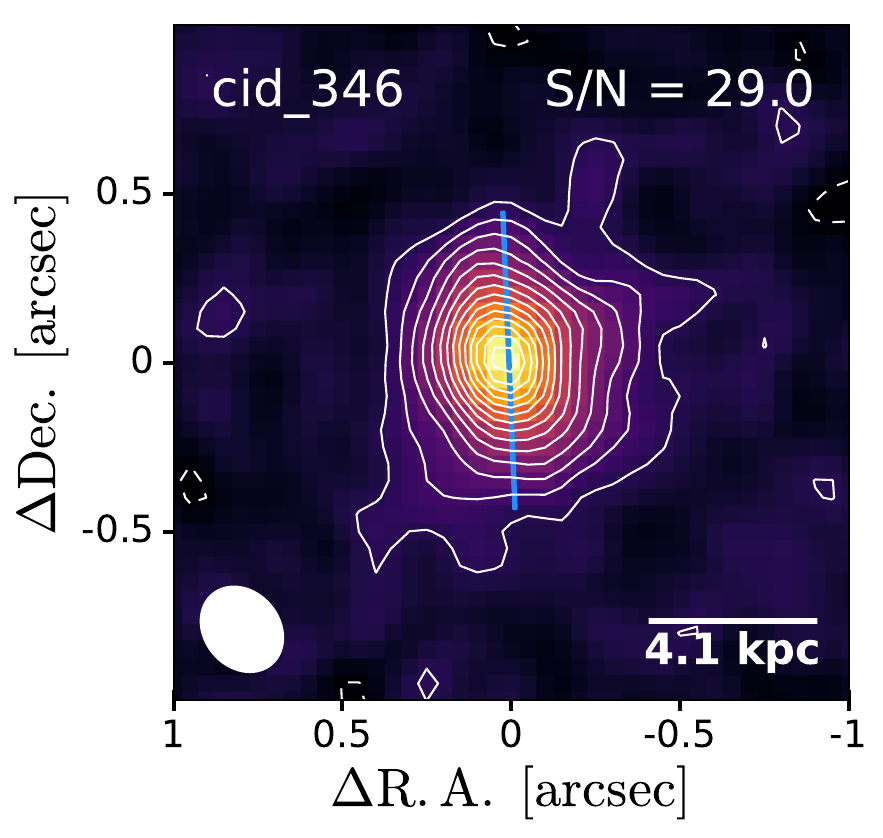}
\includegraphics[width=0.38\textwidth]{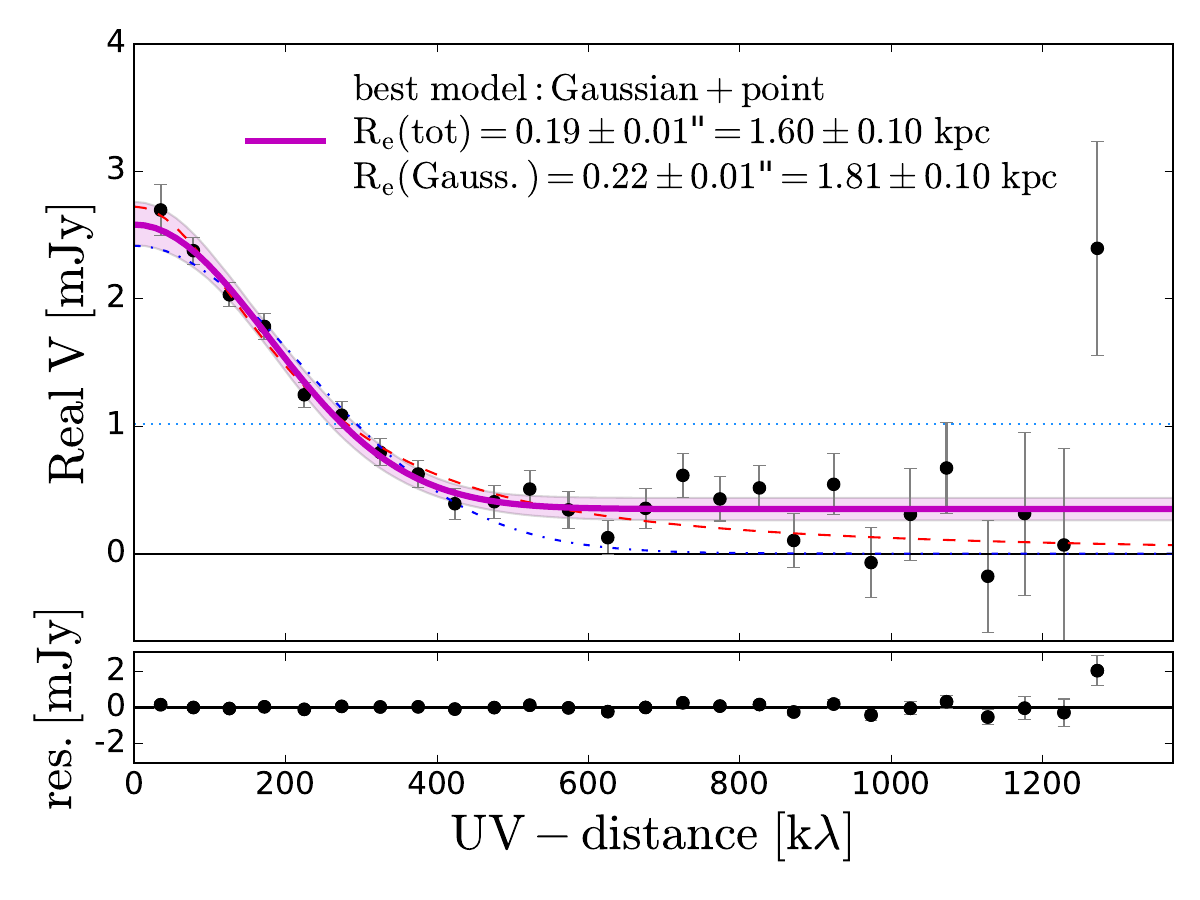}

\includegraphics[width=0.3\textwidth]{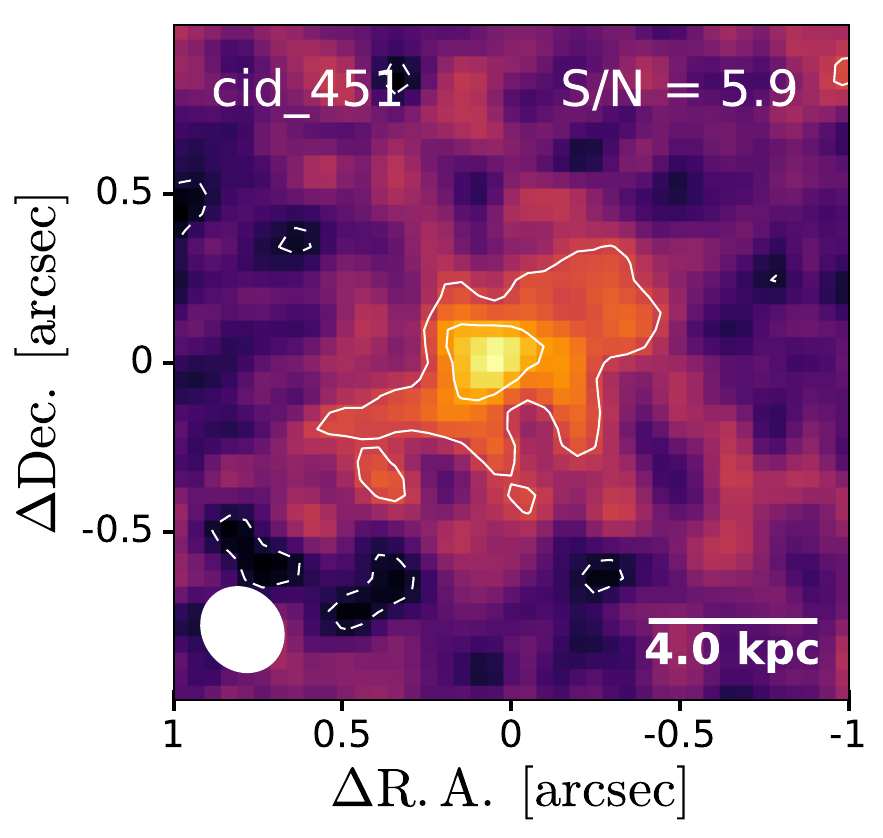}
\includegraphics[width=0.38\textwidth]{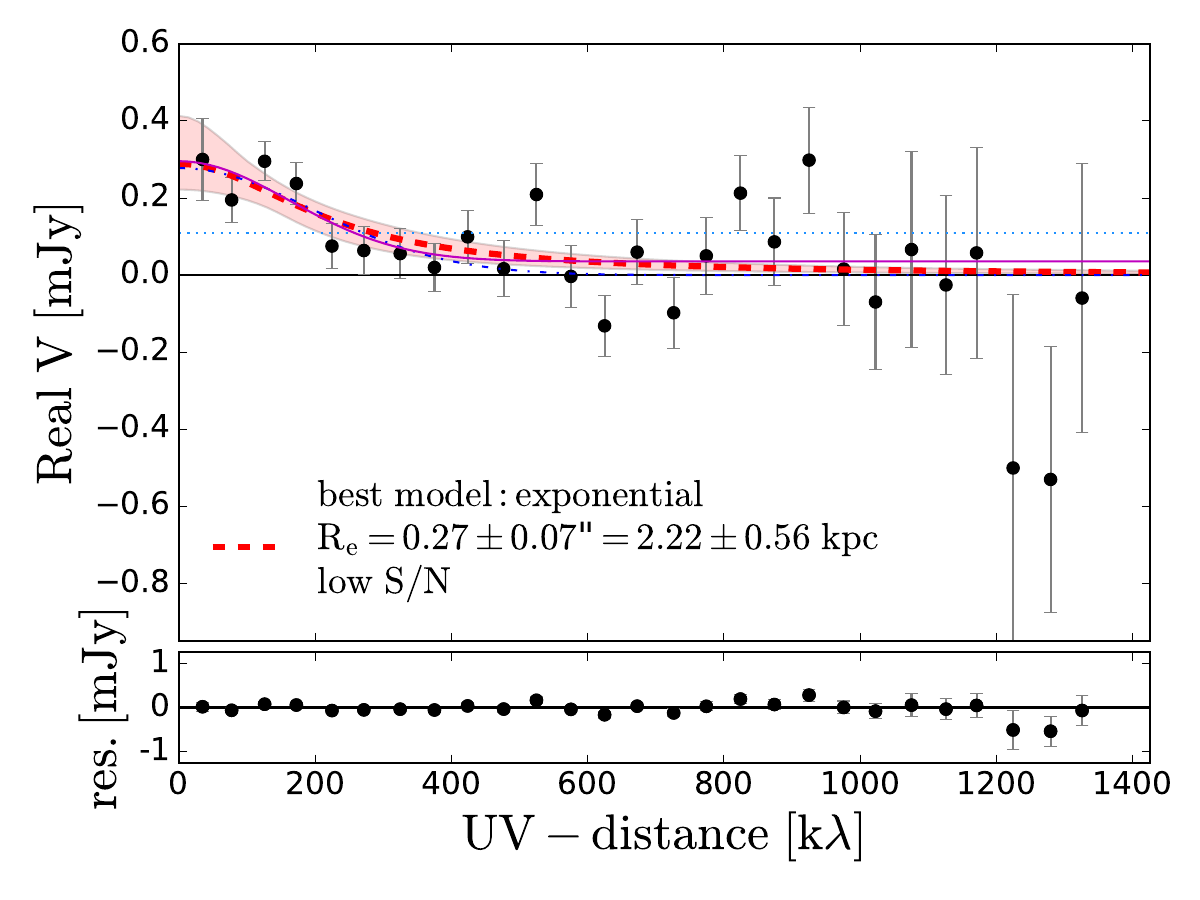}

\includegraphics[width=0.3\textwidth]{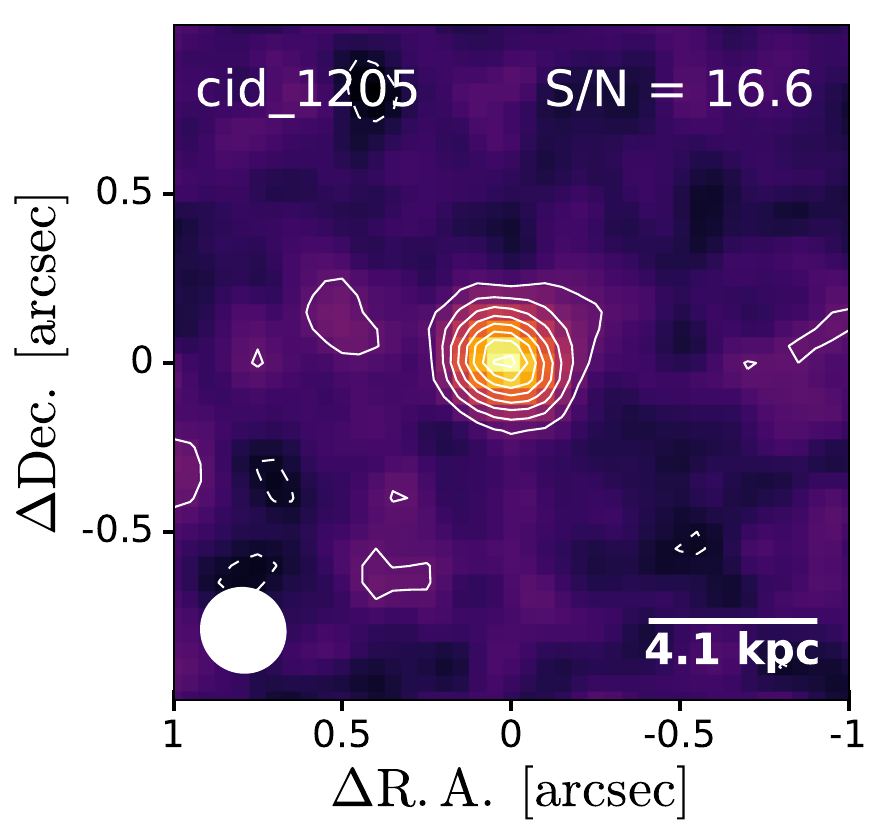}
\includegraphics[width=0.38\textwidth]{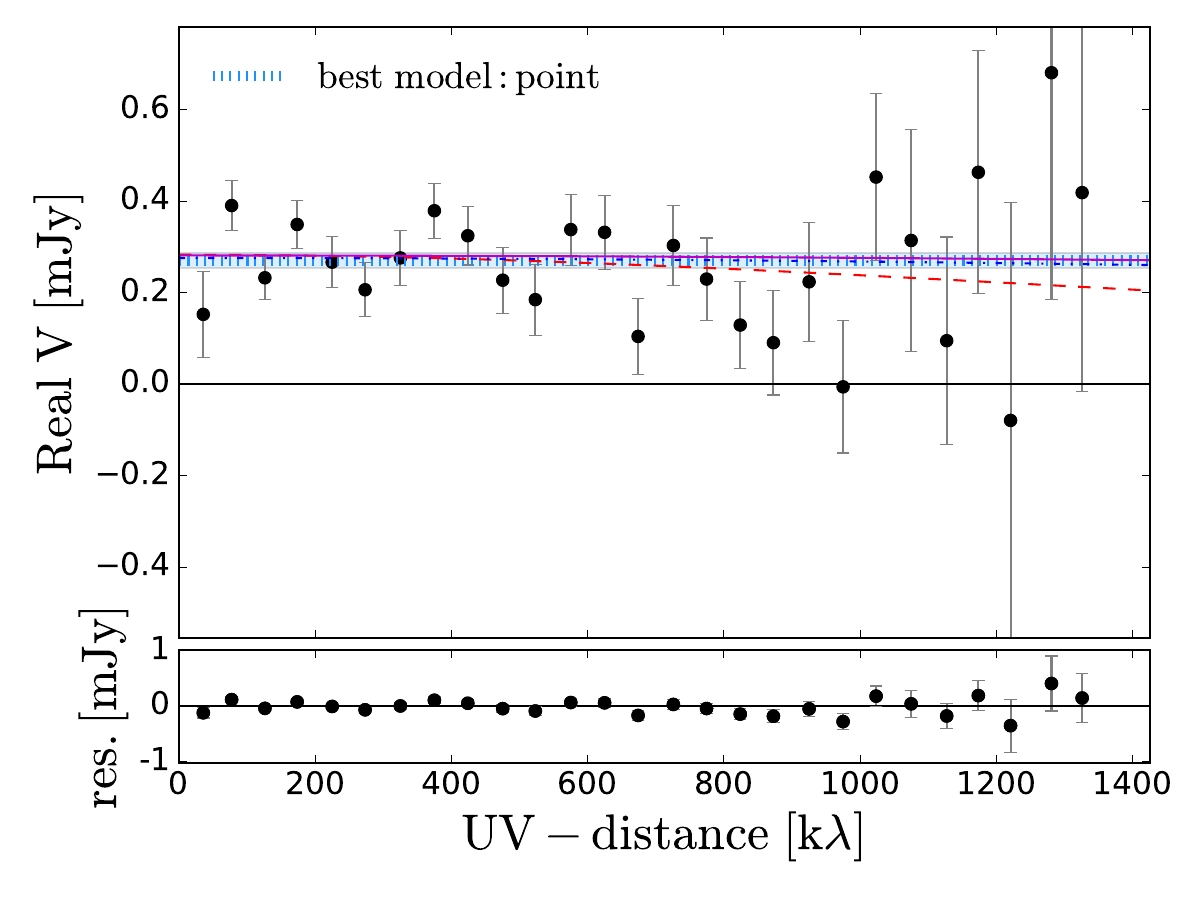}

\includegraphics[width=0.3\textwidth]{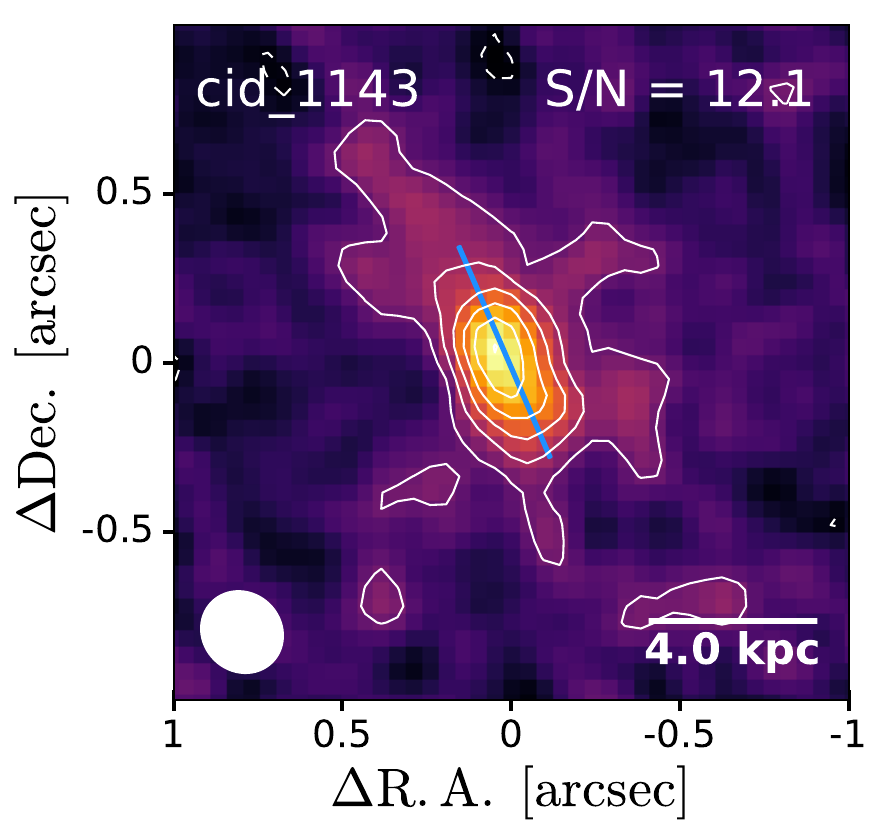}
\includegraphics[width=0.38\textwidth]{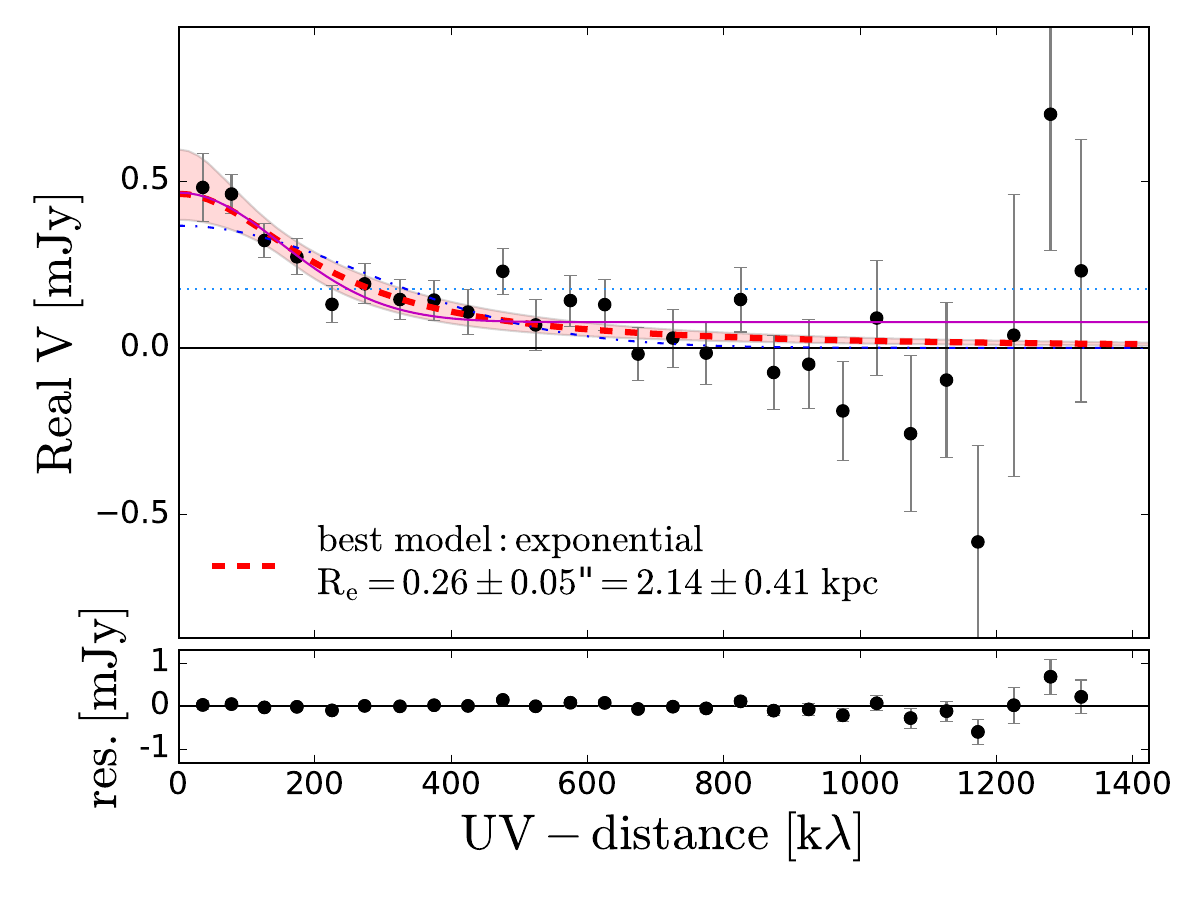}

\caption{Continued.}
\label{fig:ALMA_maps_2}
\end{figure*}

\begin{table*}
\centering
\caption{Properties of the ALMA maps and measurement of FIR sizes and flux densities at rest-frame $\sim260$~\micron. 
(1) Synthesized beam size (FWHM) of the ALMA maps created with Briggs weighting with robust=2.  
(2) RMS noise (measured from the `clean' maps). 
(3) Peak signal-to-noise in the ALMA map. 
(4) Best fit model to the $uv$-visibility data, according to the Bayesian Information Criterion (BIC). ($^{\star}$) If the peak $S/N < 8$ in the ALMA maps, we do not consider the size measurements to be reliable.
(5) Flux density at rest-frame $\sim260$~\micron, measured from the $uv$-visibilities using the best model. 
(6) Half-light radius (deconvolved from the beam) in arcsec measured from the $uv$-visibilities using the best model. For cid\_1057 the radius is not reported due to the low S/N.
(7) Half-light radius in kpc.
(8) Axial ratio measured using \uvmodelfit\ (excluding point sources). Axial ratios consistent with one are marked with~$^*$.
(9) Position angle (P.A.) measured using \uvmodelfit\ for the objects with axial ratios significantly different from one. The P.A. measures the angle of the major axis with respect to the north in anti-clockwise direction. 
(10) AGN contribution to the rest-frame 260~\micron\ ALMA flux, estimated from the SED fit.
(11) Contribution from synchrotron emission to the rest-frame 260~\micron\ ALMA flux. We model the synchrotron emission as a power law with  a spectral index $\alpha_r$. 
(12) Spectral index $\alpha_r$  used to estimate the synchrotron contribution, calculated from the 1.4, 3 and 5.5~GHz data (when available). In the cases where we could not constrain $\alpha_r$ with the available data, we assume $\alpha_r=-0.7$ (see Sec.~\ref{sec:origin}).
}
\setlength{\tabcolsep}{3pt}
\small{
\begin{tabular}{lcccccccccccc}
\hline 
ID  & Beam size & RMS  &S/N  & best model & F(260~\micron) & R$_{e}$ & R$_{e}$  & axial ratio & P.A. & $f_{AGN}$  & $f_{sync}$  & $\alpha_r$  \\
   &  [arcsec$^2$] & [$\mu$Jy beam$^{-1}$] & & & [mJy] & [arcsec] & [kpc] & & [deg.] & \% & \%  & \\ 
   & (1) & (2)  &(3)  & (4)  & (5) & (6) & (7) & (8) & (9) & (10) & (11) & (12)\\ 
  \hline \hline

X\_N\_81\_44 & 0.23$\times$0.22 & 31.44 & 22.7 & exponential & 1.13$\pm$0.16 & 0.10$\pm$0.01 & 0.83$\pm$0.10 & 1.00$\pm$0.11$^*$  & - & $<$ 1 & $<$ 1 & -0.70\\
XID36 & 0.22$\times$0.21 & 26.62 & 21.5 & exponential & 1.05$\pm$0.13 & 0.12$\pm$0.01 & 1.00$\pm$0.10 & 1.22$\pm$0.12$^*$  & - & 3 & $<$ 1 & -1.25 \\
XID419 & 0.22$\times$0.21 & 23.49 & 9.8 & Gaussian & 0.43$\pm$0.05 & 0.10$\pm$0.01 & 0.79$\pm$0.12 & 2.9$\pm$0.8& 88$\pm$5 & 4 & $<$ 1 & -0.70\\
cid\_1057 & 0.18$\times$0.16 & 16.34 & 3.6$^{\star}$ & point & 0.06$\pm$0.02 & -  & -  & - & - &3 & $<$ 1 & -0.70\\
cid\_346 & 0.27$\times$0.22 & 31.40 & 29.0 & Gauss.+point & 2.58$\pm$0.14 & 0.19$\pm$0.01 & 1.61$\pm$0.09 & 1.3$\pm$0.1& 2$\pm$5 & $<$ 1 & $<$ 1 & -0.98 \\
cid\_451 & 0.26$\times$0.23 & 16.25 & 5.9$^{\star}$ & exponential & 0.29$\pm$0.09 & 0.24$\pm$0.07 & 1.98$\pm$0.55 & 3.1$\pm$0.8& -58$\pm$5 & 6 & 21 & -0.99 \\
cid\_1205 & 0.25$\times$0.25 & 16.20 & 16.6 & point & 0.27$\pm$0.02 & $<$ 0.12 & $< $ 2.06 & - & - &$<$ 1 & $<$ 1 & -0.70\\
cid\_1143 & 0.25$\times$0.23 & 14.76 & 12.1 & exponential & 0.46$\pm$0.11 & 0.25$\pm$0.05 & 2.01$\pm$0.41 & 4.6$\pm$0.7& 23$\pm$2 & $<$ 1 & $<$ 1 & -0.70\\

\hline
\end{tabular}
}
\label{tab:ALMA_obs}
\end{table*}

\subsection{Modelling the ALMA continuum data: sizes and flux densities}
\label{sec:ALMA_sizes}
In this section, we describe how we measure the sizes and flux densities of the 260~\micron\ (rest-frame) emission. 
In particular, we are interested in comparing the 260~\micron\ sizes with FIR sizes of literature samples, to test whether AGN show different sizes compared to non-AGN.
We also want to compare the FIR sizes with the sizes of the optical emission-line regions (\Ha\ and \OIII), to investigate whether they are tracing similar spatial scales.

We tested models describing different morphologies (e.g. point source, exponential disk profile)  to extract reliable size and flux density measurements. 
 We measured the sizes both from the visibilities and in the image plane.
Measuring the sizes directly from the $uv$-visibilities has the advantage that it is not dependent on the choices made to create the images (e.g. weighting scheme, cleaning threshold, ...).
For completeness, all of the size measurements using the different methods are shown in the appendix in tables~\ref{tab:ALMA_meas_1} and \ref{tab:ALMA_meas_2}.

First we used the \casa\ routine {\tt uvmodelfit} to fit a model directly to the visibilities. The models available with {\tt uvmodelfit} are a point source, a 2D Gaussian and an elliptical disk. 
 We ran {\tt uvmodelfit} on the phase-centred calibrated measurement sets assuming a 2D Gaussian model and  using 20 iterations, which are enough to reach convergence. We note that \uvmodelfit\ measures the intrinsic sizes, deconvolved from the beam.  The results from \uvmodelfit\ also provide the information on the ratio  of the major and minor axis (see Table~\ref{tab:ALMA_obs}).

We also performed a fit on the collapsed visibilities using python outside of \casa, which gives us more freedom in  the choice of models. 
We  extracted the visibility amplitudes from the phase-centred calibrated measurements sets.
Then, we binned the data in $uv$-distance intervals of 50~k$\lambda$. We measured the average of the visibilities at each $uv$-distance, weighted by the corresponding uncertainties. The uncertainty on each mean visibility is given by the standard deviation of the mean divided by the square root of the number of visibility points in that bin. 
 In Figure~\ref{fig:ALMA_maps} we show the visibility amplitudes as a function of $uv$-distance for our sample.
 
For simplicity, we only considered symmetrical models, that is an axial ratio of one. We note that the radii derived with symmetrical Gaussian models  are consistent with the mean radii measured with \uvmodelfit. 
We tested the following models: point source, 2D Gaussian, 2D Gaussian plus a point source, and exponential profile (equivalent to a \sersic\ profile with index $n=1$).

To fit the visibilities versus $uv$-distances, we used the Markov chain Monte Carlo (MCMC) program {\tt Stan}\footnote{http://mc-stan.org/} \citep[][]{Carpenter2017}, which allows us to accurately estimate the uncertainties on the derived parameters. Specifically, we employed {\tt PyStan}\footnote{http://pystan.readthedocs.io/en/latest/http://mc- stan.org}, the python interface to {\tt Stan}. We assumed a Gaussian likelihood and use uniform priors, allowing the free parameters to vary in a large parameter range (the prior ranges are tabulated in Table~\ref{tab:priors} in the appendix).

We derived the best fit parameters by taking the median values from the marginal posterior distributions. The uncertainties are given as the 16th and 84th percentiles of the posterior distributions.
We express all sizes in terms of half-light radius (also known as effective radius) $R_e$.
For a 2D Gaussian, $R_e$ is equivalent to $0.5\times$FWHM. For the exponential profile $I(R) = \exp{(-R/a)}$, where $a$ is the scale parameter, $R_e$ is given by $1.6783\times a$. For the model with Gaussian plus point source, we calculate the radius that contains half of the light based on the total profile derived from the fit, $R_e$(tot). We also report the value of $R_e$ for the Gaussian component only, $R_e$(Gauss.).

The results from our fit on the visibilities using a Gaussian model are consistent within the uncertainties with the results from \uvmodelfit\ both in terms of sizes and flux densities. 
 Therefore, for the rest of this work we use the flux densities and size measurements from our analysis of the collapsed visibility amplitude versus $uv$-distance performed in python.

To assess the preferred model to fit our data, we use the Bayesian Information Criterion \citep[BIC;][]{Schwarz1978}:
\begin{equation}
BIC = -2\cdot \ln L + q\cdot \ln (m),
\end{equation}
where $L$ is the likelihood (i.e. the probability of the data given the parameters), $q$ is the number of free parameters of the model, and $m$ is the number of data points. The BIC considers the likelihood of the model and penalises models with a larger number of parameters. The model with the smallest BIC is the preferred model\footnote{We note that using the Akaike Information Criterion (AIC) or the Akaike Information Criterion corrected for sample size \citep{Akaike1973, Akaike1974} will give the same qualitative results. }.  Figure~\ref{fig:ALMA_maps} shows the visibilities versus $uv$-distances data with the curves showing the different models fitted. The preferred model is highlighted with a thicker line.

At least a S/N$\gtrsim10$ on the flux is required to obtain reliable size measurements \citep{Simpson2015}. 
 XID419 has S/N$=9.8$ and is therefore a borderline case. Given the good agreement in the size measured with different methods for this source, we consider the size measurements to be reliable.
 cid\_1057 and cid\_451 have very low S/N (3.6 and 5.9, respectively), therefore we do not consider the size measurements of these two targets to be reliable.
 However, for completeness we show their values throughout the figures, and we highlight them as unreliable.
For these two sources, the flux densities measured from the $uv$-visibilities are more uncertain. Thus, we also check the flux densities derived from the ALMA images created by applying tapering (with tapering size = 0.5") using \casa. The peak flux densities measured in the tapered images are in agreement (within the uncertainties) with the ones measured from the fit of the $uv$-visibilities.

 We verify that we obtain consistent results by performing the analysis also in the image plane, using the 2D Gaussian fitting routine available in the \casa\ {\tt viewer}. 
 The sizes and flux densities measured in the image plane are in agreement with the measurements done in the $uv$-plane using the Gaussian model. For more details on the comparison between different methods, see Section~\ref{sec:size_comp} of the appendix.

The sizes and flux densities measured with the different methods are reported in tables~\ref{tab:ALMA_meas_1} and \ref{tab:ALMA_meas_2}  in the appendix.The measurements done in the image plane, with \uvmodelfit,  and fitting the  $uv$-visibilities with the `preferred model' are in general agreement (see Figure~\ref{fig:comp_ALMA_sizes} in the appendix). 
 We consider as our `best measurements' the FIR sizes and flux densities derived from the fit of the $uv$-visibilities with the `preferred model' (reported in Table~\ref{tab:ALMA_obs}). 
 Given the general agreement between the different methods, the choice of the method does not affect our results.  
Nevertheless, we provide the information on the sizes measured with the different methods to ease the comparison with literature values (see Section~\ref{sec:FIR_sizes_lit}).

\subsection{FIR sizes, morphologies, and flux densities}
\label{sec:ALMA_fluxes_sizes}

\subsubsection{FIR sizes and morphologies}
\label{sec:FIR_size_morpho}

The effective radii derived from the `best fit' models for our sample are in the range 0.80-2.13 kpc, with a median of  $1.31\pm0.23$ kpc \footnote{The error on the median is calculated as the standard deviation of the measurements divided by the square root of the number of objects.}.
The sizes measured from the fit with an exponential profile are larger than the sizes derived assuming a Gaussian profile by a factor of 1.38 on average (for the extended sources). This is by construction, since the exponential profile  decreases less rapidly toward zero at larger radii than the Gaussian profile and therefore there is a larger fraction of flux at large distances. However, we note that our conclusions do not change if we choose one of these models for all sources. To facilitate comparison with literature values (where both these models are considered), we provide the results from both models in Appendix \ref{sec:size_comp}.

Half of the sources with reliable size measurements (3/6) are best described by an exponential profile according to the BIC. One source (cid\_1205) is better described by a point source and one source (XID419) by a Gaussian. 
The preferred model for cid\_346 is a Gaussian plus a point source.  The point source accounts for 13.5\% of the total flux density, according to the results of our fit. For this source, we also test a model consisting of an exponential profile plus a point source.  This model gives a point source contribution to the total flux of 4.1\%, but it is not preferred over the `Gaussian plus point source' or the `exponential profile' model, according to the BIC.
The point source component could be either due to the emission from a compact starburst or from the AGN.
From our SED fitting decomposition, the AGN component contributes only 0.15\% to the total  260~\micron\ flux density, thus it cannot account for all the flux of the observed point source. 
Therefore, a compact star-forming region may be a more likely scenario. Since we have only two objects which require a point source in the model, and in the case of cid\_346 its contribution to the total flux is marginal, we consider this as an indication that on average point source emission from AGN heated dust at this wavelength is not prominent (see Section~\ref{sec:origin} for further discussion on the origin of the FIR emission).

The ratios between the major and minor axis derived using \uvmodelfit\ for the 6/8 sources with high S/N are in the range $1-4.6$. For two sources (\XN\ and cid\_1205) the axial ratio is consistent with one. 
For XID\_419, cid\_346, and cid\_1143 the difference in the sizes of the major and minor axes show that the emission is significantly elongated in one direction (axial ratio $> 1.3$). 
For these three sources, we report also the position angle measured with  \uvmodelfit (see Table~\ref{tab:ALMA_obs}).

\subsubsection{Flux densities}
\label{sec:ALMA_fluxes}
 The flux densities measured with the different methods are  in agreement within the uncertainties (see Figure~\ref{fig:comp_ALMA_sizes}). The only exception is cid\_346, for which the \uvmodelfit\ Gaussian fit gives a slightly smaller flux (8\%) than the other methods. 
This is probably because the Gaussian model cannot well describe the point source emission in the centre, and therefore  misses part of the flux.

Our high-resolution ALMA images are potentially missing some of the more diffuse emission, and therefore the measured flux densities cannot be considered as `total' flux densities. For example, \citet{Harrison2016b} found a drop in peak flux of $18-44\%$ between a resolution of 0.8'' and 0.3''.
 To assess how much flux we are potentially missing in our ALMA data, we compare these flux measurements with other   260~\micron\ flux measurements from lower resolution data from the ALMA Archive \citep[compiled by ][]{Circosta2018} and from \citet{Scholtz2020}.  Three sources have 260~\micron\ fluxes from lower resolution data.
XID419 has an upper limit from \citet{Scholtz2020}, that is consistent with our measurement. 
For cid\_451 and cid\_1205, our flux density measurements are consistent within the uncertainties with the 260~\micron\ flux densities from the low-resolution data.
 Therefore, there is no evidence that the ALMA images are resolving out a large portion of the flux.

\subsection{SED fits including the ALMA flux densities}
\label{sec:ALMA_SED_fit}

In this section, we fit the spectral energy distributions (SEDs) of our targets to estimate the stellar masses and SFRs of their host galaxies. We used the UV-FIR photometric catalogue from \citet{Circosta2018} and we added the ALMA band-7 flux densities.
Given the large number of upper limits and the large PSF of the \Herschel\ images, including the ALMA fluxes helps to place better constraints on the properties of the host galaxies. 
Moreover, in section \ref{sec:origin} we  use these best-fit SED models to estimate the AGN contribution to the $\sim 260~\mu $m flux densities.

In addition to including the ALMA photometry, we took a conservative approach for the weakly detected or possibly contaminated \Herschel\ sources; specifically, we considered the Herschel fluxes as upper limits for  \XN, cid\_1057, cid\_1205 and cid\_1143. However, if we use  the full photometry catalogue used in \citet{Circosta2018}, even excluding the ALMA Band 7 photometry, we reach the same conclusions in terms of AGN contributions (see Section~\ref{sec:origin}).

 
We use the same SED fitting approach used by \citet{Circosta2018} for the full SUPER sample, with the only difference that we include the ALMA band 7 photometry.
\citet{Circosta2018} performed the panchromatic (UV-to-FIR) SED fitting  using the Code Investigating GALaxy Emission  \citep[\cigale;][]{Burgarella2005, Noll2009, Boquien2019}.
 This code takes into account the energy balance between the absorption by dust in the UV-optical and the corresponding re-emission in the FIR. The SED model includes three emission components: i) stellar emission; ii) emission by cold dust heated by star formation; iii) AGN emission, consisting in direct emission from the accretion disk in the UV-optical range and emission from the dusty torus peaking in the mid-infrared (MIR). For more details on the SED fitting method we refer the reader to \citet{Circosta2018}.

 The SFRs are estimated from the infra-red (IR) luminosity, obtained by integrating the best-fit template SED in the rest-frame wavelength range 8-1000~\micron, after removing the AGN contribution.  The IR luminosity is converted to SFR using the \citet{Kennicutt1998} calibration, converted from a Salpeter IMF to a \citet{Chabrier2003} IMF by subtracting 0.23~dex \citep{Bolzonella2010}\footnote{
Systematic uncertainties on the stellar masses are around 0.3~dex and can be larger for SFRs \citep[e.g. ][]{Mancini2011, Santini2015}. These uncertainties are due to the models used, degeneracies between parameters, a priori assumptions and the discrete coverage of the parameter space.}.
Figure~\ref{fig:SED_fit} in the appendix shows the SEDs with the results from the SED fitting and the measured ALMA 260~\micron\ flux densities. The SFR and stellar masses derived from these SED fits are reported in Table~\ref{tab:sample}.

We compare the SFR and stellar masses derived from the SED including the ALMA data with the values reported in \citet{Circosta2018}. The SFRs vary by less than 0.13~dex for all targets except for cid\_1205 (-0.6~dex) and cid\_1143 (-1.2~dex). 
The large difference for cid\_1143 can be explained by the fact that the \Herschel\ fluxes are all upper limits, thus the previous SFR needs to be considered as an upper limit. For cid\_1205, the \Herschel\ fluxes used to calculate the previous SFR value included also the flux from the companion galaxy, thus the previous SFR was an upper limit. 
For cid\_451, only an upper limit on the SFR ($< 125$ \Msun\ yr$^{-1}$) was reported in \citet{Circosta2018} due to the possible high contamination from synchrotron emission to the FIR fluxes. This target is classified as a radio-loud source based on the comparison between infrared and radio luminosity. 
We take advantage of the ALMA Band 3 flux measurement at rest-frame 1~mm recently presented in \citet{Circosta2021} to better constrain the contribution from the AGN synchrotron emission to the FIR fluxes (see Sec.~\ref{sec:origin}). We re-measure the SFR after subtracting the synchrotron contribution from the FIR fluxes and obtain SFR~$=48 \pm 19$ \Msun\ yr$^{-1}$. 
The stellar masses differences with respect to the previous measurements are $\leq 0.42$~dex (mean 0.14~dex). cid\_1143 shows the largest difference in stellar mass.

The right panel of Figure~\ref{fig:SUPER_sample} shows the parent SUPER sample and the subset targeted for ALMA observations in this work on the  SFR versus stellar mass ($M_*$) plane.
The ALMA targets have stellar masses in the range $\log (M_{*}/M_{\odot}) = 10.7-11.2$, with a median  $\log (M_{*}/M_{\odot}) = 10.9$. The SUPER parent sample has the same median  $\log M_{*}/M_{\odot} = 10.9$, and an interquartile range $\log (M_{*}/M_{\odot}) = 10.6-11.0$.
 The SFRs of the ALMA sample are in the range SFR$ = 8-380$ \Msun\ yr$^{-1}$, with a median SFR $= 76$ \Msun\ yr$^{-1}$. 
The SUPER parent sample has a median SFR $= 144$ \Msun\ yr$^{-1}$ and interquartile range SFR $= 63-242$ \Msun\ yr$^{-1}$.
 The ALMA targets have similar stellar masses and (slightly lower) SFRs to the objects with SFR and stellar mass measurements in the parent SUPER sample.
However, we note that only 24/39 of the SUPER targets have SFR and stellar mass measurements, and ten of these have only upper limits on the SFR, which are not taken into account in the above median and interquartile ranges.
Given the presence of many SFR upper limits in the parent sample, our targets are likely to represent the upper end of the SFR distribution of the SUPER sample.

On the SFR-$M_*$ plane (see Figure~\ref{fig:SUPER_sample}), we show also the main-sequence definition by \citet{Schreiber2015} at the average redshift of our sample ($z\sim2.3$). Most of the ALMA targets lie on the main-sequence or slightly below\footnote{We note that the main-sequence from \citet{Schreiber2015} is derived assuming a Salpeter IMF \citep{Salpeter1955}, while we assume a Chabrier IMF \citep{Chabrier2003}. However,  assuming a different IMF would systematically shift both $M_*$ and SFRs by approximately the same amount \citep{Brinchmann2004, Elbaz2007}, and therefore would not affect the shape of the main-sequence.}. 
The exception is cid\_1143, which lies clearly below the main-sequence. Despite the pre-selection of FIR detections, our ALMA targets are mainly main-sequence galaxies.  However, SUPER primarily consists of  moderate luminosity X-ray AGN which, as a population, tend to have SFR distributions slightly below the main-sequence \cite[e.g. ][]{Mullaney2015, Scholtz2018, Grimmett2020}.
 \citet{Scholtz2018} study the specific SFR (SSFR=SFR/$M_*$) distribution of a sample of 81 AGN at $z=1.5-3.2$ and for AGN with X-ray luminosities $L_{2-10~keV}> 10^{44}$~erg~s$^{-1}$, they measure the mode of the SSFR distribution  log(SSFR/Gyr$^{-1})= -0.32\pm0.16$. For the SUPER targets with SFR and stellar mass measurements, the mean is log(SSFR/Gyr$^{-1})= 0.19\pm0.09$. Thus, the SUPER sample is likely to have a distribution of SFRs skewed to higher values compared with the parent population of X-ray AGN.

\section{H$\alpha$ and [OIII] observations and analysis}
\label{sec:SINFONI}

\subsection{SINFONI observations and data reduction}

The  SINFONI Adaptive Optics (AO) assisted observations and data reduction are described in detail in \citet{Kakkad2020}, for the Type 1s, and in M.~Perna et al. (in prep.), for the Type 2s.
 Here we summarise the main information. The SINFONI observations  took place between November 2015 and December 2018 (ESO large program 196.A-0377). 
  We observe the H-band ($1.45-1.85$~\micron), which includes the rest frame optical lines \Hb\ and \OIII$\lambda \lambda$4959, 5007, and the K-band ($1.95-2.45$ \micron), which includes the \NII$\lambda \lambda$6584,6548, \Ha\ and \SII$\lambda \lambda$6716, 6731 lines. 
The average spectral resolution in the H-band and K-band is $\sim$3000 and $\sim$4000 respectively, corresponding to a channel width of $\sim 2~\AA$ and $\sim 2.5~\AA$, respectively. 
The PSF sizes of the H-band and K-band are in the range $0.27-0.52$'' and $0.15-0.46$'', respectively (the PSF sizes for each target are listed in  Table~\ref{tab:SINFONI}). 
We note that cid\_1057 is not detected in \Ha, thus for this target we only show the \OIII\ (i.e. H-band) data.

\begin{table*}
\centering
\caption{ Information about the SINFONI maps. PSF of the H and K-band SINFONI images and line fluxes measured from the integrated spectra. If the S/N of a line is smaller than 3, we report a $5\sigma$ upper limit. For cid\_1057, the S/N of the K-band image is too low to derive any information about the \Ha\ and \NII\ emission. 
(4) redshift derived from the narrow component of the \OIII\ line. (5) Ratio of the fluxes of \OIII$\lambda$5007 and \Hb\ narrow emission lines. For Type 1 AGN, this is the flux of the narrow component, i.e. the BLR component is not included. 
(6) Ratio of the fluxes of \NII$\lambda$6548 and \Ha\ narrow emission lines. For Type 1 AGN, this is the flux of the narrow component, i.e. the BLR component is not included. 
(7) Half-light radius (R$_e$) of the \OIII\ systemic emission ([-300,300] \kms\ range), deconvolved from the PSF. If R$_e$ is smaller than the PSF, we give the size of the PSF as an upper limit. 
(8) Half-light radius (R$_e$) of the \Ha\ systemic emission ([-300,300] \kms\ range), deconvolved from the PSF.
(9) Components used to fit the emission lines. `n': narrow component, `b': broad component (for the outflow), `BLR': broad line region component.
}
\begin{tabular}{lcccccccc}
\hline
ID & H-band PSF & K-band PSF & z$_{\text{[O\,\textsc{iii}]}}$ & F(\OIII)/F(\Hb) &  F(\NII)/F(\Ha) & $R_e$(\OIII) & $R_e$(\Ha) & line\\ 
   & [arcsec$^2$] &  [arcsec$^2$] & & & & [kpc] & [kpc] & components\\  
(1) & (2)  &(3)  & (4)  & (5) & (6)  & (7) & (8) & (9)\\ 
  \hline \hline

X\_N\_81\_44 & 0.27$\times$0.27 & 0.24$\times$0.24 & 2.3180 & $> 8.00$ & $< 0.10$ & $< 1.11$ & 3.48$\pm$0.08 & n, b, BLR\\
XID36 & 0.35$\times$0.35 & 0.15$\times$0.15 & 2.2578 & 10.15$\pm$0.63 & 0.69$\pm$0.12 & 1.06$\pm$0.01 & 1.57$\pm$0.01 &  n, b\\
XID419 & 0.52$\times$0.47 & 0.24$\times$0.22 & 2.1430 & $> 2.15$ & 0.91$\pm$0.05 & $< 2.07$ & 1.42$\pm$0.02 & n\\
cid\_1057 & 0.32$\times$0.30 & 0.46$\times$0.43 & 2.2099 & 2.51$\pm$0.17 & - & 1.36$\pm$0.01 & - & n, b\\
cid\_346 & 0.30$\times$0.30 & 0.30$\times$0.30 & 2.2170 & 1.58$\pm$0.20 & 0.43$\pm$0.11 & 1.99$\pm$0.03 & 2.28$\pm$0.05 & n, b, BLR \\
cid\_451 & 0.30$\times$0.28 & 0.28$\times$0.27 & 2.4434 & $> 10.59$ & 0.61$\pm$0.03 & 0.78$\pm$0.00 & 0.67$\pm$0.01 & n, b\\
cid\_1205 & 0.30$\times$0.30 & 0.30$\times$0.30 & 2.2555 & $> 1.66$ & - & 0.99$\pm$0.02 & 1.33$\pm$0.09 & n, BLR\\
cid\_1143 & 0.41$\times$0.38 & 0.30$\times$0.30 & 2.4418 & 6.91$\pm$1.97 & 0.75$\pm$0.09 & $< 1.62$ & 0.92$\pm$0.02 & n, b\\

\hline
\end{tabular}
\label{tab:SINFONI}
\end{table*}

\subsection{Astrometry registration}
\label{sec:coord}
Since one of our main goals is to compare the spatial distribution of the FIR continuum and ionised gas emission, we need to have reliable astrometry for both ALMA and the SINFONI maps.
The absolute position of the SINFONI cubes, as derived from the SINFONI pipeline, is not sufficiently accurate for our purposes.
Given the small field of view of the SINFONI images ($3\times3$~arcsec$^2$), we cannot correct the astrometry using nearby stars, since usually the target is the only visible source in the field of view. Thus, we need to rely on coordinates derived from other images.

We aligned our SINFONI data-cubes to broadband H- and K-band images of the same field. We used images from VLT/VISTA taken as part of the UltraVISTA survey for COSMOS \citep{McCracken2012} and as part of the VHS (VISTA Hemisphere Survey) for XMM-XXL \citep{McMahon2013}, and from VLT/ISAAC for CDF-S \citep{Retzlaff2010}. 
We first aligned the H/K-band images to \gaia\ astrometry \citep{GAIA2016, GAIA2018} using several stars across the fields. We then made broad-band images from the SINFONI data-cubes by collapsing them over the same wavelength range as the archival H/K-band images. We found the centroids of the images from the data-cubes and aligned these to the positions of the corresponding source in the near-infrared  VLT/VISTA or VLT/ISAAC (H/K-band) images\footnote{For two of the Type 1 sources (\XN\ and cid\_346) we use the more accurate \gaia\ position and for XID419 we use more accurate coordinates from \textit{HST}/WFC3, see appendix~\ref{app:coord}.}. 
For the uncertainties on these positions, we combined in quadrature the uncertainties from: (1) the \gaia\ coordinates ($<4$~mas), (2) the alignment of the H/K-band images with \gaia\ ($75-130$~mas), (3) the PSF of the VLT/VISTA or VLT/ISAAC K-band images (3-40~mas), (4) the 2D Gaussian fit to determine the position of sources in the VLT images (1-49~mas), and (5) the centroid position of the collapsed data-cubes ($\sim$25~mas).
This results in uncertainties on the astrometry in our SINFONI cubes of $0.03-0.14$'', corresponding to 1-3 pixels\footnote{The largest uncertainties are due to the lower resolution of the VLT/ISAAC K-band images for the CDF-S field.}.
 More details about the coordinate registration are provided in the appendix~\ref{app:coord}. The ALMA astrometry has an absolute accuracy of 2\% of the synthesized beam (ALMA Cycle 6 Technical Handbook\footnote{https://almascience.eso.org/documents-and-tools/cycle6/alma-technical-handbook}), which corresponds to $\sim 3-6$~mas for our observations and it is small enough compared to the other uncertainties.

\subsection{Spectral line fitting}
\label{sec:line_fitting}

We use the integrated spectra to derive  emission line flux ratios and investigate the main source of ionisation in our objects. For this purpose, we measure the fluxes of the \Hb, \OIII$\lambda \lambda$4959,5007, \Ha\ and the \NII$\lambda \lambda$6548,6583  emission lines.
The integrated SINFONI spectra of the \Ha+\NII\ and \Hb+\OIII\ spectral regions for our targets are shown in Figure~\ref{fig:Ha_OIII_images}.
The spectra are extracted from a circular aperture centred on the targets. The diameter of the aperture was defined to include at least $\sim95\%$ of the emission.

Up to three Gaussian components were used to fit each emission line. These are a narrow (systemic) component, a broad (outflow) component and a broad line region (BLR) component (only for the Hydrogen lines). The components used for each target are summarised in Table~\ref{tab:SINFONI}. 

For the Type 1 AGN, the \Hb+\OIII\ and \Ha+\NII\ spectral regions are fitted separately. The modelling of the \Hb+\OIII\ spectral region is described in details in \citet{Kakkad2020}. 
 The width and relative position of the systemic and outflow components of \Hb\ and \OIII\ are tied together. The broad outflow component is used in the fit only if its addition decreases the reduced $\chi^2$ value of the overall model. Specifically, for the three Type 1 in our sample: this is not required for cid\_1205  but is required for \XN\ and cid\_346\footnote{We note that in \citet{Kakkad2020} the  \Hb+\OIII\ emission lines were fitted with one component, while two components are required for the fit of the  \Ha+\NII\ emission lines (based on the reduced $\chi^2$). In order to have consistent measurements for the four emission lines, we decide to use the fit with two components also for the \Hb+\OIII\ emission lines.}.
An additional Gaussian component is used to model the broad line region (BLR) component of \Hb\ for all Type 1 sources. For the \Ha+\NII\ spectral region, we use the same number of Gaussian components as for \OIII, and an additional Gaussian component to model the BLR component of \Ha. 
Additionally, we apply the following constraints \citep[following][]{Vietri2020}: i) the widths of the broad components of \Ha\ and \NII\ are tied to the width of the broad component of \OIII, allowing variation in the parameters within the measurement errors; ii) the relative centroid locations of the narrow and broad components of \Ha\ and \NII\ are tied to the ones measured with \OIII; iii) the maximum width allowed for the narrow \Ha\ component is set by the width of the narrow \OIII\ component plus its measurement error.  Finally, empirical FeII templates from the literature are used to model the FeII emission \citep{Boroson1992, Veron-Cetty2004, Tsuzuki2006}.

For the Type 2 sources, the fitting procedure is similar to the one used for the Type 1 AGN.  The results of the emission line fits for all the Type 2 objects in the SUPER sample will be presented in M.~Perna et al. (2021, in prep.). 
The  \Ha\ and \Hb\ lines, and the \OIII, \NII\ and \SII\ doublets were modelled using Gaussian profiles. The wavelength separation between emission lines was tied according to atomic physics, and the FWHM was constrained to be the same for all the emission lines. The relative flux of the two \NII\ and \OIII\ lines was fixed to 2.99 \citep{Osterbrock2006}. 
Each spectral fit was performed with one and two Gaussian components (each centred at a given velocity and with a given FWHM). 
 The  Bayesian information criterion \citep[BIC,][] {Schwarz1978} was used to decide whether a second component was needed for the fit \citep[see e.g. ][]{Perna2019}.
All  targets require a second component except for XID419.
 We note that we use our emission-line profile fits to characterise the integrated emission-line profiles; however, in this study we do not attempt to map individual \Ha\ components associated with star formation, AGN narrow line regions and/or outflows (see Section~\ref{sec:SINFONI_maps}).

The H and K-band (continuum- and BLR-subtracted) spectra with the corresponding emission line fits are shown in Figure~\ref{fig:Ha_OIII_images}. The total spectra of the Type 1 targets (before subtracting the continuum and BLR components) are shown in Appendix~\ref{app:type1_spectra}. 
A summary of the line components used for each target is provided in Table~\ref{tab:SINFONI}.

For both Type 1 and Type 2 sources, the uncertainties on the line fluxes were estimated using a Monte Carlo approach \citep[for  details, see][]{Kakkad2020}. We define a line as detected if it has a S/N=$F/F_{err}\geq 3$. If a line is not detected, we calculate a conservative upper limit equal to the flux of a Gaussian emission line with amplitude five times the noise level (measured in a line-free region of the spectrum) and with the same width as the other detected emission lines. The $5\sigma$ upper limit corresponds to a `false negative' fraction of 2$\%$, which is the probability that a source with `true' flux higher than this upper limit is not detected.

We note that we did not correct the line fluxes for obscuration.  In most of the cases we are not able to calculate the Balmer decrement, because the \Hb\ line is not detected. 
Nevertheless, we do not consider this a major limitation since we are mainly interested in the ratio of emission lines close in wavelength, which therefore are negligibly affected by differential obscuration.

\subsection{\Ha\ and [OIII] maps}
\label{sec:SINFONI_maps}

In this section we describe the methods used to derive the maps of the narrow \Ha\ (i.e. without Broad Line Region) and \OIII\ emission. In this paper, we consider  a $\mathrm{w_{80}}$ (i.e. the width containing 80\% of the line emission) value $> 600$~\kms\ as a signature of an AGN-driven outflow \citep{Kakkad2020}. 
Following this definition, we consider as non-outflowing emission a velocity range of 600~\kms, centred on the peak of the \OIII\  (or \Ha) emission line. 
For the two targets with a two component fit to the \OIII\ emission line, this definition covers the bulk of the narrower component and excludes the broader, blue-shifted component. 
We note that it is possible that the outflow component is contributing also in the central velocity channel ($v =[-300, 300]$~\kms), especially in the sources with a strong outflow component.
We chose this definition because we prefer to adopt a consistent definition across all targets and for both emission lines (\Ha\ and \OIII). Moreover, a definition which is as independent as possible from the modelling of the emission line profile is highly preferred. However, it is inevitable that the resulting \Ha\ maps of the Type 1s will depend on the method used to subtract the BLR component.

To create the maps, we use the continuum-subtracted data-cubes. For the \Ha\ maps, we also subtract the BLR component (for the Type 1) and the \NII\ components. 
 For \XN, we note that the broad component of \Hb\ is much more prominent than the broad component detected in \OIII\ (see Figure~\ref{fig:spectra_Type1} in the appendix). We suspect that this component is a residual component belonging to the BLR. This argument is supported by the fact that the spatial distribution of the \Hb\ broad component is similar to the spatial location of the BLR. 
 Therefore, for this object we consider the broad \Hb\ and \Ha\ components as part of the BLR and we subtract them from the data-cube. 
 We use the peak of the modelled  \OIII$\lambda 5007$ line profile to derive the systemic redshift and set the zero velocity (see Sec.~\ref{sec:line_fitting}). 
We create the \OIII\ and \Ha\ maps by collapsing the spectra over the selected velocity channels for each spaxel. 

We note that our approach to map the \Ha\ emission line is different from the method applied in other resolved studies of the \Ha\ emission of Type 1 AGN \citep[e.g. ][]{Carniani2016}, where the residuals of the fit (of their AGN-dominated components) were used to trace \Ha\ emission due to star formation. After subtracting the emission line fitting model from our data, we do not find any significant residual at the position of \Ha, both in the maps and  in the integrated spectra (see residual panels in Figure~\ref{fig:Ha_OIII_images}). Thus, we cannot use this method to map the \Ha\ emission from star formation. We defer to a future work a detailed investigation of the spatially-resolved narrow \Ha\ emission distribution and kinematics for the few Type 1 sources where the S/N are sufficiently high for such analyses. 
In Section~\ref{sec:BPT} we  use emission line diagnostics to investigate the main mechanism responsible for the ionisation of \Ha\ (star formation or AGN) in our objects.
 However, due to the limited S/N of our observations, we cannot perform a BPT analysis in a spatially resolved manner.

Figure~\ref{fig:Ha_OIII_images} shows the contours of the \Ha\ and \OIII\ emission in the central 600~\kms\ channels. 
For every map, we measure the size and the centroid (`peak position') of the emission by fitting a 2D Gaussian using the python package \curvefit. The \Ha\ and \OIII\ sizes are reported in Table~\ref{tab:SINFONI}. We use these measurements in Sec.~\ref{sec:ALMA_SINFONI_comp}, where we compare the sizes and positions of the \Ha\ and \OIII\ emissions with the location of the FIR emission.
 The uncertainties on the  positions are calculated by adding in quadrature the uncertainties on the coordinates registration (see Section~\ref{sec:coord}) and the uncertainties due to the 2D Gaussian fit (mean uncertainty $\sim$ 2 mas). The uncertainties on the position are in the range 0.04-0.15''.

\begin{figure*}
\includegraphics[width=0.215\textwidth]{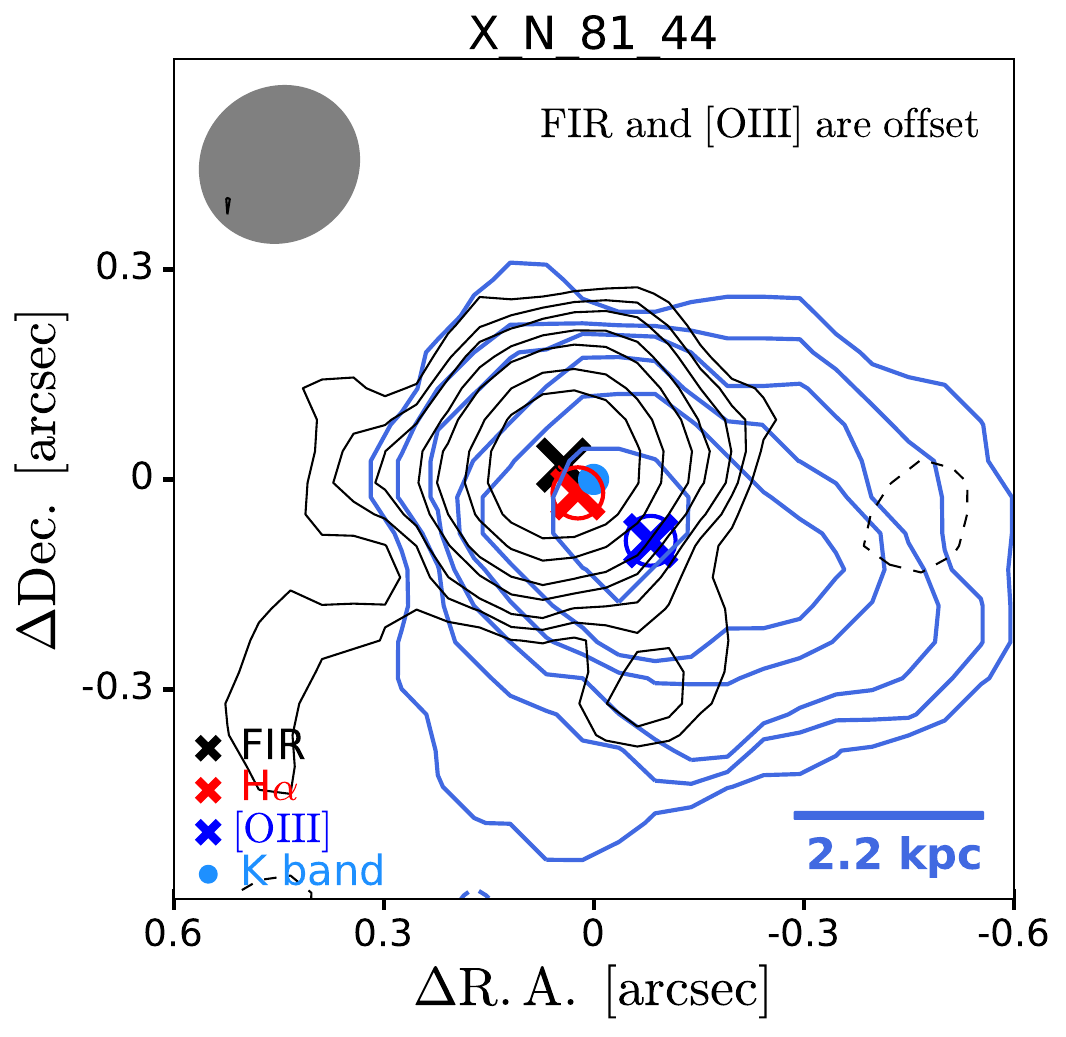}
\includegraphics[width=0.27\textwidth]{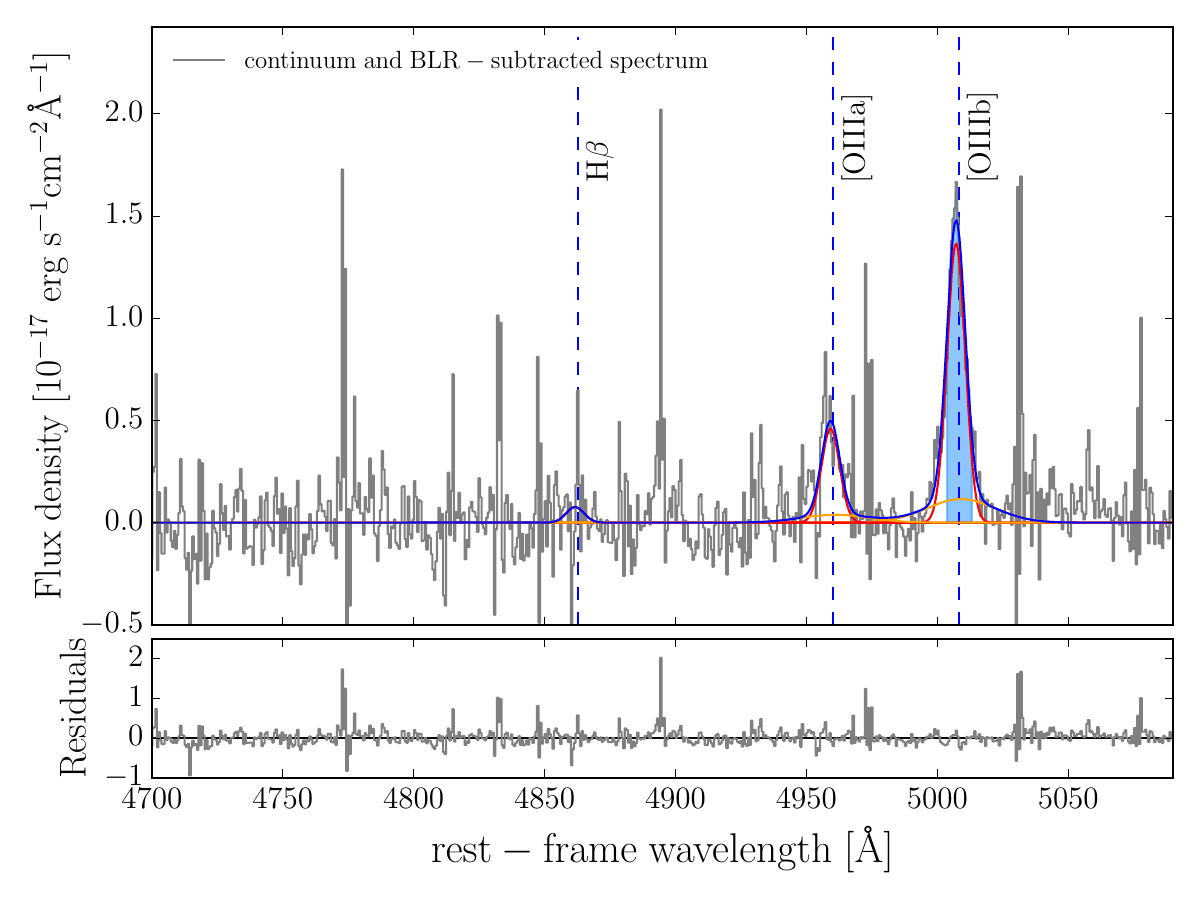}
\includegraphics[width=0.215\textwidth]{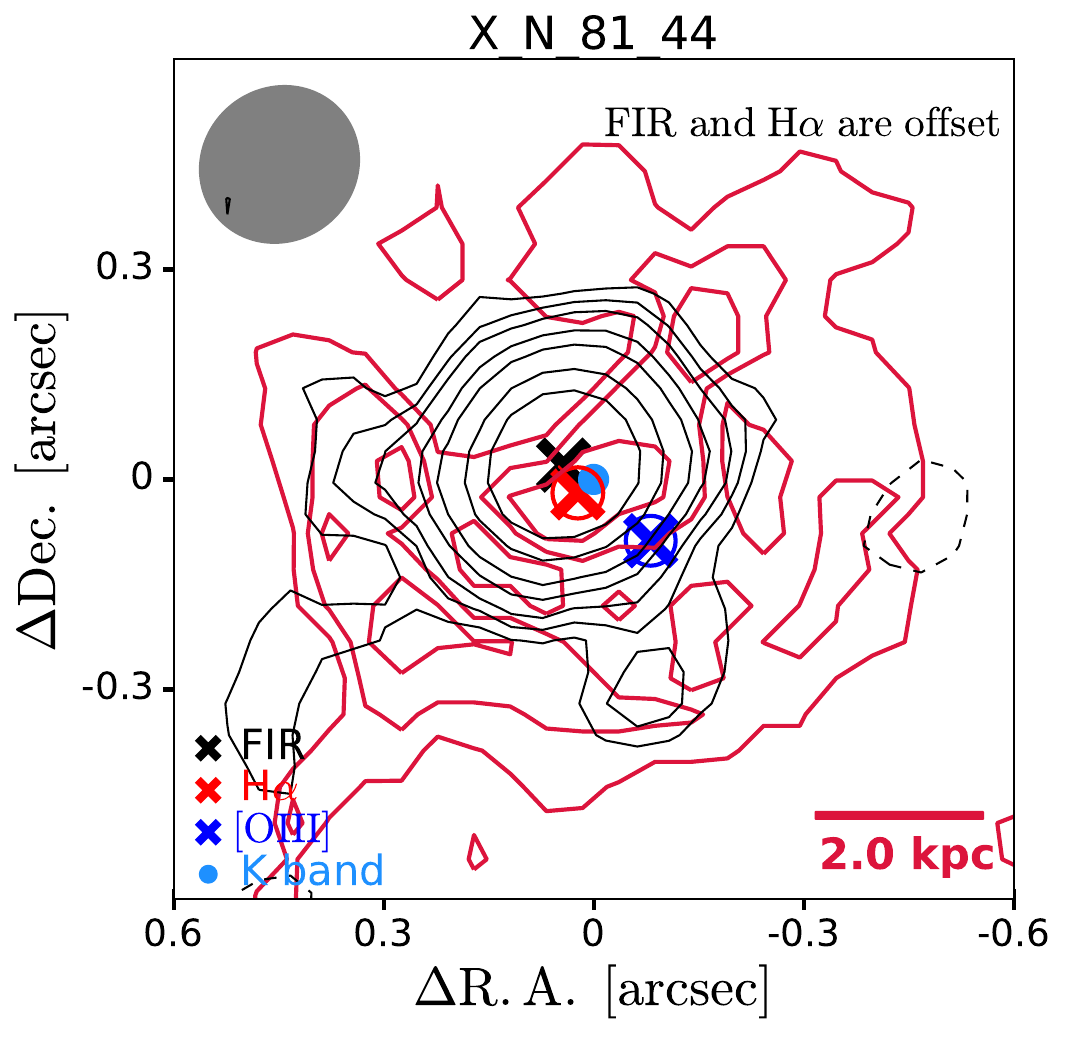}
\includegraphics[width=0.27\textwidth]{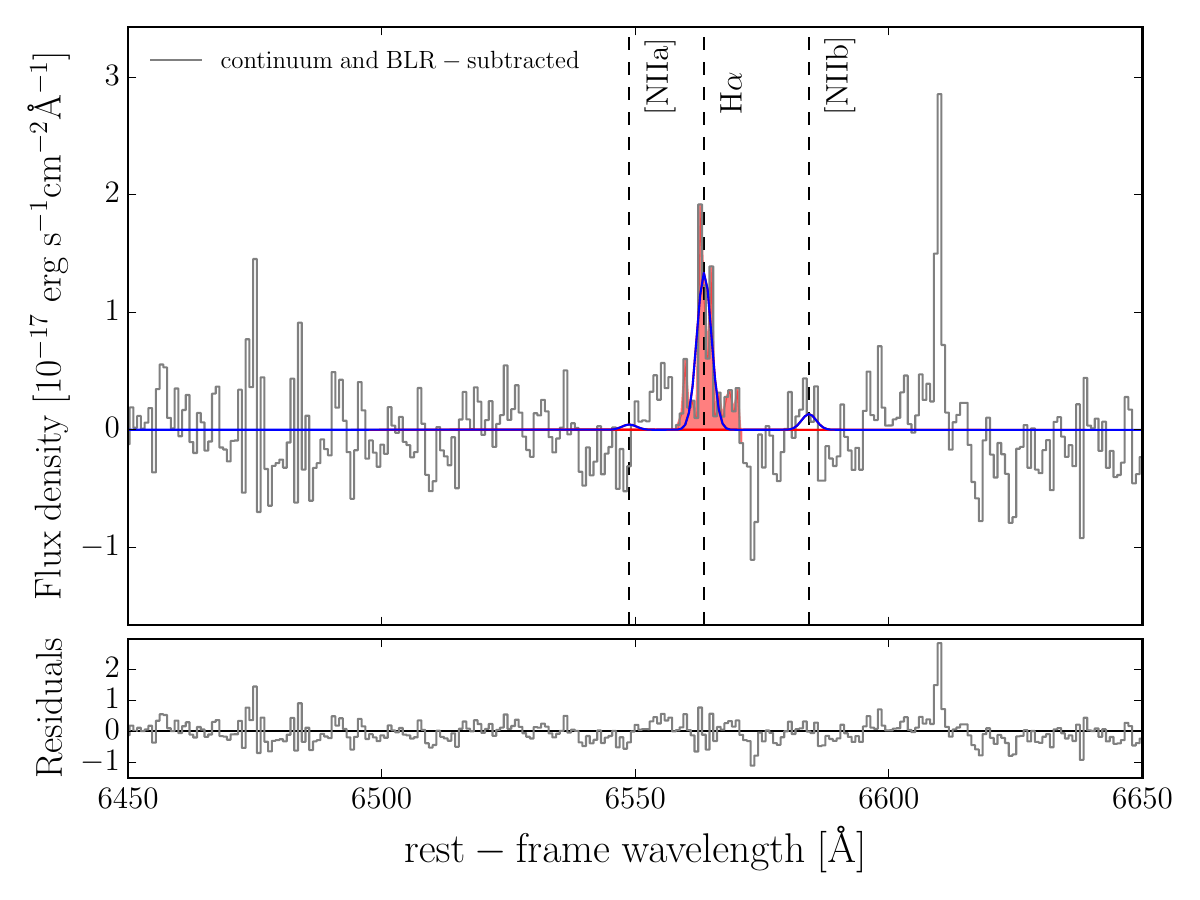}

\includegraphics[width=0.215\textwidth]{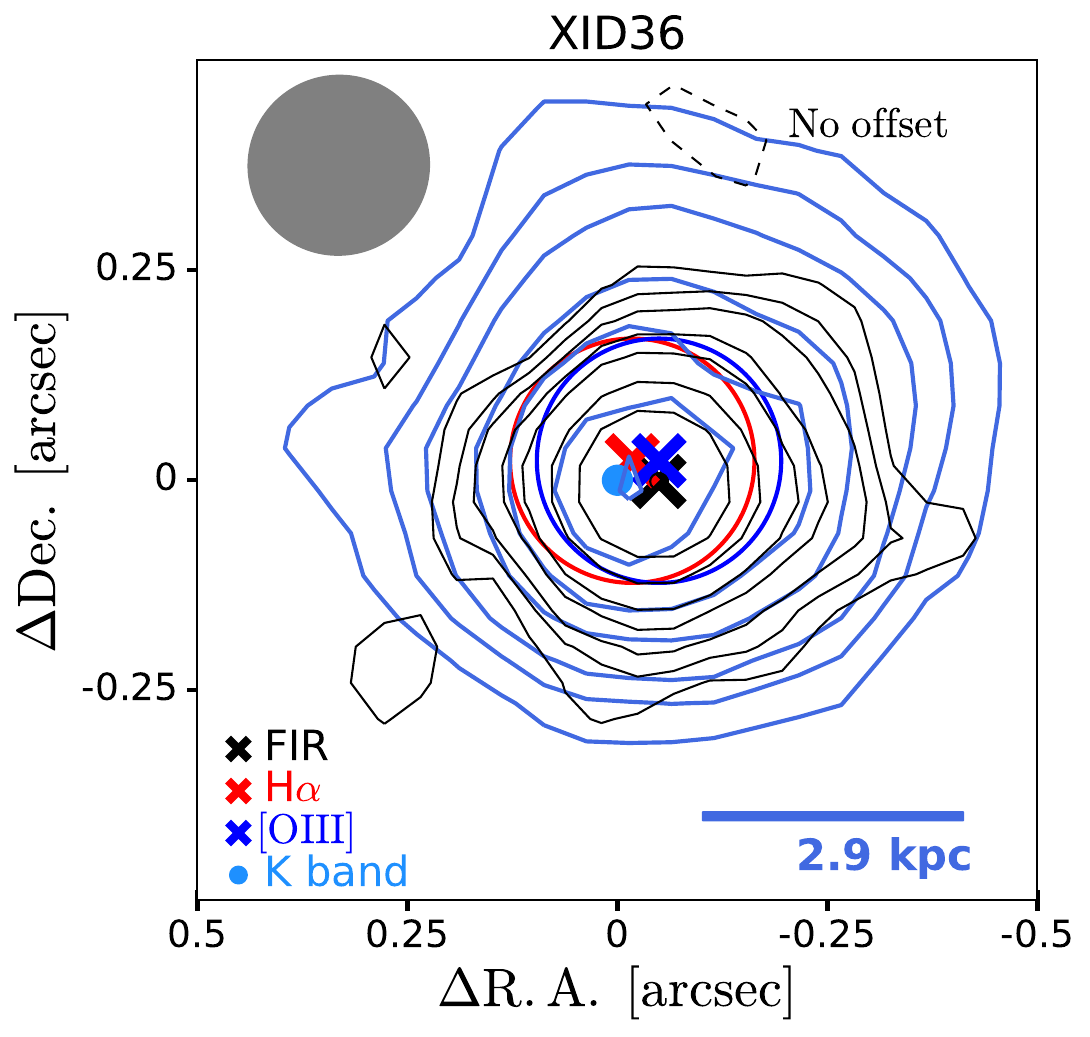}
\includegraphics[width=0.27\textwidth]{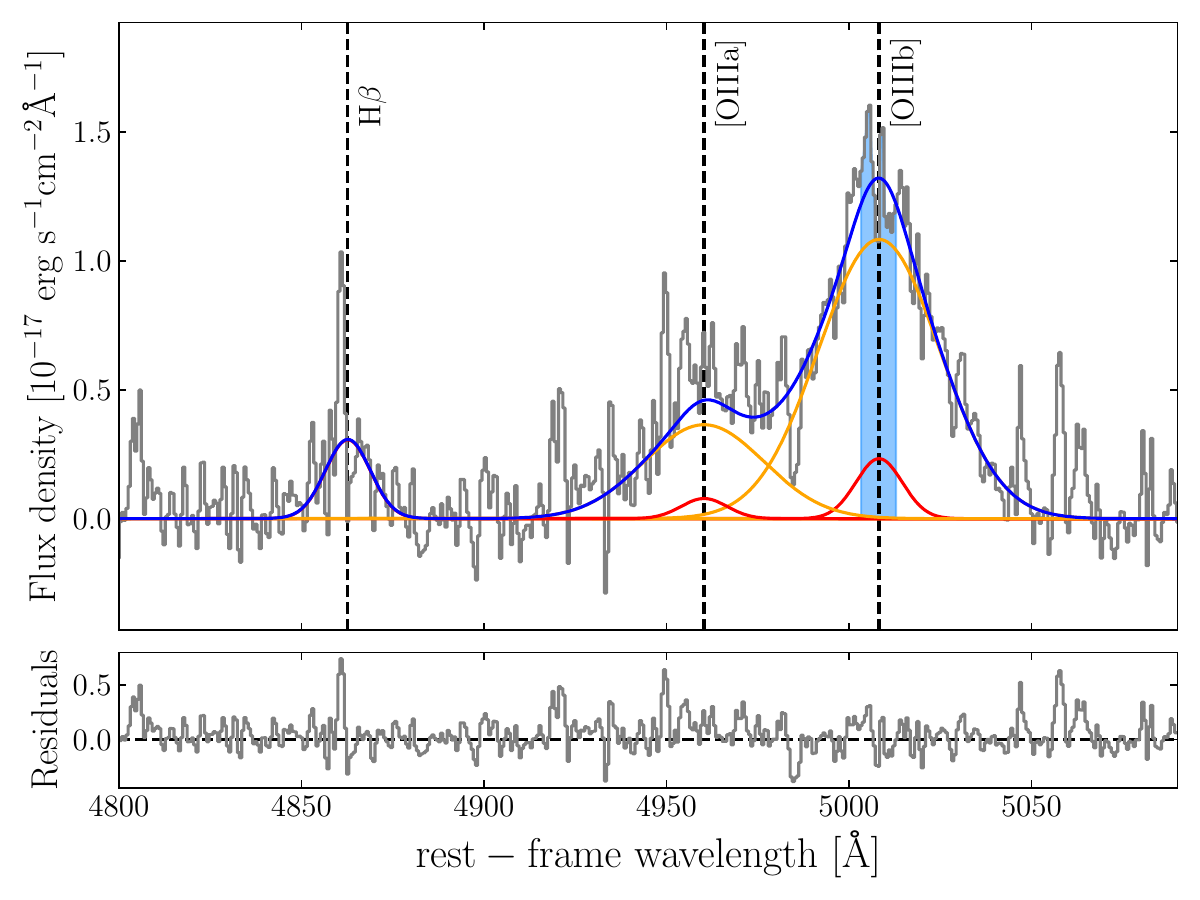}
\includegraphics[width=0.215\textwidth]{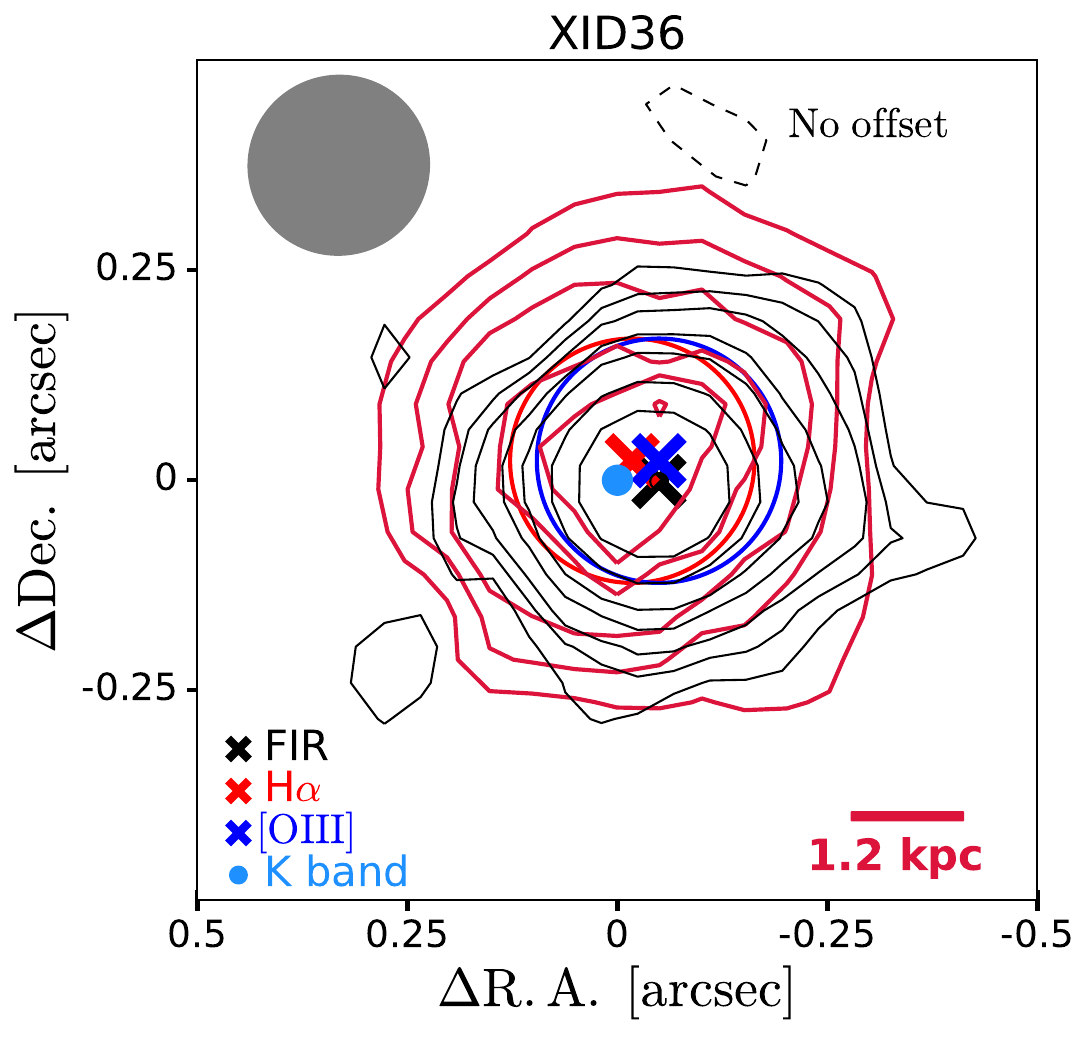}
\includegraphics[width=0.27\textwidth]{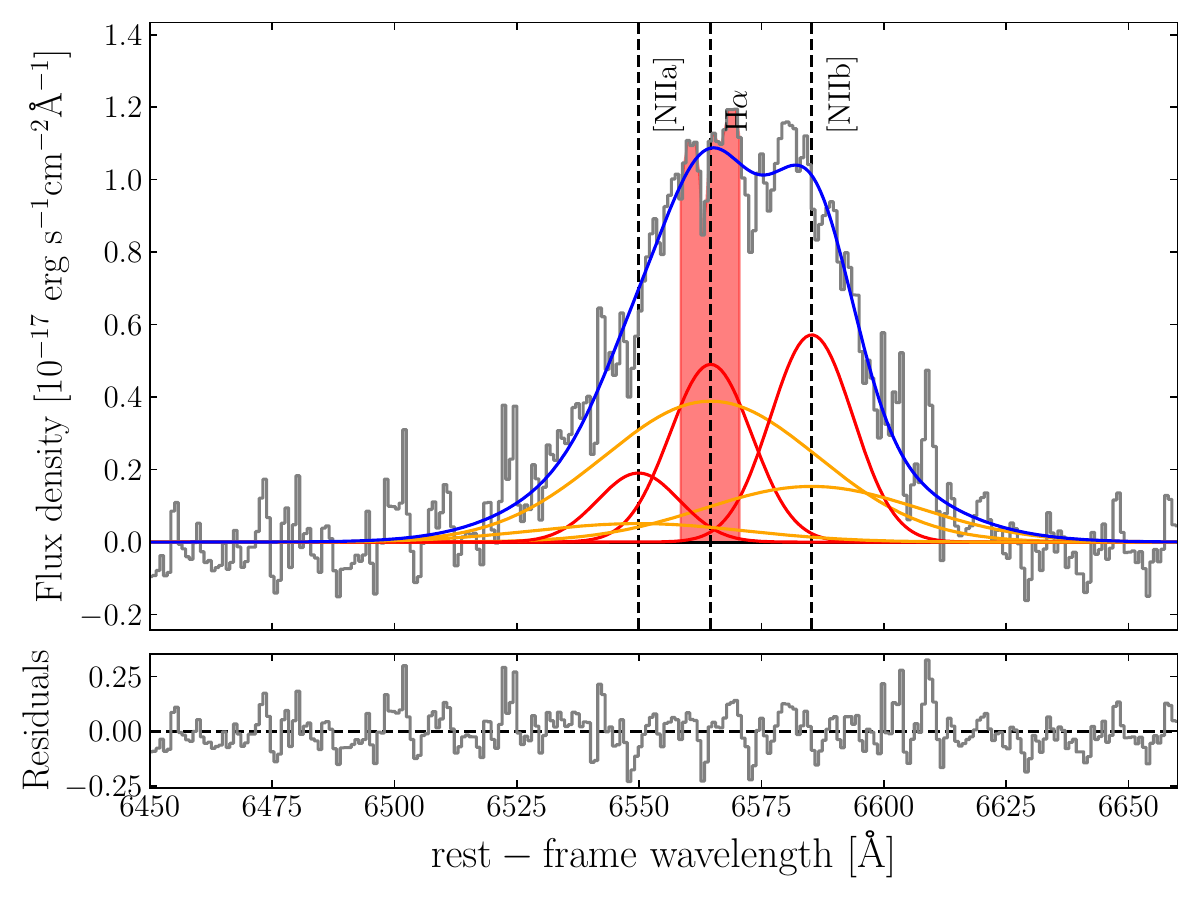}

\includegraphics[width=0.215\textwidth]{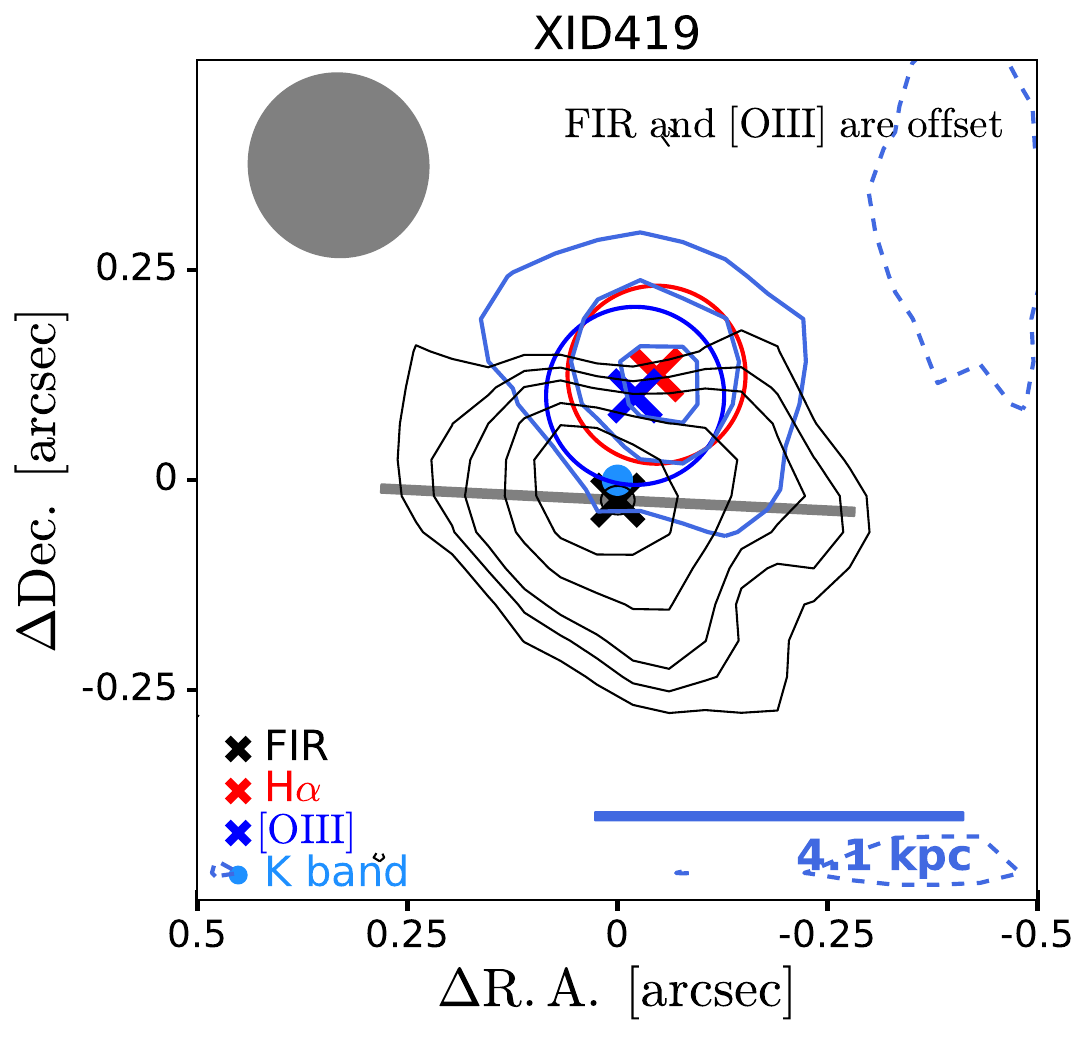}
\includegraphics[width=0.27\textwidth]{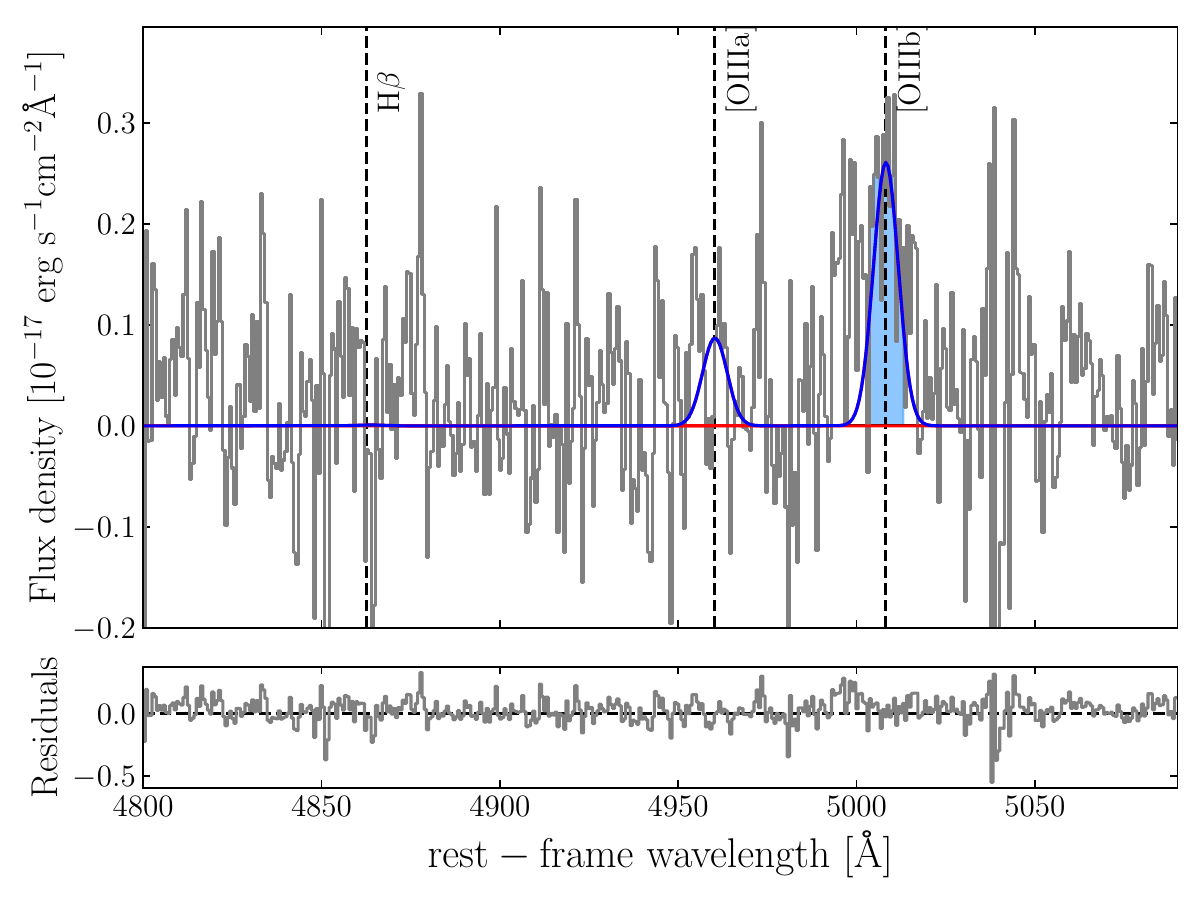}
\includegraphics[width=0.215\textwidth]{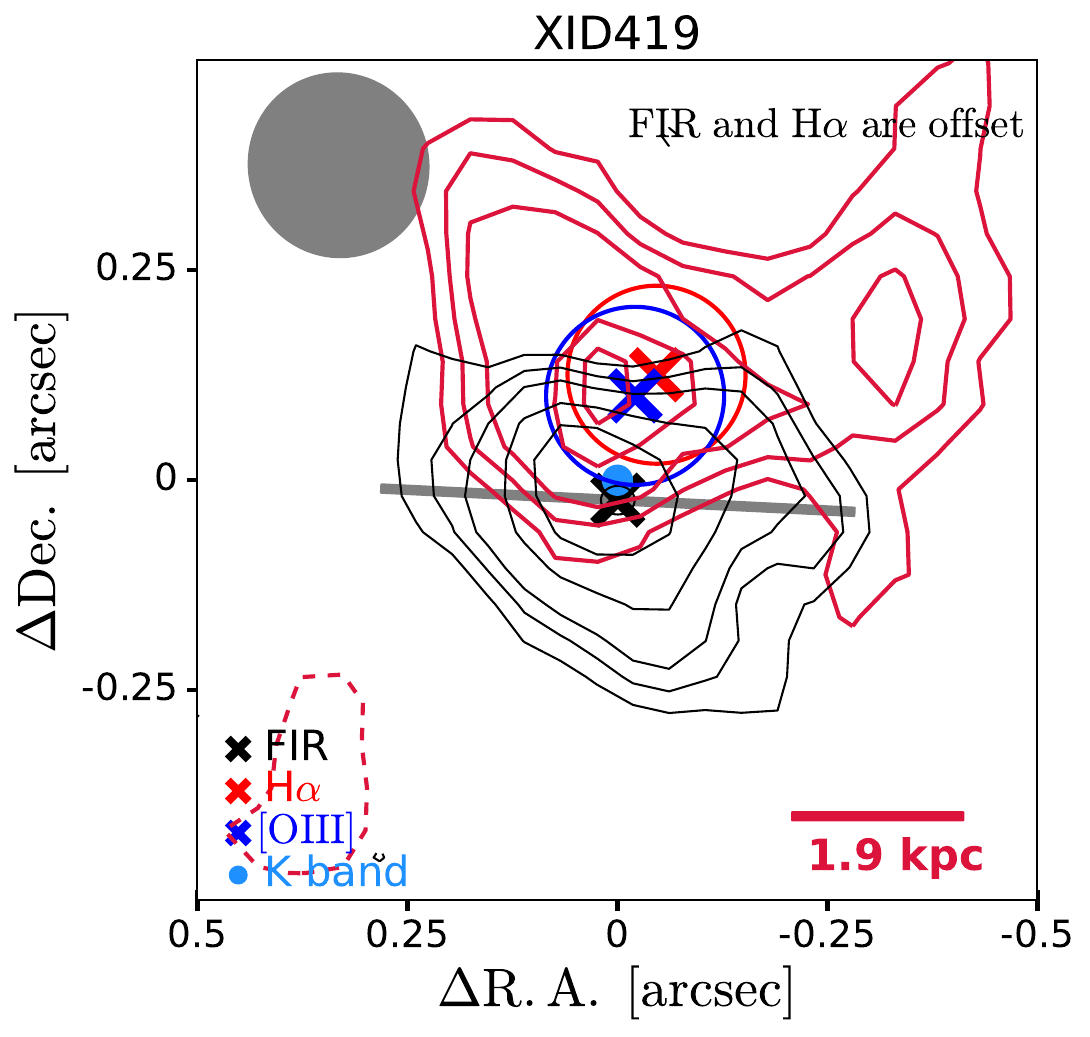}
\includegraphics[width=0.27\textwidth]{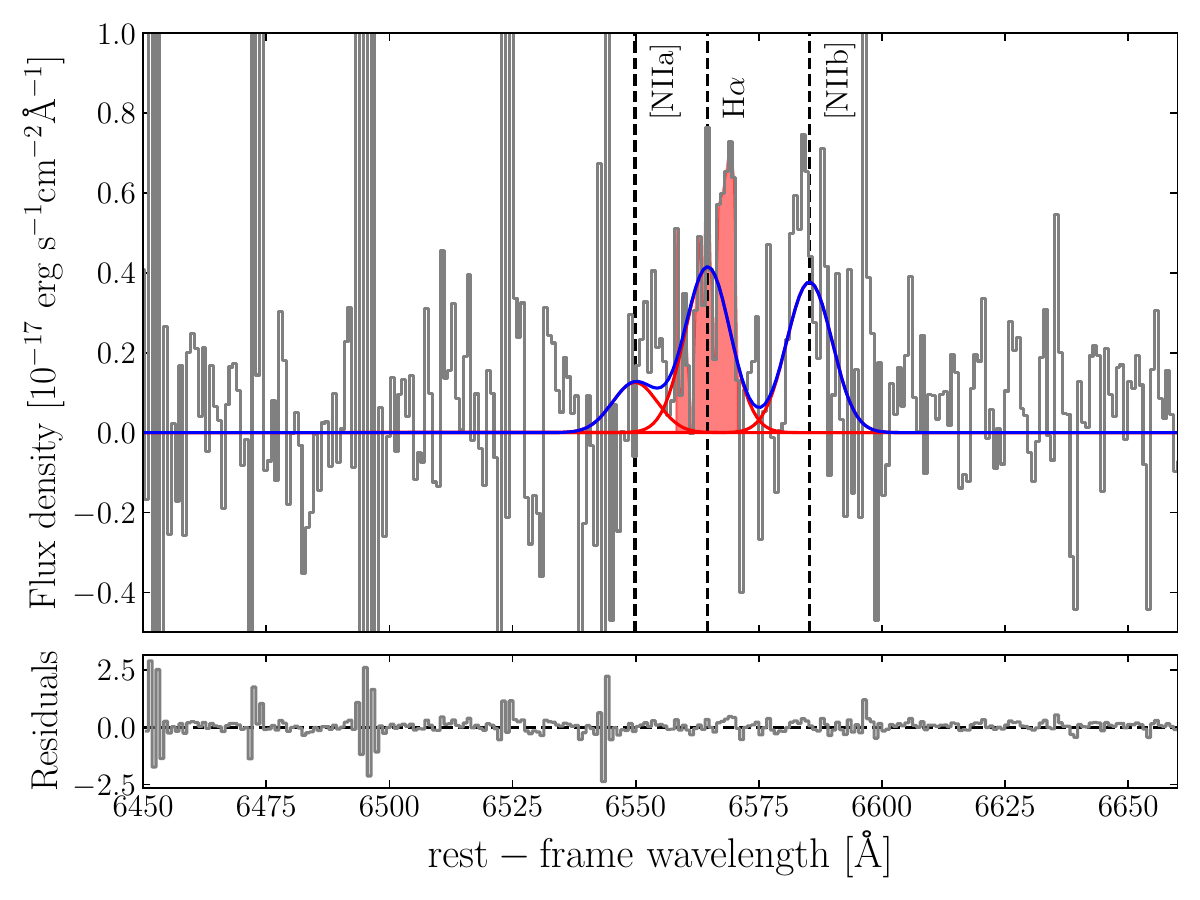}

\includegraphics[width=0.215\textwidth]{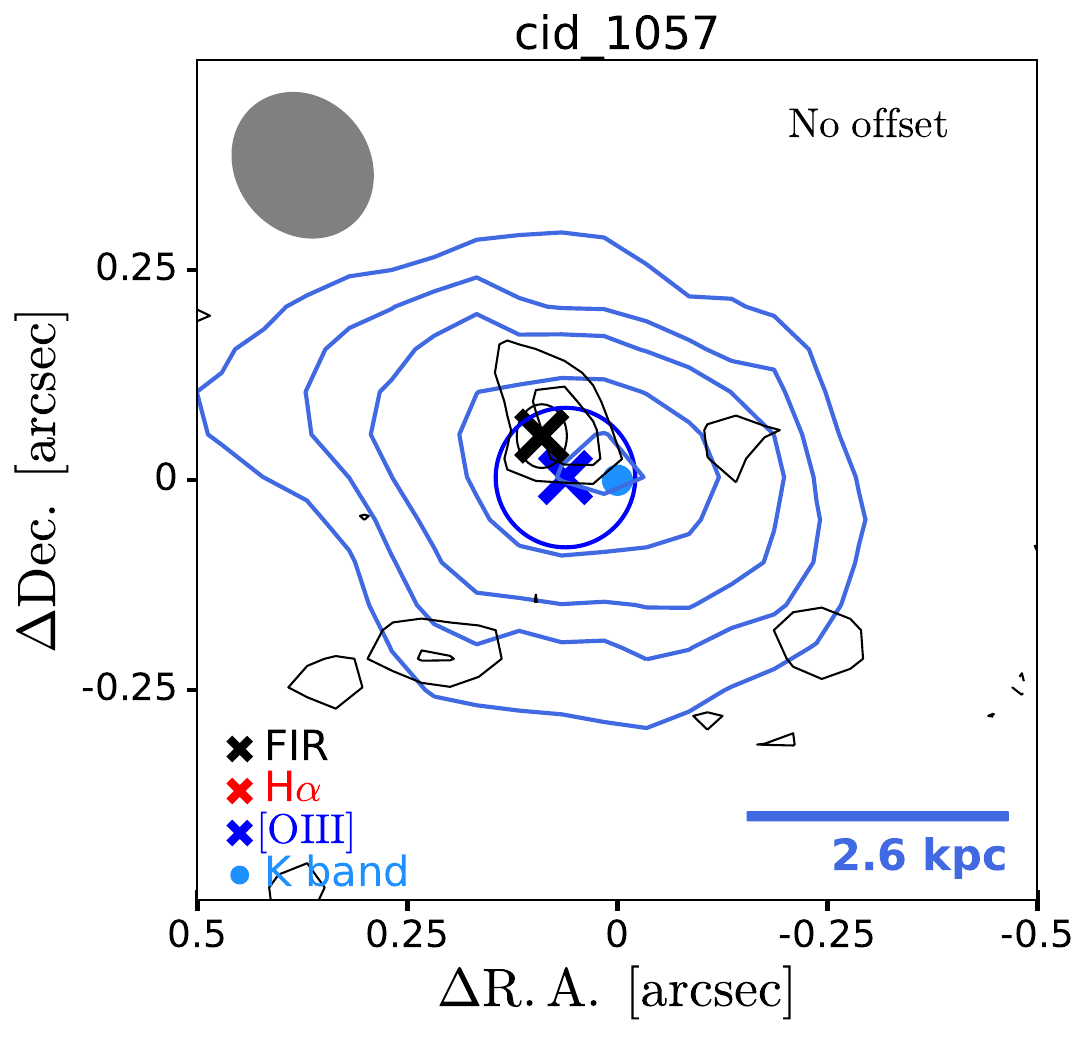}
\includegraphics[width=0.27\textwidth]{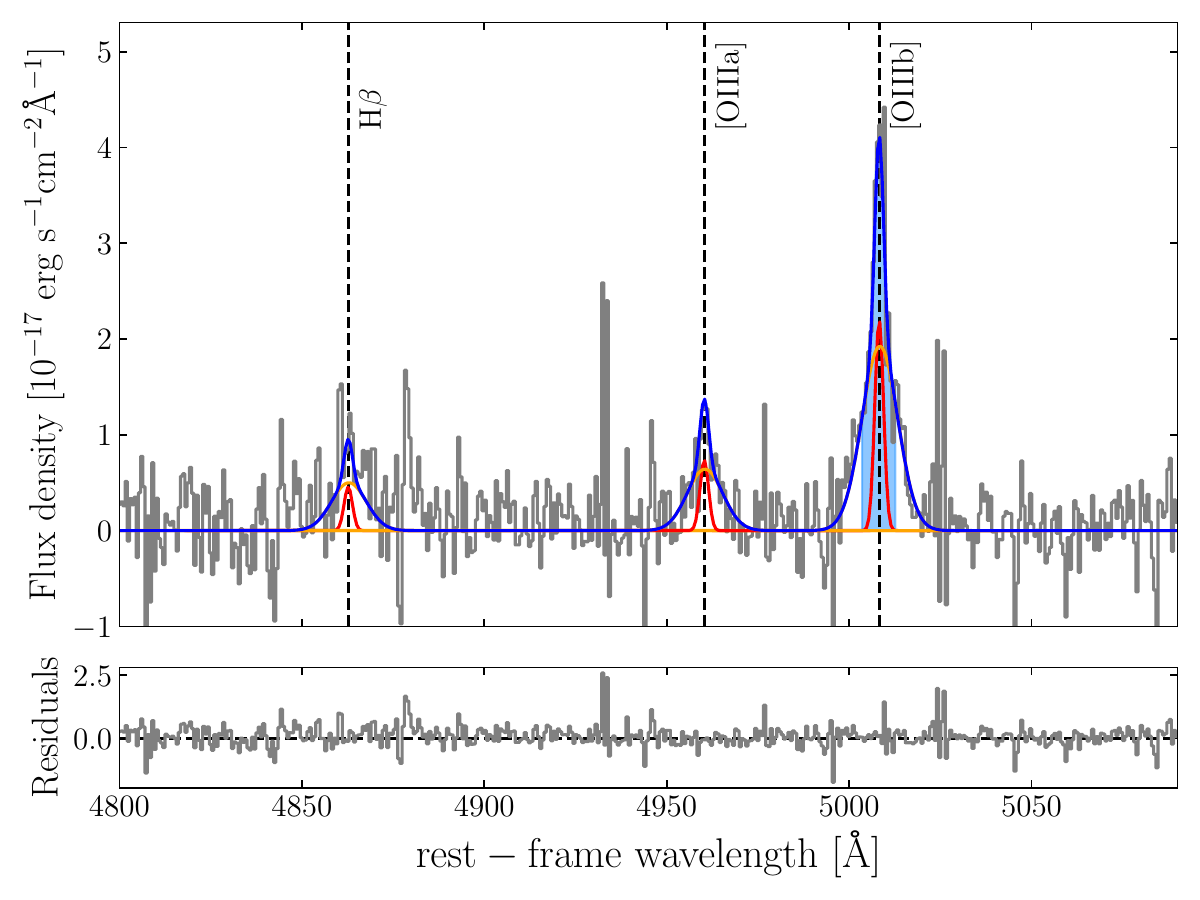}

\includegraphics[width=0.215\textwidth]{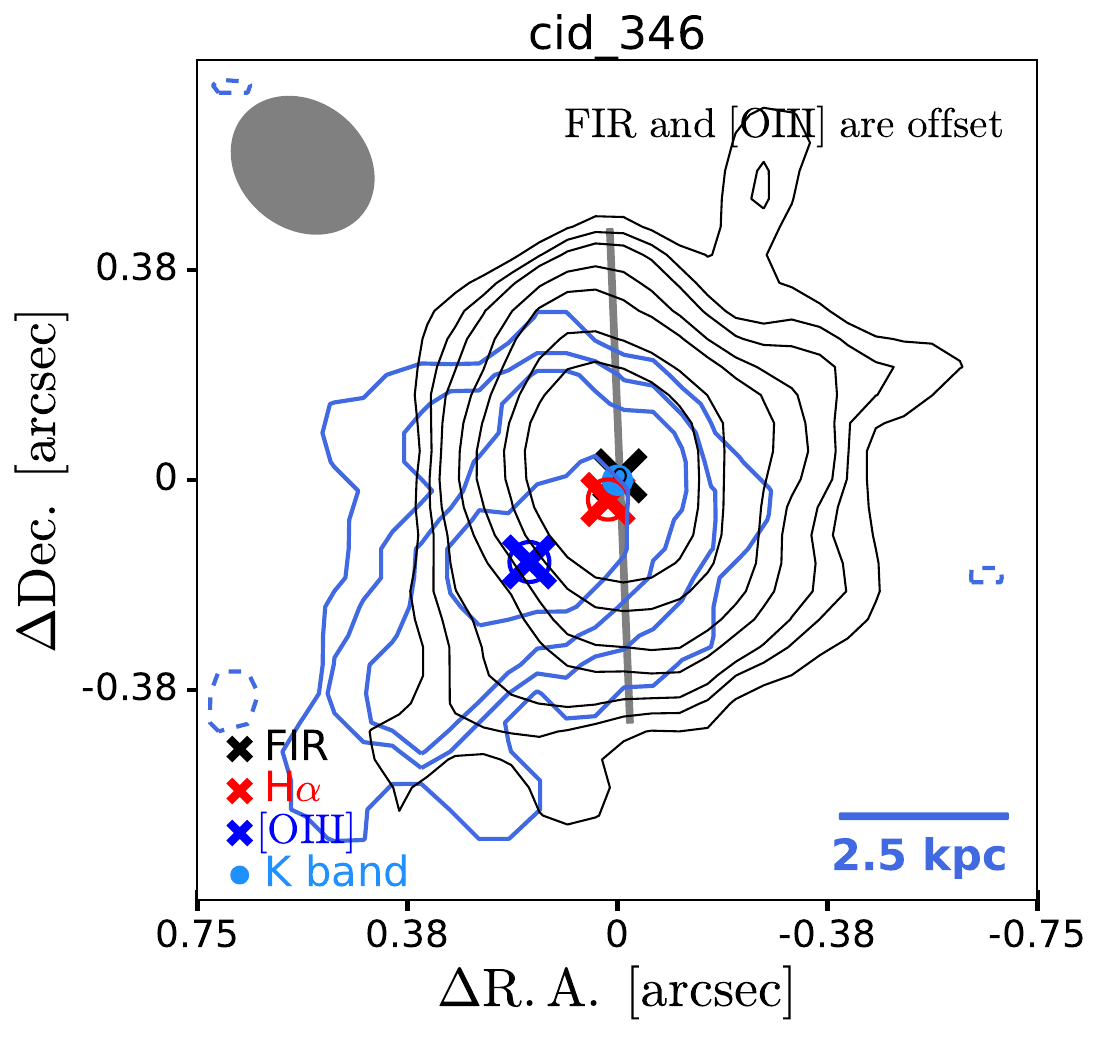}
\includegraphics[width=0.27\textwidth]{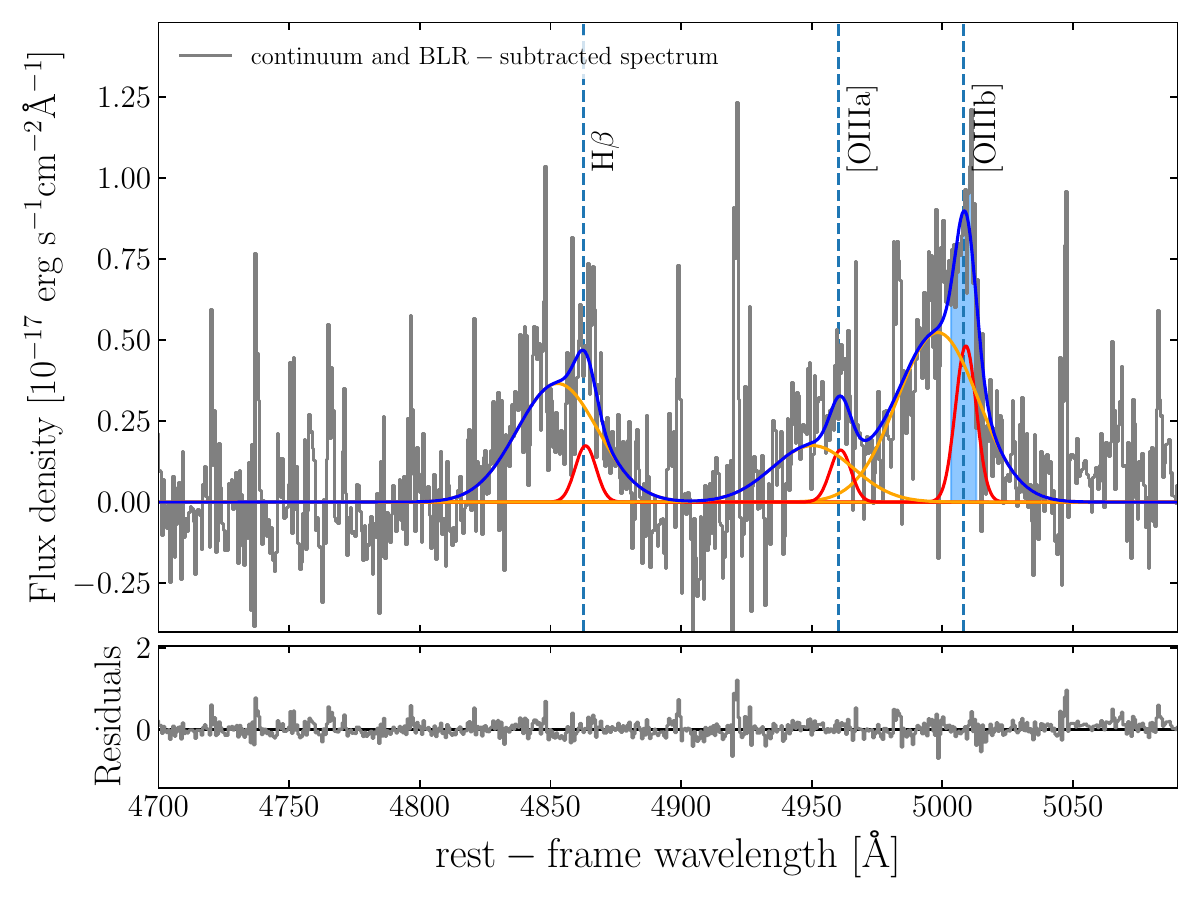}
\includegraphics[width=0.215\textwidth]{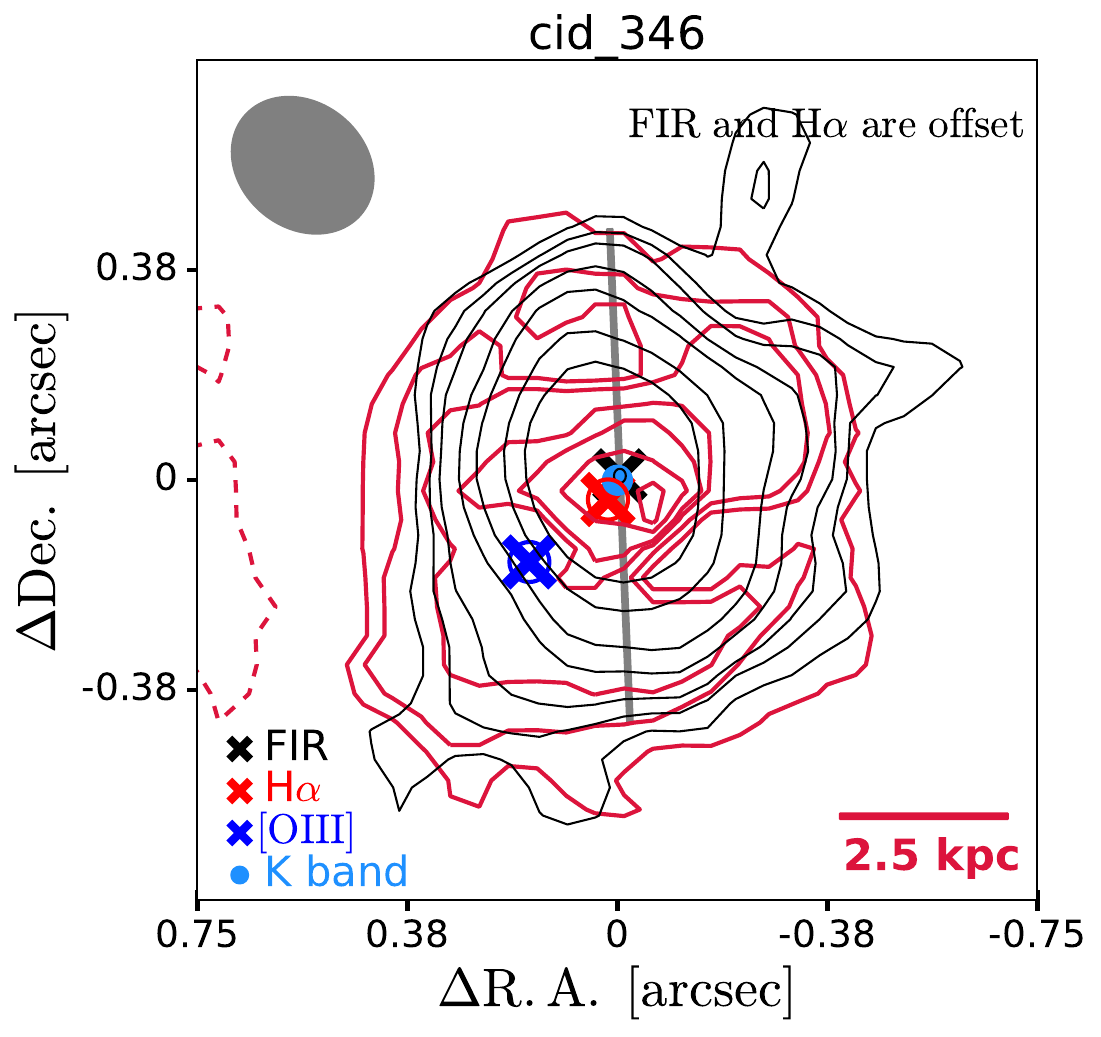}
\includegraphics[width=0.27\textwidth]{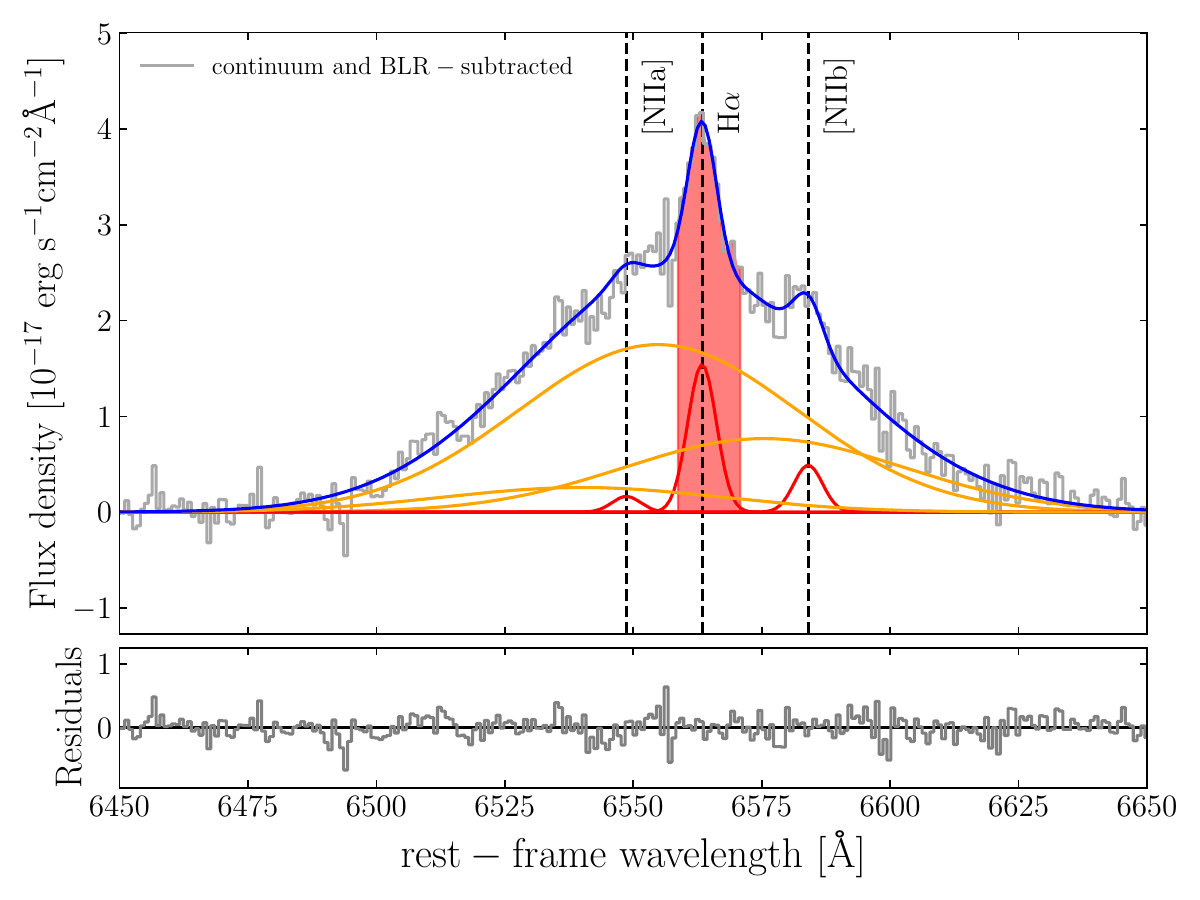}

\caption{\OIII\ and \Ha\ continuum-subtracted maps (where available), with the corresponding spectra. For the Type 1 AGN (\XN, cid\_346 and cid\_1205),the broad line region component is also subtracted.
The  \OIII\ and \Ha\ maps were created by integrating the spectrum over the [-300, 300] \kms\ velocity range, with respect to the centre of the line (see shaded regions on spectra). 
The FIR (black), \OIII\ (lightblue), and \Ha\ (red)  emission contours are shown at the 2, 3, 4, 6, 8, 12, and 16$\sigma$ levels. 
 Dashed lines indicate the negative -2$\sigma$ contours.  The crosses show the position of the centroid of the FIR (black), \OIII\ (blue) and \Ha\ (red) emission with the respective uncertainties (circles). 
 The lightblue point shows the centroid position of the optical continuum (i.e. K-band observed-frame). In all maps, the grey ellipses show the size of the ALMA  beam, while the scale-bars give the size of the PSF of the line emission maps. The grey bar shows the position angle along the major axis of the FIR emission, when it can be reliably determined (see Section~\ref{sec:FIR_size_morpho}).
\textit{Column~1:} 
Contours of the \OIII\ emission  (lightblue) and FIR (rest-frame) 260~\micron\ continuum emission (black).
\textit{Column~2:} Continuum-subtracted (integrated)
 \OIII +\Hb\ spectra. The blue curve shows the total fit  to the H$\beta$, \OIII$\lambda$4959 and \OIII$\lambda$5007 lines. The magenta and orange curves show the narrow and broad components.
 \textit{Column~3:} Contours of the narrow \Ha\ flux map (red) FIR (rest-frame) 260~\micron\ continuum emission (black). 
 \textit{Column~4:} Same as column 2, but for the
 continuum-subtracted (integrated) \Ha +\NII\ spectra. The blue curve shows the total fit to the H$\alpha$, \NII$\lambda$6548 and \NII$\lambda$6584 lines. 
 }
\label{fig:Ha_OIII_images}
\end{figure*}

\begin{figure*}\ContinuedFloat
\centering

\includegraphics[width=0.215\textwidth]{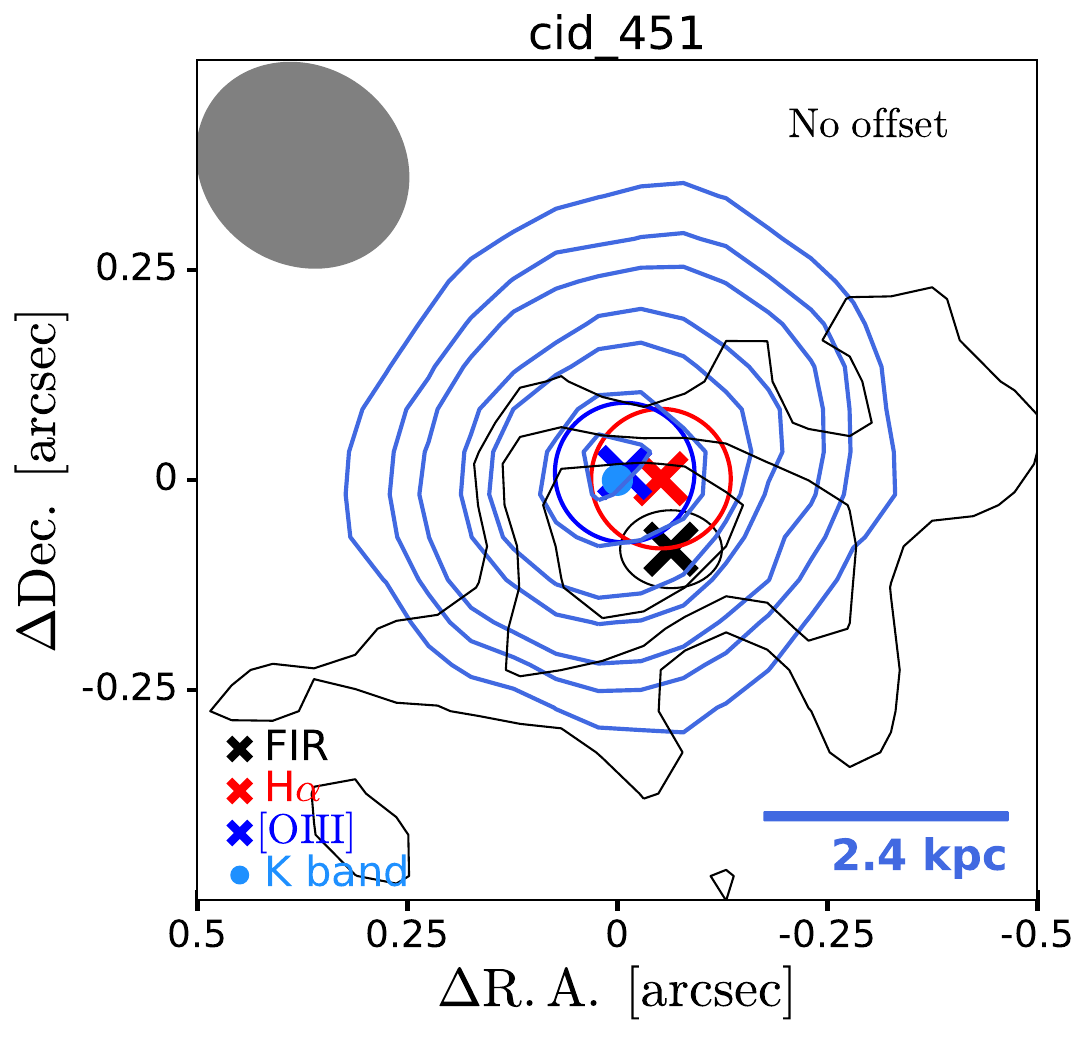}
\includegraphics[width=0.27\textwidth]{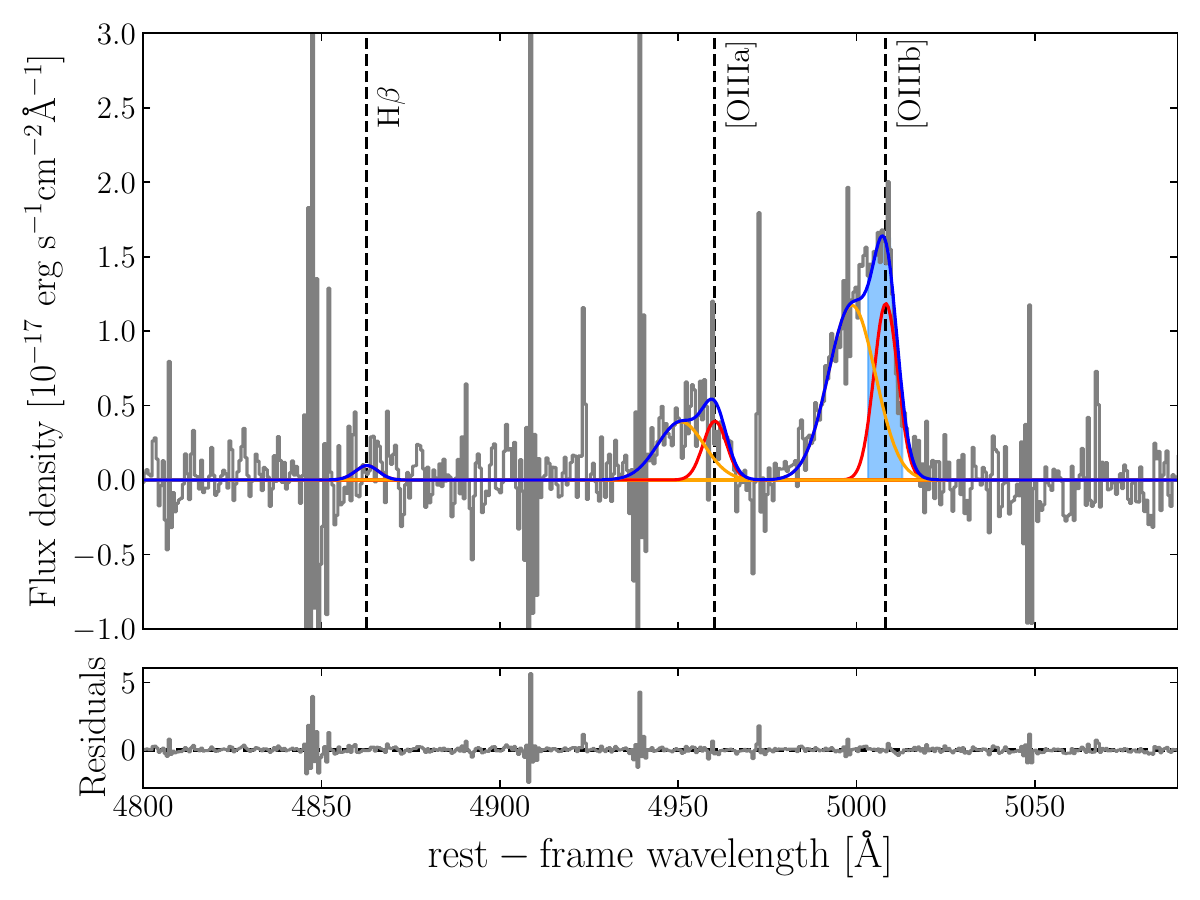}
\includegraphics[width=0.215\textwidth]{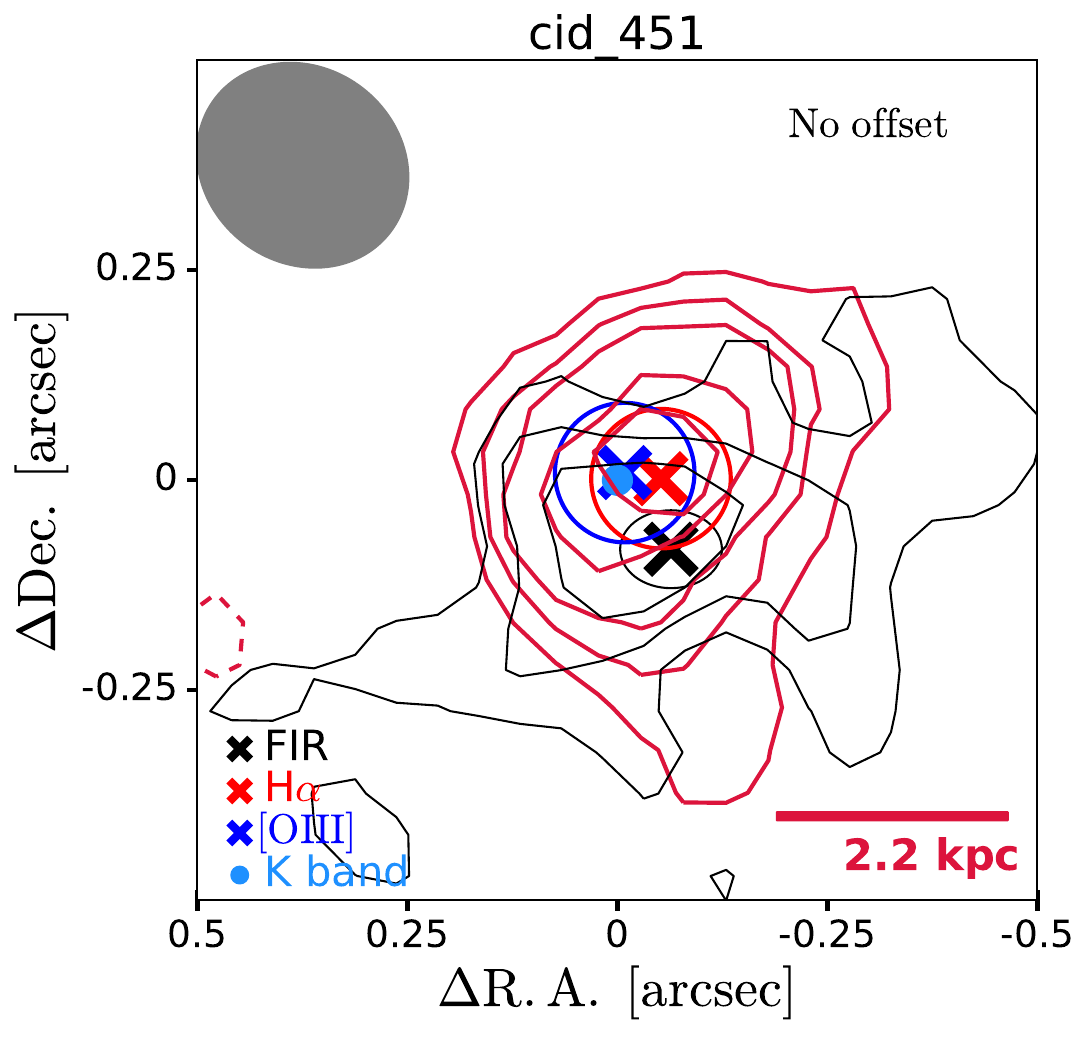}
\includegraphics[width=0.27\textwidth]{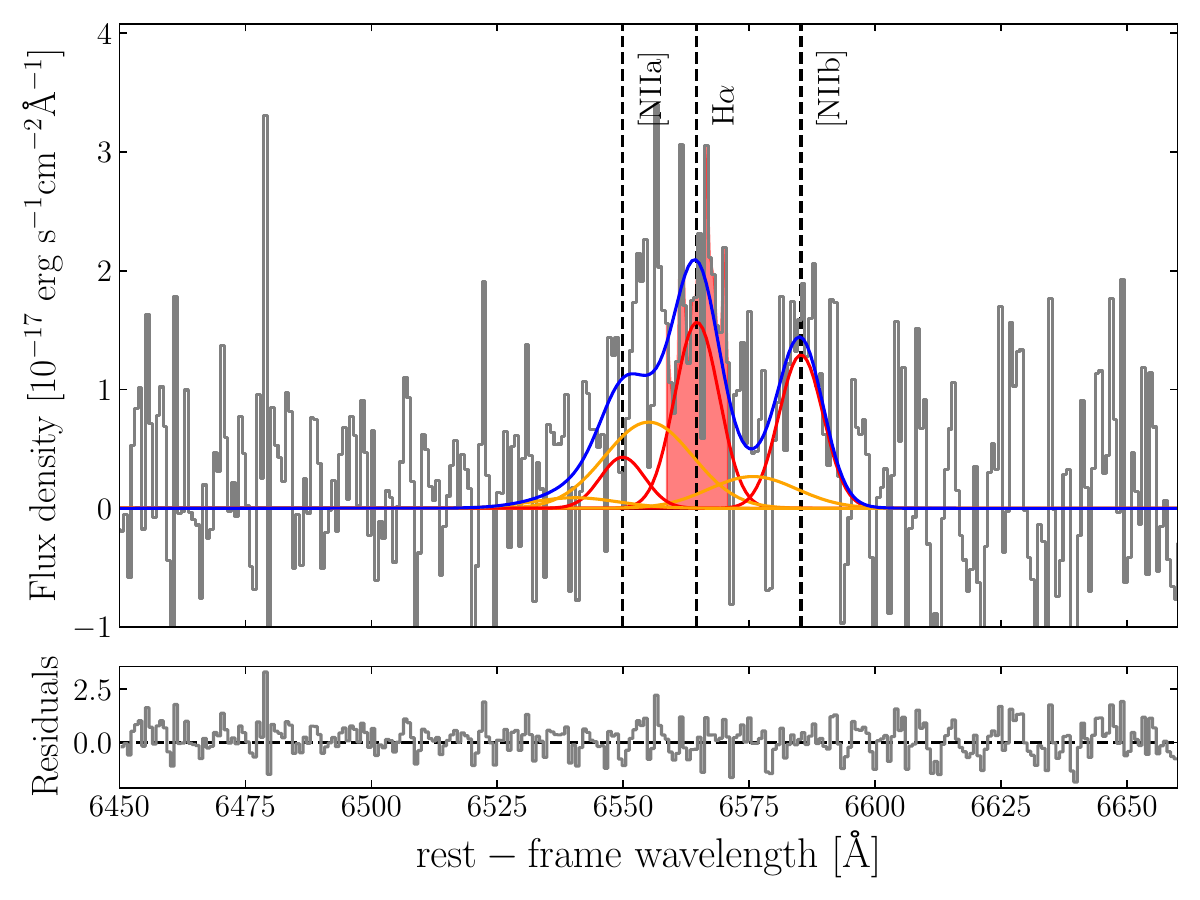}

\includegraphics[width=0.215\textwidth]{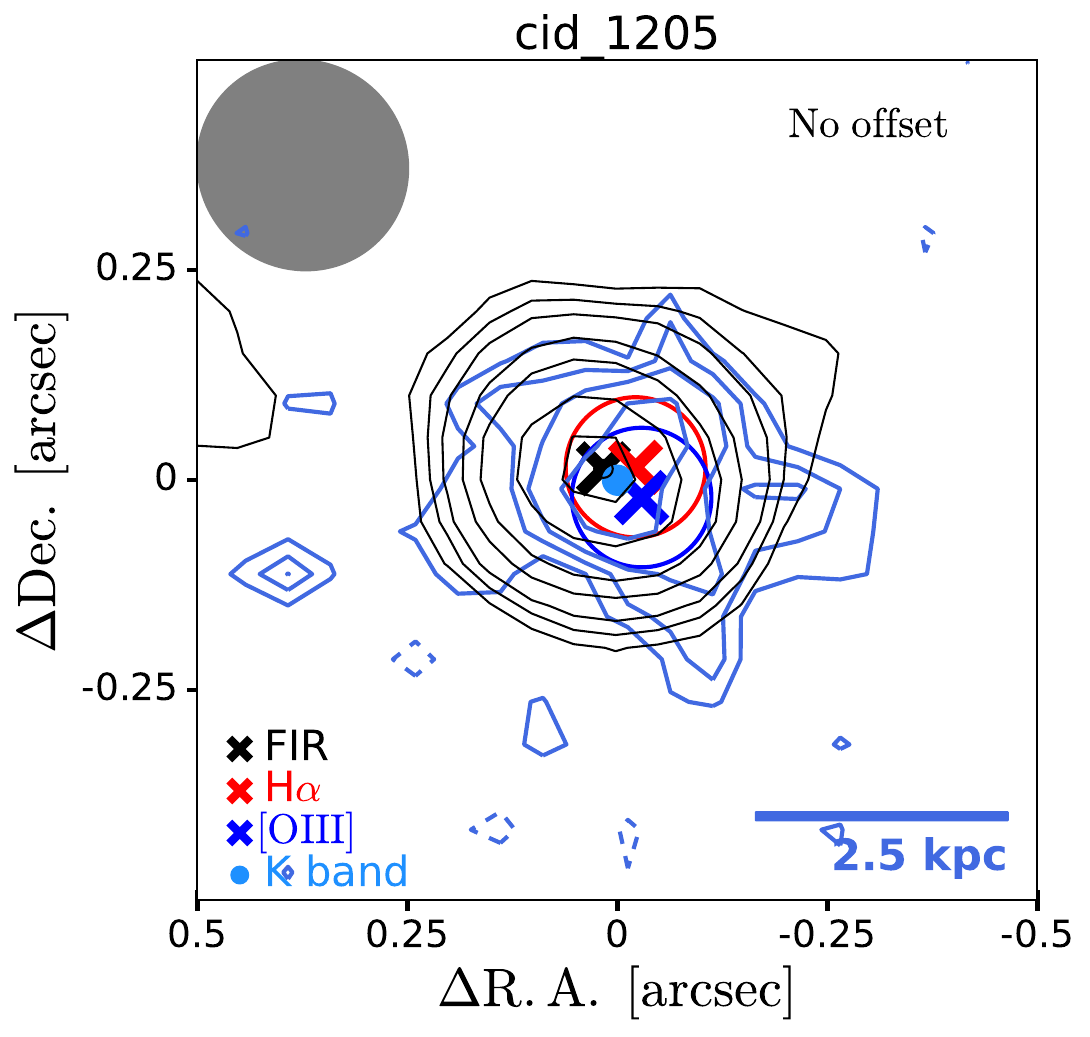}
\includegraphics[width=0.27\textwidth]{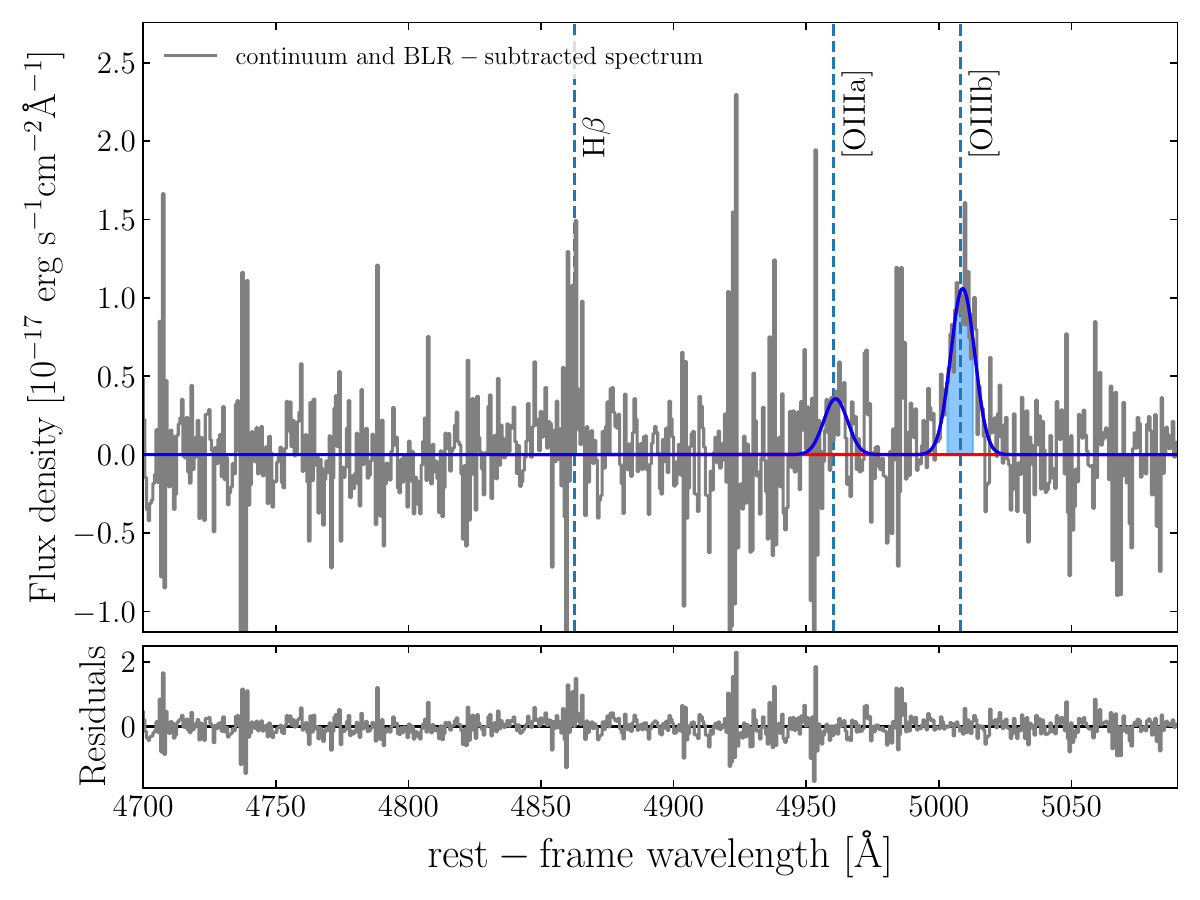}
\includegraphics[width=0.215\textwidth]{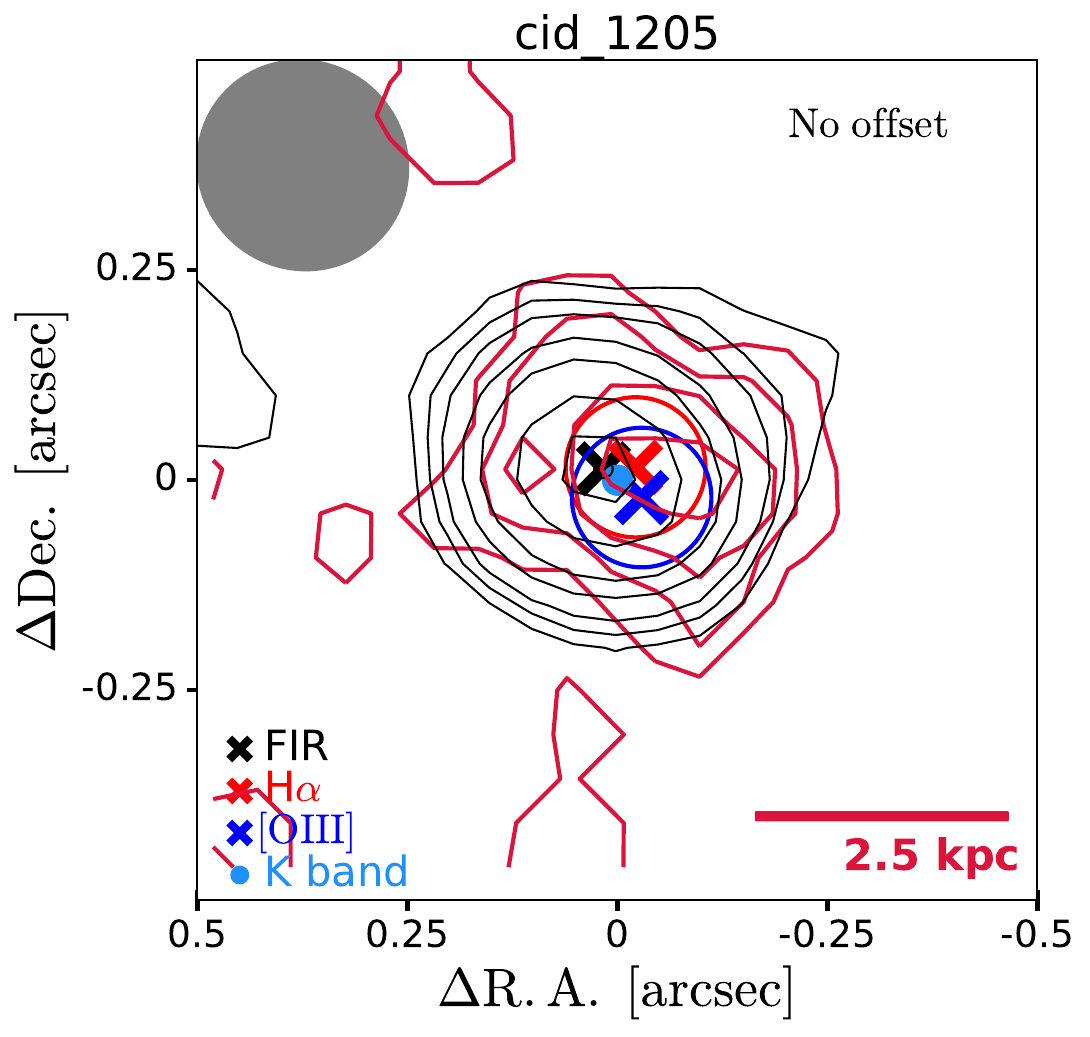}
\includegraphics[width=0.27\textwidth]{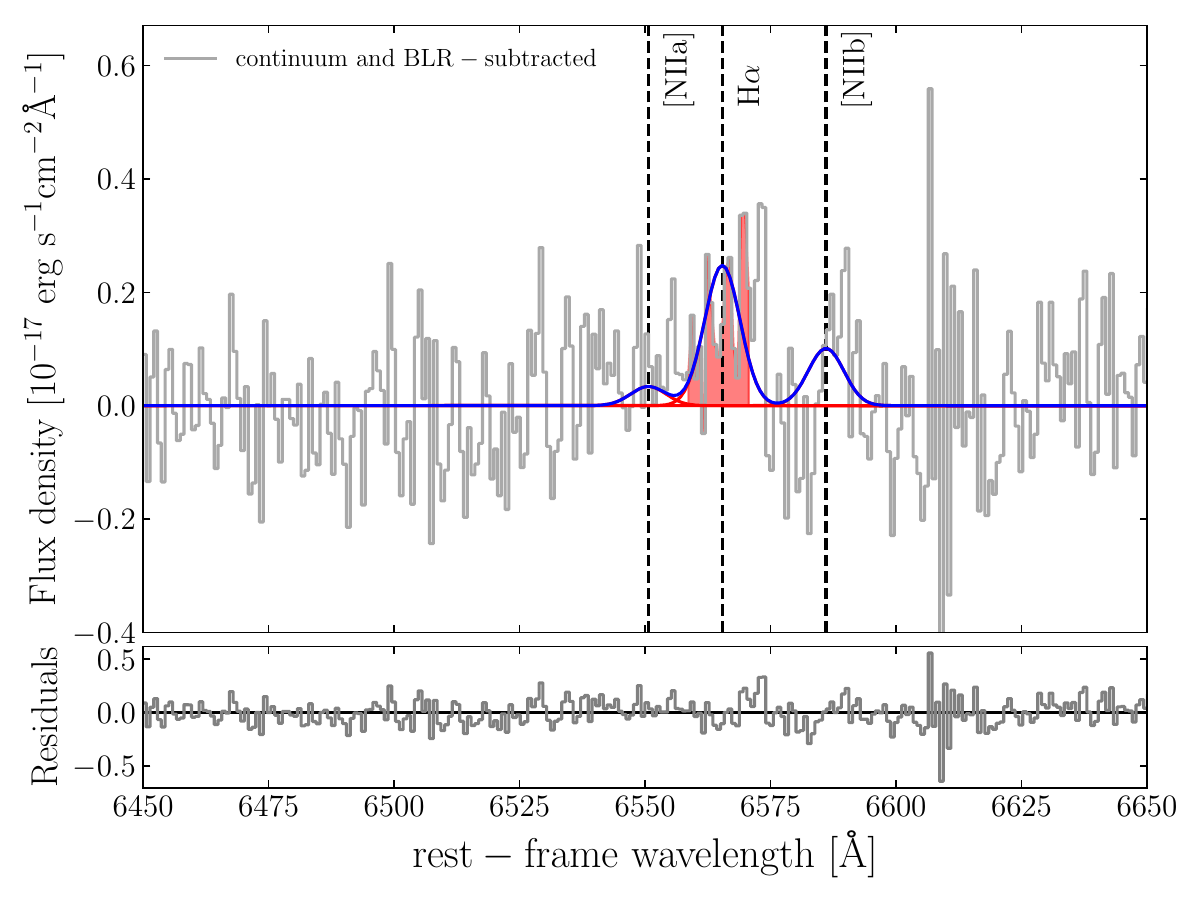}

\includegraphics[width=0.215\textwidth]{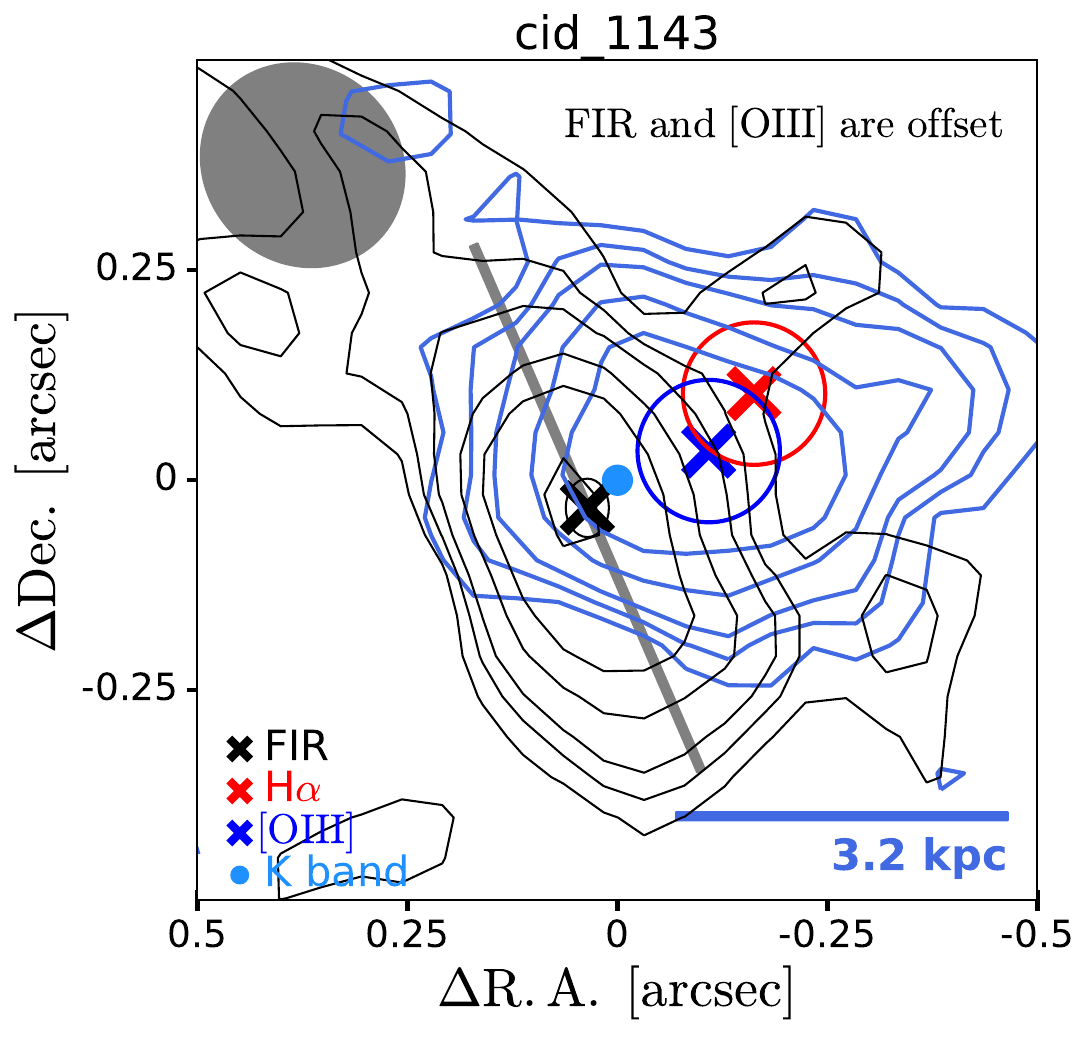}
\includegraphics[width=0.27\textwidth]{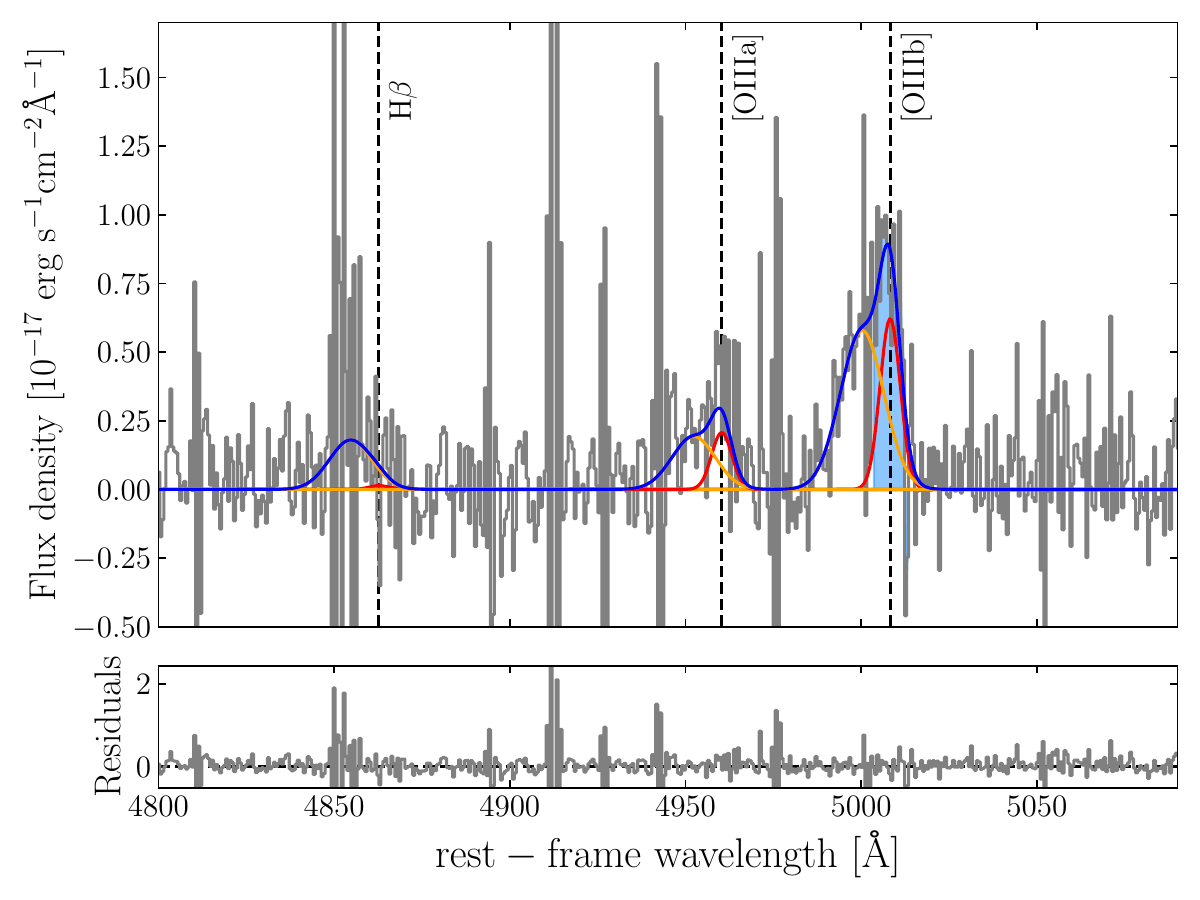}
\includegraphics[width=0.215\textwidth]{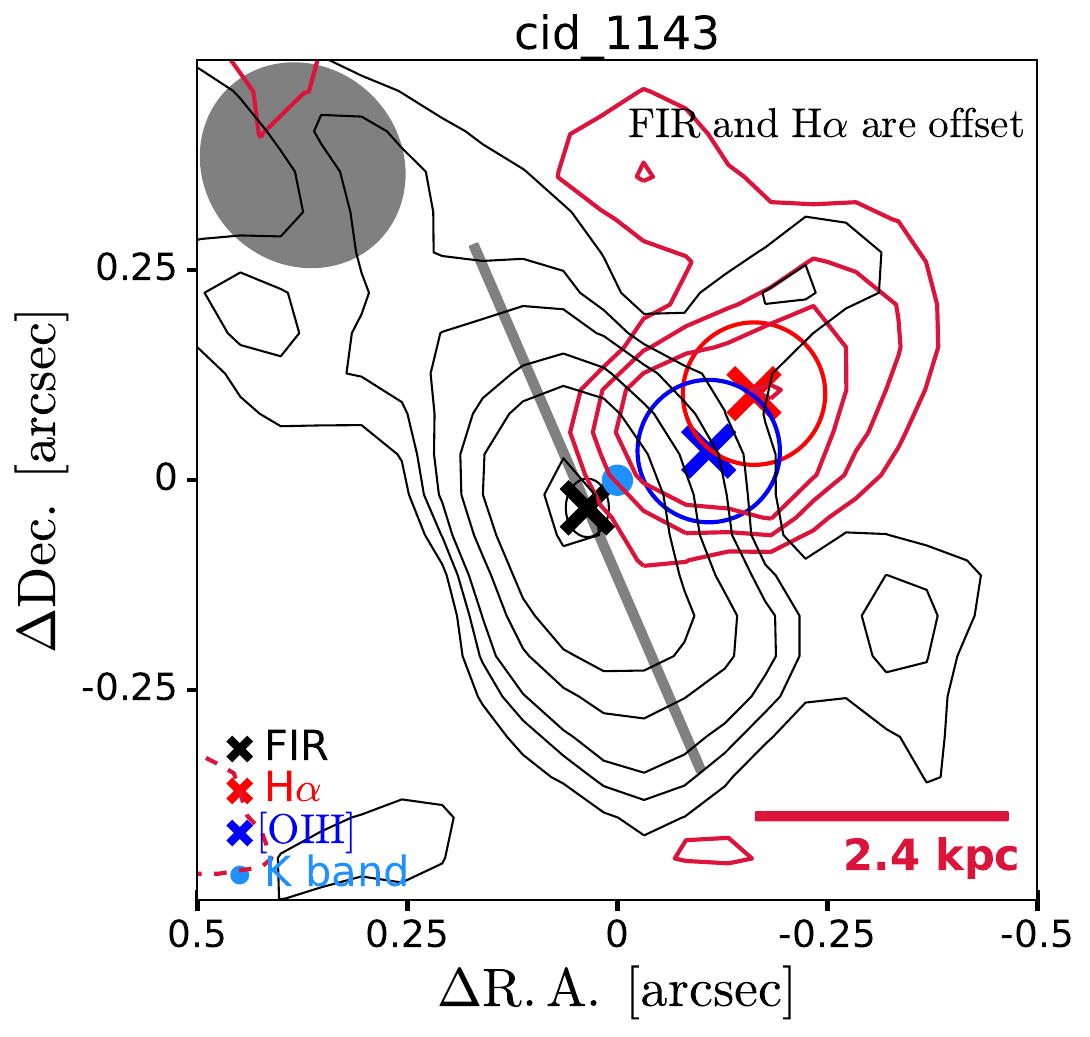}
\includegraphics[width=0.27\textwidth]{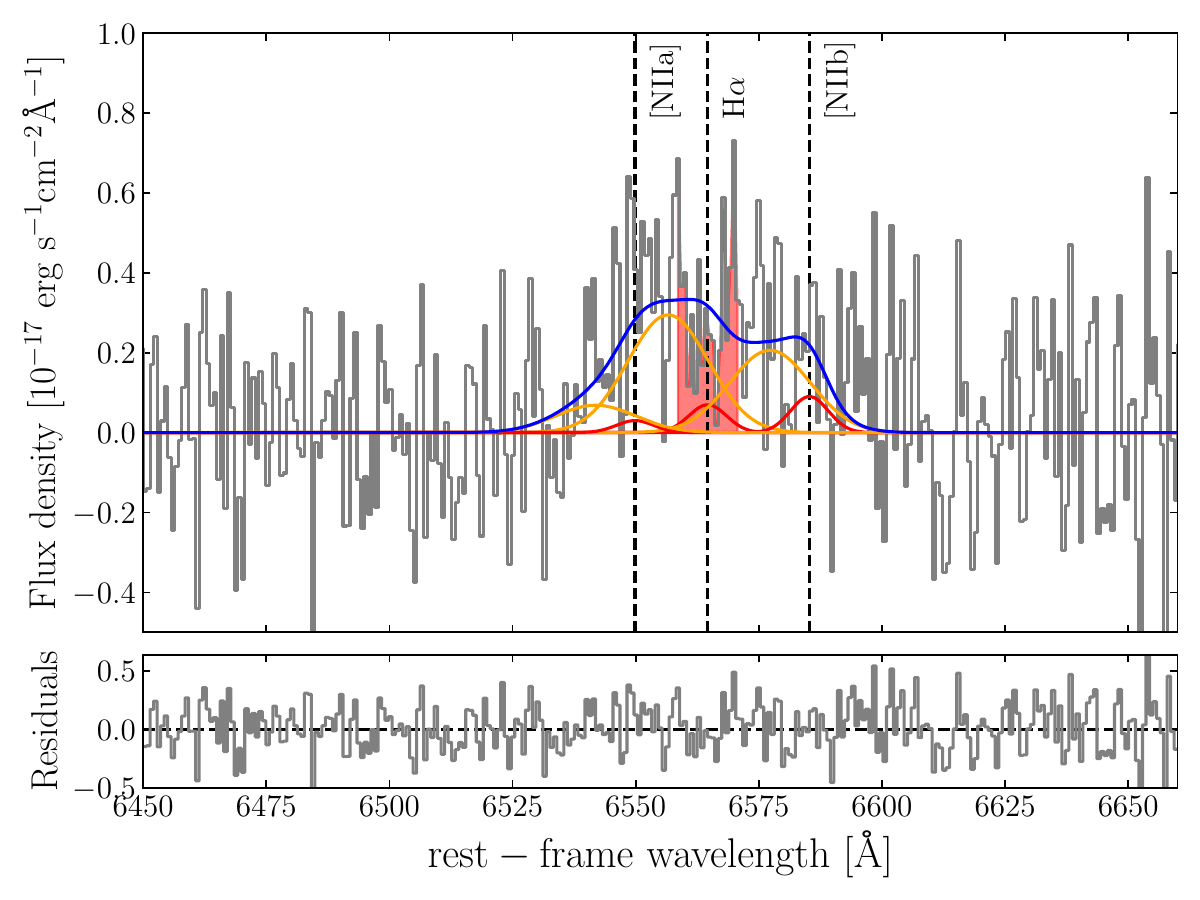}

\caption{Continued.}
\label{fig:Ha_OIII_images_2}
\end{figure*}

\section{Results and discussion}
\label{sec:results}
In this section, we first investigate the origin of the (rest-frame) 260~$\mu$m emission in our targets, to assess the relative importance of the physical processes responsible for the FIR emission (Section~\ref{sec:origin}).
 Then, we compare the FIR size of our sample with other AGN hosts and non-AGN galaxies at similar redshift from the literature, to test whether our sample has similar sizes to the general population of $z\sim2$ galaxies  (Section~\ref{sec:FIR_sizes_lit}). Finally, we compare the spatial distribution of the FIR emission with the ionised gas distribution (Section~\ref{sec:FIR_Ha_OIII}), as well as with the position of the ionised outflows (Section~\ref{sec:outflows}).

\subsection{Origin of the FIR ALMA Band 7 emission}
\label{sec:origin}

We compare the distribution of dust-obscured star formation with the distribution of ionised gas and corresponding ionised outflows in our targets. For this goal, we use the rest-frame FIR images from our ALMA 260~\micron\ maps. However, it is important to first assess the physical processes that are responsible for the 260~\micron\ emission in our sample, to test whether this emission can be used as a reliable tracer of dust obscured star formation.

The rest-frame $\sim$~260~\micron\ emission (corresponding to the observed 870~\micron\ emission) can have different origins. The three main sources of emission in the FIR are: 1) dust heated by star formation, 2) dust heated by the AGN, and 3) AGN synchrotron emission \citep[e.g. ][]{Falkendal2019}.
The observed emission is likely a combination of the three processes above, nevertheless in this section we attempt to estimate their fractional contribution, and to determine the dominant source of emission in our eight ALMA targets.\\

\noindent \textit{Dust heated by the AGN}: We consider first a diagnostic to estimate the AGN contribution in the FIR that does not rely directly on our specific SED fits.
 Following \cite{Stanley2018}, this method focuses on the ratio of the (observed-frame) FIR (870$\mu$m) to MIR (24$\mu$m) flux. AGN can have a stronger MIR emission compared to star-forming galaxies, due to emission from hot dust in the torus \citep[e.g. ][]{Pier1992, Lacy2004, Donley2012, Stern2012}. Therefore, looking at the (observed-frame) flux ratio $F_{870\mu m}/F_{24\mu m}$, it is possible to assess whether the SED is dominated by AGN emission in the $24-870$~\micron\ regime.  
 
\begin{figure}
\centering
\includegraphics[width=0.47\textwidth]
{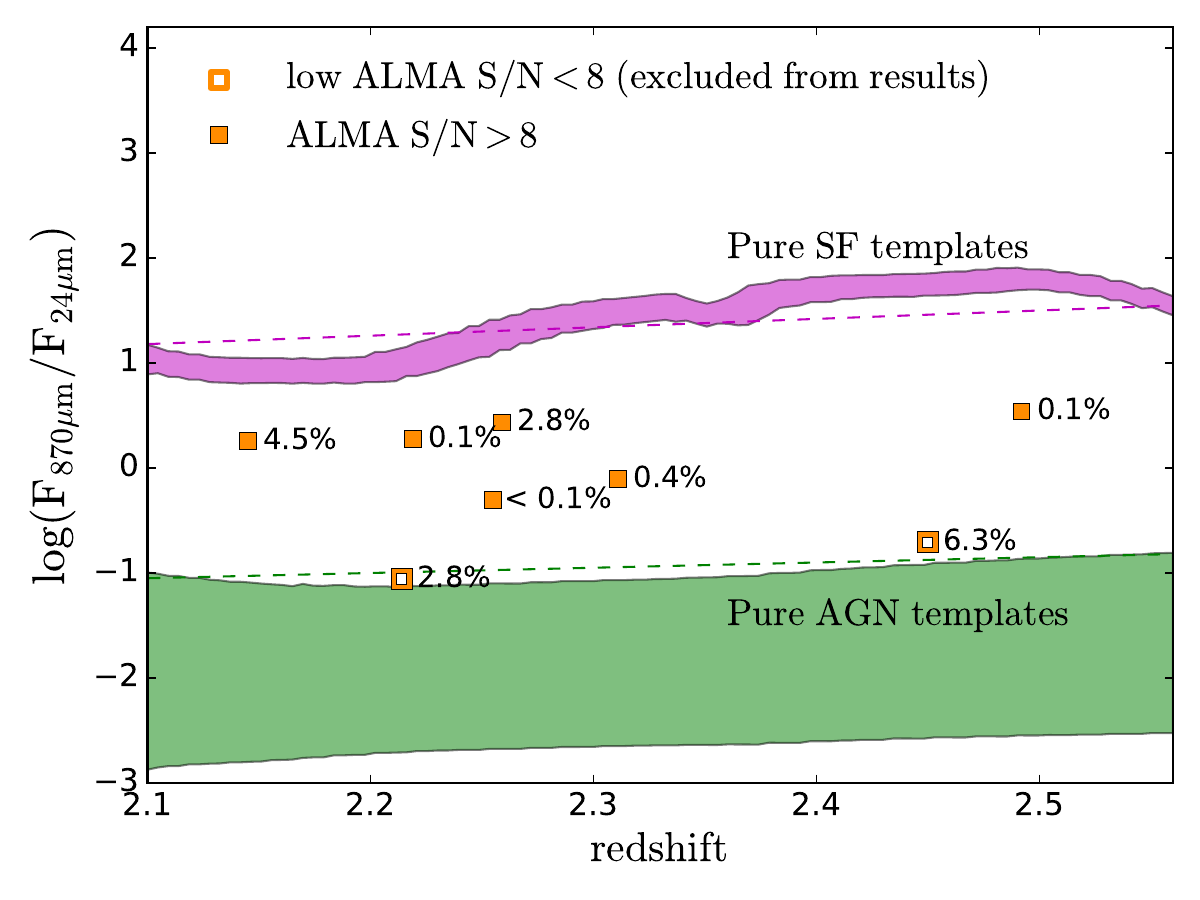}
\caption{Ratio of (observed-frame) 870$\mu$m to 24$\mu$m flux as a function of redshift for our targets calculated using the measured F(870$\mu$m) from the ALMA maps. The two sources with peak S/N$<8$ in the ALMA maps are marked with empty symbols (cid\_1057 and cid\_451).
 The numbers show the percentage AGN contribution at 870~\micron\ (observed-frame) estimated from the modelled SEDs (see Section~\ref{sec:origin}). The dashed lines show the median flux ratio as a function of redshift for star-forming (SF) galaxies templates (magenta) and AGN templates (green) from \citet{Stanley2018}.
 The shaded areas show the range of flux ratios obtained from the SF (magenta) and AGN (green) templates used to fit our ALMA targets.
}
\label{fig:Stanley2018}
\end{figure}

Figure~\ref{fig:Stanley2018} shows the $F_{870\mu m}/F_{24\mu m}$ flux ratios as a function of redshift for our sample. To compute the flux ratios, we use the 24~\micron\ \textit{Spitzer}/MIPS flux densities from the photometric catalogue compiled by \citet{Circosta2018} and our measurements of the  870~\micron\ ALMA flux densities.
 On Figure~\ref{fig:Stanley2018}, we show the range of flux ratios obtained from the SED fit of our objects by \citet{Circosta2018} using AGN models from \citet{Fritz2006} and star-forming galaxies (SF) templates from \citet{Dale2014}.
We also show  the curves corresponding to the average pure SF template and the average AGN only template from \citet{Stanley2018}. They use AGN templates from \citet{Mullaney2011} and SF templates from \citet{Mullaney2011} and \citet{Silva1998}. 

All our targets are in the region between the SF and AGN templates, suggesting that the (observed-frame) $24-870$~\micron\ emission is produced by a mixture of SF and AGN.
 The only two targets which are close to the AGN-dominated region of this diagram are cid\_1057 and cid\_451. These two targets are those with low S/N in the 870~\micron\ data (S/N$<8$). 
 This is probably reflecting the uncertainty in the measured  $F_{870\mu m}$ fluxes and the higher fraction of flux that may be resolved out in the ALMA observations, rather than suggesting a higher AGN contribution. 

To provide a more quantitative estimate of the contribution of the different physical processes to the rest-frame 260~\micron\ (observed-frame 870~\micron) flux density of our targets, we also analyse the results of the UV-to-FIR SED fitting performed with \cigale\ including our ALMA Band 7 photometric points  (see Section~\ref{sec:ALMA_SED_fit}). The MIR to FIR spectral range is modelled with two main components: emission by cold dust heated by SF and AGN emission due to the dusty torus peaking in the MIR. 
 To estimate the AGN contribution to the 260~\micron\ flux density, we consider the ratio between the flux density of the AGN component at 260~\micron\ derived from the SED best-fit model and that measured from our ALMA data.

\begin{figure}
\centering

\includegraphics[width=0.5\textwidth]{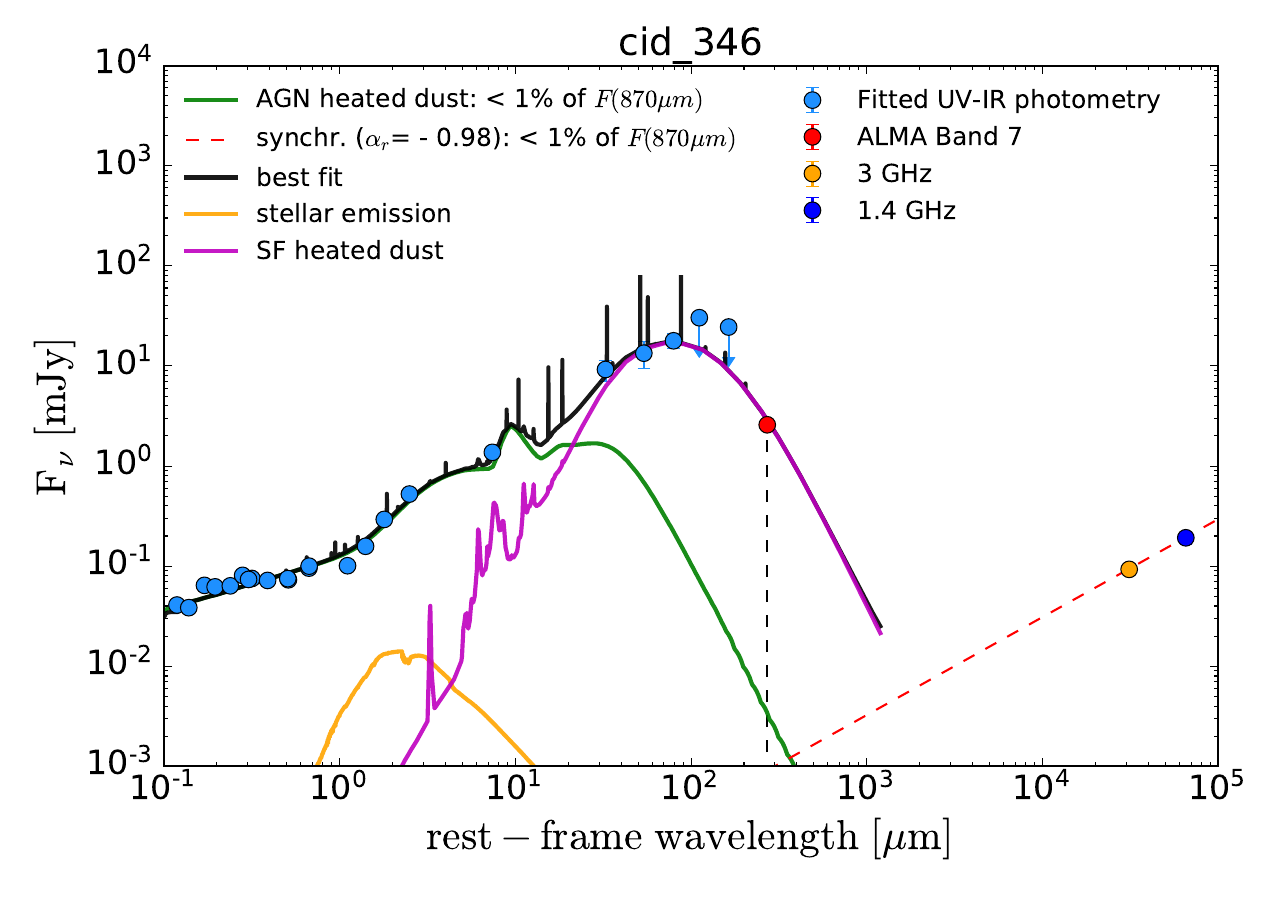}
\caption{Example rest-frame spectral energy distribution (SED) of one of our targets (cid\_346). The light blue data points represent the UV-IR photometry used for the SED fit. 
The arrows indicate 3$\sigma$ upper limits. The red  point shows our ALMA Band 7 flux measurement (included in the fit). 
The blue and orange points show the radio fluxes at 1.4 and 3~GHz, respectively. 
The solid curves show the different components of the SED modelling performed with \cigale. 
To estimate the maximum contribution of synchrotron emission to the rest-frame 260~\micron\ flux density, we parametrised this emission as a power law with spectral index $\alpha_{r}=-1.25$ (dashed red line), derived based on the 1.4 and 5.5~GHz data. The estimated contributions to the 260~\micron\ flux from dust heated by the AGN and from synchrotron emission are also shown.
}
\label{fig:SED_fit_example}
\end{figure}

 In Figure~\ref{fig:SED_fit_example}, we show as an example the SED of one target (cid\_346). The SEDs of the other targets  are shown in Figure~\ref{fig:SED_fit} in the appendix.
The AGN contribution to the total flux at 260~\micron\ ranges from $<1\%$  to 6.3\% (for cid\_451), with a mean of 2.1\%. Therefore, the results of the SED fitting decomposition suggest that the 260~\micron\ emission is dominated by star formation rather than heating from the AGN in all our targets.
 For the two targets with low S/N in the 870~\micron\ data (cid\_1057 and cid\_451), which are close to the AGN-dominated region of the diagnostic diagram from \citet{Stanley2018}, the SED fitting would imply that the AGN only contributes 2.8\% and 6.3\%, respectively, to the 260~\micron\ flux.
 The \cigale\ SED fitting code does not provide the uncertainties on the best-model SED. We therefore consider the uncertainties on the AGN fraction in the FIR range (8-1000~\micron), derived from the posterior distribution of the \cigale\ SED fit. 
These uncertainties are in the range 3-10\%. Thus, even adding these uncertainties, the AGN contribution remains small ($<16$\%).
 We note that the SED fit was performed including our ALMA high-resolution flux densities. We test that excluding the ALMA points from the SED fit, the inferred AGN contributions vary by less than 3\%.
Additionally, we test that using a difference SED fitting code \citep[AGNfitter,][]{Calistro2016}, the AGN contributions remain small ($<7\%$).

 We also consider a third method to estimate the AGN contribution in the FIR. The AGN re-processed emission depends on the model assumed to describe the torus, thus different models could potentially lead to different estimates. For the SED fitting, \citet{Circosta2018} use the models from  \citet{Fritz2006}.
As an alternative approach, we model the torus emission using a modified black-body (MBB) following the methodology described by \citet{Lani2017}. Using the AGN bolometric luminosity ($L_{bol}$) and the measured dust sizes ($R_e$), we apply Eq.~2 from \citet{Lani2017} to estimate the temperature that the dust heated by the AGN would have at a distance $R_e$ from the AGN:
\begin{equation}
    \left( \frac{T [K]}{1500}\right) ^{\frac{4+\beta}{2}}=   \frac{1}{R_e[pc]} \left( \frac{L_{bol} [erg/s]}{10^{46}} \right)^{\frac{1}{2}}.
\label{eq:Lani}
\end{equation}
We note that this equation assumes optically thin dust and we assume a dust emissivity index $\beta=1.8$ for this test. For our sample, the  temperatures of the AGN component (estimated using equation~\ref{eq:Lani})  are in the range  $68-153$~K.
Then, for each target we create a MBB with the estimated temperature and we scale this model to match the photometric data. In this way, we can estimate the maximum AGN contribution at rest-frame 260~\micron\ assuming a MBB model. Since this model is constrained only in the MIR range ($\sim5-50$~\micron), we can apply it only to the six targets with enough photometric coverage in the MIR (i.e. excluding  cid\_1057 and cid\_1143).
The AGN contributions to the ALMA band 7 photometry estimated with this method are in the range $0.5-20\%$. These percentages are consistent within the uncertainties with the results from the \cigale\ SED fitting. 
We consider more accurate the results  from the \cigale\ SED fitting, since this method uses a larger wavelength range to constrain the contribution from the AGN, including the near-infrared (NIR) photometry. Nevertheless, it is re-assuring that also using a different model to reproduce the torus emission returns consistent results.

Although we are limited by photometric coverage at long wavelengths in most of the SEDs, which makes the exact percentage of the AGN contribution uncertain, all our analyses indicate that the contribution from dust heated by the AGN to the 260~\micron\ emission, and hence to our maps produced from the ALMA data, is negligible.
We note that some studies based on radiative transfer models have highlighted the possibility that AGN could contribute significantly to the heating of diffuse warm dust on host-galaxy scales \citep{Schneider2015, Duras2017, Viaene2020, DiMascia2021, McKinney2021}. This emission is not related to the torus emission and therefore it is not taken into account in our SED fitting decomposition. This AGN contribution is more  important in the central region  \citep[$\lesssim100$ pc,][]{Viaene2020}, but in more luminous AGN it could also affect larger scales. This effect on large scales is more likely to appear on extremely dusty and infrared-luminous objects, such as in infrared and submm selected galaxies \citep{McKinney2021}, or in very luminous quasars \citep[ $L_{bol} > 10^{47}$ erg s$^{-1}$,][]{Duras2017}, which is not the case of our sample.
Detailed radiative transfer modelling would be required to test this effect on our sample, but this is beyond the scope of this work.\\

\noindent \textit{Synchrotron emission}:
In AGN, synchrotron emission can also contribute to the FIR flux, in particular in sources classified as `radio loud' \citep[e.g. ][]{Dicken2008, Falkendal2019}.
 All our targets have  flux measurements or 3$\sigma$ upper limits at 1.4~GHz and/or 3~GHz obtained with the Very Large Array (VLA). The radio fluxes are reported in \cite{Circosta2018}. Briefly, the two sources from E-CDF-S (XID36 and XID419) have  1.4~GHz fluxes from \cite{Miller2013}. XID36 has an additional flux measurement at 5.5~GHz taken with the Australia Telescope Compact Array (ATCA) from \citet{Huynh2012}. \XN\ has a 1.4~GHz flux upper limit from the VLA's FIRST survey \citep{Becker1995}. The sources from COSMOS have flux measurements from the 3~GHz VLA-COSMOS project \citep{Smolcic2017}. Two targets (cid\_346 and cid\_451) have also a measurement at 1.4~GHz from \citet{Schinnerer2007}. 
 
 To estimate the contribution from synchrotron emission to the ALMA Band 7 flux (rest-frame $\sim$260~\micron), we extrapolate from the radio fluxes using a power-law model \citep{Dicken2008}. 
For the two sources that have two flux measurements in the radio (XID36 and cid\_346), we use the slope between the two points to determine the spectral index $\alpha_{r}$ ($F_{\nu} \propto \nu^{\alpha_{r}}$), assuming no significant variability in the radio at the two frequencies. 
For cid\_346, we measure a spectral index $\alpha_r=-0.98$ and the contribution from synchrotron emission is 0.03\%. For XID36 we measure a spectral index $\alpha_r=-1.25$ and a contribution of 0.06\%. 
 For cid\_451, which is classified as a radio loud AGN, the 3~GHz flux is higher than the 1.4~GHz flux.
 Using the ALMA Band 3 flux at $\sim 100$~GHz from \citet{Circosta2021} and the 3~GHz flux, we estimate $\alpha_r=- 0.99$ and a contribution of 21\%.  
We note that since this source has low S/N, it is not included in our spatially resolved analyses. For the targets with only one flux measurement in the radio, we assume $\alpha_r=-0.7$, which is the median value measured for AGN from \citet{Smolcic2017}. For these targets, the contribution from synchrotron emission is small ($< 1\%$).
 We note that  even assuming a conservative value of $\alpha_r=0$, which is towards the most extreme value measured for AGN \citep{Dicken2008, Smolcic2017}, the synchrotron contribution is small ($< 34\%$ for cid\_1057 (a source with low S/N in the ALMA map) and $< 14\%$ for the other targets).

We conclude that in our ALMA sample most of the emission at rest-frame 260~\micron\ is due to dust heated by star formation, with contribution from dust heated by the AGN $\leq 6$\% and synchrotron contribution $\leq 21$\%. If we consider only the six targets with high S/N in the 260~\micron\ maps, the contribution from AGN-heated dust is $\leq 4$\% and the synchrotron contribution is $< 1$\%. The estimated contributions for each target are tabulated in Table~\ref{tab:ALMA_obs}.

\subsection{FIR size comparison with other samples from the literature}
\label{sec:FIR_sizes_lit}

 In this section, we compare the FIR sizes of our sample with other samples of galaxies at similar redshift from the literature. We also investigate if there is any difference in the size of galaxies with and without an AGN.
 
 From our analysis, we find that the FIR effective radii derived from the fit with the `preferred models' for our sample are in the range $0.80-2.01$~kpc, with a mean of $1.36\pm0.21$ kpc.
 These sizes are comparable to previous measurements presented from ALMA data for $z\sim1-3$ AGN and star-forming galaxies  \citep{Barro2016, Hodge2016, Fujimoto2017, Gullberg2019, Scholtz2020, Chen2020}.
 
\begin{figure*}
\centering
\includegraphics[width=0.8\textwidth]{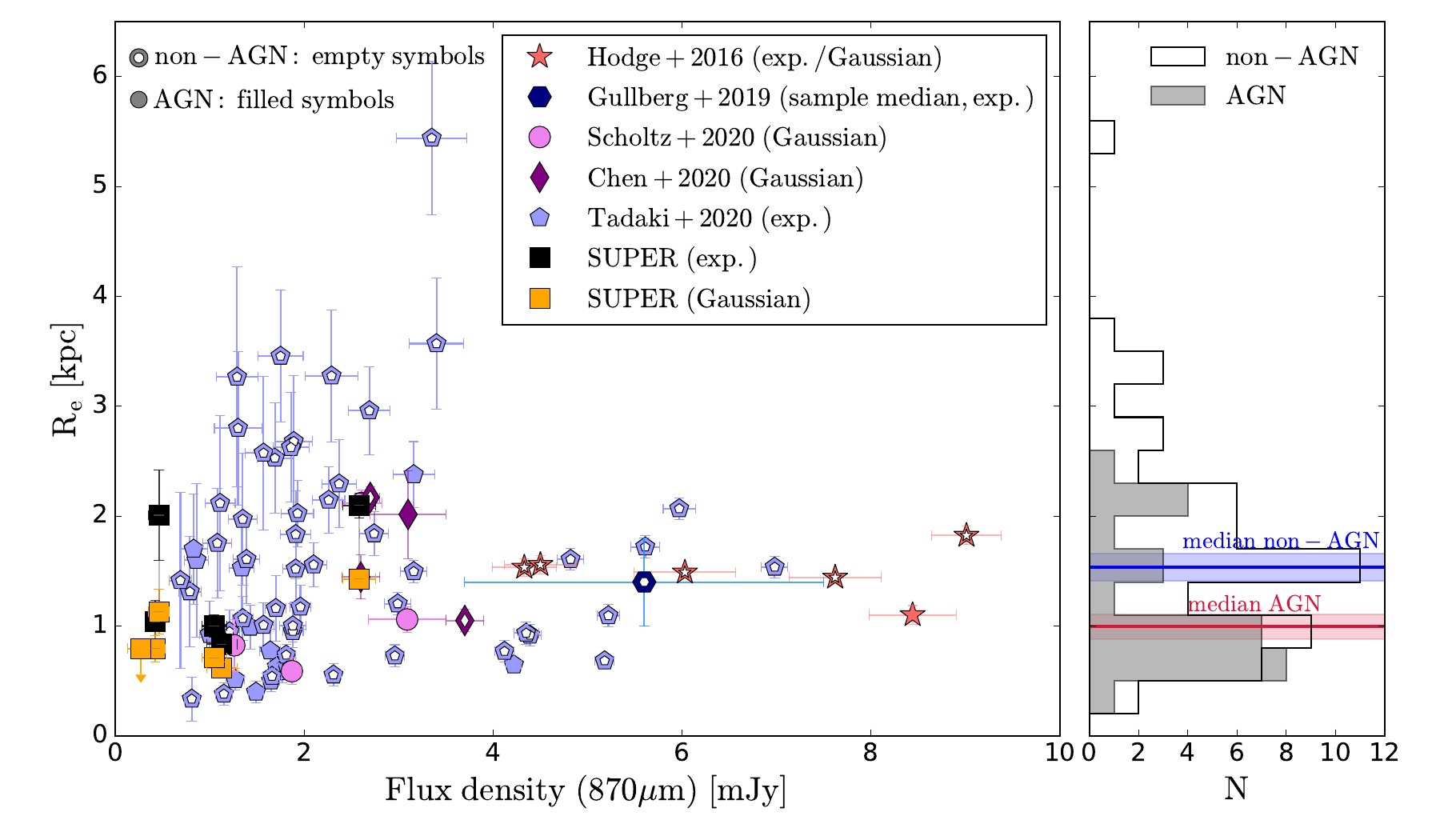}
\caption{ FIR effective radii versus observed-frame 870~\micron\ flux density for our targets and other samples from the literature  at similar redshift ($z \sim 2$), derived from ALMA observations at similar resolution ($\sim 0.2$''). The effective radii are measured assuming a Gaussian profile for the \citet{Scholtz2020} and \citet{Chen2020} samples; an exponential profile  for the \citet{Gullberg2019} and \citet{Tadaki2020} sample; and a \sersic\ profile with index $n$ between Gaussian and exponential for the \citet{Hodge2016} sample. For the  \citet{Gullberg2019} sample, we show the median value of the sample.  For SUPER, we show the radii measured from the fit of the visibilities assuming both a Gaussian and an exponential profile. Filled symbols are used for AGN and empty symbols for galaxies with no AGN signature. The \citet{Gullberg2019} sample is a mixture of AGN and non-AGN. The histograms on the right show the distributions of FIR effective radii for AGN (filled histogram) and star-forming galaxies with no AGN (empty histogram), with the corresponding medians. }
\label{fig:comp_sizes}
\end{figure*}

 We investigate in more detail how our sizes compare to other samples in Figure~\ref{fig:comp_sizes}, where we plot size as a function of observed-frame 870~\micron\ flux density. We show the 6/8 sources with high S/N in our sample. For the sake of a more direct comparison to literature, we show the sizes from our exponential and Gaussian fits (see Section~\ref{sec:FIR_size_morpho}; Appendix~\ref{sec:size_comp}), noting that these are consistent with the sizes derived from our `preferred' models.

For this comparison, we identify five samples of star-forming galaxies (SFGs) and AGN with i) similar redshifts to our sample ($z\sim1.5-2.5$) and ii) FIR sizes measured from high-resolution (< 0.5'') ALMA Band 7 continuum data: \citet{Hodge2016}, \citet{Gullberg2019},  \citet{Scholtz2020}, \citet{Chen2020}, and \citet{Tadaki2020}. We consider only the sources detected with S/N$>8$, for which it is possible to derive reliable sizes.
 In Table~\ref{tab:literature_samples} we summarise the main properties of the literature samples (sample size, redshift, resolution and RMS sensitivity of the ALMA maps, and the model used to derive the FIR sizes). 

\begin{table*}
\centering
\caption{Properties of the literature samples used to compare the FIR sizes (see Section~\ref{sec:FIR_sizes_lit}).
(1) Number of objects considered in our comparison.
(2) Redshift of the sources.
(3) RMS sensitivity of the ALMA observed-frame 870~\micron\ maps.
(4) Resolution of the ALMA observed-frame 870~\micron\ maps.
(5) Model used to measure the FIR sizes: exponential (exp.) or Gaussian.
(6) Mean FIR size (half-light radius) derived from the 870~\micron\ data.}
\setlength{\tabcolsep}{6pt}
\begin{tabular}{lccccccc}
\hline 
Reference & N & z & RMS  & resolution & model & $<R_e>$  \\
   &  & &[mJy beam$^{-1}$] & [arcsec] & & [kpc]   \\ 
   & (1) & (2)  &(3)  & (4)  & (5) & \\ 
  \hline \hline
  This work & 6 & 2.1-2.5  & 0.02-0.04  & 0.16-0.27 & exp. (Gaussian) & 1.31$\pm$0.23 (0.93$\pm$0.23)$^{a}$\\
\citet{Hodge2016} & 6 & 1.5-2.5  & $\sim$0.064 & $\sim$0.16 & exp./Gaussian$^{b}$ & 1.48$\pm$0.08 \\
\citet{Gullberg2019} & 153 & 1.5-5.8 (median 2.9) & 0.09-0.34 & $\sim$0.18 &  exponential & 1.20$\pm$0.4 \\
\citet{Scholtz2020} & 4 & 1.4-2.6  & 0.02-0.69 & 0.16-0.28 & Gaussian & 0.97$\pm$0.24\\
\citet{Chen2020} & 4 & 1.5-2.5  & 0.03-0.07  & 0.17-0.25 & Gaussian & 1.67$\pm$0.22\\
\citet{Tadaki2020} & 62 & 1.9-2.6  & $\sim$0.06  & 0.20-0.30 & exponential & 1.56$\pm$0.12\\
\hline
\end{tabular}
$^{(a)}$For our work, we report the mean $R_e$ obtained using both the exponential (exp.) and the Gaussian model.
$^{(b)}$\citet{Hodge2016} fit the sample assuming a \sersic\ profile with index $n$ free. The six SMGs with $z=1.5-2.5$ have FIR profiles between Gaussian and exponential (\sersic\ index $n=0.5-1$).
\label{tab:literature_samples}
\end{table*}

The first sample consists of 16 sub-millimetre galaxies (SMGs) from the ALESS survey \citep{Hodge2013, Karim2013}, whose FIR sizes are reported in \citet{Hodge2016}. We consider only the SMGs in the redshift range $z=1.5-2.5$ (six objects). Based on X-ray data \citep{Wang2013}, one object (ALESS17.1) is confirmed as AGN. We note that two of the targets have X-ray luminosities $L_X(0.5-8keV) > 10^{42.2}$ erg/s, for which it is ambiguous if the emission is due to an AGN or to star formation.

The second sample is presented by \citet{Gullberg2019}. They measure the FIR sizes from  exponential fits for 153 SMGs from the ALMA SCUBA-2 UDS survey \citep[AS2UDS; ][]{Stach2019}.
Using MIR diagnostics, \citet{Stach2019} identified one third of the sample as AGN, one third as non-AGN, with the final third having insufficient MIR photometry for this classification.
 Using stacking analyses, \citet{Gullberg2019} show that their measured individual sizes underestimate the true sizes by $\sim 50\%$ due to the relatively low RMS (median of 0.3~mJy/beam compared to our mean 0.02~mJy/beam). Therefore, in Figure~\ref{fig:comp_sizes}, we just show their derived median size which accounts for this effect.

The third sample is presented in \citet{Scholtz2020} and consists of eight X-ray AGN at $z = 1.4-2.6$ from the KMOS AGN Survey at High-redshift (KASHz)\footnote{We note that \citet{Scholtz2020} reported the effective radii as $R_e$=$\sigma$=FWHM/2.355. For our comparison, we convert their size measurements to $R_e$=FWHM/2, to be consistent with our definition. }. 
We exclude from the comparison two targets with low-resolution ALMA data (beam size $> 0.5$'') and two targets with S/N$< 8$. 

Moreover, we consider the sample of six SMGs in the redshift range $z = 1.5-2.5$ presented by \citet{Chen2020}. Three objects are classified as X-ray AGN, one as infrared AGN, and two as non-AGN.  Two targets are also  present in the sample from \citet{Hodge2016}  and have consistent size measurements, thus we consider them only once in our comparison.

Finally, we consider the sample of 62 massive (\Mstar$> 10^{11}$ \Msun) star-forming galaxies at redshift $z = 1.9-2.6$ from \citet{Tadaki2020}. X-ray data for this sample are available from \citet{Luo2017} and \citet{Kocevski2018}. Based on a threshold of intrinsic X-ray luminosity  $L_{X}> 10^{42}$ erg s$^{-1}$, 13 objects are classified as AGN.
We note that this sample combines high-resolution ALMA observations (0.2'') with a compact array configuration (maximum recoverable scale $\sim6.7$''). Therefore, their observations are more sensitive to large scales than the other samples, which only use the high-resolution configuration.

The mean effective radius obtained with the exponential fit for our sample ($1.30\pm0.22$ kpc) is consistent with the mean sizes from \citet[][]{Hodge2016} ($1.48\pm0.08$~kpc), \citet{Gullberg2019} ($1.2\pm 0.4$~kpc), and \citet{Tadaki2020} ($1.56\pm 0.12$~kpc), that also used the exponential fit.
Using the Gaussian fit, we obtained a mean radius of $0.91\pm0.11$~kpc, which is consistent with \citet{Scholtz2020} ($1.14\pm0.29$ kpc) and smaller (by $>3\sigma$) than the mean size from \citet{Chen2020} ($1.67\pm0.22$~kpc), both obtained assuming a Gaussian profile. 
However, we note that our sample has a smaller range of flux densities ($F_{870\mu m} =0.3-2.6$ mJy) compared to these other samples ($F_{870\mu m} =0.7-9.0$ mJy), which include also SMGs, that is galaxies selected because of their high submm flux. 
 If we limit the comparison to objects with $F_{870\mu m}< 2.6$~mJy, the mean size derived, $1.44\pm0.11$~kpc, is consistent with the mean size of our sample (measured with the exponential fit).

We also separate the sample in AGN and non-AGN (see open and filled symbols in Figure \ref{fig:comp_sizes}), to investigate whether there is a difference in their FIR sizes. 
We consider the sizes measured assuming a Gaussian profile, when possible, in order to have a consistent method for all samples. We note that for the \citet{Tadaki2020} sample this is not possible, since the sizes have been measured only with the exponential profile. 
However, we note that if we were to use the sizes measured with the exponential profile fit, we would obtain the same qualitative results. We exclude from this analysis the sample from \citet{Gullberg2019}, for which we only have the `median' size of the sample that includes both AGN and non-AGN.

 The mean size for AGN is $1.16\pm0.11$ kpc,  while the mean size for non-AGN is $1.69\pm0.13$ kpc (4$\sigma$ difference).  Applying the two-samples Kolmogorov-Smirnov (KS) test, we find a $p-$value$=0.03$, meaning that the distributions of sizes for the AGN and non-AGN are significantly different ($p-$value$<0.05$).
 We note that if we consider the sizes measured only with exponential or only with Gaussian profiles, the KS test finds a significant difference between the sizes of AGN and non-AGN with $p-$values of $0.02$ and $0.01$, respectively.
 The AGN sample has a  low mean flux density ($F_{870\mu m}= 2.0\pm0.3$~mJy) compared with the non-AGN sample ($F_{870\mu m}= 2.8\pm0.2$~mJy).  Therefore, it is possible that the difference in FIR sizes is partly related to the difference in flux density, and not to the presence of an AGN.
 To test this, we limit the samples to objects with flux densities $< 4$~mJy, and we find that the difference between AGN and non-AGN becomes even larger: the mean size for AGN is $1.18\pm0.12$ kpc and for non-AGN is $1.80\pm0.16$ kpc ($p-$value of the KS test is 0.03).
 The AGN sample has a stellar mass distribution slightly skewed to lower stellar masses (30\% of objects have $M_\star<10^{10}$ \Msun, median $M_\star=10^{11.07}$ \Msun), compared with the non-AGN sample (3\% of objects have $M_\star<10^{10}$ \Msun, median $M_\star=10^{11.21}$ \Msun). If we consider only objects with $M_\star>10^{11}$ \Msun, we still find a significant difference in the mean size of AGN ($1.06\pm0.14$ kpc) and  non-AGN ($1.70\pm0.10$ kpc).

As mentioned before, the observations of \citet{Tadaki2020} are more sensitive to large scales than the other samples, and therefore they may be more effective in recovering larger and more diffuse dust emission. This could introduce a bias in our comparison. To avoid it, we compare the sizes of AGN and non-AGN within the \citet{Tadaki2020} sample. 
  We note that there is no significant difference in the stellar mass or SFR distribution of the AGN and non-AGN within this sample. Using the KS test, we find that the FIR sizes of AGN are significantly smaller than the ones of non-AGN ($p-$value~$=0.03$). To take into account the uncertainties in the size measurements, we apply a Monte Carlo approach. For each source, we draw 100 samples from a Gaussian distribution centred on the measured size and with standard deviation equal to the uncertainty on the size. Then we perform the KS test on the drawn samples. In 79\% of the samples, the KS test gives a $p-$value$<0.05$, meaning that the difference is significant at the 79\% level.

Some previous studies have also found smaller FIR sizes in galaxies hosting AGN. 
For example, \citet{Chang2020} also find smaller sizes in obscured IR-selected AGN compared to non-AGN at $z\sim1$. Furthermore, \citet{Lutz2018} find smaller FIR sizes in nearby ($z<0.06$) X-ray selected AGN than in non-AGN at the same FIR luminosity. 
Under the assumption that the dust is spatially coincident with the gas reservoir, their interpretation is that a compact configuration of dust and gas favours the accretion to the central SMBH.
However, another interpretation is that the AGN are contributing significantly to the heating of the diffuse dust in the central region of the galaxy, resulting in more concentrated FIR emission in AGN host galaxies \citep[e.g. ][]{Schneider2015, Viaene2020, DiMascia2021, McKinney2021}. On the other hand, \citet{Ni2021} recently reported a relation between the black hole accretion rate and the compactness of the host galaxy using optical/NIR imaging. They also interpret this relation as a link between the black hole growth and the central gas density.

The significance of  the difference in dust size between AGN and non-AGN presented in this paper is still limited by the sample size and by the lack of homogeneity in the samples (although we have tried to minimise these as much as possible). In order to confirm our result, it would be important to have a larger AGN and control samples well matched in SFR, stellar mass, redshift and with similar FIR sensitivity and resolution.

\subsection{ionised gas and  FIR emission}
\label{sec:FIR_Ha_OIII}

\subsubsection{Origin of the \Ha\ emission}
\label{sec:BPT}

The \Ha\ emission is sometimes used to trace the spatial distribution of unobscured star formation in AGN host galaxies. Previous studies have performed a variety of analyses to try to minimise the challenges in using \Ha\ maps to investigate the impact of AGN outflows on star formation \citep{Cano-Diaz2012, Cresci2015, Carniani2016}. These challenges include the fact that part of the \Ha\ flux could be due to the ionising radiation of the AGN itself, and therefore special care should be taken if we want to use it as a star formation rate tracer (e.g. \citealt{Scholtz2020}). In addition, \Ha\ can suffer from significant obscuration \citep[see also][]{Kewley2013a, DAgostino2019a}. 
Therefore, if we want to use \Ha\ as a SFR tracer we need to determine what is the dominant contributor to the line emission and whether we can identify a component of the \Ha\ emission that can be used to trace star formation in our sample.

\begin{figure}
\centering
\includegraphics[width=0.45\textwidth]
{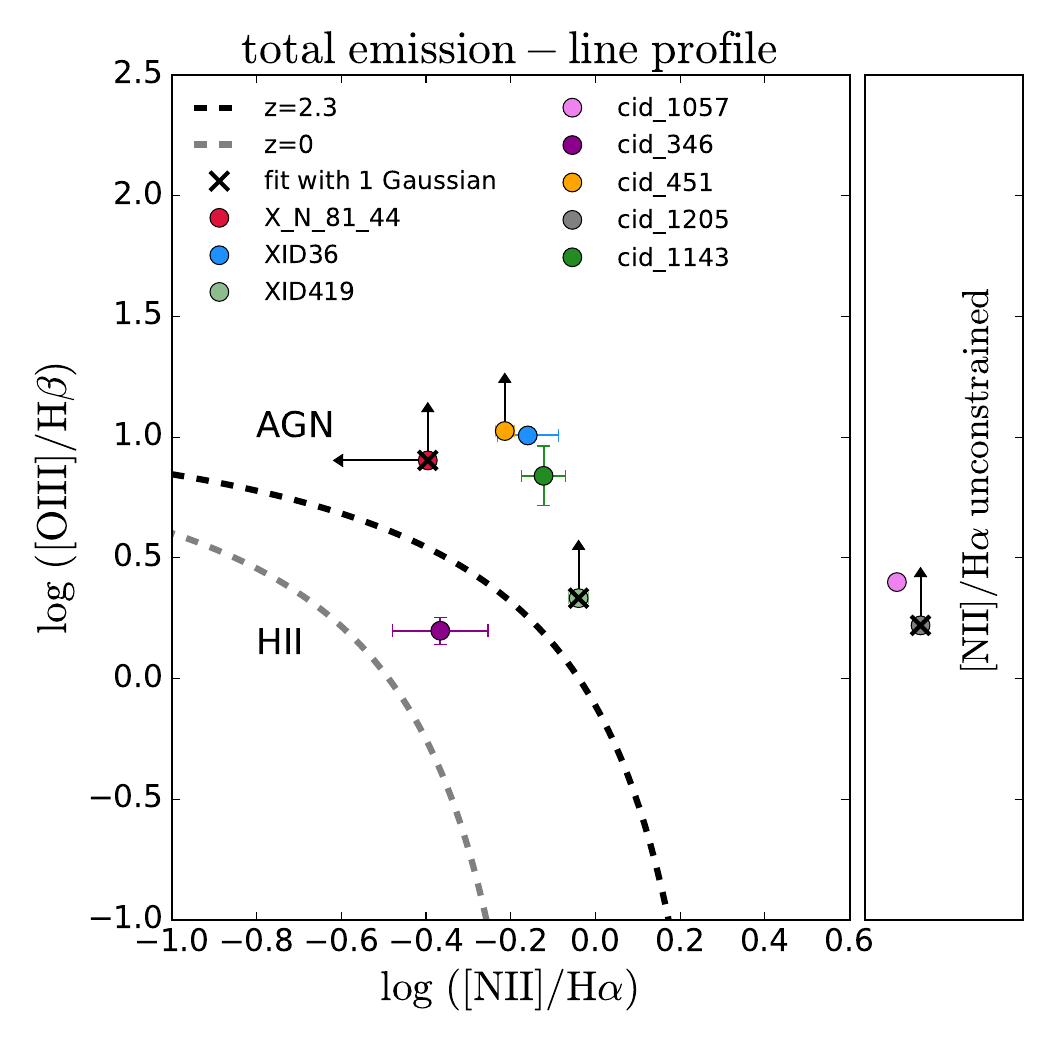}
\includegraphics[width=0.45\textwidth]
{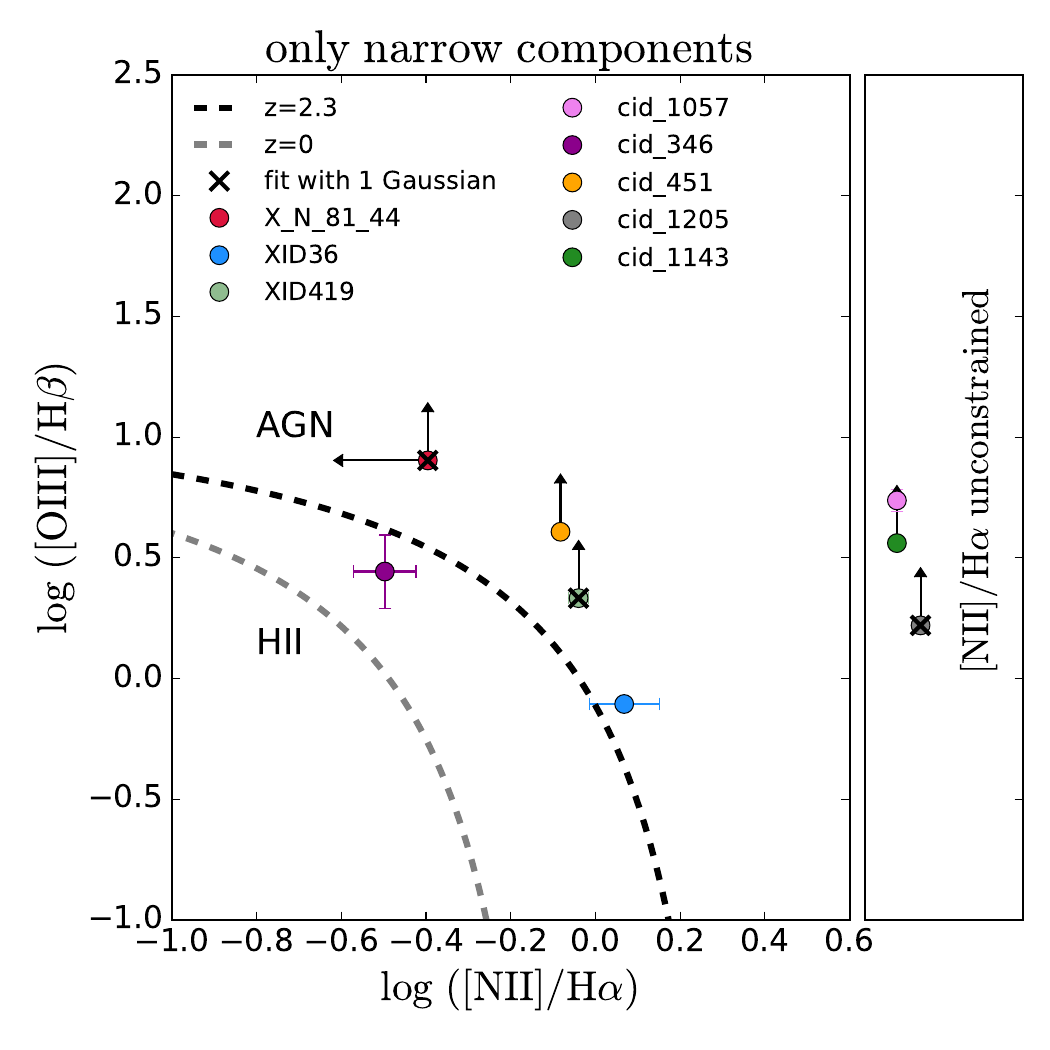}
\caption{Emission lines diagnostic diagrams. 
 The curves show the separation between star-forming galaxies and AGN at redshift $z=0$ (grey) and $z=2.3$ (black) from \citet{Kewley2013a}. 
  In the cases where one line is not detected (S/N$<3$), the line ratio is shown with an arrow indicating a 5$\sigma$ upper (or lower) limit.
In the cases where both the \Ha\ and \NII\ lines are not detected, the \NII/\Ha\ line ratio is unconstrained (see box on the right).
\textit{Upper panel:} The line ratios are measured from the total emission line profile. \textit{Lower panel:} The line ratios are calculated using the fluxes of only the narrow Gaussian component. For the three targets for which the emission lines were fitted with only one Gaussian component (marked with black crosses), the line ratios are the same as in the upper panel. 
   }
\label{fig:BPT}
\end{figure}

Unlike \cite{Carniani2016}, we were unable to identify any residual emission beyond the components associated with the AGN (i.e. the broad-line regions, AGN narrow line regions and outflows, see Section~\ref{sec:SINFONI_maps}). Therefore, we may expect the \Ha\ emission we have mapped to be dominated by AGN-related processes. To verify this, in Figure~\ref{fig:BPT} we use a BPT-like diagram \citep{Baldwin1981} which uses the optical line ratios \OIII/\Hb\ and \NII/\Ha\ to separate galaxies depending on the different source of ionisation (HII region or AGN).

The black dashed curve is the separation between star formation and `composite' (AGN/star-forming) at the mean redshift of our sample ($z=2.3$) from \citet{Kewley2013b}. 
As a reference, we also show as a grey dashed curve the separation at $z=0$ \citep{Kauffmann2003}. 
 The upper panel shows the line ratios measured from the total line profiles (narrow and broad component), while the lower panel shows the line ratios of only the narrow components.
 In the cases where \Hb\ is not detected, based on a detection threshold of S/N$>3$, we derive a 5$\sigma$ lower limit for the \OIII/\Hb\ line ratio (see Sec.~\ref{sec:line_fitting}). Similarly, for the cases where \NII\ is not detected, we show the 5$\sigma$ upper limit for the  \NII/\Ha\ line ratio. 
In some targets, both the \Ha\ and \NII\ lines are not detected, therefore the \NII/\Ha\ line ratio is unconstrained (see box on the right).

Most of our objects lie above the division line when we consider the line ratios measured from the total profile (see upper panel of Figure \ref{fig:BPT}). This means that the emission is not dominated by star formation. This is  not surprising, since these are X-ray selected AGN. For two objects, the \NII/\Ha\ line ratio is unconstrained. If we assume that they have a similar \NII/\Ha\ line ratio as the other objects, they would be in the AGN region. The only target which lies in the HII region is cid\_346.

 The targets remain in the same region of the BPT diagram also when considering only the narrow component (see lower panel of Figure \ref{fig:BPT}). This again is not surprising given the fact that our targets are bright AGN. Therefore, even if we were to consider the narrow component separately, it would still be dominated by AGN emission in most of our galaxies.  The only exception is cid\_346, which lies in the HII part of the diagram both when considering the total profile and when considering only the narrow component.
 For this target, we do not see a large difference in the spatial distribution of the `narrow' map and the map that contains both the narrow and broad components. Thus, our main conclusions would not change if we were to consider only the narrow \Ha\ component for cid\_346.

In summary, the BPT analysis suggests that \Ha\ emission is dominated by AGN ionisation in most of our sample.
As discussed in the next section, the central position of \Ha\ and \OIII\ are in in agreement for most targets (see Figure~\ref{fig:comp_coord_SINFONI}). 
 Our sources lie within the scatter of the \OIII -X-ray luminosity relation derived by \citet{Kakkad2020} using data from SUPER and from X-ray selected AGN at $z\sim1.1-2.5$ from the KASHz survey \citep{Harrison2016a}. This  suggests that the \OIII\ emission is mainly ionised by the AGN, as traced by the X-rays.
Consequently, the similar spatial distribution of \Ha\ and \OIII\ supports the idea that also the \Ha\ emission is predominantly ionised by the AGN.

For this analysis, we have considered spatially integrated spectra, but there may be spatial variations in the line ratios and some of the emission is probably more AGN dominated (e.g.  towards the centre). Unfortunately, with the current resolution we are not able to investigate spatial variations or de-couple AGN and star formation contributions as done in \citet{Cano-Diaz2012}, \citet{Cresci2015} and  \citet{Carniani2016}. 
Additionally, PSF-smearing could spread the AGN emission from the central spaxels to the other spaxels,  contaminating the \Ha\ emission at larger scales, which may be originally dominated by star formation.

Summarising, we are not able to use the \Ha\ emission as a good indicator of short term ($\sim10$~Myr) star formation in our sample. The FIR emission can provide information on the spatial location of the dust-obscured star formation, but in general it can trace SFR up to longer timescales \citep[up to 100 Myr,][]{Kennicutt2012}. As an alternative to \Ha, star formation on short timescales can be traced using rest-frame UV observations, which however could suffer from AGN contamination and dust-obscuration as well.

\subsubsection{Comparison of spatial distribution of FIR, optical, H$\alpha$, and  [OIII] emission}
\label{sec:ALMA_SINFONI_comp}

In this section we compare the spatial distribution of the FIR, optical, \Ha\ and \OIII\ emission.
In Figure \ref{fig:Ha_OIII_images}, we show the FIR continuum contours, with the  \OIII\  and \Ha\ contours overlaid, created using the `central' 600 \kms\ wide maps. We measure the positions of the FIR, \Ha\ and \OIII\ emission by fitting a 2D Gaussian to the images using the python routine \curvefit\ (see Sec.~\ref{sec:SINFONI_maps}).
The positions of the centroids are plotted as crosses on the images, with circles indicating the corresponding 1$\sigma$ uncertainties. The position of the rest-frame optical continuum (see Section~\ref{sec:coord}) is also shown in lightblue.

\begin{figure*}
\centering
\includegraphics[width=0.99\textwidth]
{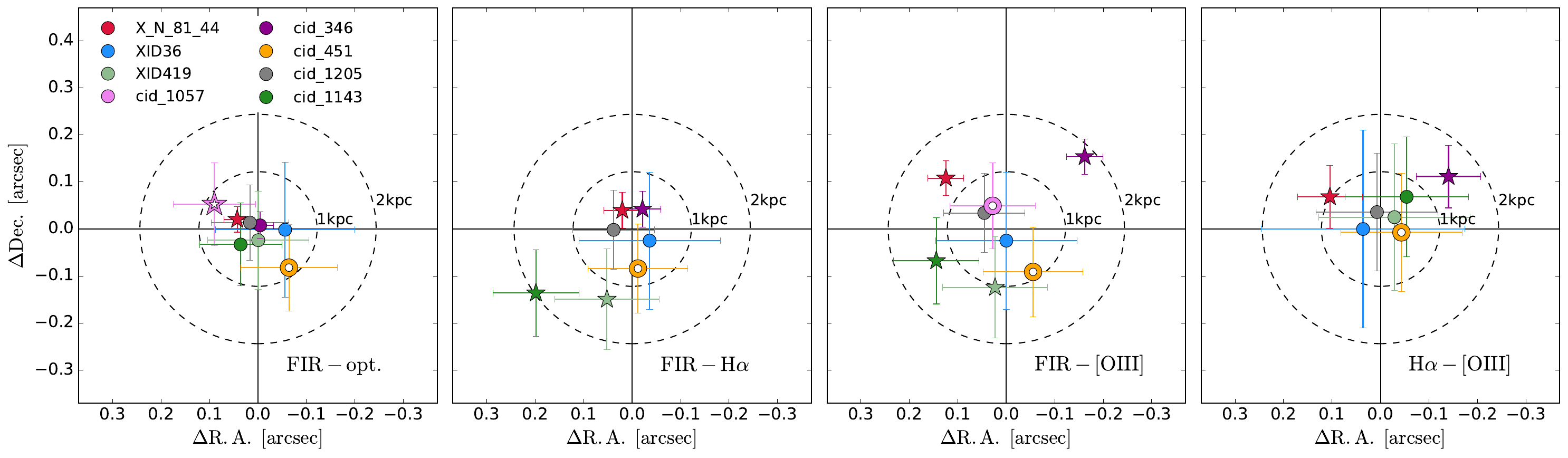}
\caption{Comparison of  the centroid positions of the 2D Gaussian fit to the images of (from left to right): FIR and optical continuum,  FIR and \Ha\ emission, FIR and \OIII\ emission, and \Ha\ and \OIII\ emission.
 The coordinates are derived from the centroid of a 2D Gaussian fit to the images. The \Ha\ and \OIII\ images were created by integrating the spectra over the [-300,300] \kms\ velocity range. The uncertainties on the position combine the uncertainties on the coordinate registration, on the Gaussian fit and the pixel size. The objects with a significant offset (i.e. difference in R.A. or Dec. larger than the $1\sigma$ uncertainty) are shown as star symbols in each plot, while the ones with no significant offset are shown as circles. The dashed circles show the offset corresponding to 1~kpc and 2~kpc at the median redshift of the sample.
The two objects with ALMA low S/N$ < 8$ are shown as empty symbols (cid\_1057 and cid\_451). 
The position of the \Ha\ and \OIII\ centroids are in fairly good agreement, while the FIR centroids show larger offset to the centroid positions of the ionised gas (both \Ha\ and \OIII ).
}
\label{fig:comp_coord_SINFONI}
\end{figure*}

 In  Figure~\ref{fig:comp_coord_SINFONI} we directly compare the difference in position between the FIR continuum, optical continuum,  \Ha\ and \OIII\ emission. The two targets with low S/N in the ALMA maps are marked with empty symbols. 

The rest-frame optical continuum emission is in general agreement with the position of the FIR continuum (see left panel of Figure~\ref{fig:comp_coord_SINFONI}). 
 The offset between optical and FIR continuum is smaller than the uncertainties for all targets, with the exception of cid\_1057 and \XN.  cid\_1057 has low S/N in the ALMA maps (S/N$=3.6$), therefore, the FIR position is not well constrained. 
\XN\  shows an offset larger than the uncertainties, but still small ($< 0.05$'').
 In the Type 1s, the optical continuum is dominated by the AGN emission, while in the Type 2s it is probably tracing the stellar disk, since the central AGN is obscured. The good alignment between the FIR continuum emission and the optical continuum suggests that the FIR is aligned with the host galaxy position.
 Unfortunately, we do not have other information about the host galaxy morphology. For three sources, there are HST/WFC3 images available,  but the images show only strong point sources, so we cannot derive any information about the host galaxy morphology.

For the majority of the sources with reliable FIR positions (4/6),  the \Ha\ and \OIII\ centroid positions are not consistent with the FIR position (difference larger than 1$\sigma$, see Figure~\ref{fig:comp_coord_SINFONI}). 
 The offsets between the FIR and \Ha\ position are in the range $0.3-1.9$~kpc, with a mean offset of $0.8\pm0.2$~kpc. The offsets between the FIR and \OIII\ are in the range $0.2-1.8$~kpc, with a mean offset of $1.0\pm0.2$~kpc. Two objects have an offset $> 1$~kpc, both for \Ha\ and \OIII: XID419 and cid\_1143.
We compare our results with the work by \citet{Scholtz2020}, who measured the offset between the FIR emission and the \Ha\ emission in a sample of eight AGN at redshift $1.4-2.6$. They found projected offsets between \Ha\ and FIR in the range $0.8-2.8$~kpc, with a mean offset of $1.4\pm0.6$ kpc. The offsets found by \citet{Scholtz2020} are a bit larger but consistent with our findings. However, we note that their rest-frame optical observations have a lower spatial resolution (FWHM PSF $\sim0.6-1$'') compared to the AO observations presented here.

To summarise, we find that the centroid of the FIR emission is on average offset ($> 1\sigma$, see star symbols in Fig.~\ref{fig:comp_coord_SINFONI})  from the \OIII\ and \Ha\ centroids. However, we note that there can still be significant overlap between the emissions even if the centroids are offset. One possible explanation of the offset is that the dust and the ionised gas have different locations. 
Another possibility is that the outflow component is contributing substantially also in the central velocity  channel ($v =[-300,300]$~\kms). In the next Section \ref{sec:outflows}, we compare the position of the \OIII\ emission in the central velocity channel with the emission in the blue-shifted and red-shifted channels ($v < -300$~\kms and $v>300$~\kms). For XID36 and cid\_451, the morphology and position of the \OIII\ emission is similar in the three velocity channels, which suggests that the outflow component may contribute significantly also in the central velocity channels. For the other targets, the morphology and position of the emission in the three channels are different, suggesting that the central \OIII\ channel is not strongly contaminated by the outflow component.

It is also possible that the dust is obscuring part of the \Ha\ and \OIII\ emission, especially where the peak of the dust emission is located.
 Unfortunately, in most of our objects, we do not have a measurement of the Balmer decrement, because of the low S/N of the \Ha\ and \Hb\ emission lines.  However, even the Balmer decrement could underestimate the level of obscuration, because  in the most obscured regions the ionised gas emission may be totally obscured by dust \citep[see ][]{Chen2020}. 
The fact that the \Ha\ and FIR emission are not co-spatial has implications if we want to use the \Ha\ emission to trace the total star formation in the host  galaxies. 
First, the assumption that \Ha\ emission is dominated by star formation is likely not to be true in most of our cases (see Section~\ref{sec:BPT}).
Second, even if we could derive the extinction correction for the \Ha\ emission, we would still not be able to recover the total SFR of the galaxy since we would not be sampling the same region as that covered by the FIR emission \citep[e.g. ][]{Brusa2018}.

 In Figure~\ref{fig:Ha_OIII_images}, which presents the \Ha\ and \OIII\ maps, we also show the position angle along the major axis of the FIR emission for the three targets where it can be reliably determined (XID419, cid\_346, and cid\_1143). The \Ha\ and \OIII\ emission are offset from the FIR emission in a direction roughly perpendicular to the FIR major axis. This could be indicative of AGN ionisation cones, extending perpendicular to the plane of the galaxy \citep[e.g. ][]{Crenshaw2010a, Venturi2017, Venturi2018}. 
This would favour a scenario where a significant fraction of the ionised gas has a different locations with respect to the FIR emission.  However, we cannot rule out that dust is obscuring part of the ionised gas emission.

The \Ha\ and \OIII\ positions are in general in agreement with each other, with offsets in the range $0.3-1.5$~kpc and mean offset $0.6\pm0.1$~kpc (see fourth panel in Figure~\ref{fig:comp_coord_SINFONI}). Only two objects (\XN\ and cid\_346) show a \Ha-\OIII\ offset larger than the uncertainties. In \XN\, the morphologies of \Ha\ and \OIII\ emission are also different, extending in two  almost perpendicular directions.  In cid\_346 instead the \Ha\ distribution is similar to the FIR distribution, while \OIII\ is more extended  in the S-E direction. 
 We note that cid\_346 is the only object in our sample which is classified in the HII region according to the emission line diagnostic diagram (see Section~\ref{sec:BPT}). The offset between \Ha\ and \OIII\ could be a result of the fact that they are ionised by different mechanisms (AGN for \OIII\ and star formation for \Ha).

\begin{figure}
\centering
\includegraphics[width=0.45\textwidth]{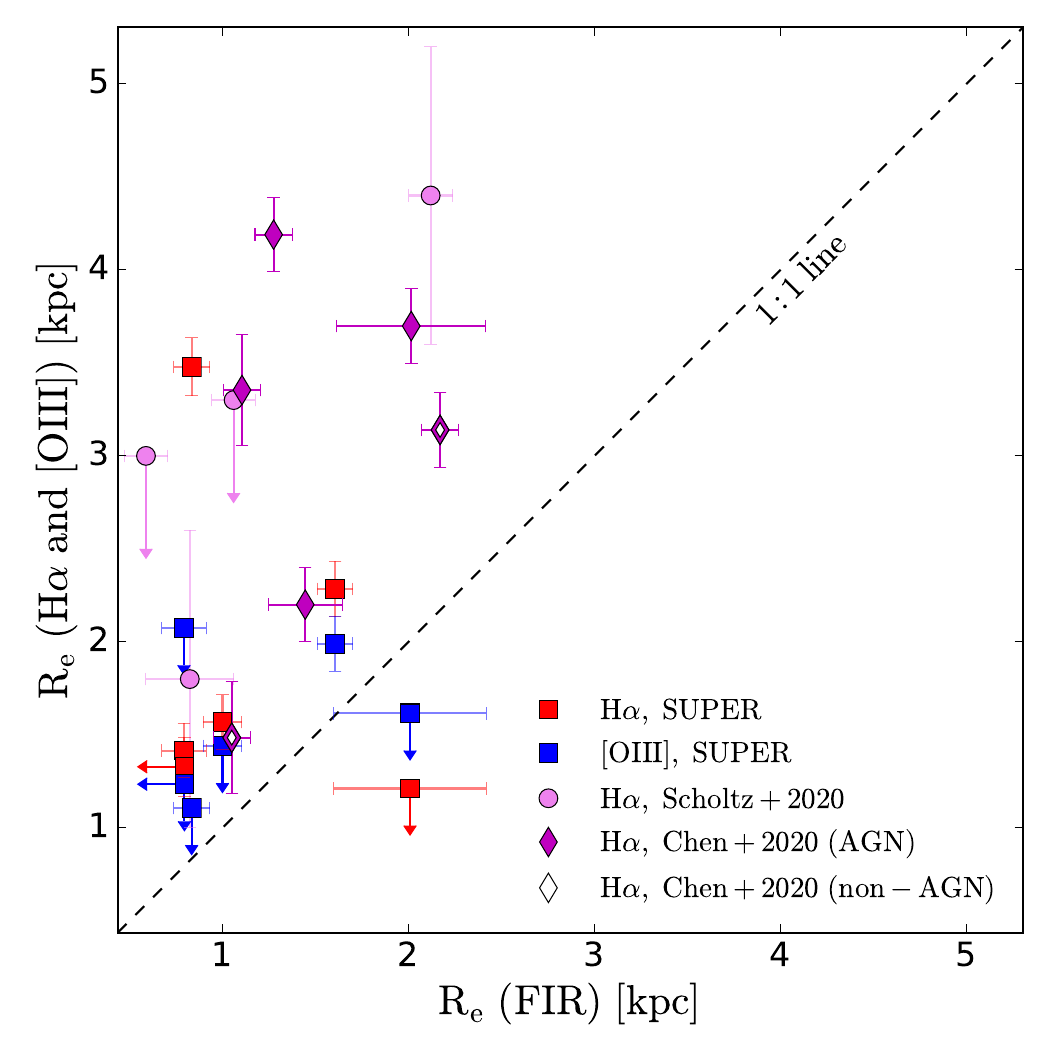}
\caption{Comparison of the size of the FIR emission with the sizes of the \Ha\ (red) and \OIII\ (blue) emission. The effective radii of the FIR emission are measured from the best fit on the visibilities. The \Ha\ and \OIII\ sizes are measured by fitting a 2D Gaussian to the images. If the measured size is smaller than the beam, we show the point as an upper limit. The two objects with low S/N ($< 8$) in the ALMA data are not shown (cid\_1057 and cid\_451).
The FIR and \Ha\ sizes of the sample of $z\sim 2$ AGN from \citet{Scholtz2020} and from \citet{Chen2020} are shown with violet circles and magenta diamonds, respectively. The white diamonds are non-AGN from \citet{Chen2020}.
On average, the FIR sizes are smaller than the \Ha\ and \OIII\ sizes by a factor of $\sim2$.
}
\label{fig:comp_size_SINFONI}
\end{figure}

Following  \citet{Chen2020} and \citet{Scholtz2020}, we also compare the sizes of the FIR, \Ha, and \OIII\ emission in Figure~\ref{fig:comp_size_SINFONI}. We use the FIR sizes derived with the best model fit of the $uv$-visibilities. In general, the \Ha\ and \OIII\ sizes are comparable or larger than the FIR sizes. The only exception is cid\_1143, for which the upper limits on the \Ha\ and \OIII\ sizes are smaller that the FIR sizes. We note that the FIR emission of this object is significantly elongated (axis ratio $=4.6$). 
A possible interpretation is that the dust emission is tracing the stellar disk of this galaxy (seen edge-on), and the ionised gas is more extended out of the galaxy plane.
The mean ratios are $R_e$(\Ha)/$R_e$(FIR) = $1.87\pm0.45$ and $R_e$(\OIII)/$R_e$(FIR) = $1.50\pm0.23$.  
Larger \Ha\ sizes compared to FIR sizes have also been observed by \cite{Scholtz2020}, \cite{Chen2020} and \citet{Tadaki2020}, who find mean ratio $R_e$(\Ha)/$R_e$(FIR) of $3.1\pm0.6$, $2.1\pm0.3$, and $2.3$ respectively.
 These studies find on average larger \Ha\ sizes (and larger $R_e$(\Ha)/$R_e$(FIR) ratios)  compared to our observations. The difference could be due to the fact that our AO observations are missing part of the more extended flux \citep{ForsterSchreiber2018b}.
Additionally, in our case we cannot exclude that the FIR sizes are underestimated because the high-resolution ALMA observations are not sensitive to more diffuse emission.

The larger sizes of the ionised gas emission compared to the FIR sizes could be due to different reasons.
One possibility is that the  ionised gas is more extended because of the AGN, that ionises the gas to larger distances \citep[e.g. ][]{Scholtz2020}. 
 However, larger \Ha\ sizes compared to the FIR have been observed also in star-forming galaxies not hosting  AGN \citep{Chen2020, Tadaki2020}. In $z\sim2$ star-forming galaxies, the optical continuum is found to be systematically larger than the FIR sizes by a factor of $2-3$, suggesting that the FIR emission is tracing a compact star-burst region \citep[e.g.][]{Barro2016,  Tadaki2017a, Fujimoto2017, Fujimoto2018, Elbaz2018, Calistro2018, Lang2019, Puglisi2019}.
It is interesting to note that \citet{Popping2021} use simulations to show  that the larger rest-frame optical sizes are due to higher dust-obscuration in the centre of galaxies which artificially increases the derived sizes in this band, and that the FIR emission is not intrinsically more compact than the stellar distribution.

\subsection{ionised outflows and star formation}
\label{sec:outflows}

In this section, we qualitatively compare the spatial distribution of the \OIII\ outflows and star formation. 
 As we have shown in Section~\ref{sec:origin}, the rest-frame 260~\micron\ emission in our targets is mostly due to star formation,  while we cannot rely on the \Ha\ emission as a star formation indicator.  Thus, here we focus on the spatial comparison of the outflows and FIR emission.

Following \citet{Kakkad2020}, we define as outflow the \OIII\ emission with absolute velocities $> 300$~\kms\ with respect to the zero velocity (see Section~\ref{sec:SINFONI_maps}).
In Figure~\ref{fig:OIII_maps} we show the \OIII\ maps in three velocity channels: blue-shifted emission < -300~\kms, central emission [-300, 300]~\kms, and red-shifted emission > 300~\kms. We create the maps by collapsing the continuum subtracted spectra over the selected velocity channels for each spaxel. 
We check that our conclusions are not sensitive to the choice of the threshold adopted to select the blue- or red-shifted emission (i.e. absolute velocities $>300$~\kms). By selecting only the extreme velocity wings (e.g. $>600$~\kms), the qualitative conclusions remain the same.

All our targets apart from cid\_1205 show blue-shifted emission (above 4$\sigma$), in  agreement with the analyses of \citet{Kakkad2020} on the Type 1 targets.  We note that the \OIII\ data-cube of cid\_1205 presents an artefact (horizontal stripe) south of the target. After removing the artefact, we do not see significant emission in the blue-shifted map, but we cannot exclude that a faint outflow could be undetected. However, since the \OIII\ spectrum of this target shows only a weak emission at velocities $<-300$~\kms, we do not expect a strong \OIII\ outflow in this target.

From the \OIII\ line profiles, we can see that the blue-wing is generally more prominent than the red-wing. This effect can be due to dust that is obscuring the receding side of the outflow \citep[e.g. ][]{Bae2016}.  
 The more extended blue-shifted emissions ($> 4.5$~kpc) are detected in  cid\_1143, and cid\_1057. The red-shifted emission is fainter and detected above 5$\sigma$ in only 5/8  objects (\XN, XID36, cid\_1057, cid\_451 and cid\_1205).  These objects show a significantly extended ($\sim 2.5$~kpc) red-shifted emission.

The shape of the outflow is certainly also a function of the viewing angle. In XID36 and cid\_451, the blue-shifted emission is symmetric (not elongated) and the shape is similar to the emission in the central velocity channel. A possible interpretation is that the outflow is aligned in the direction of our line-of-sight. On the contrary, in cid\_1143 and cid\_1057 the outflow has an extended and biconical shape which is suggesting that our line of sight is almost perpendicular to it. In particular, in  cid\_1143, the direction of the outflow is perpendicular to the major axis of the FIR emission. 
If the stellar disk is oriented as the dust emission, this may suggest that the outflow is moving along the path of least resistance, which is perpendicular to the disk  \citep{Gabor2014}.

Under the assumption that the FIR emission is a reliable tracer of the SFR, we do not see evidence that the obscured star formation  is suppressed or disturbed in the location of the outflow. 
We discuss this result further in the next section.

\begin{figure*}
\centering

\includegraphics[width=0.228\textwidth]{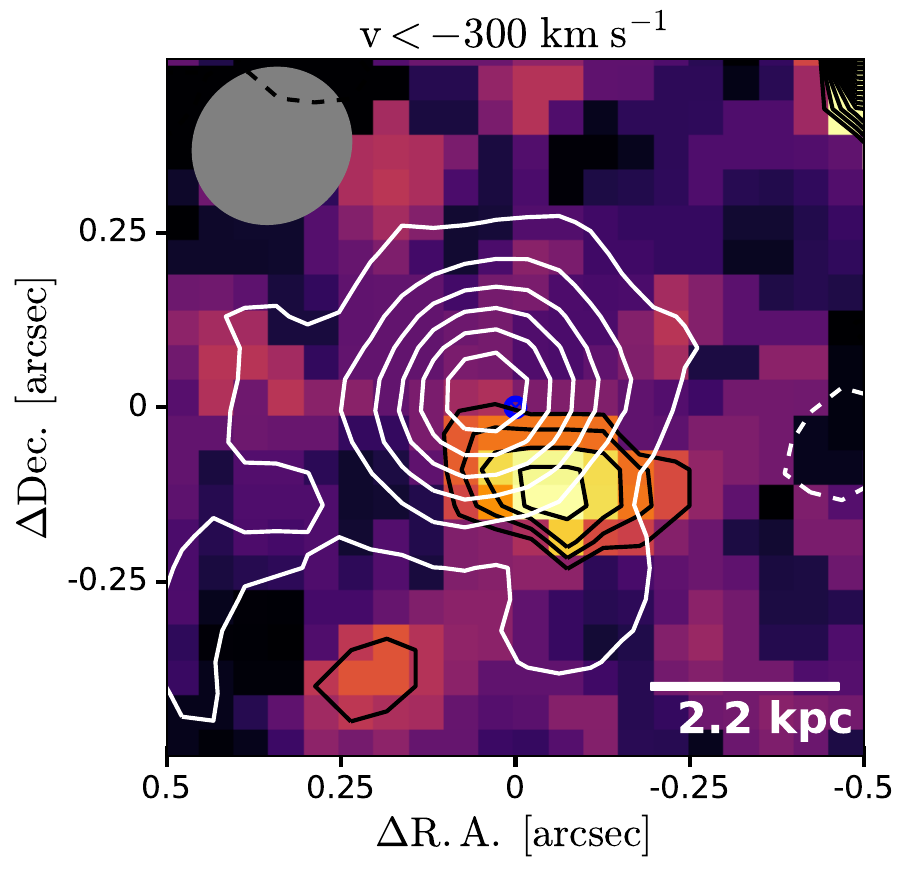}
\includegraphics[width=0.2\textwidth]{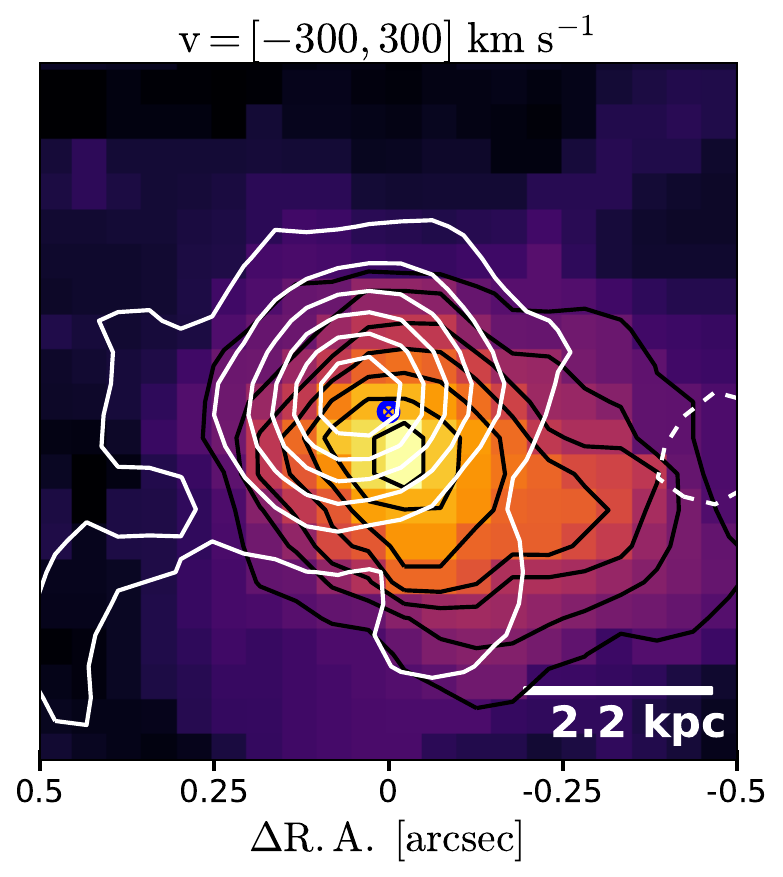}
\includegraphics[width=0.2\textwidth]{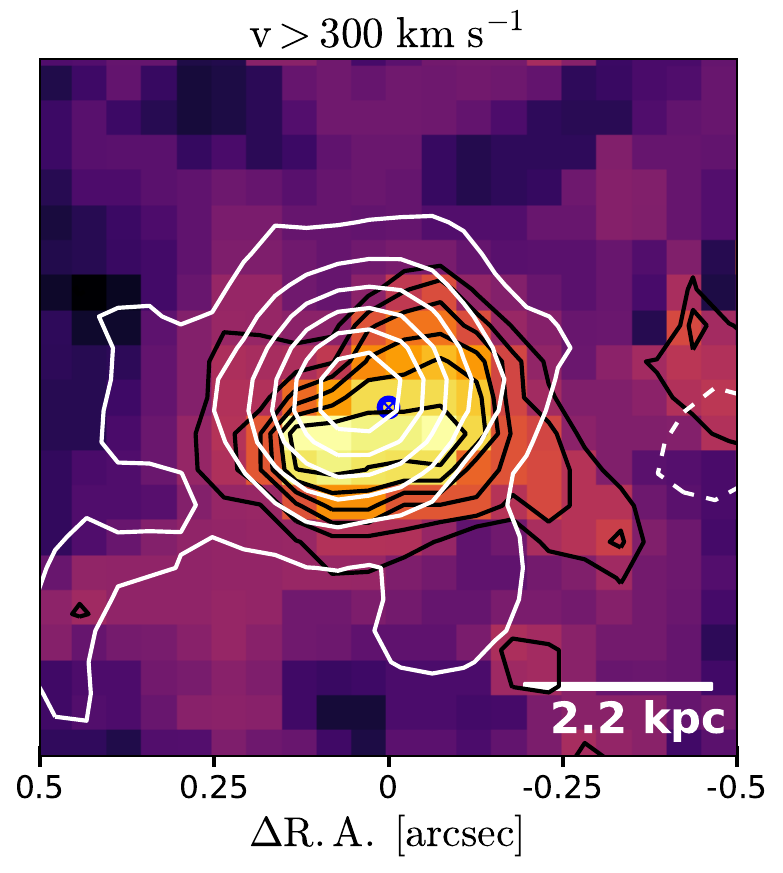}
\includegraphics[width=0.32\textwidth]{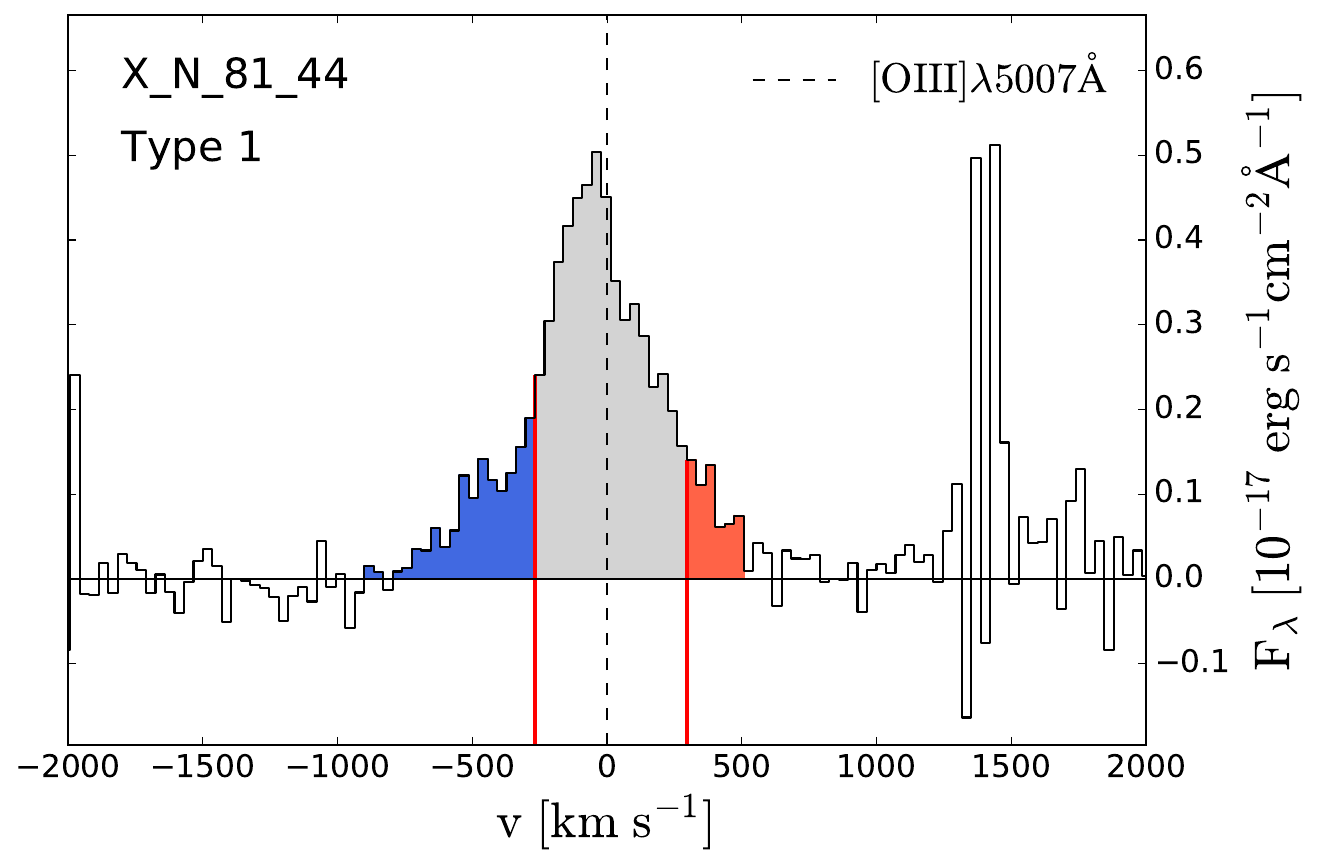}

\includegraphics[width=0.228\textwidth]{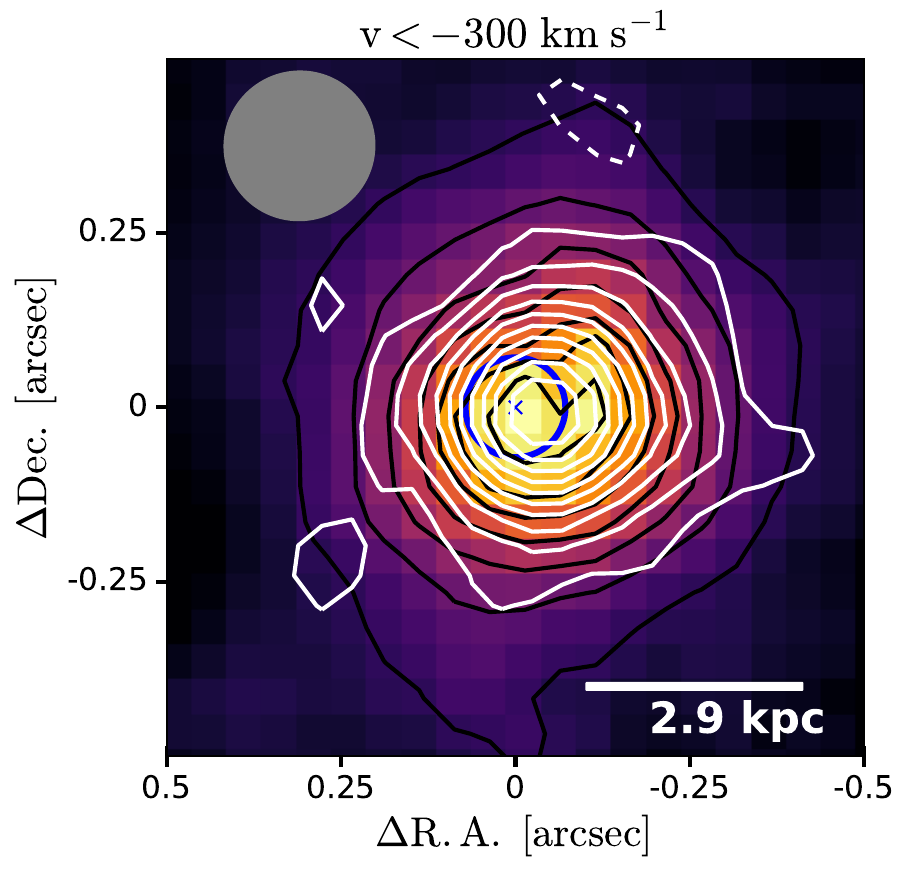}
\includegraphics[width=0.2\textwidth]{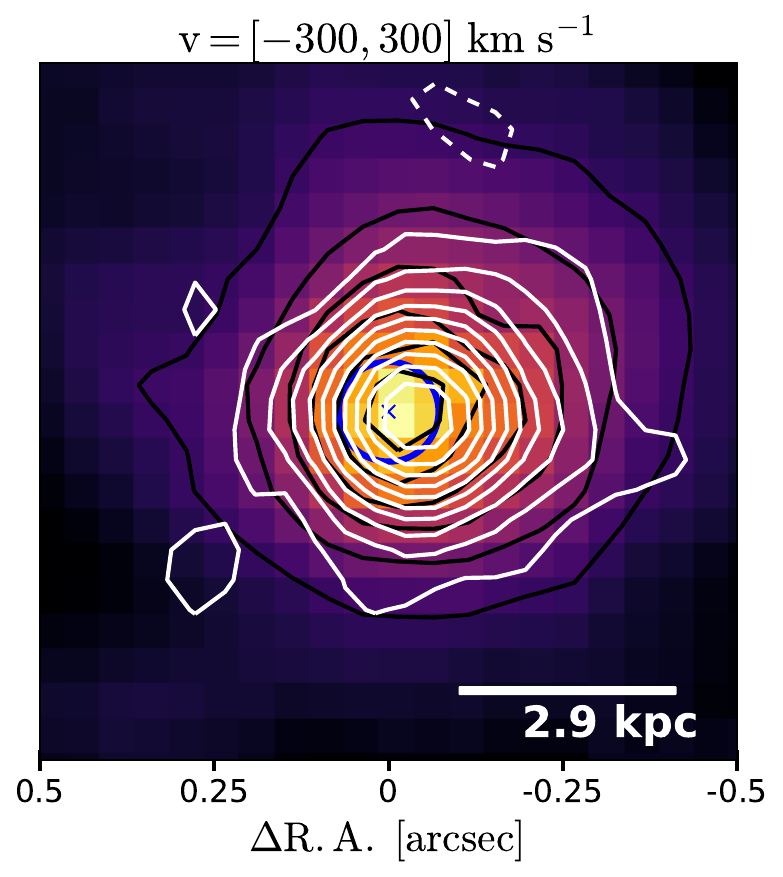}
\includegraphics[width=0.2\textwidth]{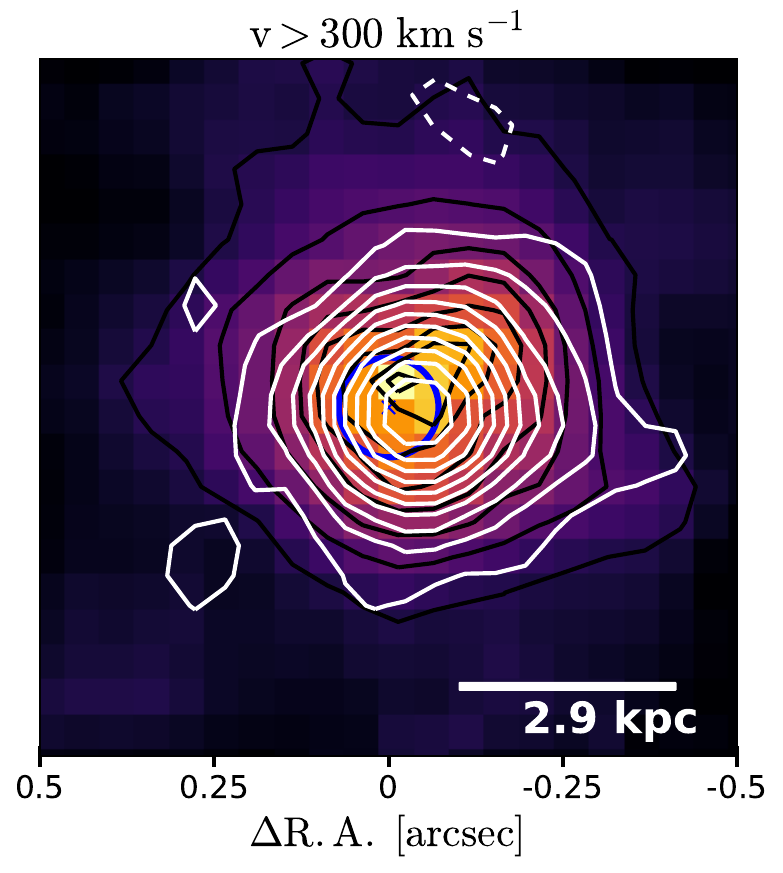}
\includegraphics[width=0.32\textwidth]{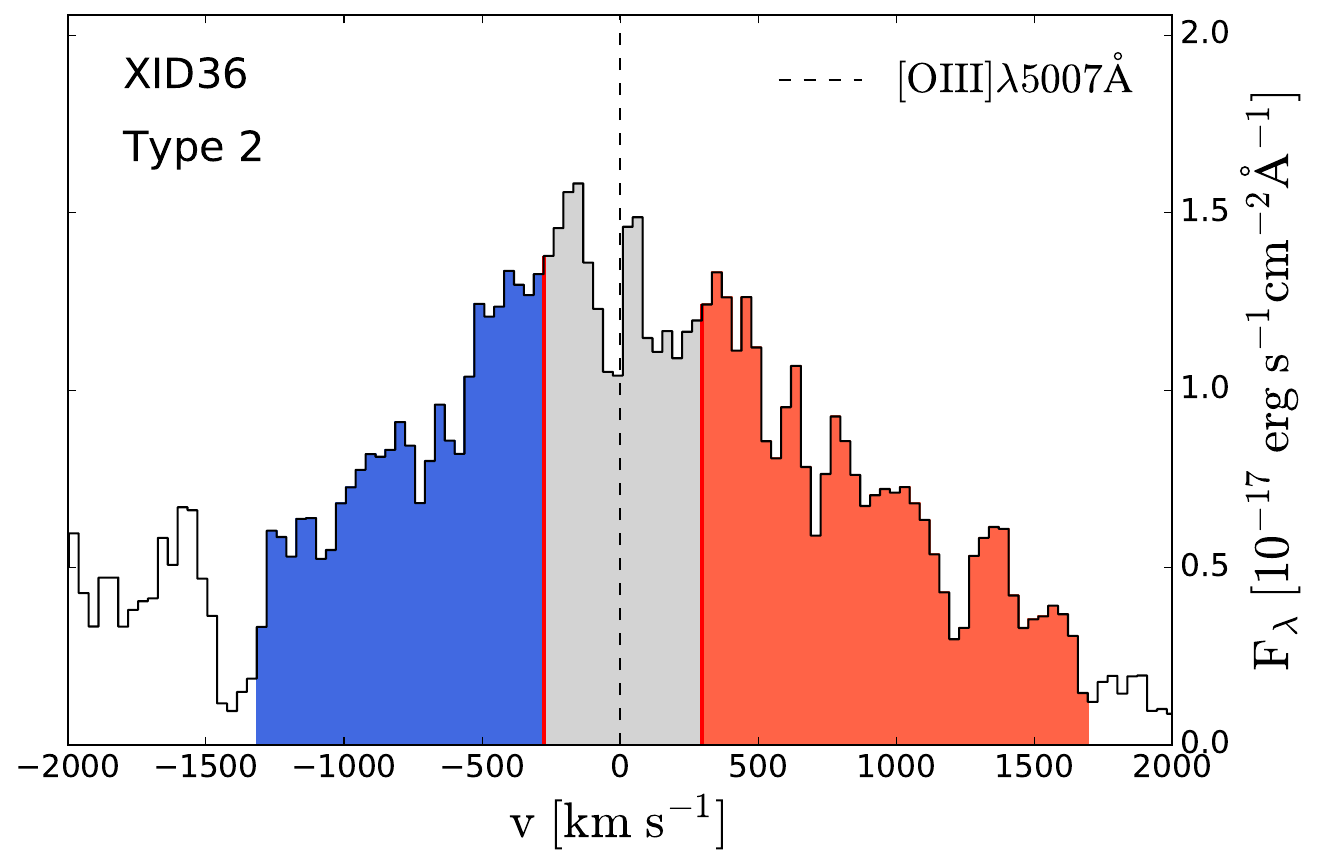}

\includegraphics[width=0.228\textwidth]{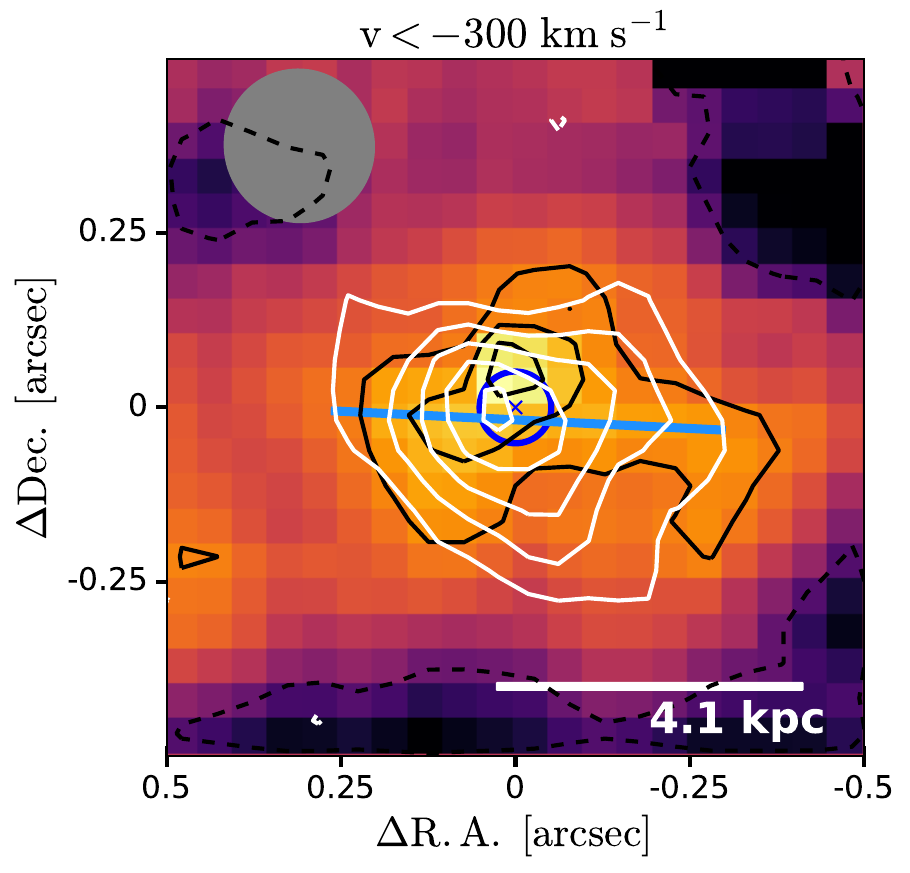}
\includegraphics[width=0.2\textwidth]{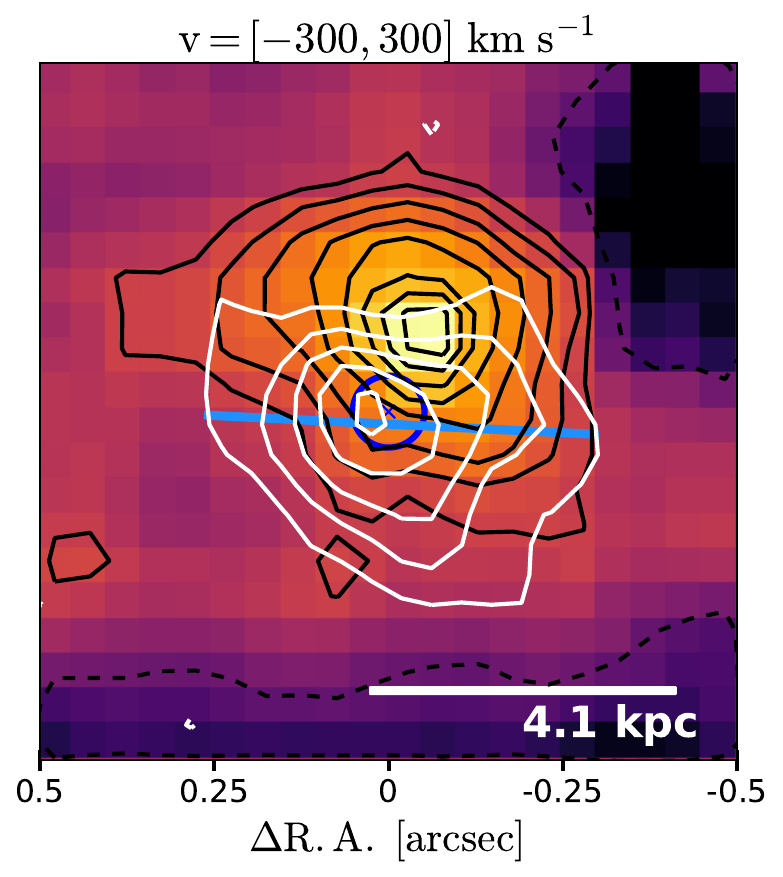}
\includegraphics[width=0.2\textwidth]{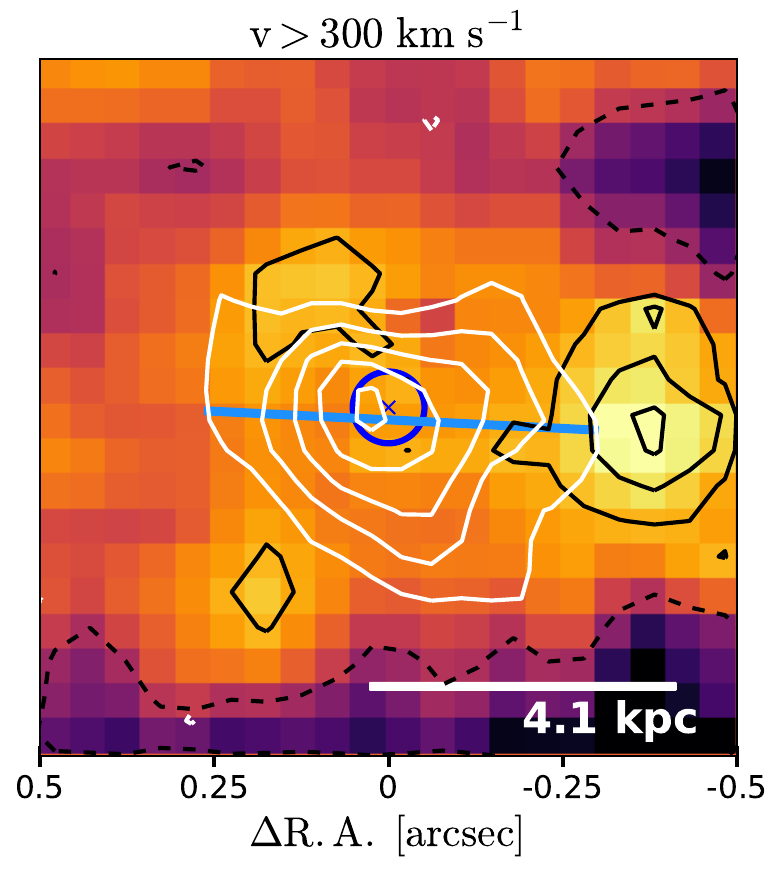}
\includegraphics[width=0.32\textwidth]{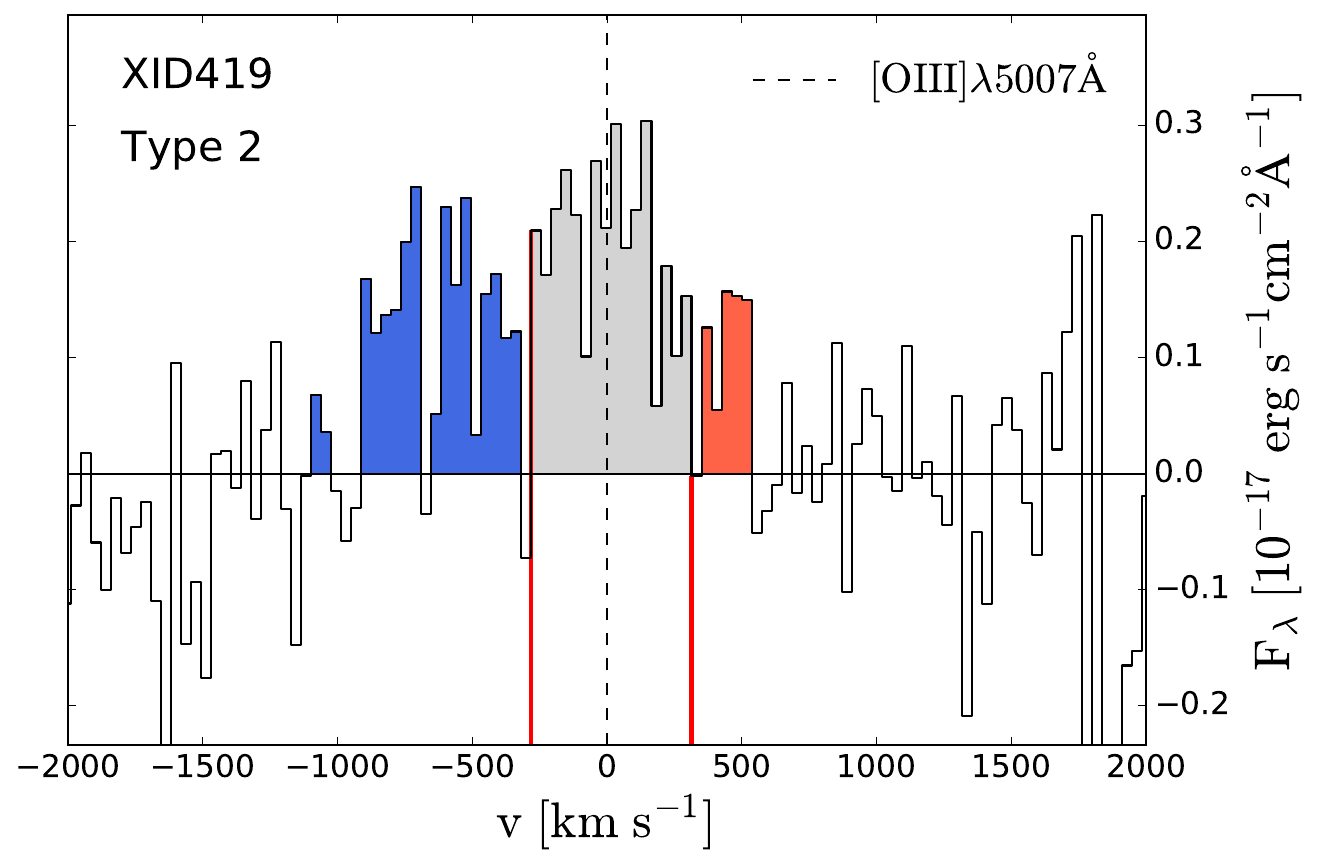}

\includegraphics[width=0.228\textwidth]{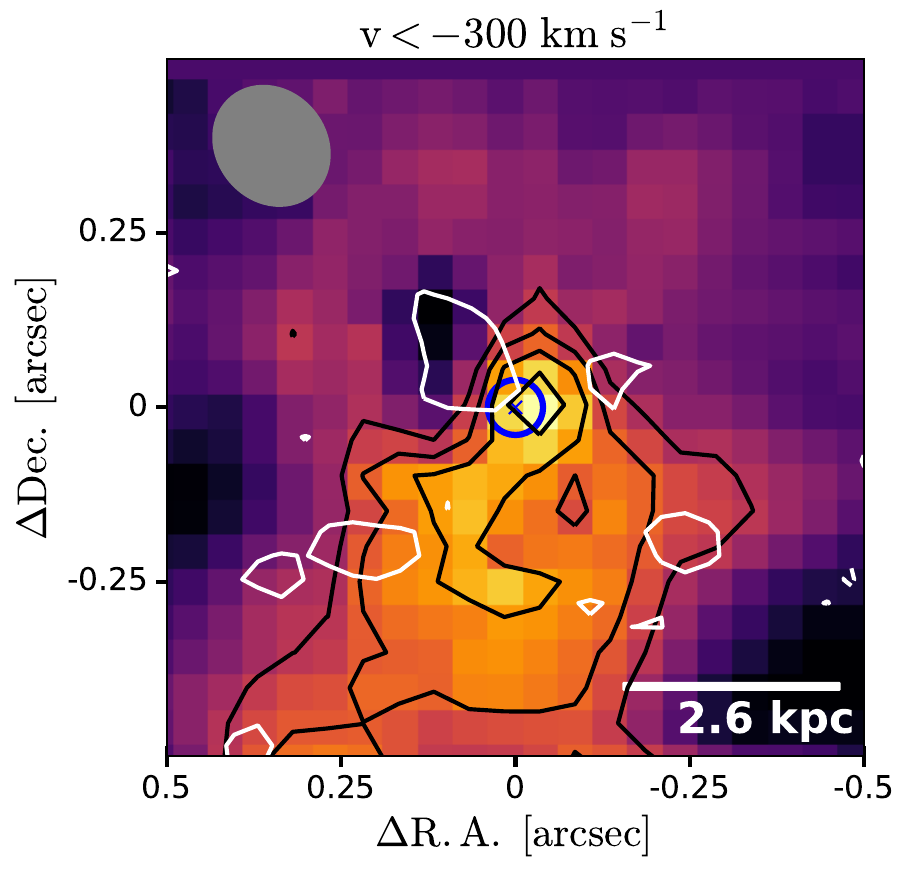}
\includegraphics[width=0.2\textwidth]{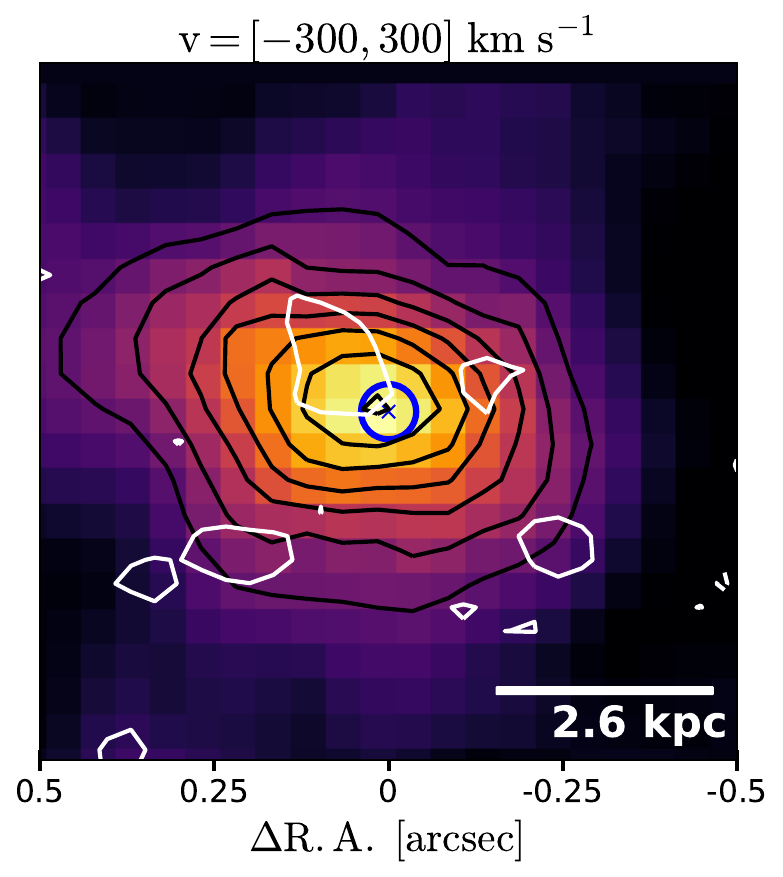}
\includegraphics[width=0.2\textwidth]{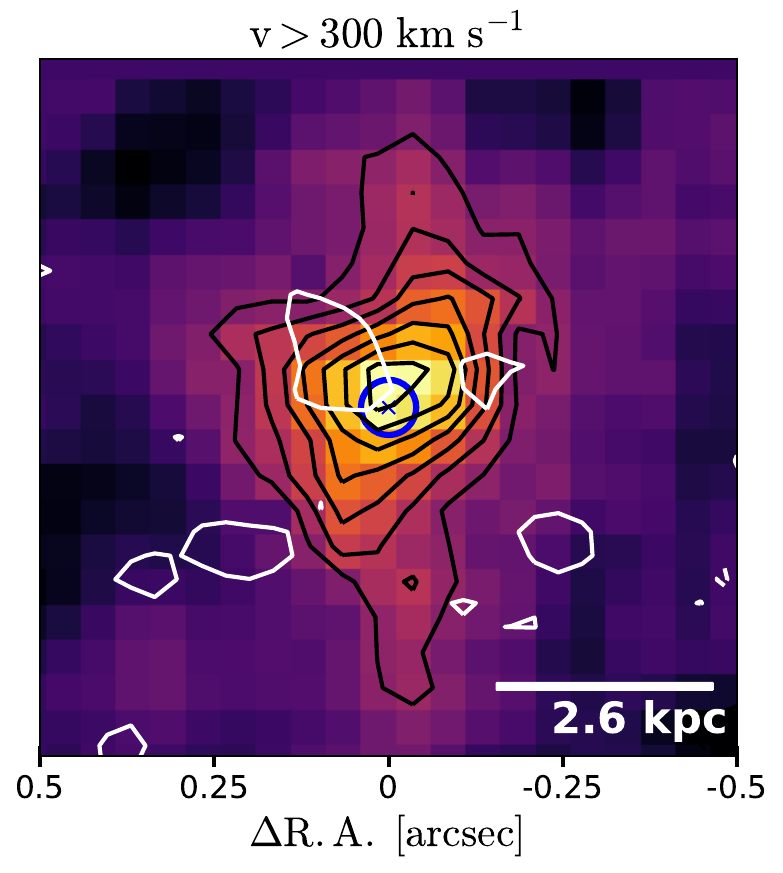}
\includegraphics[width=0.32\textwidth]{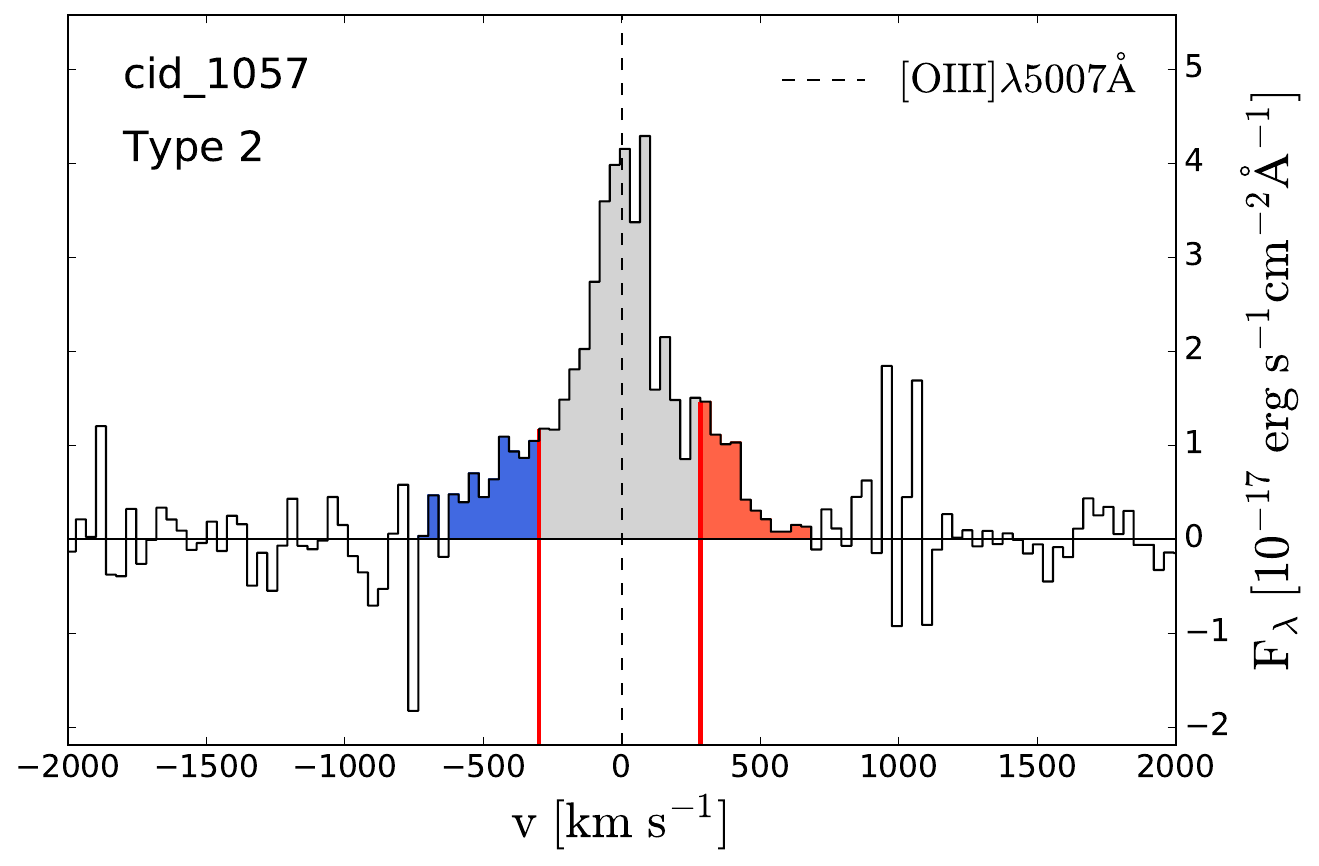}

\includegraphics[width=0.228\textwidth]{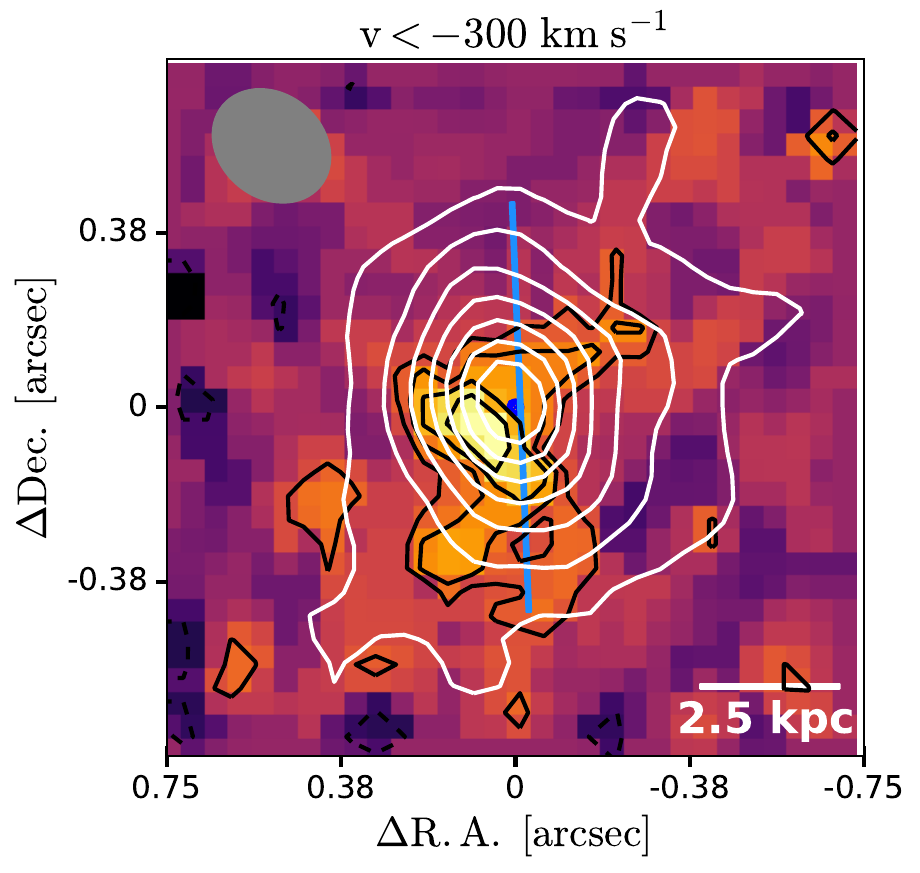}
\includegraphics[width=0.2\textwidth]{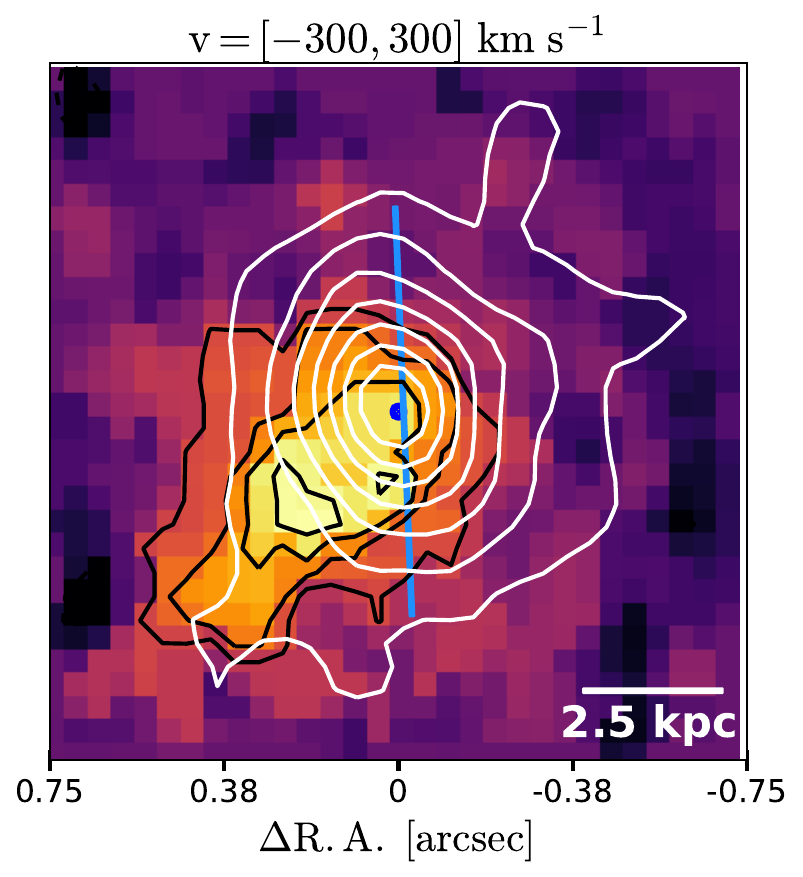}
\includegraphics[width=0.2\textwidth]{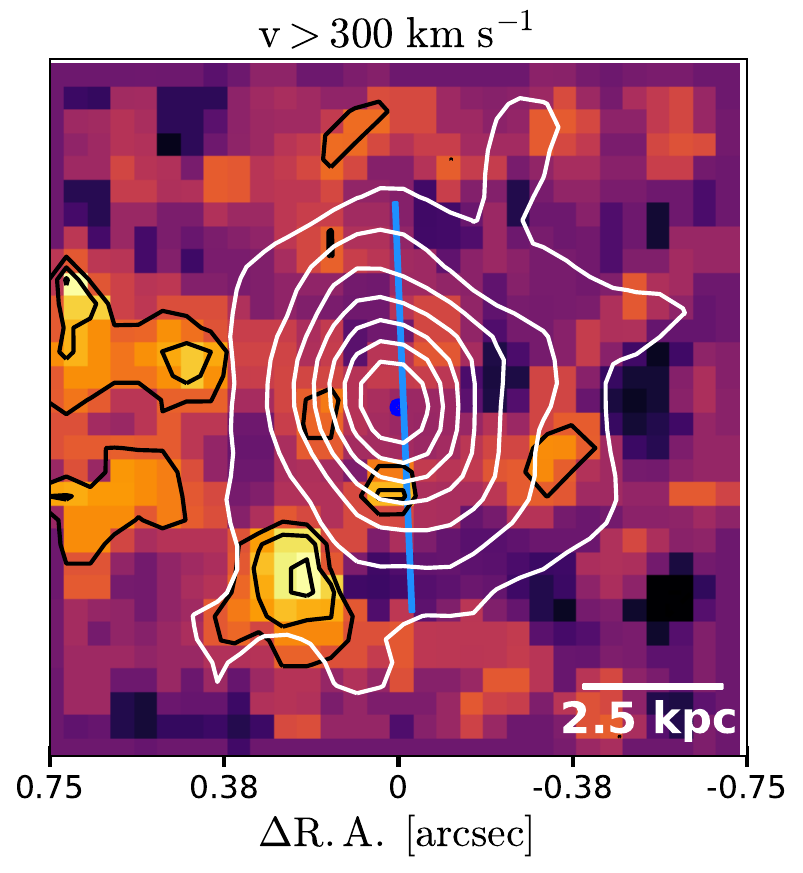}
\includegraphics[width=0.32\textwidth]{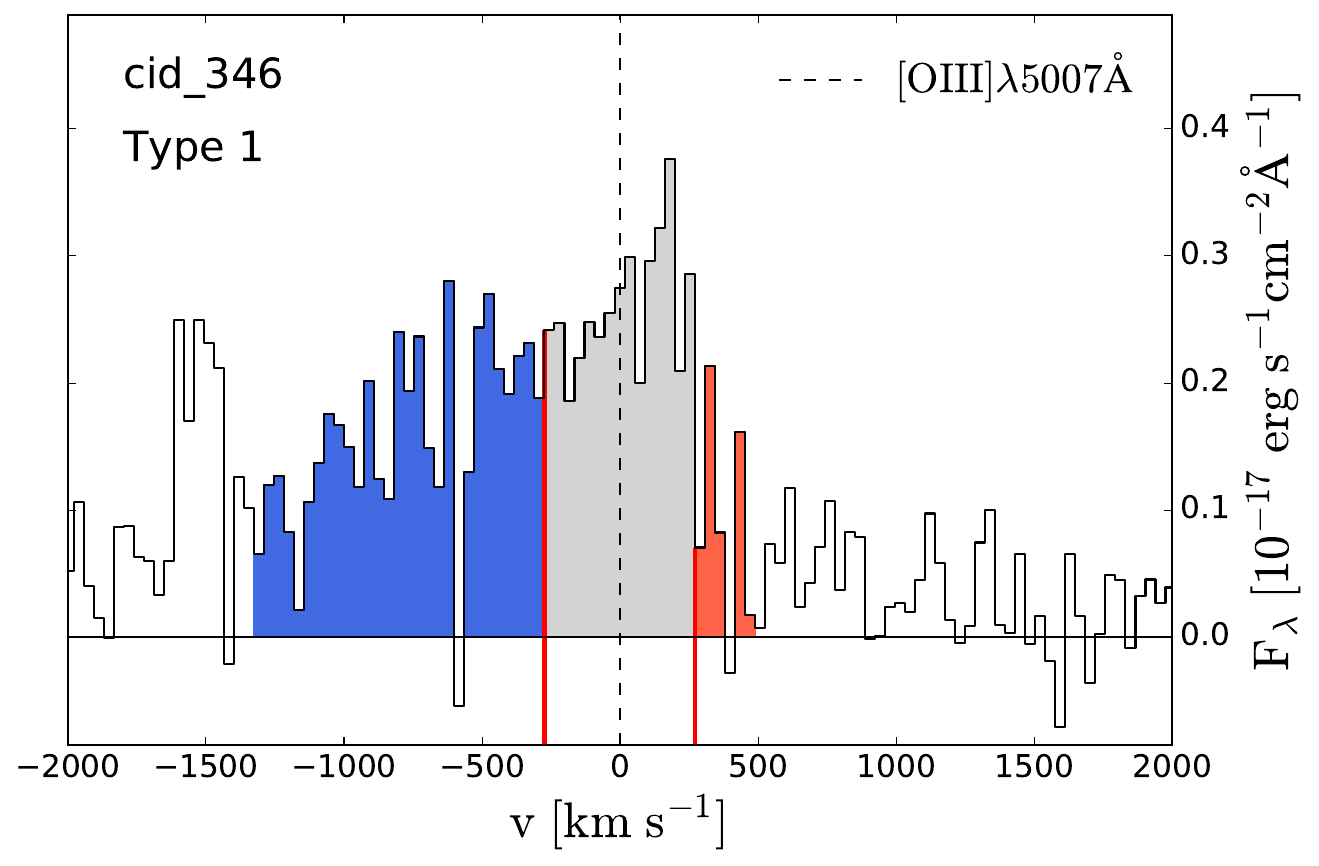}

\caption{Maps of \OIII\ emission in three different velocity channels: blue-shifted emission (< -300 \kms, \textit{first column}), central channel ([-300, 300] \kms, \textit{second column}), red-shifted emission (> 300 \kms, \textit{third column}). The interval over which the spectrum has been integrated is indicated on the right plot. In black are the \OIII\ emission contours  starting from 2$\sigma$ and increasing in intervals of 1$\sigma$ (and in intervals of 2$\sigma$ for XID36, cid\_451 and cid\_1143, to improve presentation).  In white are the FIR emission contours, starting from 2$\sigma$ and increasing in intervals of 2$\sigma$ (or intervals of 4$\sigma$ for \XN\ and cid\_346, to improve presentation). Negative -2$\sigma$ contours are shown with dashed curves. 
 The grey ellipse shows the size of the ALMA beam, while the white scale-bar shows the size of the PSF of the \OIII\ image in kpc.
 The blue cross and circle show the position and uncertainty of the optical continuum.
 The lightblue bar indicates the position angle along the major axis of the FIR emission, when it can be reliably determined (see Section~\ref{sec:FIR_size_morpho}).
\textit{Fourth column:} spectrum around the \OIII $\lambda 5007$ emission line. The coloured areas show the spectral regions over which the emission was integrated to create the three images on the left.}
\label{fig:OIII_maps}
\end{figure*}

\begin{figure*}\ContinuedFloat
\centering

\includegraphics[width=0.228\textwidth]{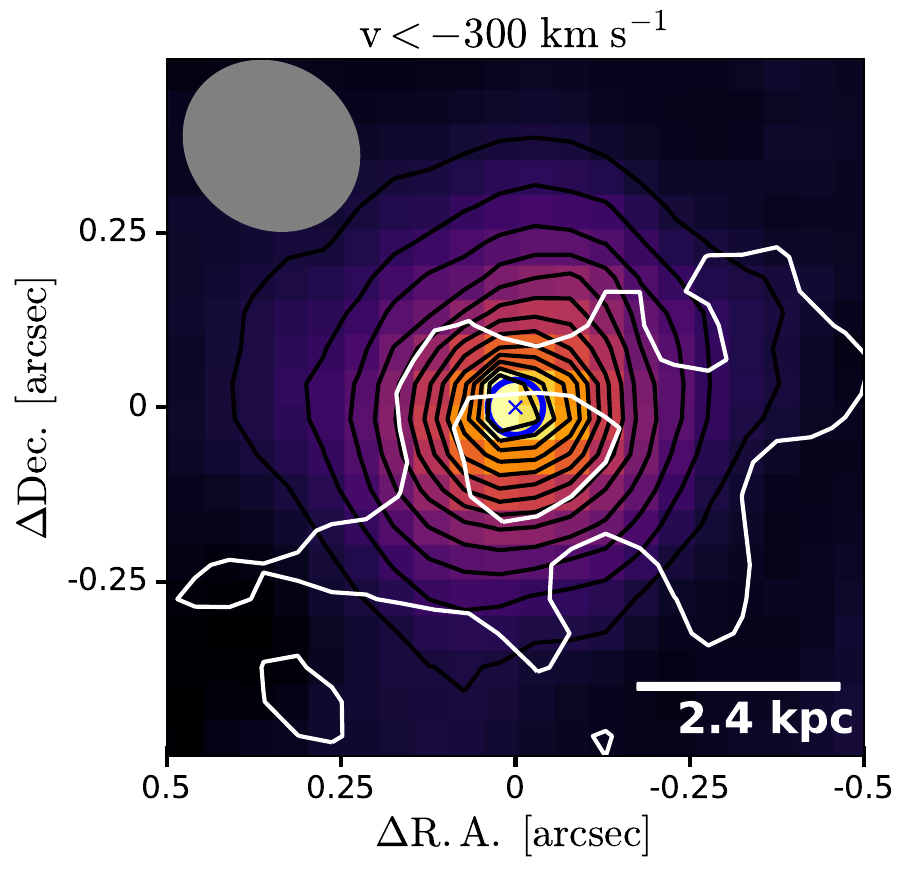}
\includegraphics[width=0.2\textwidth]{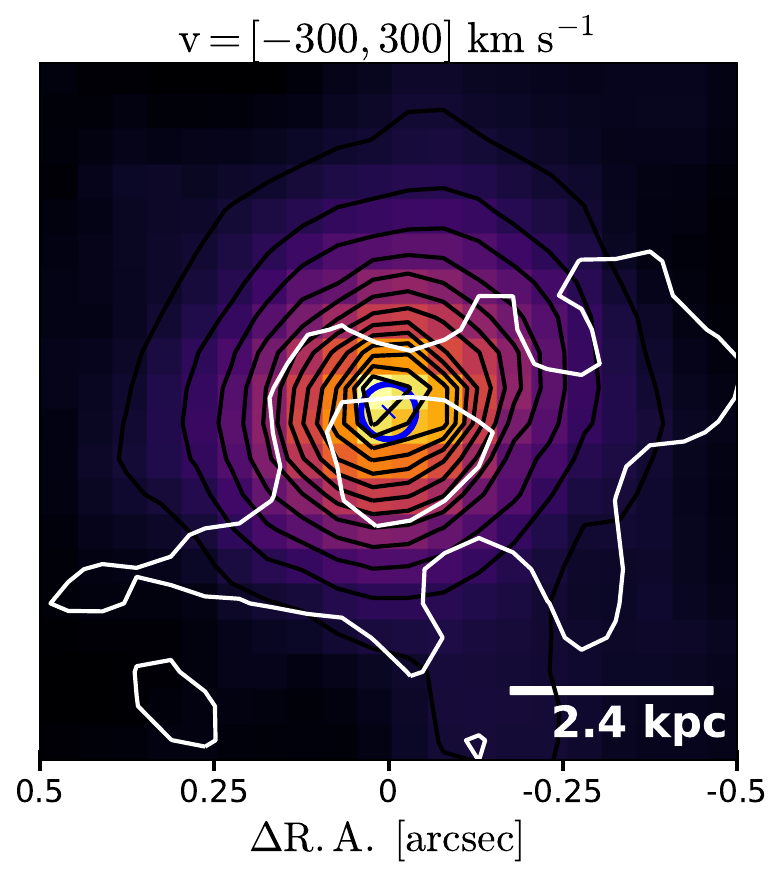}
\includegraphics[width=0.2\textwidth]{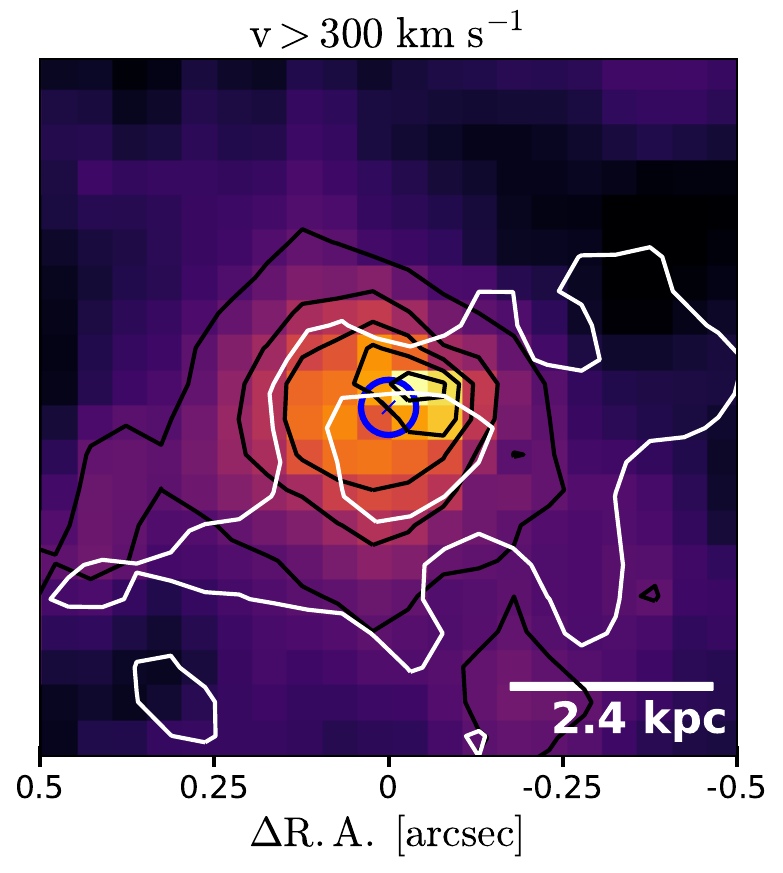}
\includegraphics[width=0.32\textwidth]{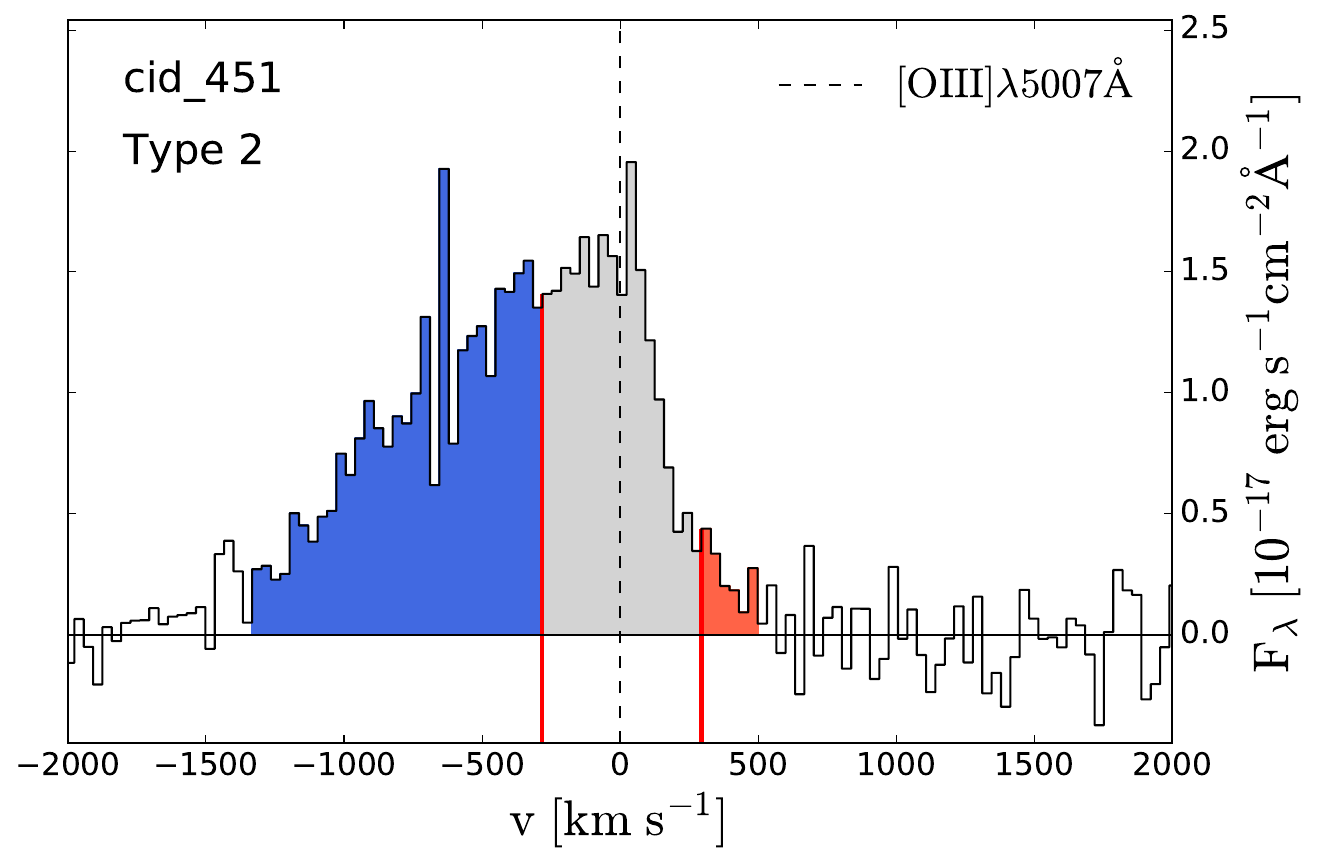}

\includegraphics[width=0.228\textwidth]{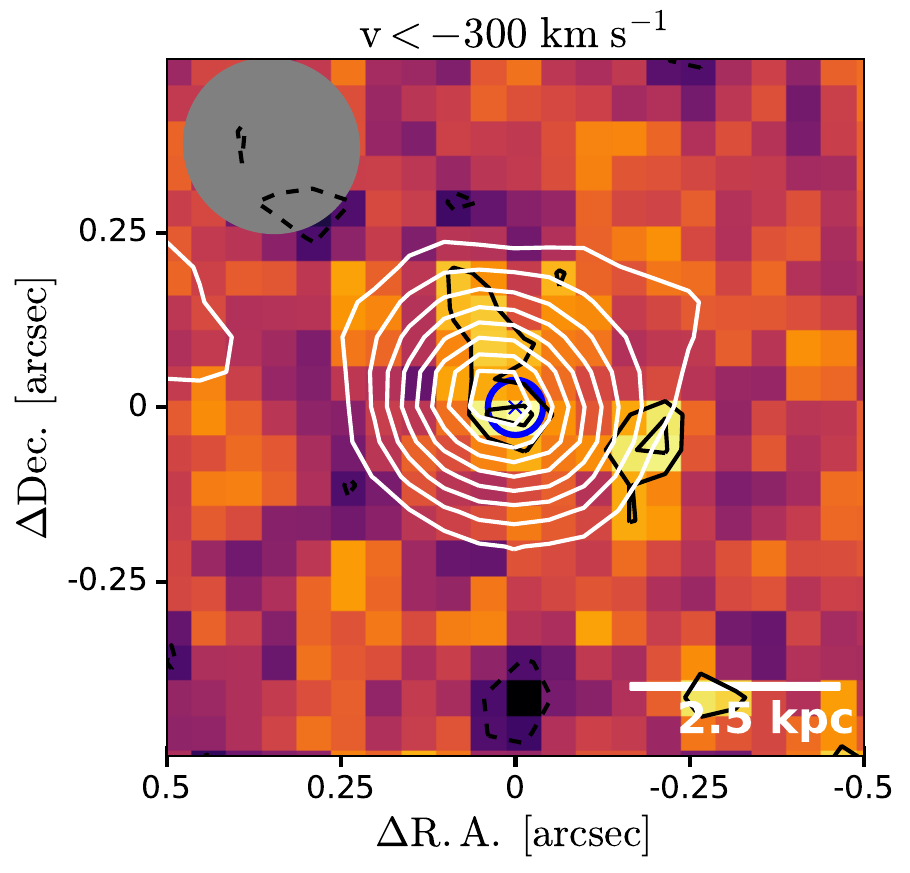}
\includegraphics[width=0.2\textwidth]{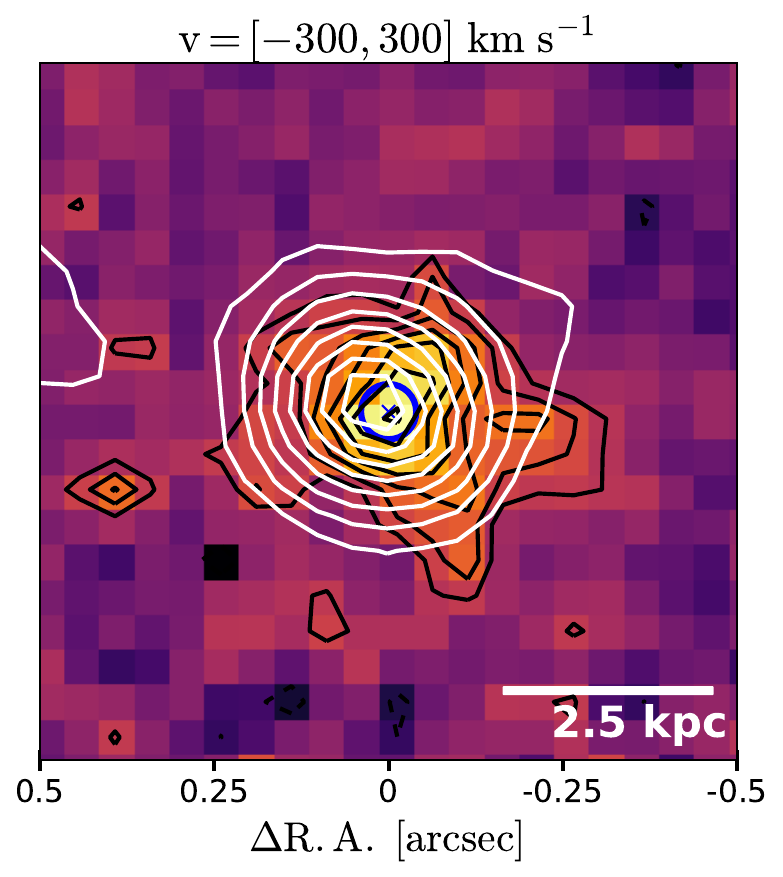}
\includegraphics[width=0.2\textwidth]{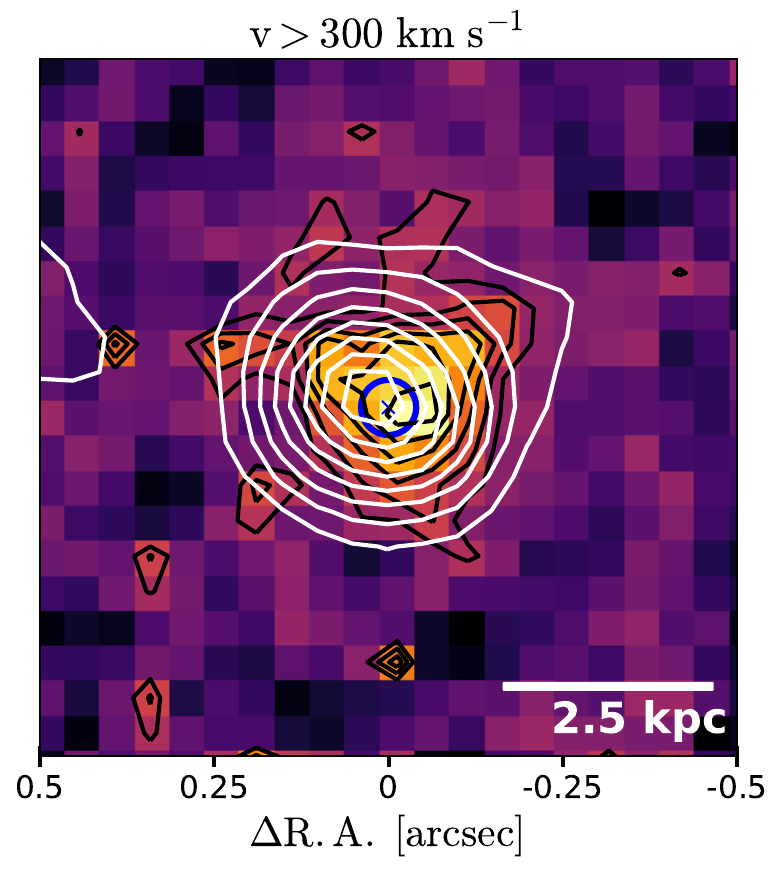}
\includegraphics[width=0.32\textwidth]{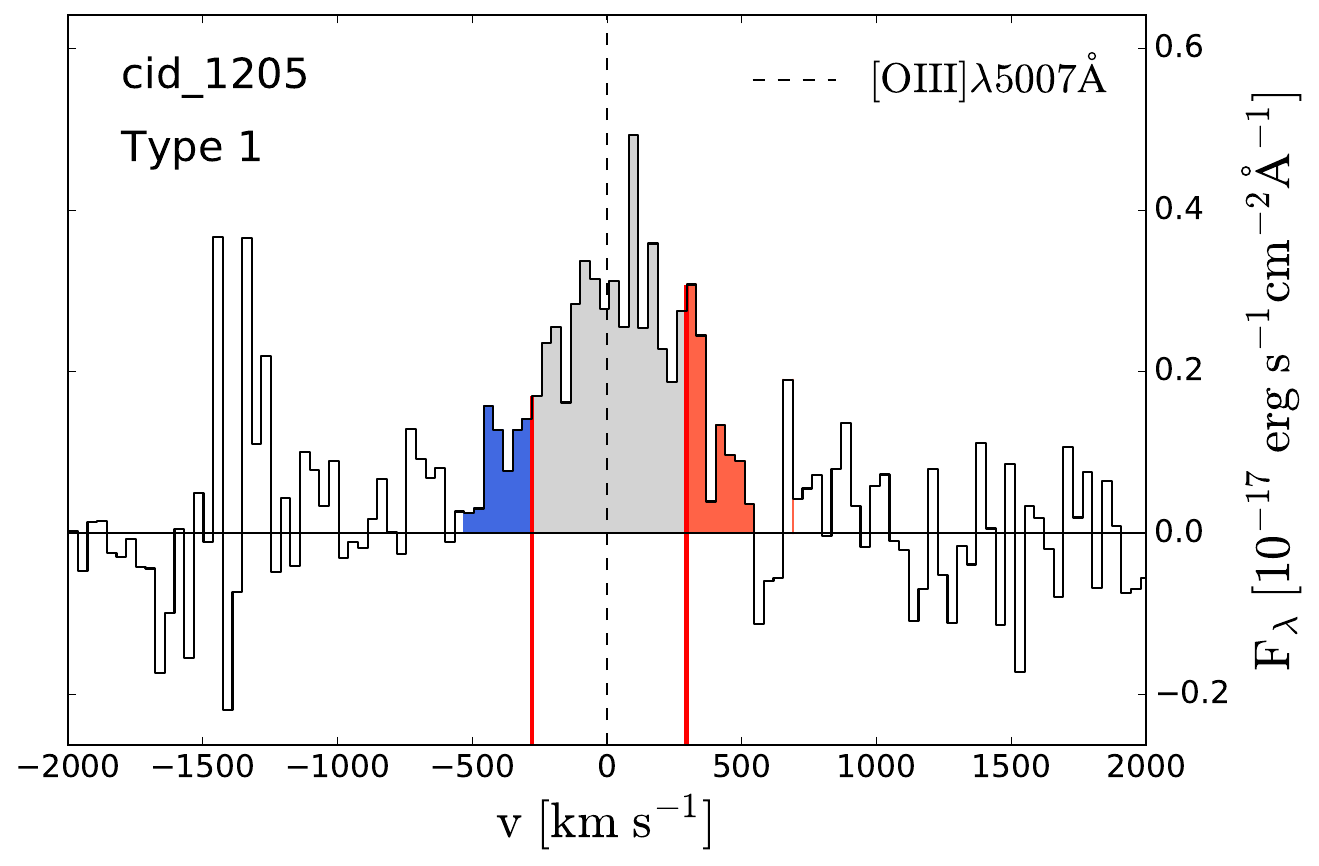}

\includegraphics[width=0.228\textwidth]{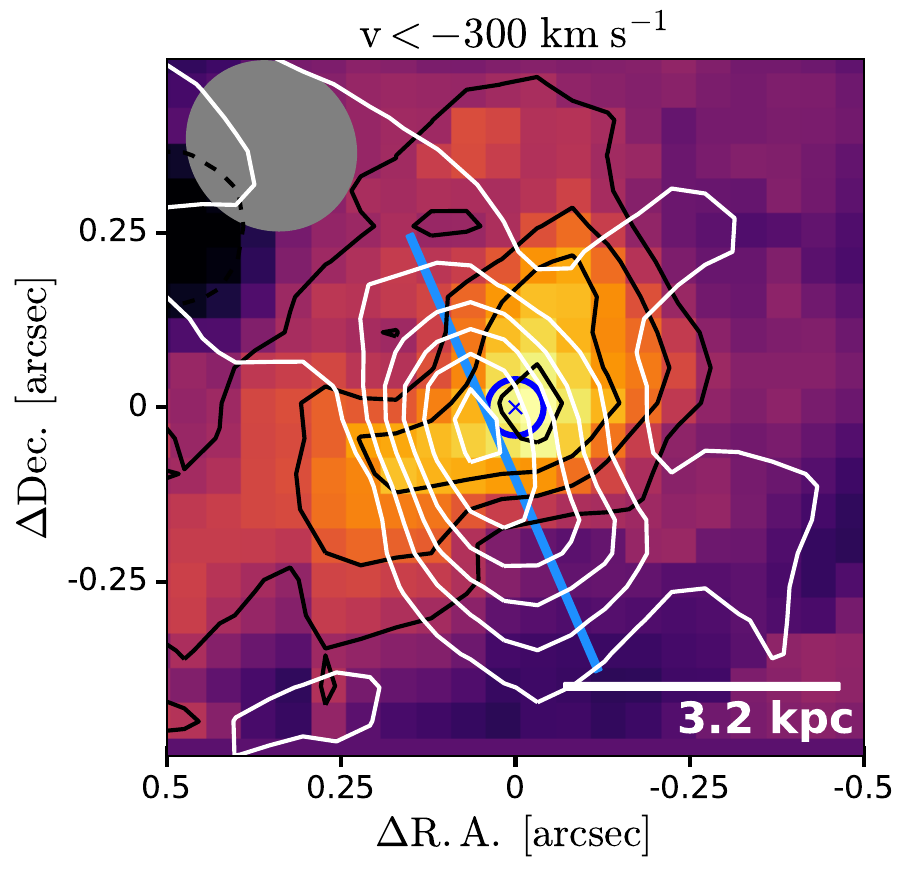}
\includegraphics[width=0.2\textwidth]{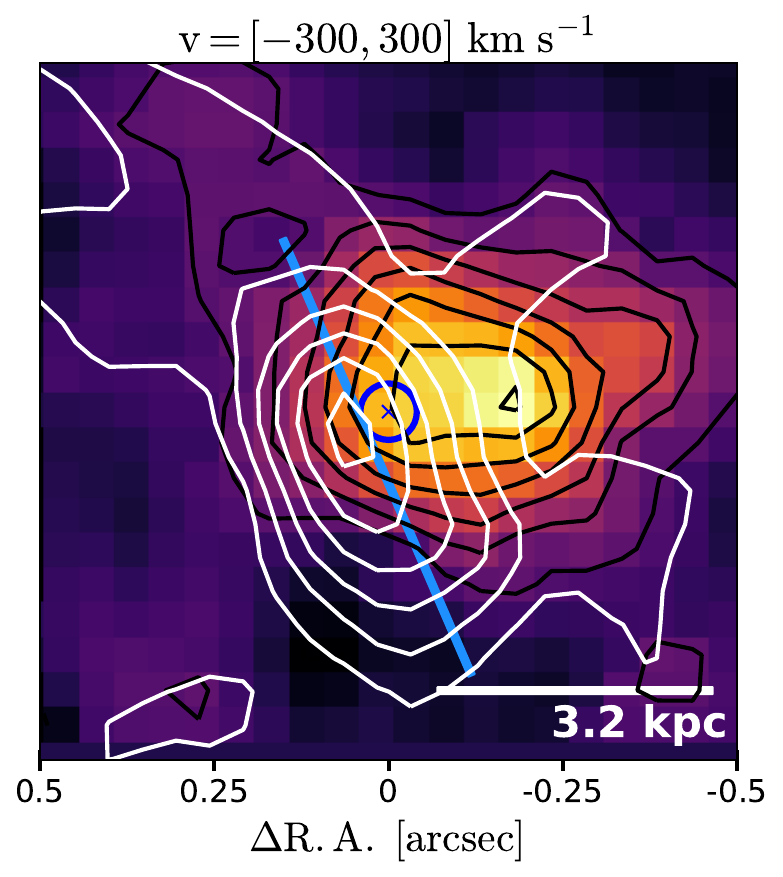}
\includegraphics[width=0.2\textwidth]{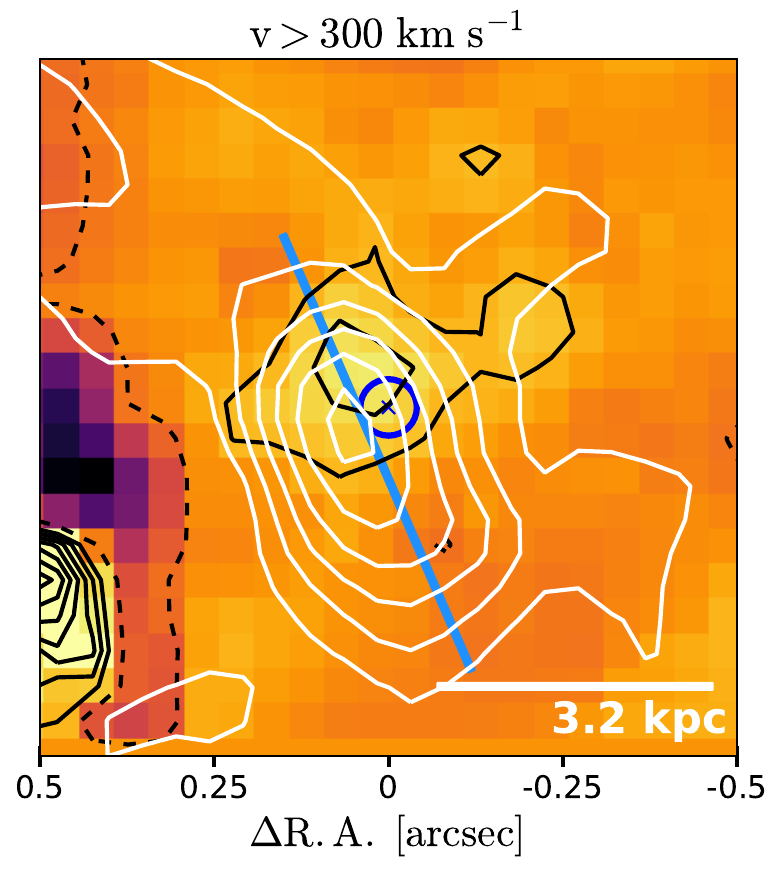}
\includegraphics[width=0.32\textwidth]{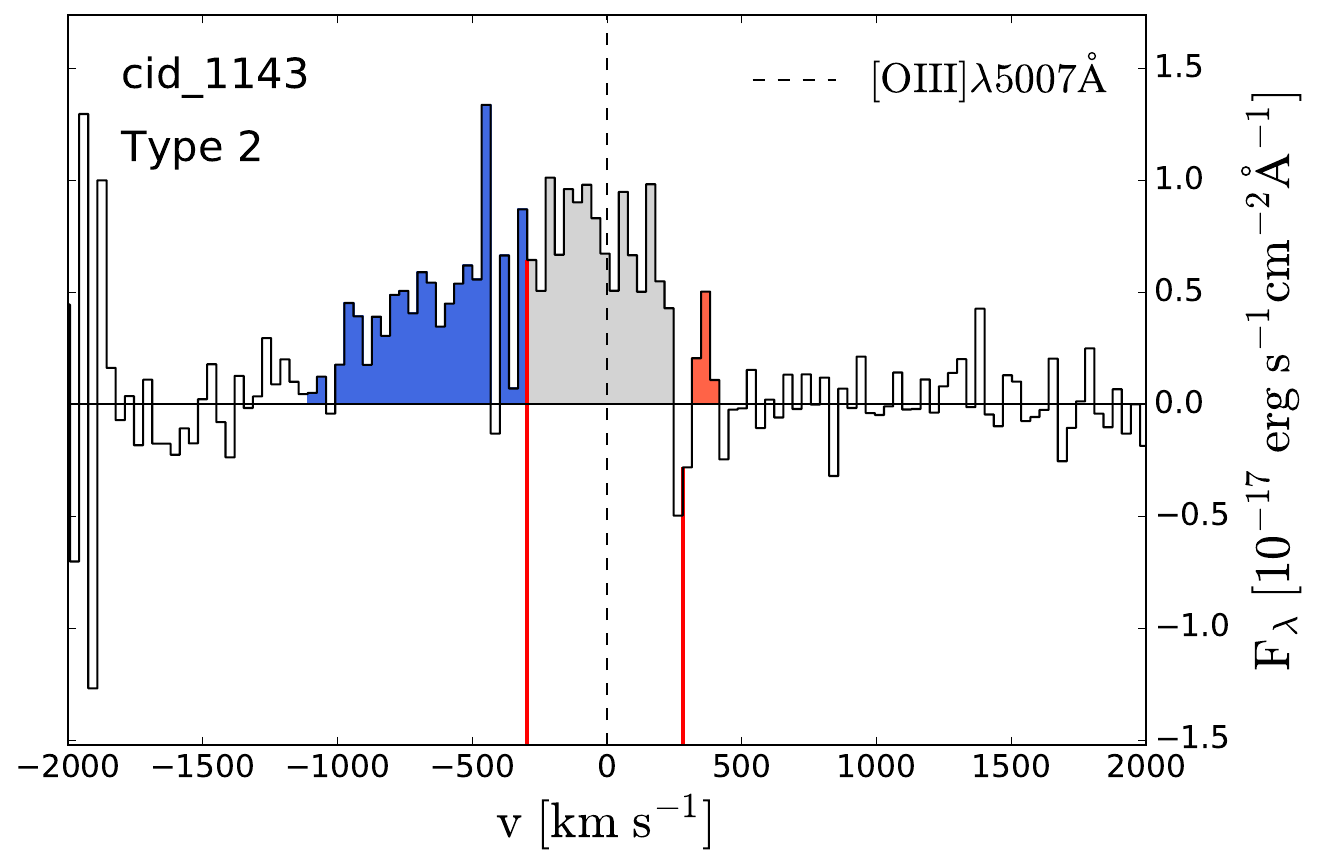}

\caption{Continued.}
\label{fig:OIII_maps_2}
\end{figure*}

\subsection{Implications of our results}
\label{sec:implication}
The main goal of this study is to trace with high spatial resolution the star formation in the host galaxies of AGN with powerful outflows, to assess the impact that these may have on the mass build-up of the galaxies.

Since our targets are powerful AGN, we expected the \Ha\ maps to be contaminated by AGN emission. Our BPT analysis confirms that for all but one target (cid\_346) the \Ha\ emission is AGN dominated. Thus, we cannot use \Ha\ as a star formation tracer in our sample and we have to rely only on the FIR, which traces obscured star formation.

In this study, we use the monochromatic FIR emission at $\sim260$~\micron\ to trace obscured star formation. This relies on the assumption that spatial variations of the shape of the FIR SED (due to changes in the dust properties, as for example the dust temperature and emissivity index $\beta$) are negligible. Under this assumption, the single FIR image can be translated into a map of the FIR luminosity and, consequently, of the obscured SFR. In order to test how spatial variations in the dust temperature may affect our results, we would need spatially resolved FIR observations at other wavelengths.

As described Section~\ref{sec:outflows}, we do not find any evidence of the ionised outflows instantaneously regulating the star formation in their host galaxies, for example, through cavities or `holes' in the FIR emission.
Similar results are reported by \citet{Balmaverde2016}, who study a sample of quasars at $z<1$ and find that quasars with strong outflows do not show lower SFR than those with weaker outflows.
Other spatially-resolved studies of AGN at $z\sim1.4-2.6$ also find no evidence of instantaneous star formation suppression due to the ionised outflows \citep[][]{Scholtz2020, Scholtz2021}.

When looking for evidence of AGN feedback on star formation, we need to consider the timescales of the star formation tracer we are using.
If the impact of the AGN on the level of star formation is on short timescales, then a star formation rate tracer too sensitive to the past star formation history of the galaxy may dilute the feedback signature.
The FIR is commonly believed to trace star formation on timescales of  $\sim100$~Myr. However, some studies find no significant difference in the SFR measured from the FIR and from \Ha\ in star-forming galaxies \citep[e.g. ][]{Lam2013, Dominguez-Sanchez2014, Rosario2016}, which is believed to measure SFR on shorter timescales ($\sim10$~Myr). 
 This would suggest that the FIR  traces similar star formation timescales as the \Ha\ emission. However, this result may depend on the star formation history of the individual  galaxies. 
In this context, it would then be very interesting to have an alternative and reliable tracer of the star formation on short timescales. Standing the limitations of using \Ha\ or the UV emission for AGN hosts, an alternative could be to use the mid-infrared PAHs features which would be accessible at these redshifts with JWST.

On the other hand, it is also possible that the outflow needs some time to produce an effect on the star formation. Thus, there is a delay between the moment the outflow is visible and when the star formation decreases.
Episodes of AGN activity happen on shorter timescales compared to timescales of star formation variations  \citep{Gabor2014, Hickox2014}. 
 The outflow may need more time to remove a substantial quantity of gas from the host galaxy and cause a suppression of star formation. 
Simulations by \citet{Gabor2014} show that AGN-driven outflows do not cause instantaneous quenching, but they may remove a substantial amount of gas on long timescales ($\gtrsim$~Gyr).
 Recently, \citet{Costa2020} found that AGN feedback acts in two modes in their simulations: a rapid mode that removes dense gas from the nucleus, and a slower mode that  prevents halo gas accretion. The first mode alone will not cause a decrease in the star formation, but the combination of the two modes will suppress star formation in the long-term.
 Some observations show that high-redshift AGN host galaxies may have lower molecular gas fractions, the fuel for star formation, compared to mass-matched star-forming galaxies at high redshift \citep[e.g. ][] {Kakkad2017, Circosta2021}. Over time, this should be visible as an impact on the star formation itself.

Moreover, we need to consider also the physical scales that we are sampling in this study. We do not see signs of star formation suppression at the scales probed by our observations ($\geq2$~kpc), but it is possible that the impact of the outflows is only visible at smaller scales. Outflows may influence star formation only in a small region of the galaxy, on scale $< 1$~kpc \citep[e.g. ][]{Croft2006, Alatalo2015, Cresci2015a, Querejeta2016, Rosario2019, Shin2019, Husemann2019}.

There is also the possibility that the interaction between the outflow and the ISM is limited. We observe that, at least in two targets (cid\_346 and cid\_1143),  the ionised gas is located preferentially perpendicular to the dust. This may indicate that the outflow is propagating following the path of least resistance. Simulations have shown that AGN-driven outflows propagate preferably away from the plane of the galaxy, avoiding dense gas regions in the galactic disk \citep{Gabor2014, Costa2014, Mukherjee2016}.

To interpret our results, we also need to consider how representative of the parent population our sample is.
 Due to the requirement to have FIR detections, our targets tend to have higher SFRs compared to the average of the parent SUPER sample, in which there are many SFR upper limits (see Figure~\ref{fig:SUPER_sample}).  Additionally, our sample has a higher SSFR compared to the average of the parent population of X-ray AGN at $z\sim2$ \citep{Scholtz2018}.
It is possible that if we were to target AGN with lower SFRs, we would be more likely to observe localised suppression of star formation, since the total star formation level is lower.
To test this hypothesis, it  would require deep ALMA continuum observations  to get spatially resolved maps of the dust continuum emission for the whole SUPER sample.

\section{Summary and conclusions}
\label{sec:conclusions}
We present ALMA Band 7 observed-frame 870~\micron\ (rest-frame $\sim 260$~\micron) continuum high-resolution ($\sim$0.2'', corresponding to $\sim$2~kpc) observations of eight X-ray AGN at redshift $z\sim$2 from the SUPER sample.
The ALMA targets were selected from the parent SUPER sample based on  photometric detections in the FIR (i.e. observed wavelength $24-870$ \micron) and \OIII\ detections in the SINFONI IFS maps. The selected sample has a range of bolometric luminosities which is representative of the parent sample ($L_{bol}= 10^{44.9}-10^{46.8}$ erg s$^{-1}$). They have SFRs in the range $8-380$ $M_\odot$ yr$^{-1}$ and  most of the targets lie near the star formation main-sequence, with the exception of cid\_1143 that lies  below (see Figure~\ref{fig:SUPER_sample}).
The main conclusions of this work are:
\begin{itemize}

\item We detect 6/8 of our targets  with S/N$>8$ in the ALMA rest-frame 260~\micron\ continuum maps. The rest of the conclusions are based on these six high S/N targets. We model the 260~\micron\ data in the visibilities versus $uv$-distance space using different models (point source, Gaussian, exponential profile, Gaussian+point source). From the `best fit' models, we measure flux densities in the range $0.27-2.58$~mJy and half-light radii  in the range $R_e= 0.83-2.01$ kpc (median $1.31\pm0.23$~kpc)  (Section~\ref{sec:ALMA_fluxes_sizes}). \\

\item   From the SED decomposition, we find that the contribution of AGN-heated dust to the total rest-frame 260~\micron\ emission is $\leq 4$\%  in our sample.
The contribution from synchrotron emission from AGN is small in most of the targets ($< 1$\%). 
We conclude that the main contribution to the 260~\micron\ flux is due to dust heated by star formation (see Section~\ref{sec:origin}).\\

\item We compare the FIR sizes of our sample with other samples of SFGs and AGN at similar redshift, for which the FIR sizes are measured from ALMA (observed-frame) 870~\micron\ observations of similar resolution and sensitivity (see Section~\ref{sec:FIR_sizes_lit}, Figure~\ref{fig:comp_sizes}). Our sample is in agreement with literature samples of redshift-matched star-forming and AGN host galaxies.
Across these samples, we find that the mean FIR size of AGN ($R_e = 1.16\pm 0.11$~kpc) is smaller than the mean FIR size of non-AGN ($R_e = 1.69\pm 0.13$~kpc). A possible interpretation is that a compact dust/gas configuration favours the accretion to the central SMBH.\\

\item We use the redshift-dependent BPT diagnostic diagram from \citet{Kewley2013a,Kewley2013b} to identify the main process responsible for gas ionisation in our sample (Figure~\ref{fig:BPT}). 
 Most of our objects lie in the AGN region, with the exception of cid\_346 (see Section~\ref{sec:BPT}).
We find a good agreement between the spatial distribution of \Ha\ and \OIII\ in most of the targets (see Section~\ref{sec:ALMA_SINFONI_comp}, Figure~\ref{fig:Ha_OIII_images}). This suggests that the same mechanism responsible for the \OIII\ emission, that is the AGN, is also partly responsible for the \Ha\ emission. 
Only two objects (\XN\ and cid\_346) show a significant offset between the peak of \Ha\ and \OIII\ emission, which could be a sign that there are different contributions from the ionisation sources or it could be due to differential obscuration. 
In summary, across our sample we do not identify any \Ha\ emission that we can confidently use to trace star formation. \\

\item We observe  different spatial distributions of ionised gas and FIR emission (Figure~\ref{fig:Ha_OIII_images}, Figure~\ref{fig:comp_coord_SINFONI}). Specifically, in most of our targets (4/6) there is a significant offset between the central position of the ionised gas emission and the FIR emission ($0.4-1.9$~kpc).
We also find that the ionised gas emission tends to be larger than the FIR emission by a factor of $\sim1.7$ (Figure~\ref{fig:comp_size_SINFONI}). 
Most strikingly, in two sources (cid\_1143 and cid\_346) we observe that the ionised gas is perpendicular to the dust emission. All of these observations provides further evidence that dust and ionised gas emission are not directly associated in our sample, with the dust most likely to be tracing the host galaxy and the ionised gas most likely to be tracing the AGN emission-line regions.\\

\item Comparing the position of the \OIII\ blue-shifted emission with the FIR emission (Figure~\ref{fig:OIII_maps}), we find that the FIR emission is unaffected by the ionised outflow. 
Assuming that the FIR is tracing the obscured star formation, we see no evidence for star formation suppression due to the ionised outflows at the scale probed by our observations ($\sim 2$ kpc).
However, we do not have a resolved map of an alternative and reliable tracer of only recent star formation.
 Additionally, the outflow may need longer timescales to significantly affect star formation.
  In at least two objects (cid\_346 and cid\_1143), the outflow is almost perpendicular to the direction of the major axis of the FIR emission, which can be an indication that the outflow propagates following the path of least resistance.
\end{itemize}

We do not observe any evidence in this study for ionised outflows directly influencing star formation. However, one limitation is that our sample may be biased to relatively high SFRs.  To confirm that the ionised outflows do not have an impact on star formation, we need to obtain spatially-resolved FIR maps also for the SUPER targets with lower SFRs. 
Furthermore, future IFS observations with higher spatial resolution (e.g. with ELT/HARMONI) will help to get spatially-resolved emission-line ratio diagnostics that will allow us to map and identify the ionisation sources of \Ha\ in a spatially-resolved way.

\begin{acknowledgements}

We thank the anonymous referee for carefully reading the paper and providing constructive comments.
 We thank  Ian Smail and Bitten Gullberg for helping us to understand the data of the AS2UDS sample. 
 I.L. acknowledges support from the Comunidad de Madrid through the Atracci\'on de Talento Investigador Grant 2018-T1/TIC-11035.
 M.P. is supported by the Programa Atracci\'on de Talento de la Comunidad de Madrid via grant 2018-T2/TIC-11715.  C.C.C. acknowledges support from the Ministry of Science and Technology of Taiwan (MOST 109-2112-M-001-016-MY3).
 D.M.A. and D.J.R. acknowledge support from STFC (ST/T000244/1).
 A.P. gratefully acknowledges financial support from STFC through grants ST/T000244/1 and ST/P000541/1. 
 G.V. acknowledges financial support from Premiale 2015 MITic (PI B. Garilli).

This paper is based on observations collected at the European organisation for Astronomical Research in the Southern Hemisphere under ESO programme 196.A- 0377.
This paper makes use of the following ALMA data: ADS/JAO.ALMA$\#$2018.1.00992.S. ALMA is a partnership of ESO (representing its member states), NSF (USA) and NINS (Japan), together with NRC (Canada), MOST and ASIAA (Taiwan), and KASI (Republic of Korea), in cooperation with the Republic of Chile. The Joint ALMA Observatory is operated by ESO, AUI/NRAO and NAOJ.
Based in part on data products produced by TERAPIX and the Cambridge Astronomy Survey Unit on behalf of the UltraVISTA consortium.
 Based on observations obtained as part of the VISTA Hemisphere Survey, ESO Progam, 179.A-2010 (PI: McMahon). 
This work has made use of data from the European Space Agency (ESA) mission {\it Gaia} (\url{https://www.cosmos.esa.int/gaia}), processed by the {\it Gaia} Data Processing and Analysis Consortium (DPAC,
\url{https://www.cosmos.esa.int/web/gaia/dpac/consortium}). Funding for the DPAC has been provided by national institutions, in particular the institutions participating in the {\it Gaia} Multilateral Agreement.

The {\tt Starlink} software \citep{Currie2014} is currently supported by the East Asian Observatory.
This research has made use of the NASA/IPAC Extragalactic Database (NED) which is operated by the Jet Propulsion Laboratory, California Institute of Technology, under contract with the National Aeronautics and Space Administration.
This research made use of Astropy, a community-developed core Python package for Astronomy \citep{astropy}, {\tt Matplotlib} \citep{Hunter2007} and {\tt NumPy} \citep{VanDerWalt2011}.  
This research used the {\tt TOPCAT} tool for catalogue cross-matching \citep{Taylor2005}, the {\tt Stan} interface for Python {\tt PyStan} \citep{pystan}, and {\tt APLpy}, an open-source plotting package for Python \citep{Robitaille2012}.

\end{acknowledgements}

%
%
\bibliographystyle{aa} 
\bibliography{SUPER_paper.bib}


\begin{appendix} 

\section{Models used to fit the FIR profile}
To assess the morphology of the FIR emission, we fit the visibilities versus $uv$-distances using the following models: point source, Gaussian, Gaussian+point source, and exponential profile. We Fourier-transformed the models to perform the fit in the visibilities versys $uv$-distance plane. The models in the visibilities versus $uv$-distance plane are defined as:
\begin{itemize}
\item \textit{Point source}: a point source is represented as a constant model as a function of $uv$-distance:
\begin{equation}
f(x, F) = F ,
\end{equation}
where  $F$ is the flux density in mJy and  $x$ is the $uv$-distance in units of k$\lambda$, where $\lambda$ is the wavelength of the observation.\\

\item \textit{Gaussian profile}: the Fourier transform of a Gaussian model is also a Gaussian defined as:
\begin{equation}
f(x, F, \sigma_{uv}) = F\cdot \exp\left({-\frac{1}{2}\left(\frac{x}{\sigma_{uv}} \right) ^2}\right),
\end{equation}
where  $F$ is the flux density in mJy and $\sigma_{uv}$ is the scale parameter. $\sigma_{uv}$ is related to the effective radius $R_e$ (in radians) as:
\begin{equation}
R_e \text{[rad]} = \frac{FWHM}{2} = \frac{2.355}{2}\cdot \sigma = \frac{2.355}{2}\cdot \frac{1}{2\pi\sigma_{uv}\cdot10^{3}},
\end{equation}
where the factor $10^3$ is necessary to convert from k$\lambda$ to $\lambda$. The effective radius in arcsec is obtained as $R_e \text{[arcsec]} =R_e \text{[rad]} \frac{180\cdot 3600}{\pi}  $.\\

\item \textit{Gaussian+point source}: this model combines the previous two models:
\begin{equation}
f(x, F_{Gauss.}, \sigma_{uv}, F_{point}) = f_{Gauss}(x, F_{Gauss}, \sigma_{uv})+ F_{point},
\end{equation}
where $F_{Gauss.}$ and $F_{point}$ are the flux densities of the Gaussian and point source components, respectively. The total flux density is given by $F(tot) = F_{Gauss.}+ F_{point}$.\\

\item \textit{Exponential profile}: the exponential profile is modelled as :
\begin{equation}
f(x, F, \sigma_{uv}) = F \cdot \frac{\sigma_{uv}}{\sigma_{uv}^2+x^2},
\end{equation}
where $\sigma_{uv}$ is related to the effective radius $R_e$ (in radians) as:
\begin{equation}
R_e \text{[rad]} = 1.6783 \cdot \sigma = 1.6783 \cdot \frac{1}{2\pi\sigma_{uv}\cdot10^{3}}.
\end{equation}
\end{itemize}

The priors used for the Bayesian fitting are reported in Table~\ref{tab:priors}.

\begin{table}
\centering
\caption{Range of priors used for the fit of the visibilities vs. $uv$-distances. $F$ is the flux density and $\sigma_{uv}$ is the scale parameter.}
\begin{tabular}{lccc} 
\hline
 Model & parameter & priors \\ \hline 
 
point & $F$ & $0-1000$~mJy \\
Gaussian& $F$ & $0-1000$~mJy \\
 & $\sigma_{uv}$ & $0-10000$ k$\lambda$  \\
Gaussian+point & $F$(Gauss) & $0-5$~mJy \\
 & $\sigma_{uv}$ & $0-10000$ k$\lambda$ \\
 & $F$(point) & $0-5$~mJy \\
exponential & $F$ & $0-1000$~mJy \\
 & $\sigma_{uv}$ & $0-20000$ k$\lambda$\\
\hline 
\end{tabular}
\label{tab:priors}
\end{table}

\section{Comparison of FIR sizes and flux densities derived using different methods}
\label{sec:size_comp} 
\begin{figure*}
\includegraphics[width=0.9\textwidth]{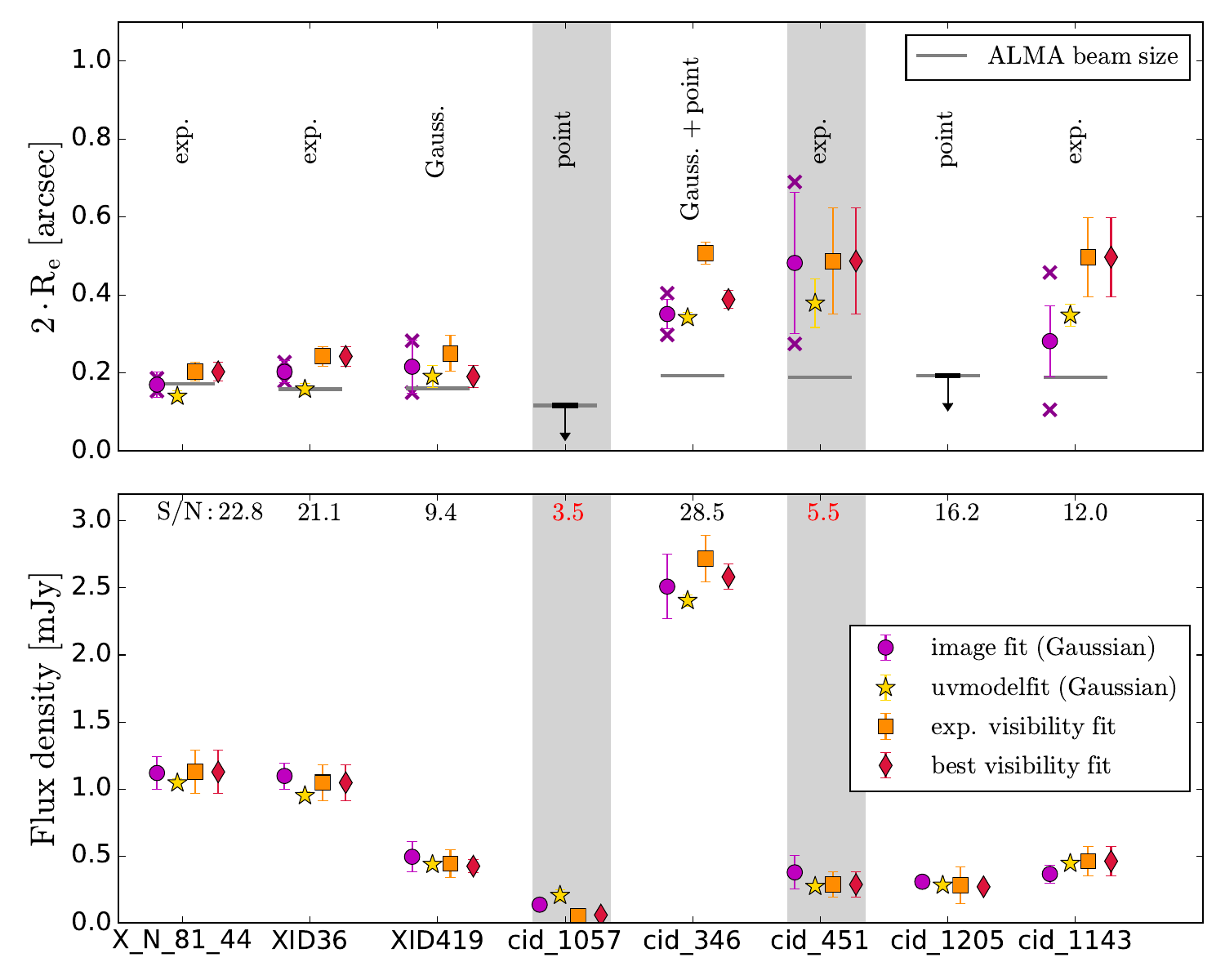}

\caption{Comparison of the FIR sizes and flux densities derived using different methods:  fit of a 2D Gaussian on the image plane (circles), fit of a 2D Gaussian using {\tt uvmultifit} (stars), fit of the visibilities assuming an exponential profile (squares), and fit of the visibilities with the `preferred model' according to the BIC (diamonds). The preferred model for each object is written on the top part of the figure. 
\textit{Upper panel}: Size measurements expressed as two times the effective radius ($R_{e}$). The grey horizontal lines represent the size of the ALMA beam for each object. 
For the 2D Gaussian fit on the image plane, the two magenta crosses represent the sizes of the major and minor axis. For point sources, we show the upper limit corresponding to the beam size. \textit{Bottom panel:} integrated flux densities measured with different methods. The peak signal-to-noise (S/N) of the ALMA images is written on the top part of the panel. 
The results for the two objects with S/N too low to obtain reliable measurements (cid\_1057 and cid\_451) are highlighted with background grey shading.}
\label{fig:comp_ALMA_sizes}
\end{figure*}

In Figure~\ref{fig:comp_ALMA_sizes}, we compare the FIR sizes and and flux densities of the rest-frame 260~\micron\ ALMA emission, measured using four different methods: 1) fit on the image assuming a 2D Gaussian profile, 2) fit with \uvmodelfit\  assuming a 2D Gaussian profile, 3) fit of the $uv$-visibilities assuming an exponential profile (equivalent to a \sersic\ profile with $n=1$), and 4) fit of the $uv$-visibilities with the `preferred' model according to the BIC (point, Gaussian, exponential or Gaussian+point model, see Section~\ref{sec:ALMA_sizes}). For the 2D Gaussian fit on the image plane, we show two points representing the sizes of the major and minor axis (in violet) and the mean value in magenta. For \uvmodelfit, we show the mean value between the major and minor axis.
For cid\_1057 and cid\_451, the S/N is very low (3.6 and 5.9, respectively), therefore we do not consider their size  measurements to be reliable. These two sources are highlighted with a grey band in Figure~\ref{fig:comp_ALMA_sizes}.

In general, the sizes measured with different methods are in agreement within the uncertainties. We note that the sizes measured with the exponential profile are larger than the sizes measured with the Gaussian profile (factor of 1.38 on average) by construction, since the exponential profile does not decrease rapidly towards zero at larger radii and thus considers a larger amount of flux at large radii.

The flux densities are mostly insensitive to the method used. The only notable differences are for cid\_1057 and cid\_346. cid\_1057 has a very low S/N, thus the flux measurements are not very reliable. For cid\_346, the Gaussian fit with \uvmodelfit\ measures a lower flux than the other methods. The rest-frame 260~\micron\ emission of this galaxy is better described by a Gaussian+point source or by an exponential profile, therefore the Gaussian model cannot fit well the central flux peak and underestimates the total flux. 

 In Tables~\ref{tab:ALMA_meas_1} and \ref{tab:ALMA_meas_2}, we provide all the measurements obtained with the different methods. We note that when we compare our measurements to literature values in Section~\ref{sec:FIR_sizes_lit}, we used the sizes obtained with the same method used in the literature.

\begin{sidewaystable}
\caption{ALMA rest-frame 260~\micron\ (observed-frame 870~\micron) flux densities (F), effective radii (R$_e$), and ratio between major and minor axes obtained with two methods: 1) fit on the image (created with natural weighting) and 2) using \uvmodelfit. For both methods we assume a 2D Gaussian profile.
For the two point sources (cid\_1057 and cid\_1205), the sizes are not reported.
($^{\star}$) If the peak S/N < 8 in the ALMA maps, we do not consider the size measurements to be reliable.}
\begin{tabular}{lc|ccccccccccc}
\hline
Method: & & & image fit &  & & \uvmodelfit &  \\

ID & S/N & F & R$_{e}$ & axial ratio & F & R$_{e}$ & axial ratio  
 \\ 
   & & [mJy]  & [mas] & &
     [mJy]  & [mas] & \\  
  \hline \hline
 
X\_N\_81\_44 & 22.7 & 1.18$\pm$0.10 & 98$\pm$18 & 1.23$\pm$0.32 & 1.05$\pm$0.03 & 70$\pm$5 & 1.00$\pm$0.11  \\
XID36 & 21.5 & 1.01$\pm$0.08 & 99$\pm$13 & 1.30$\pm$0.25 & 0.95$\pm$0.03 & 79$\pm$5 & 1.22$\pm$0.12 \\
XID419 & 9.8 & 0.45$\pm$0.06 & 112$\pm$22 & 1.60$\pm$0.46 & 0.44$\pm$0.03 & 96$\pm$14 & 2.88$\pm$0.80  \\
cid\_1057$^{\star}$ & 3.6 & 0.29$\pm$0.13 & - & - & 0.21$\pm$0.04 & - & - \\
cid\_346 & 29.0 & 2.61$\pm$0.17 & 186$\pm$17 & 1.27$\pm$0.16 & 2.41$\pm$0.04 & 171$\pm$5 & 1.31$\pm$0.06 & \\
cid\_451$^{\star}$ & 5.9 & 0.35$\pm$0.08 & 244$\pm$66 & 1.91$\pm$0.76 & 0.27$\pm$0.02 & 190$\pm$36 & 3.15$\pm$0.83 & \\
cid\_1205 & 16.6 & 0.29$\pm$0.02 & - & - & 0.28$\pm$0.01 & -& -  \\
cid\_1143 & 12.1 & 0.56$\pm$0.07 & 222$\pm$36 & 2.49$\pm$0.60 & 0.45$\pm$0.02 & 174$\pm$18 & 4.60$\pm$0.70 & \\

\hline
\end{tabular}
\label{tab:ALMA_meas_1}
\end{sidewaystable}

\begin{sidewaystable}
\caption{ALMA rest-frame 260~\micron\ (observed-frame 870~\micron) flux densities (F) and effective radii (R$_e$), obtained through the fit on the visibilities using python, assuming different symmetric (i.e. axis ratio = 1) models. The model with the smallest BIC (Bayesian Information Criterion) value is the preferred model. The stars ($^{\star}$) mark the targets with the peak S/N < 8 in the ALMA maps, for which we do not consider the size measurements to be reliable. For point sources, we tabulate only the results of the point source model. For extended sources, we do not tabulate the result of the fit with point source model, since it gives a poor fit to the data and underestimate the fluxes.
 $^{a}$For the Gaussian+point model, we report: the total flux F(tot)=F(point)+F(Gaussian), the flux of the point source F(point), the half-light radius of the total profile R$_{e}$(tot), the half-light radius of the Gaussian component R$_{e}$(Gauss). 
$^{b}$For XID36 and cid\_451, the difference in BIC between the exponential and Gaussian model is $<2$, therefore it is not clear than one model is performing better than the other.}
\setlength{\tabcolsep}{4pt}
\begin{tabular}{lccccccccccccccc}
\hline
Model: & \multicolumn{2}{c}{point} & &  Gaussian &  &  & exponential & &  \multicolumn{4}{c}{Gauss.+point $^{a}$}    & \\
ID &  
F & BIC  &
F & R$_{e}$ & BIC &
F & R$_{e}$ & BIC & 
F(tot) & F(point) & R$_{e}$(tot) & R$_{e}$(Gauss) & BIC & 
 preferred model \\ 
  & [mJy]  &  &
   [mJy]  & [mas] & &
   [mJy]  & [mas] & &
   [mJy]  & [mJy] & [mas] & [mas]  & &
      &\\  
  \hline \hline

X\_N\_81\_44 & - & - & 1.06$\pm$0.05 & 75$\pm$7 & 80.73 & 1.13$\pm$0.16 & 102$\pm$12 & 74.57 & 1.13$\pm$0.19 & 0.29$\pm$0.10 & 84$\pm$20 & 111$\pm$20 & 77.75 & exponential\\
XID36 & - & - & 0.97$\pm$0.05 & 86$\pm$7 & 51.46 & 1.05$\pm$0.13 & 122$\pm$13 & 50.02 & 0.98$\pm$0.13 & 0.08$\pm$0.06 & 90$\pm$11 & 96$\pm$11 & 55.44 & exp./Gaussian$^{b}$\\
XID419 & - & - & 0.43$\pm$0.05 & 96$\pm$14 & 42.34 & 0.44$\pm$0.10 & 125$\pm$23 & 45.77 & 0.42$\pm$0.08 & 0.02$\pm$0.02 & 97$\pm$17 & 101$\pm$17 & 47.23 & Gaussian \\
cid\_1057$^{\star}$ & 0.06$\pm$0.02 & 51.69 & - & - & - & - & - & - & - & - & - & - & - & point \\
cid\_346 & - & - & 2.41$\pm$0.08 & 173$\pm$7 & 79.40 & 2.72$\pm$0.17 & 254$\pm$14 & 42.91 & 2.58$\pm$0.14 & 0.35$\pm$0.05 & 194$\pm$11 & 219$\pm$11 & 38.77 & Gauss.+point \\
cid\_451$^{\star}$ & - & - & 0.28$\pm$0.05 & 194$\pm$42 & 45.88 & 0.29$\pm$0.09 & 244$\pm$68 & 45.46 & 0.29$\pm$0.08 & 0.04$\pm$0.02 & 212$\pm$53 & 236$\pm$53 & 47.28 & exp./Gaussian$^{b}$\\
cid\_1205 & 0.27$\pm$0.02 & 43.71 & - & - & - & - & - & - & - & - & - & - & - & point \\
cid\_1143 & - & -& 0.37$\pm$0.04 & 140$\pm$25 & 54.79 & 0.46$\pm$0.11 & 249$\pm$51 & 47.25 & 0.47$\pm$0.10 & 0.08$\pm$0.03 & 221$\pm$77 & 257$\pm$77 & 53.57 & exponential\\

\hline
\end{tabular}
\label{tab:ALMA_meas_2}
\end{sidewaystable}

\section{Astrometry of the SINFONI data}
\label{app:coord}
In this Section, we explain the details of the registration of the astrometry of the SINFONI data-cubes.
As explained in Section~\ref{sec:coord}, the absolute position of the SINFONI cubes, as derived from the SINFONI pipeline, is not sufficiently accurate for our purposes. The small field of view of the SINFONI images (3$\times$3 arcsec$^2$) does not allow us to correct the astrometry using nearby stars, since usually the target is the only visible source in the field of view. Therefore, we have to derive the absolute coordinates from other images.

\noindent \textbf{Reference coordinates:}
We use H-band and K-band images with a large field of view (3$\times$3 arcmin$^2$) to determine the reference coordinates of our targets that we later use to register the astrometry of the SINFONI data-cubes. 
We use K-band and H-band images from the VLT/VISTA and VLT/ISAAC. 

To determine the coordinates from the H/K-band images, we apply the following procedure.
 First, we align the H/K-band images to the \gaia\ DR2 catalogue, using the Graphical Astronomy and Image Analysis Tool (GAIA) that is part of the {\tt Starlink} software  \citep{Currie2014}. 
 There are at least five objects in common between the \gaia\ DR2 catalogue and each image that allow us to accurately align the images to the \gaia\ astrometry\footnote{There is an exception: cid\_1205 that has only 3 sources in common between the GAIA catalogue and the H/K-band image. For this object the H/K-band position is in perfect agreement with the FIR position measured from the ALMA map, therefore we consider the H/K-band coordinates to be reliable.}.
 Then, we determine the centroid position of our target in the `astrometry corrected' image by fitting a 2D Gaussian to the source. 
 
 We apply this procedure to both the H-band and K-band images. 
The offsets between the coordinates derived from the H-band and K-band are smaller than one pixel ($< 0.07$''), therefore we can assume that the H-band and K-band emission peak at the same position. 
 We decide to use the coordinates derived from the K-band corrected images to register the astrometry of both the H-band and K-band SINFONI images. 

We compare the K-band coordinates with the \gaia\ coordinates for the two objects detected in \gaia.
For cid\_346, the K-band coordinates agree very well with the \gaia\ coordinates (offset 7.3 mas).
For \XN, the K-band coordinates are shifted by 0.05''  to the west with respect to the coordinates from \gaia. Given the low resolution of the K-band image for this target (FWHM PSF 1.04''), we decide to use the coordinates from \gaia\ for this source. 
We note that these two targets are Type 1 AGN, and are dominated by the point source emission of the AGN across the optical and near-infrared bands.
Finally, we note that XID419 is only marginally detected in the K-band image (peak S/N$ < 6$) and thus it is difficult to determine its position. Therefore for this target we rely on the coordinates derived from the \textit{HST}/WFC3 images reported in \citet{Scholtz2020}.

\noindent \textbf{Registration of the SINFONI images:} 
 We use these coordinates to register the position of the peak of the emission in the SINFONI data cubes. 
 The emission of the H/K-band filters is dominated by the continuum, but there is also some contribution from the emission lines.
  
For the Type 1 AGN, we check that the position of the continuum and the position derived by collapsing the total SINFONI data-cubes are in agreement, both in the H- and K-band (offset $<0.02''$).

For the Type 2, the continuum is significantly detected in the SINFONI maps only in one object (XID36). For this target, we test that the position of the continuum and the position derived by collapsing the total data-cube are in agreement (offset $<0.007''$), both in the H- and K-band.
For the other targets, we use the SINFONI spectra to estimate the relative contribution of continuum and emission lines to the total emission in the H/K band VISTA and ISAAC filters. The continuum contribution is $>70 \% $ for all targets.

\noindent \textbf{Uncertainties on the coordinate registration:}
The typical uncertainty on the \gaia\ coordinates is $\leq$~4 mas. 
The precision of the alignment of the K-band images with the \gaia\ images is about half pixel (75~mas for COSMOS and 130~mas for CDF-S).
 The uncertainties from the 2D Gaussian fit of the VISTA or ISAAC K-band images are in the range 1-49 mas  (median 4~mas). 
 We also consider the uncertainties due to the size of the VISTA or ISAAC K-band PSF (0.78-1.04''), calculated as PSF/($2\times$S/N) following \citet{Condon1997}, which are in the range  3-40~mas.
 The uncertainties on the position of the emission of the SINFONI cubes is $\sim$ half pixel (25 mas).

  To estimate the total uncertainties of the derived K-band coordinates, we added in quadrature all the above uncertainties. These uncertainties are dominated by the pixel size of the K-band images. 
In summary, the uncertainties on the derived SINFONI astrometry for our sample are in the range 0.03-0.14''. 

\section{SINFONI spectra of Type 1 AGN}
\label{app:type1_spectra}
Figure~\ref{fig:Ha_OIII_images} in Section~\ref{sec:SINFONI} shows the BLR-subtracted spectra that we used in the analysis. In this Section, we show the total SINFONI emission line spectra (including the BLR component) of the three Type 1 AGN in our sample (\XN, cid\_346, cid\_1205).

\begin{figure*}
\centering

\includegraphics[width=0.4\textwidth]{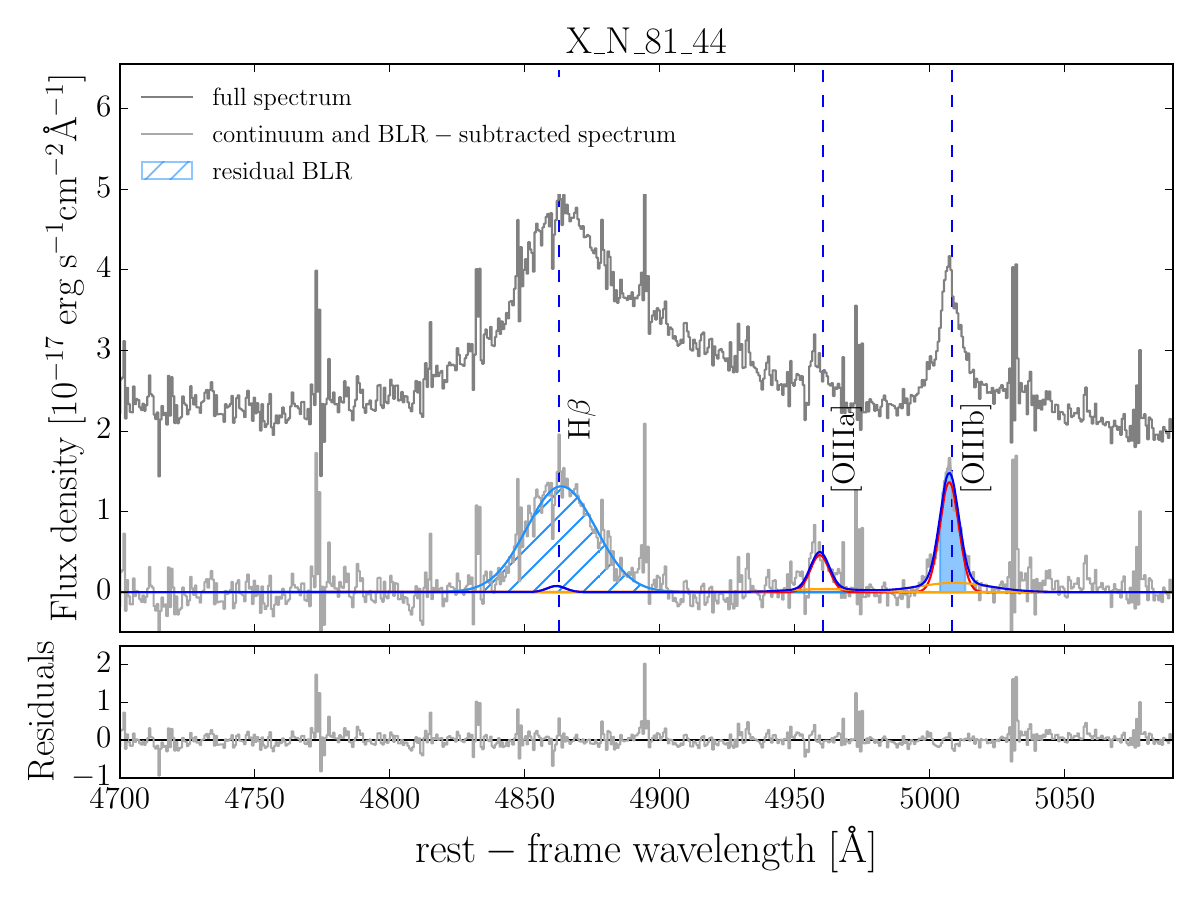}
\includegraphics[width=0.4\textwidth]{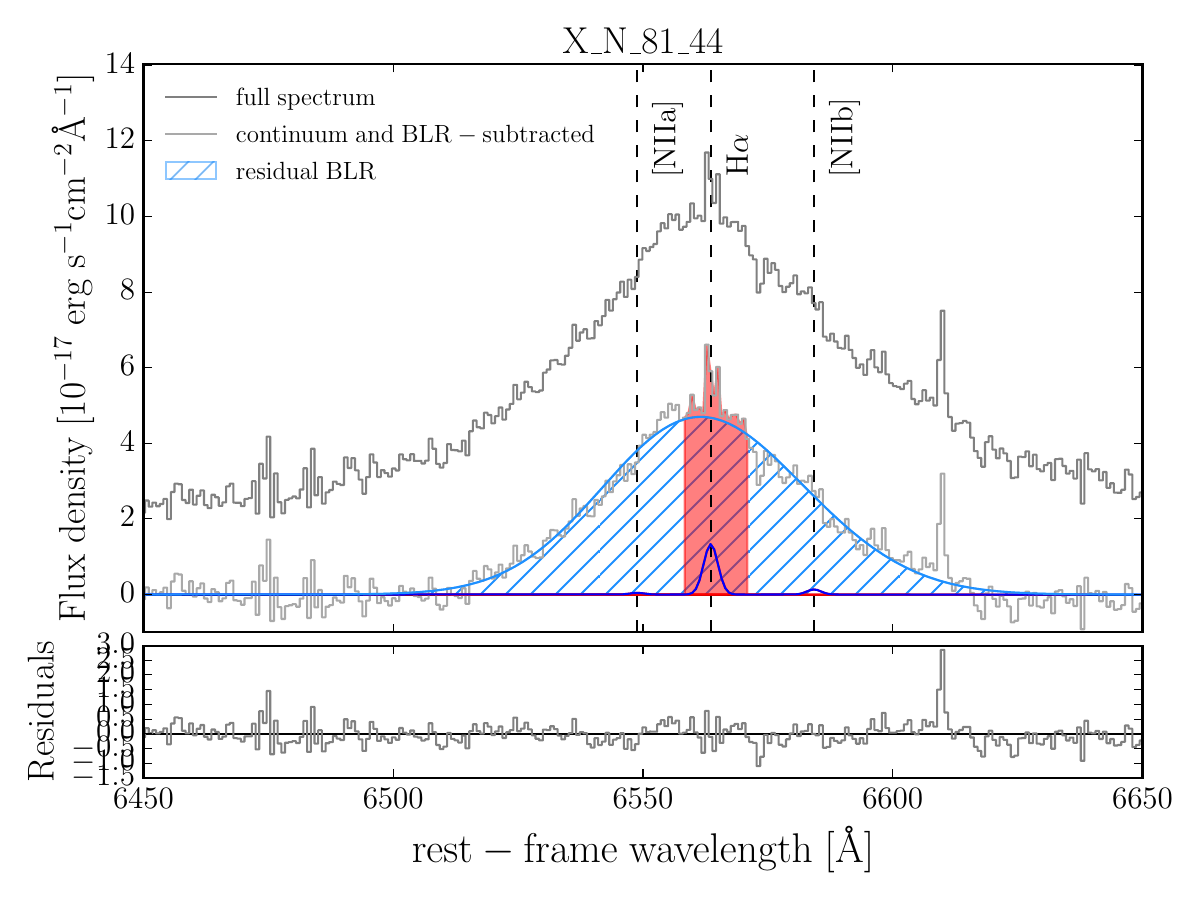}

\includegraphics[width=0.4\textwidth]{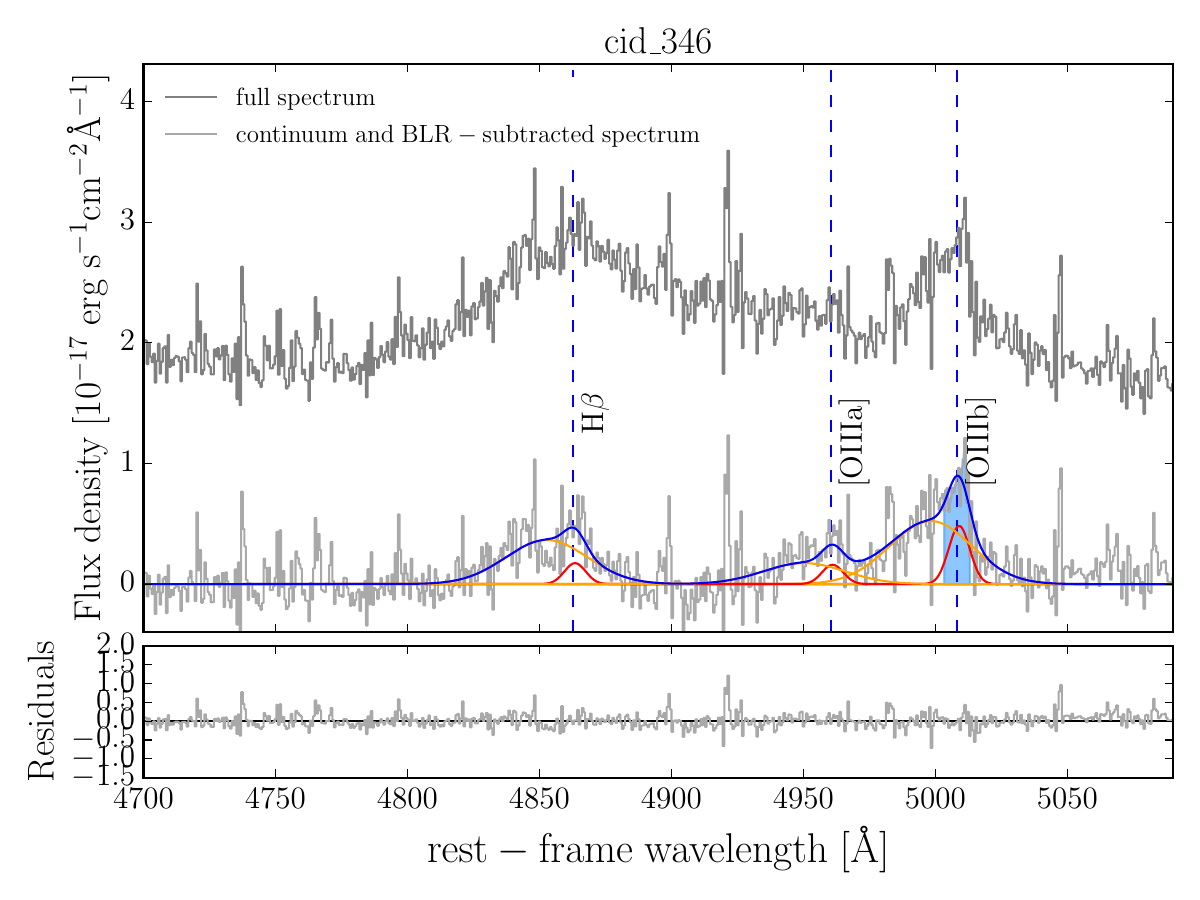}
\includegraphics[width=0.4\textwidth]{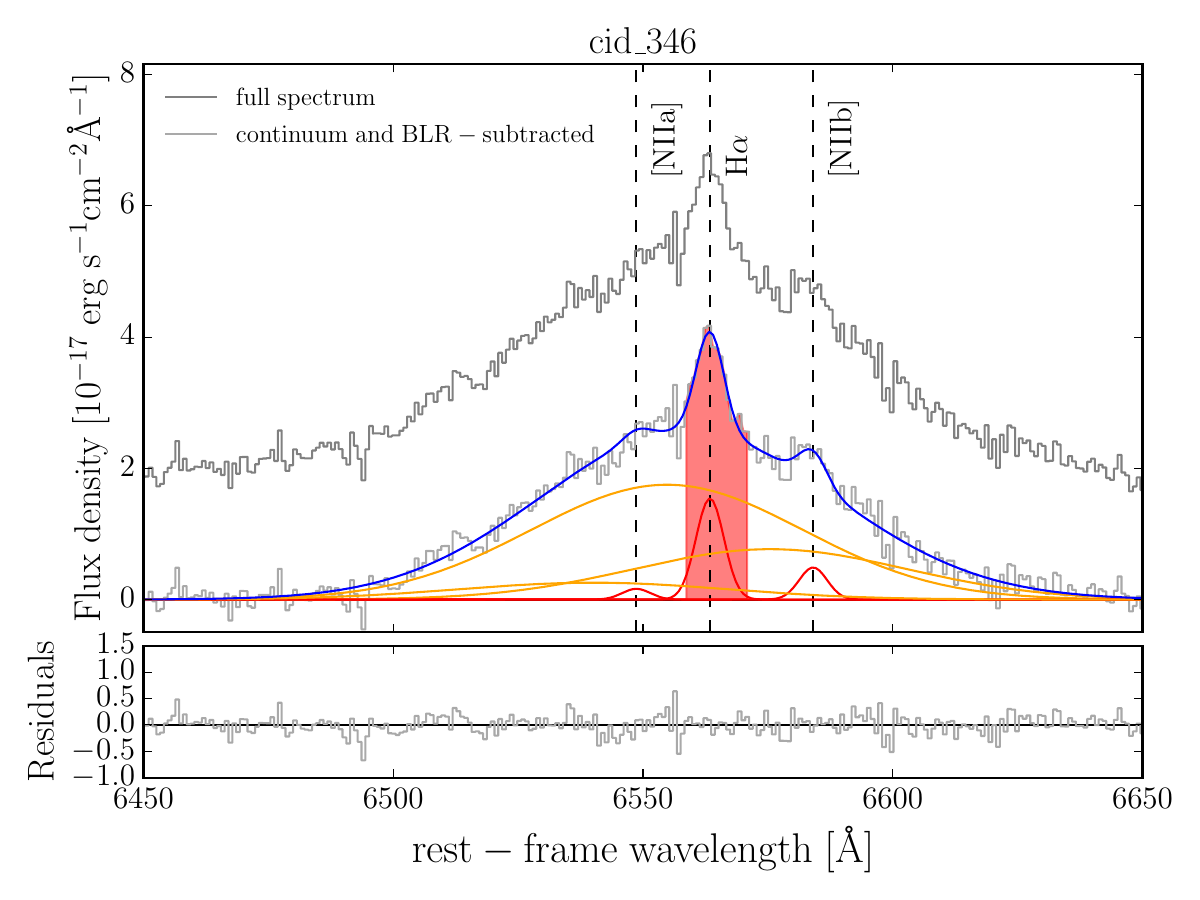}

\includegraphics[width=0.4\textwidth]{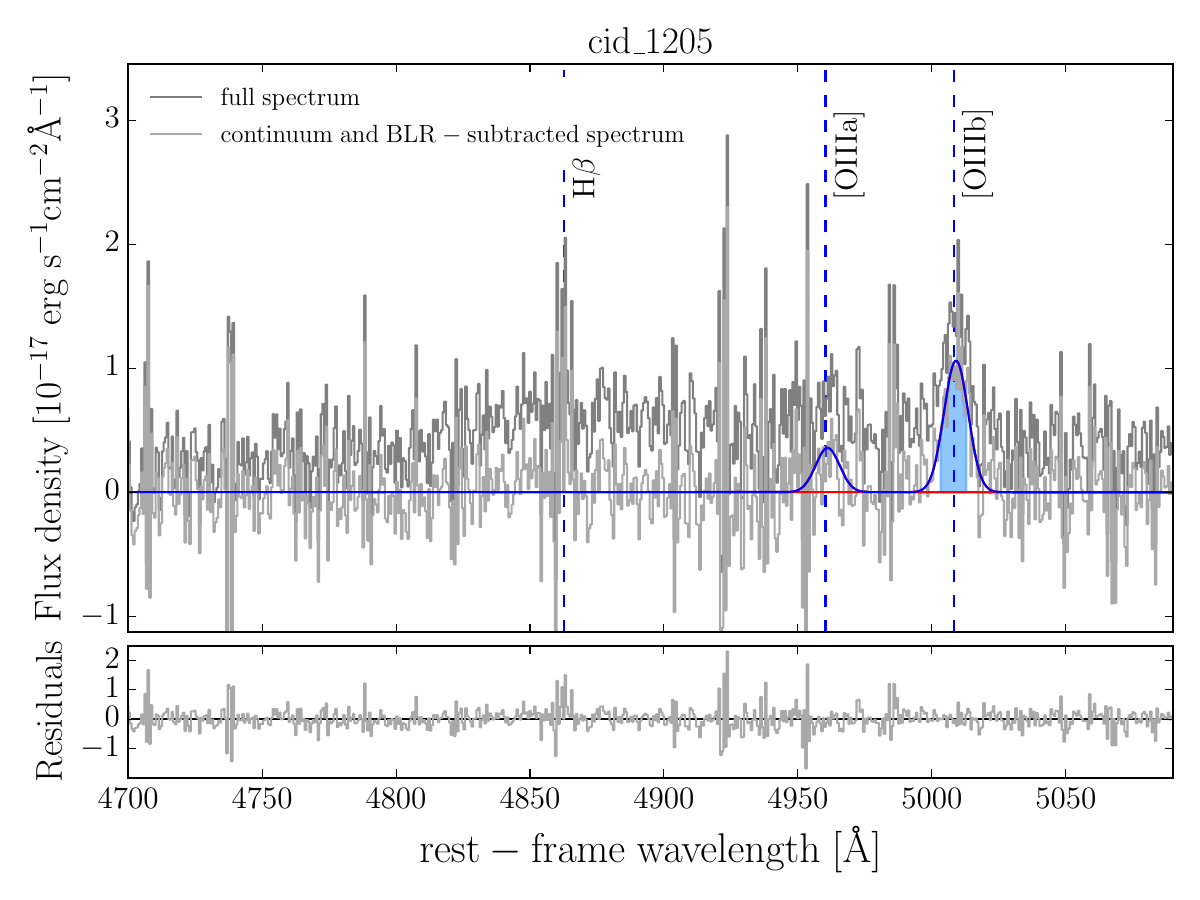}
\includegraphics[width=0.4\textwidth]{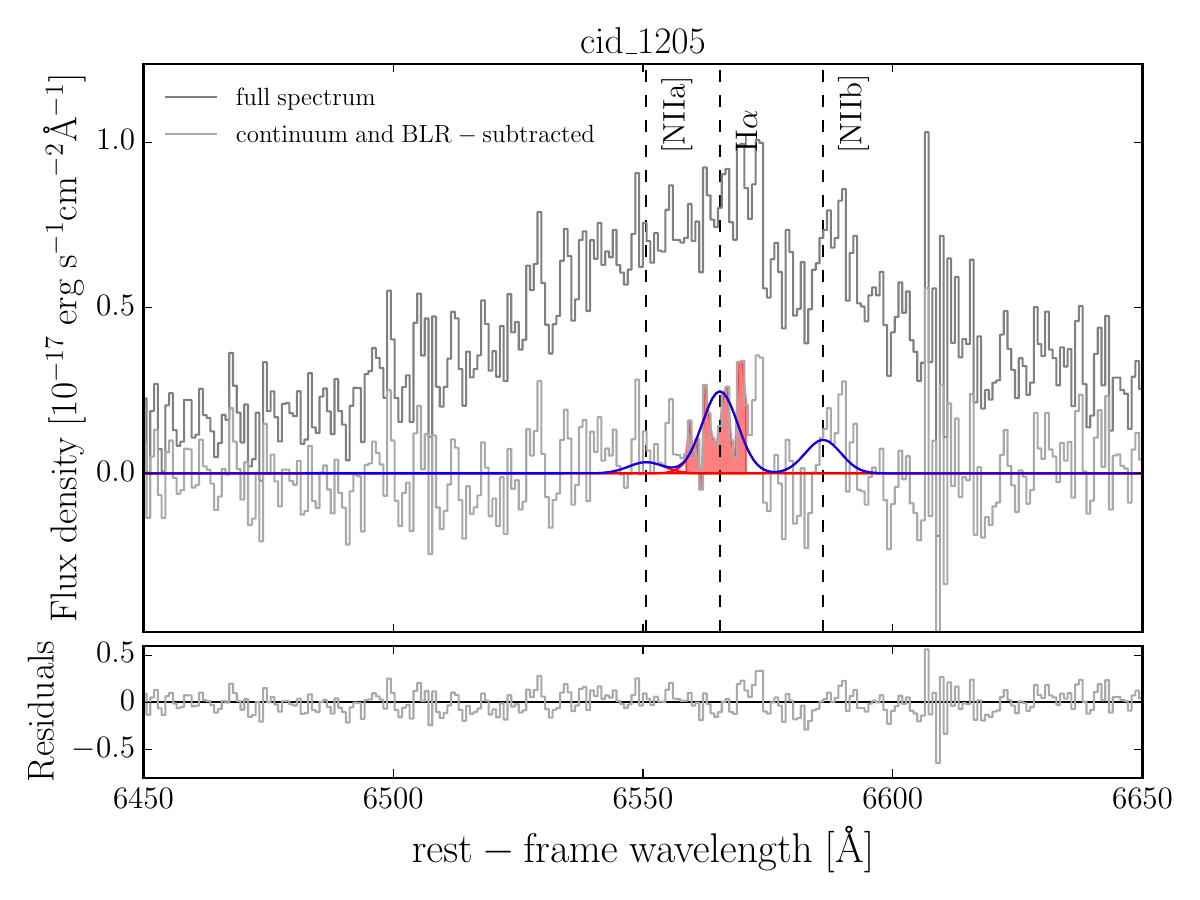}

\caption{Integrated spectra of Type 1 AGN in our sample:  \OIII +\Hb\ spectra (left) and \Ha +\NII\ spectra (right).
 The continuum- and BLR-subtracted spectra are  shown in light-grey. 
The total spectra, before subtracting the continuum and broad line region (BLR) emission, are shown in dark grey.
The blue curves show the total fit to the H$\beta$, \OIII$\lambda$4959 and \OIII$\lambda$5007 lines (left panel) and to the  H$\alpha$, \NII$\lambda$6548 and \NII$\lambda$6584 lines (right panel). 
The red and orange curves show the narrow and broad components, respectively.
For \XN, the broad components of the \Hb\ and \Ha\ are considered as part of the BLR.
 }
\label{fig:spectra_Type1}
\end{figure*}

\section{Spectral energy distributions  (SEDs)}

\begin{figure*}
\centering
\includegraphics[width=0.4\textwidth]{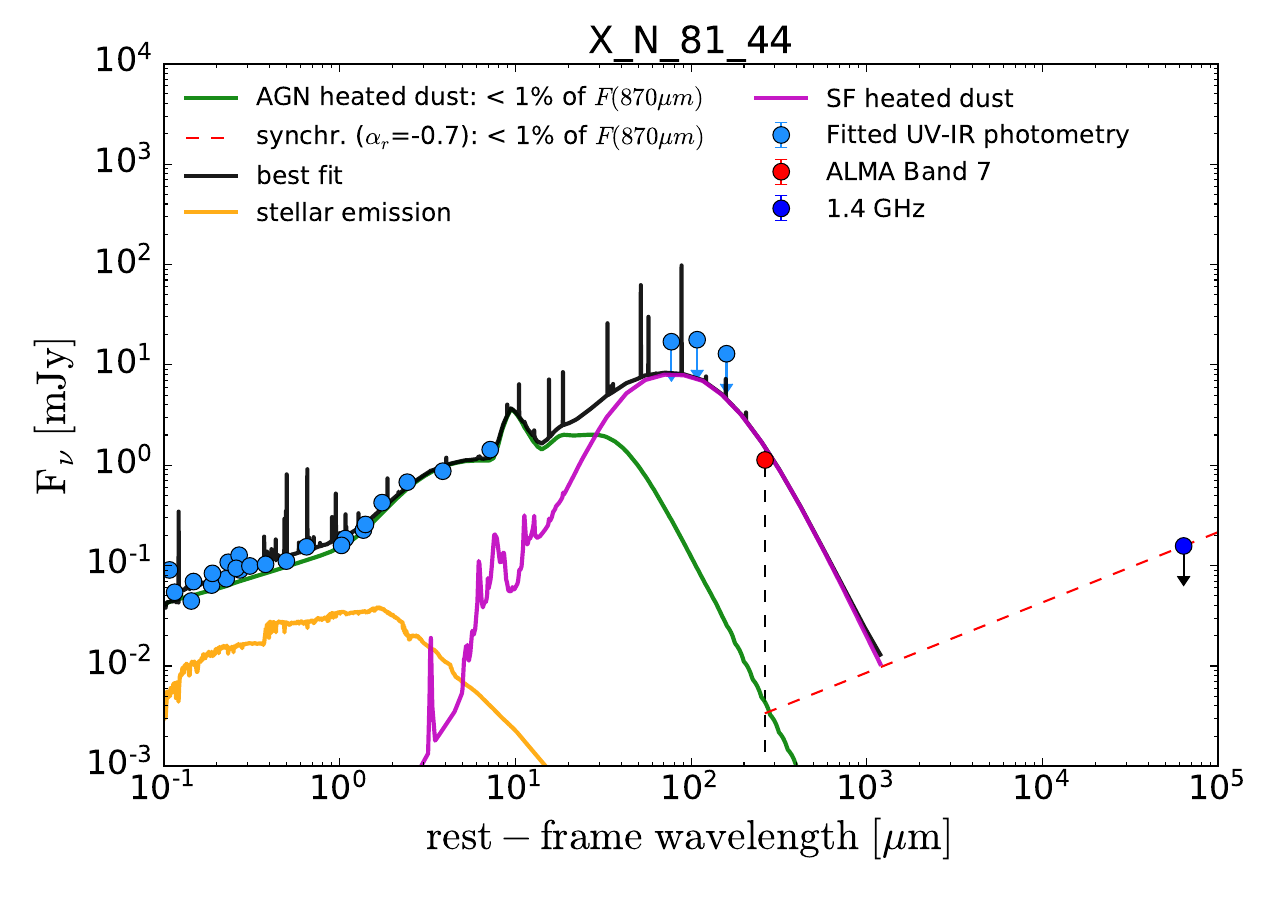}
\includegraphics[width=0.4\textwidth]{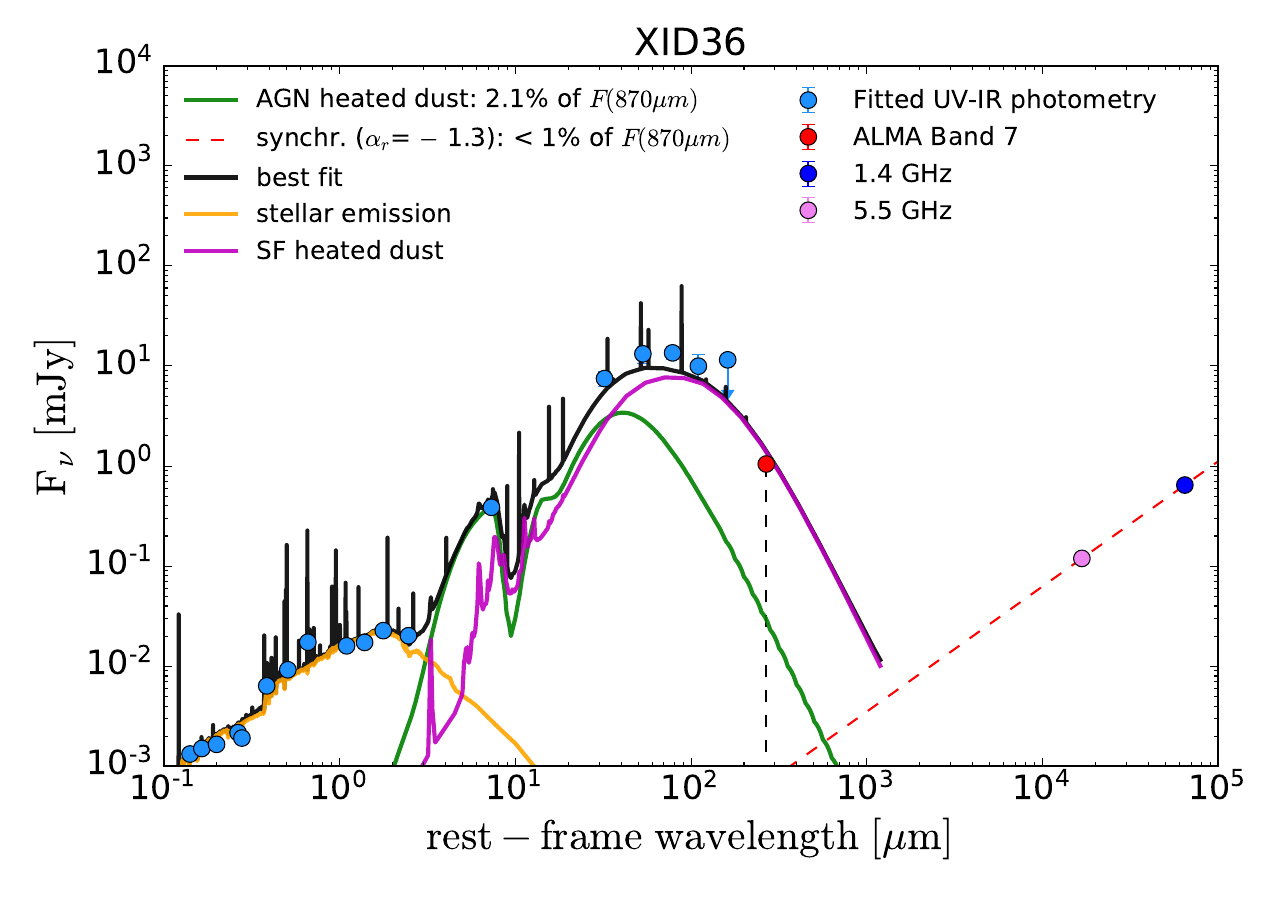}
\includegraphics[width=0.4\textwidth]{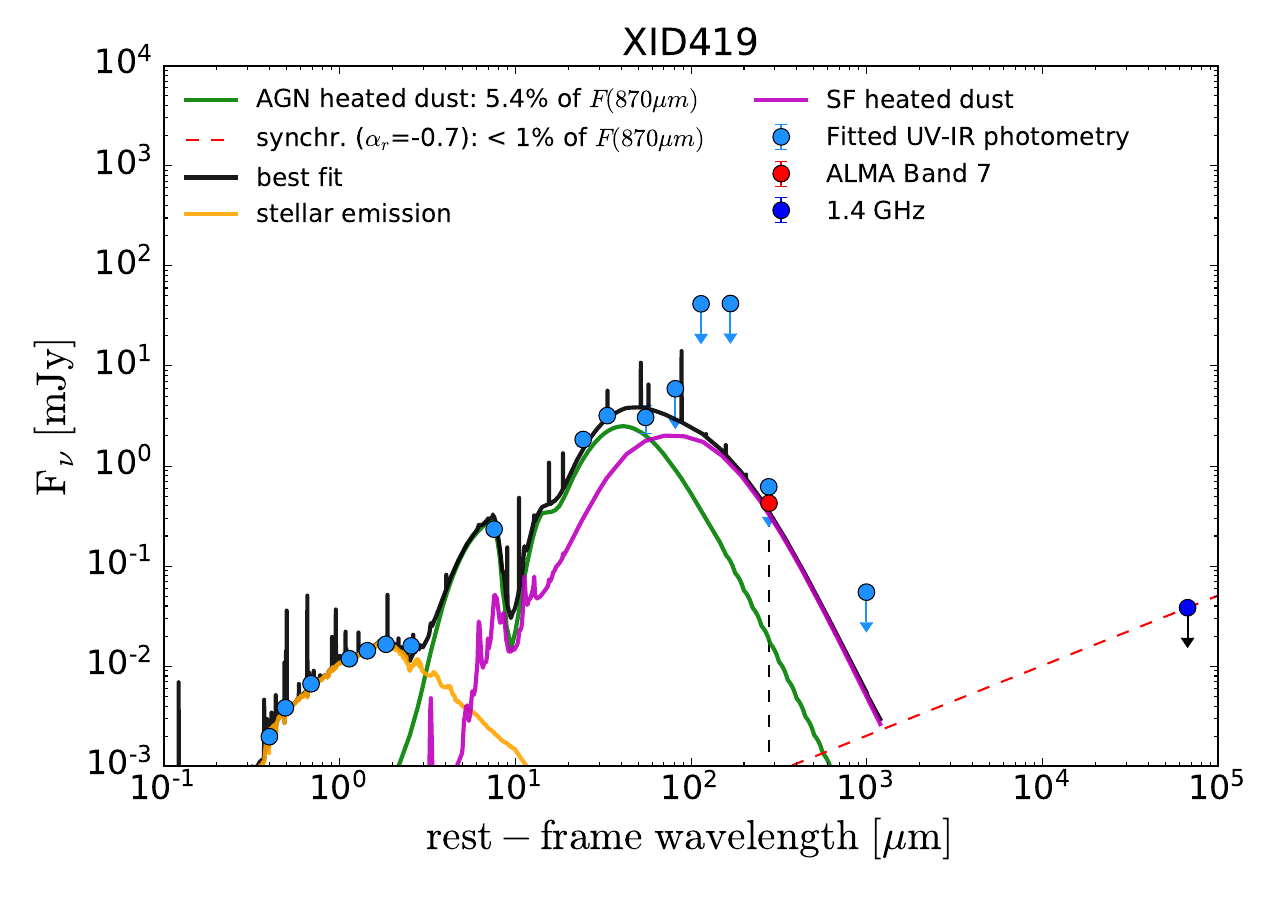}
\includegraphics[width=0.4\textwidth]{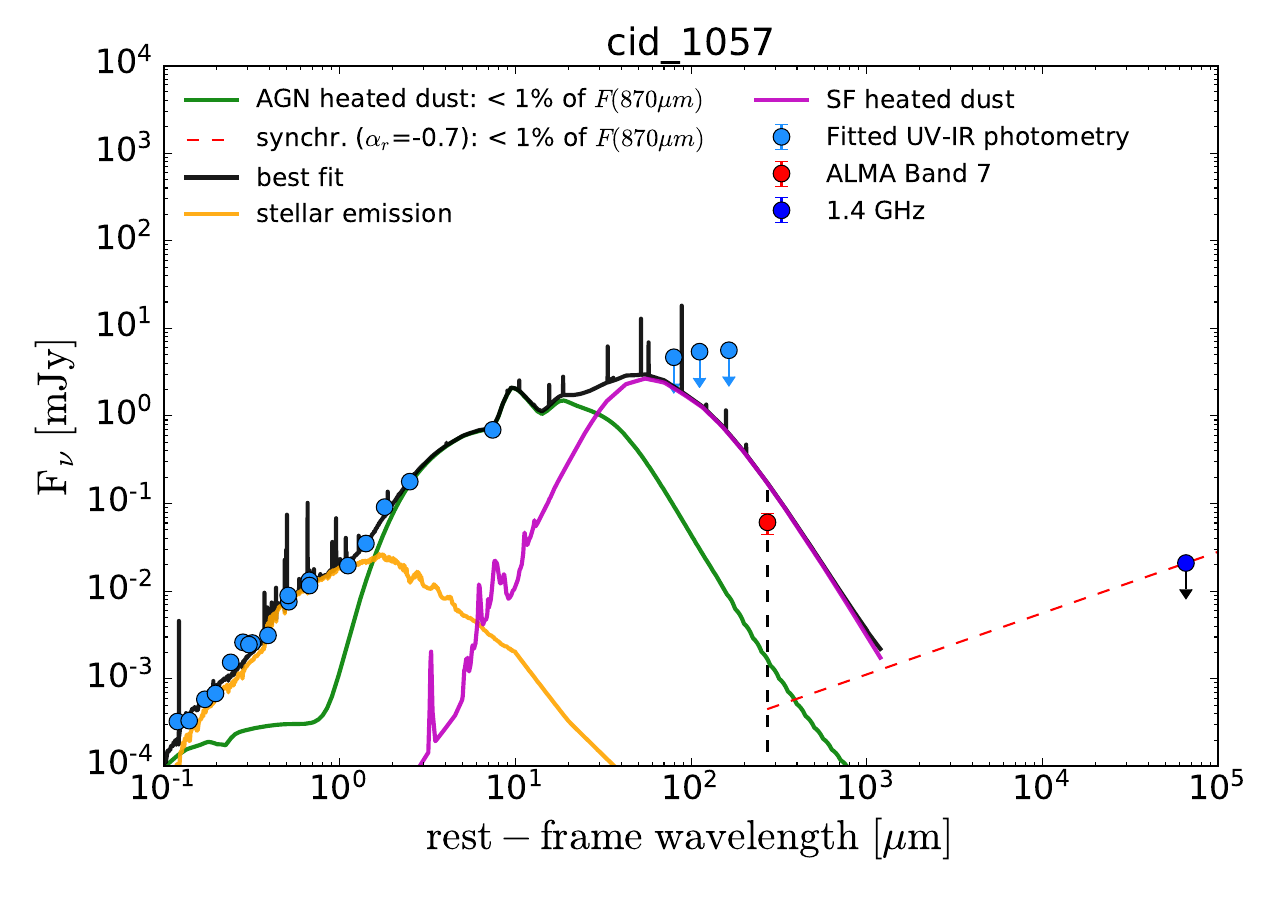}
\includegraphics[width=0.4\textwidth]{Figures/SED_plots/cid_346_SED_fit_cigale_paper_alpha07_UL_with_ALMA.pdf}
\includegraphics[width=0.4\textwidth]{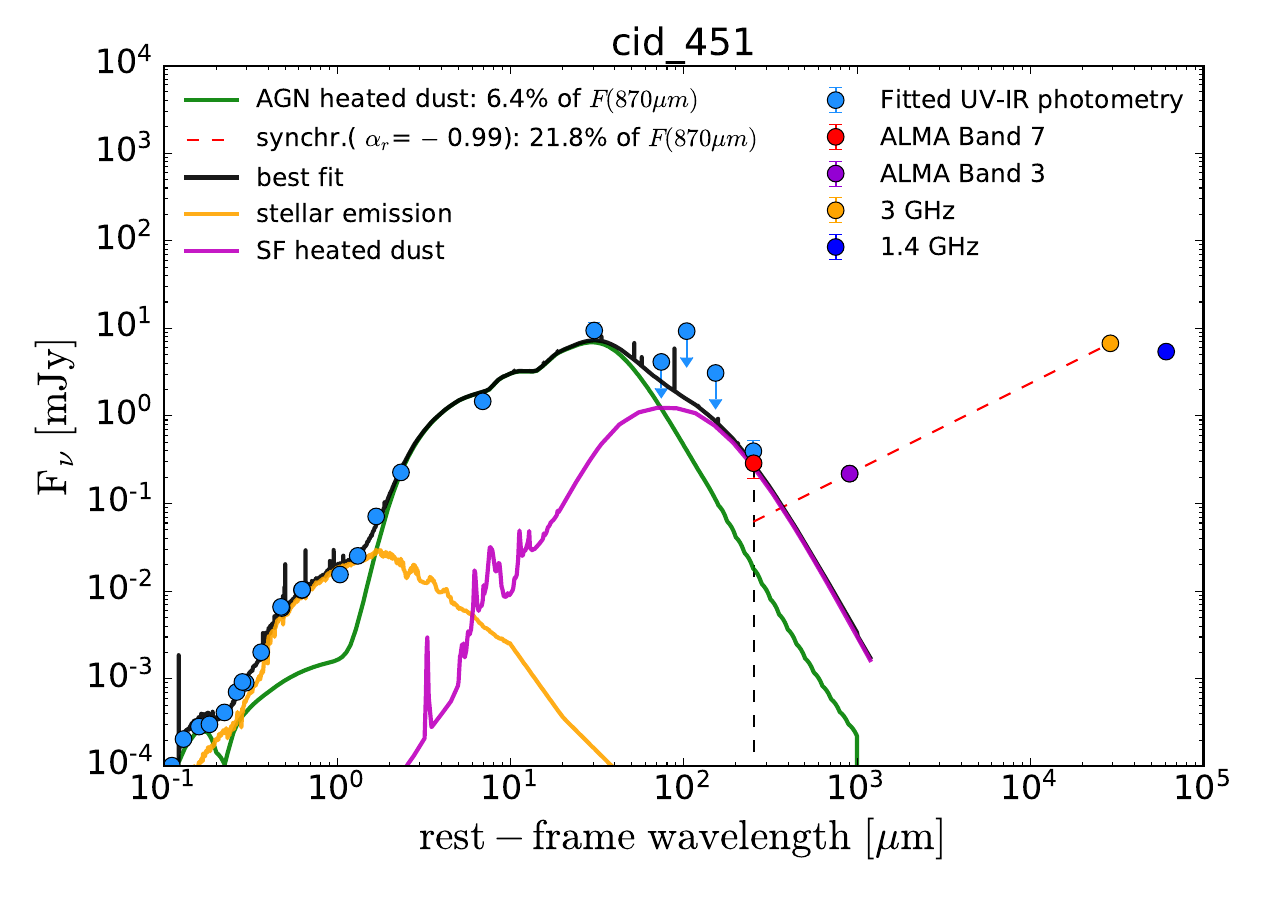}
\includegraphics[width=0.4\textwidth]{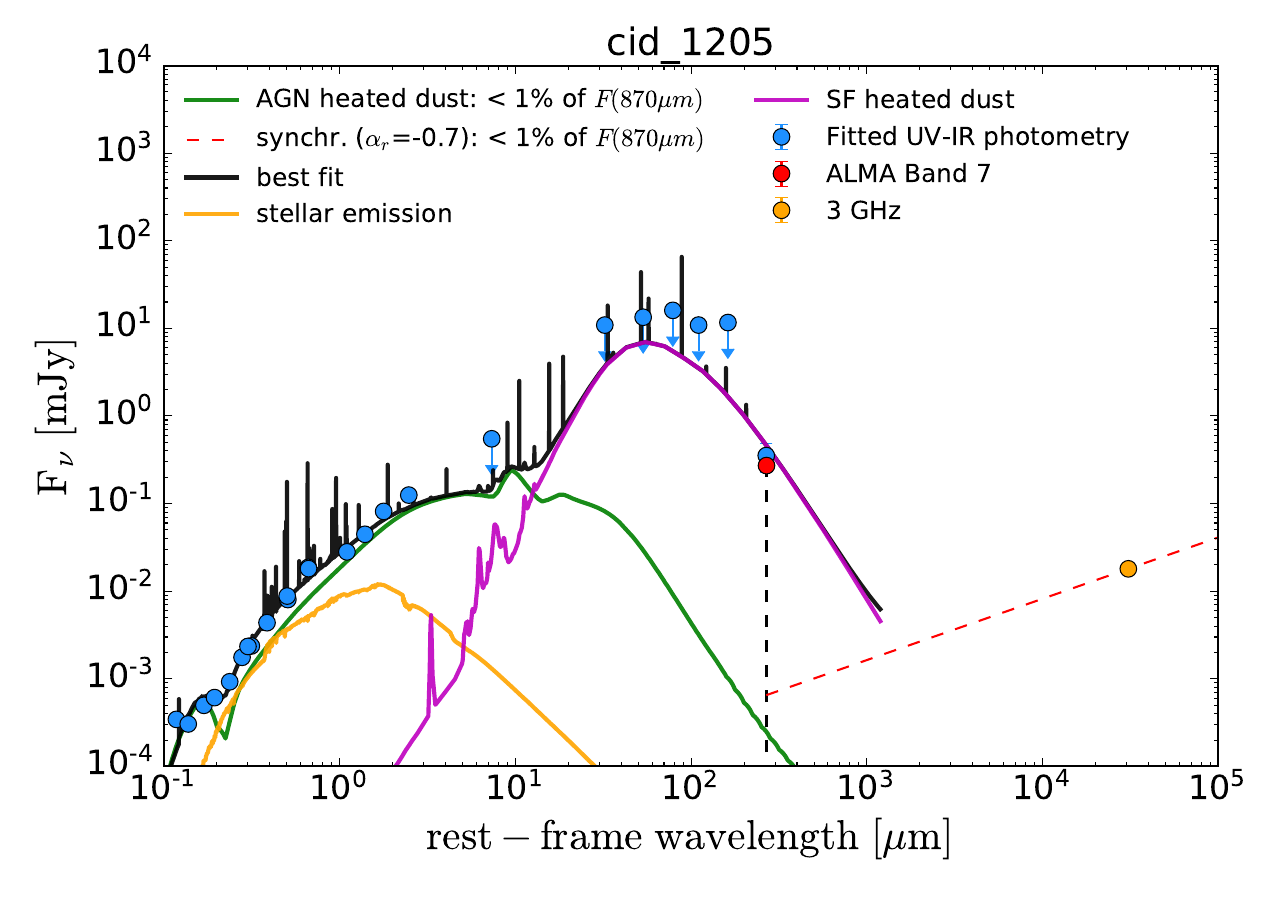}
\includegraphics[width=0.4\textwidth]{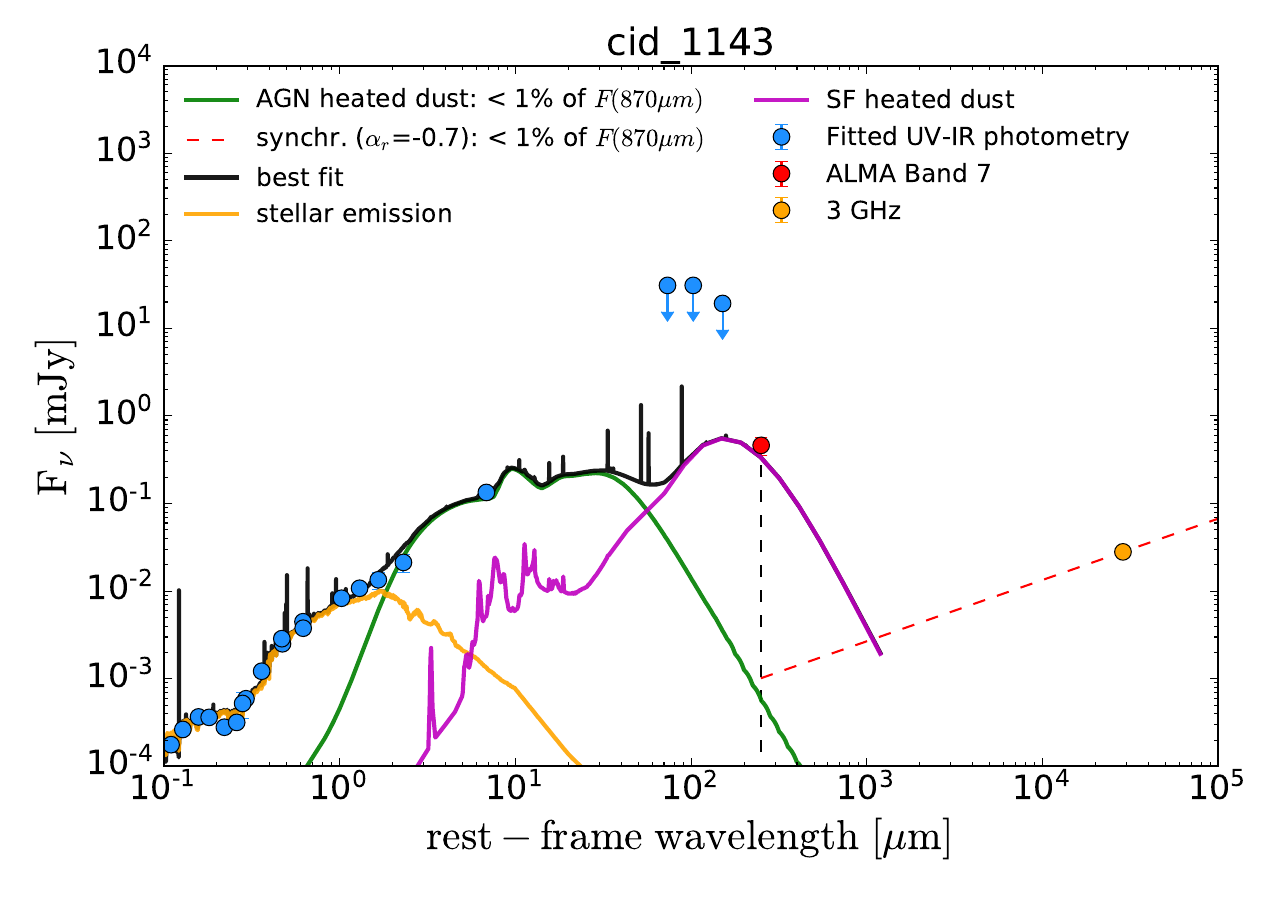}
\caption{\small Rest-frame spectral energy distribution (SED) of our sample. The light blue data points represent the UV-IR photometry that was used for the SED fitting  \citep{Circosta2018}. The red point shows our ALMA Band 7 flux measurements, which was also included in the fit.
The violet point in cid\_451 shows the ALMA Band 3 flux from \citet{Circosta2021}.
The blue, orange, and pink points show the radio fluxes at 1.4, 3, and 5~GHz, respectively. The arrows indicate 3$\sigma$ upper limits. The solid curves show the results of the SED fitting with \cigale: in black is the total best-fit model (including the contribution the from nebular emission component), in orange  the dust-attenuated stellar emission, in magenta the emission from dust heated by star formation and in green the emission from dust heated by the AGN. 
To estimate the maximum contribution of synchrotron emission to the rest-frame 260~\micron\ flux density we parameterized this emission as a power law with spectral index $\alpha_{r}$ (dashed red line), normalised at 3~GHz (10cm) or 1.4~GHz (21cm), depending on the available radio data. For the galaxies with radio fluxes in two bands, we derived $\alpha_{r}$ based on the two fluxes. 
 On the plot we show the estimated contribution (in percentage) from dust heated by the AGN and from synchrotron emission  to the rest-frame 260~\micron\ ALMA Band 7 flux, estimated from the total SED template. }
\label{fig:SED_fit}
\end{figure*}

In Figure \ref{fig:SED_fit}, we show the spectral energy distributions of our targets (rest-frame wavelength range $0.1$~\micron$-10$~cm), together with the best fit model from \citet{Circosta2018} (see description in Section~\ref{sec:sample}). We use these models to predict the percentage contribution due to dust heated by the AGN at  260~\micron\ (rest-frame). We also show the  synchrotron emission contribution at 260~\micron, predicted based on the available radio photometry. The different sources that can contribute to the 260~\micron\ flux are discussed in Section~\ref{sec:origin}.

\end{appendix}

\end{document}